\documentclass{report}
\usepackage{times}
\usepackage{fancyhdr}
\usepackage{epsf}
\newcommand{\dd}{\mbox{d}}
\newcommand{\DD}{\mbox{D}}
\newcommand{\bbox}[1]{\mbox{\boldmath $#1$}} 
\newcommand{\tfrac}[2]{{\textstyle\frac{#1}{#2}}}

\def\overlay#1#2{\ifmmode%
\setbox0=\hbox{$#1$}%
\setbox1=\hbox to\wd0{\hss$#2$\hss}\else%
\setbox0=\hbox{#1}%
\setbox1=\hbox to\wd0{\hss#2\hss}\fi%
#1\hskip-\wd0\box1 }
\newcommand{\lesssim}{\mbox{\raisebox{-.3ex}{\,$\stackrel{<}{\sim}$\,}}}
\newcommand{\gtrsim}{\mbox{\raisebox{-.3ex}{\,$\stackrel{>}{\sim}$\,}}}

\begin{document}
\pagenumbering{roman}
\clearpage{\pagestyle{empty}\cleardoublepage}
\title{\bf Boulevard \\ of \\ Broken
Symmetries\footnote{\normalsize 
Dedicated to the memory of Anna M.~Konings.  }}  
\author{Adriaan M.J. Schakel \\ Institut f\"ur Theoretische Physik \\
Freie Universit\"at Berlin \\ Arnimallee 14, 14195 Berlin \\ e-mail:
schakel@physik.fu-berlin.de 
}
\begin{abstract}
Effective theories of quantum liquids (superconductors and superfluids
of various types) are derived starting from microscopic models at the
absolute zero of temperature.  Special care is taken to assure Galilei
invariance.  The effective theories are employed to investigate the
quantum numbers carried by the topological defects present in the phases
with spontaneously broken symmetries.  Due to topological terms induced
by quantum fluctuations, these numbers are sometimes found to be
fractional.  The zero-temperature effective theories are further used to
study the quantum critical behavior of the liquid-to-insulator
transition which these systems undergo as the applied magnetic field,
the amount of impurities, or the charge carrier density varies.  The
classical, finite-temperature phase transitions to the normal state are
discussed from the point of view of dual theories, where the defects of
the original formulation become the elementary excitations.  A
connection with bosonization is pointed out. \vspace{2.cm} 
\end{abstract}
\maketitle
\pagestyle{fancyplain}
\renewcommand{\chaptermark}[1]%
     {\markboth{Chapter \thechapter \hspace{.1cm} \ #1}{}}
\renewcommand{\sectionmark}[1]%
     {\markright{\thesection \hspace{.1cm} \ #1}}
\lhead[\thepage]{\bf\let\uppercase\relax\rightmark}
\rhead[\bf\let\uppercase\relax\leftmark]{\thepage}
\cfoot{}
\newcommand{\TheAuthor}{}
\newcommand{\Author}[1]{\renewcommand{\TheAuthor}{#1}}
\rfoot[\TheAuthor]{}
\lfoot[]{\TheAuthor}
\Author{\copyright \bf Amstex}
%
%
%
%
%
\tableofcontents
\listoffigures
\listoftables
\chapter*{Preface\label{chap:pref}\markboth{Preface}{Preface}}
The title of this report, Boulevard of Broken Symmetries, is chosen to
indicate that the (condensed matter) systems we will be discussing in
these pages have a spontaneously broken symmetry in common\footnote{The
true story behind this title is the following.  Every late summer during
the eighties, Amsterdam was visited by a traveling theater group.  They
built up their tents in a central park of the city surrounded by the
Concert Building and the major museums (and also by the high school this
author attended).  The name of this enterprise, marking the end of the
summer and its fantasies, was called ``Boulevard of Broken Dreams''}.
The notion of a spontaneously broken symmetry is one of the paradigms of
modern physics.  It plays a central role in today's understanding of
many startling phenomena known in condensed matter physics, statistical
physics, high-energy physics, and cosmology.  A spontaneously broken
symmetry indicates the presence of a global symmetry which is not
apparent in the state the system is in. The symmetry is not lost, but
implemented in a nontrivial way.  If the symmetry involved is a
continuous one (and the dimensionality is larger than two), the breaking
is accompanied by the occurrence of gapless modes.  Due to their
gaplessness, they are the dominant elements in an effective description
of the system with broken symmetry valid at low energy and small
momentum.

This will be the subject of the first chapter.  The topic has a long
history, what is relatively new is how the effective description can be
reconciled with Galilei invariance---one of the symmetries governing the
nonrelativistic world of condensed matter physics.  More specific, we
will derive the effective theories at the absolute zero of temperature
of classical hydrodynamics, of a superconductor both in the
weak-coupling as well as in the strong-coupling limit, of superfluid
$^3$He, and of a bosonic superfluid.  We also consider the behavior of a
superconductor and the bosonic superfluid at finite temperature.
Throughout this report we employ the powerful apparatus of quantum field
theory, mostly in the functional-integral formulation.  In this
approach, the effective theories are obtained by integrating out certain
degrees of freedom contained in the microscopic description of the
quantum system under consideration.

The occurrence of gapless modes is not the only manifestation of a broken
continuous symmetry.  A closely related one is the appearance of topological
defects.  These objects, which are often of paramount importance to
understand the physical properties of a system with a broken symmetry, can
have peculiar quantum numbers associated with them. This will be the subject
of Chap.\ \ref{chap:indq}.  We shall consider a superfluid $^3$He film,
one-dimensional metals, a two-dimensional model exhibiting an exotic
mechanism which leads to superconductivity, and free electrons confined to
move in a plane.  Again using the functional-integral approach to quantum
field theory, we shall see how so-called topological terms arise in the
effective theories after the fermionic degrees of freedom contained in the
microscopic models are integrated out.  These terms are the cause of the
peculiar quantum numbers which are expected to arise in these systems.

The next chapter, Chap.\ \ref{chap:dual}, deals with so-called dual
theories.  Duality is one of the hottest topics in contemporary condensed
matter physics.  Quite generally, a dual description of a system refers to a
description in terms of a different set of variables than the original set.
The systems we will consider in this chapter---a superfluid $^4$He film, a
superconducting film, and a bulk superconductor, all at finite
temperature---have vortex solutions in common.  In the original formulation,
these topological defects are described as singular objects.  In the dual
theories discussed in this chapter, these objects are instead described in a
nonsingular fashion by field theory.  The dual theories are employed to
discuss the (equilibrium) phase transition these systems undergo as a
function of temperature.  It will also be pointed out in this chapter that
the concept of duality in two dimensions is closely related to
bosonization---a powerful computational tool in this reduced dimensionality.

In the last chapter, Chap.\ \ref{chap:qpt}, we discuss quantum phase
transitions.  These are phase transitions, taking place close to the
absolute zero of temperature, which are driven not by temperature, but
by some other parameter in the system.  In contrast to equilibrium
transitions at finite temperature, time is important in quantum phase
transitions.  This naturally leads to the use of quantum field theory to
describe these.  Recent experiments on quantum phase transitions in
various two-dimensional systems have raised some very interesting
questions.  It is generally believed that the concept of duality will
play a decisive role in answering these.  We will be concerned in this
chapter with 2nd-order quantum phase transitions.  In addition to a
diverging correlation length, which is common for 2nd-oder equilibrium
transitions at finite temperature, these transitions also have a
diverging correlation time.  We will discuss in detail the universality
class defined by repulsively interacting bosons which undergo a
so-called superfluid-to-Mott-insulating phase transition.  We will
consider the system both in the absence and in the presence of disorder.
We then continue to describe the quantum phase transition which a
quantum-Hall liquid undergoes in the absence of disorder as the applied
magnetic field varies.  We shall discuss scaling theory and consider some
relevant experiments.
\section*{Acknowledgments}
The author wishes to express his gratitude to Professor H. Kleinert for
many interesting discussions and helpful advise.

This work was performed as part of a scientific network supported by the
European Science Foundation, an association of 62 European national
funding agencies (see network's URL, http://defect.unige.ch/defect.html).

\chapter*{Notation\markboth{Notation}{Notation}}
\label{chap:not}
We adopt Feynman's notation and denote a spacetime point by $x=x_\mu =(t,{\bf
x})$, $\mu = 0,1, \cdots,d$, with $d$ the number of space dimensions, while
the energy $k_0$ and momentum ${\bf k}$ of a particle will be denoted by
$k=k_\mu = (k_0,{\bf k})$.  The time derivative $\partial_0 =
\partial/\partial t$ and the gradient $\nabla$ are sometimes combined in a
single vector $\tilde{\partial}_\mu = (\partial_0, -\nabla)$.  The tilde on
$\partial_\mu$ is to alert the reader for the minus sign appearing in the
spatial components of this vector.  We define the scalar product $k \cdot x
= k_\mu x_\mu = k_\mu g_{\mu \nu} k_\nu = k_0 t - {\bf k} \cdot {\bf x}$,
with $g_{\mu \nu} = {\rm diag}(1,-1, \cdots,-1)$ and use Einstein's summation
convention.  Because of the minus sign in the definition of the vector
$\tilde{\partial}_\mu$ it follows that $\tilde{\partial}_\mu a_\mu =
\partial_0 a_0 + \nabla \cdot {\bf a}$, with $a_\mu$ an arbitrary vector.

Integrals over spacetime are denoted by
$$
\int_{x} = \int_{t,{\bf x}} = \int \dd t \, \dd^d x,
$$
while those over energy and momentum by
$$
\int_k = \int_{k_0,{\bf k}} = \int \frac{\dd k_0}{2 \pi}
\frac{\dd^d k}{(2 \pi)^d}.
$$
When no integration limits are indicated, the integrals are assumed to be over
all possible values of the integration variables.

Natural units $\hbar = c = k_{\rm B} = 1$ are adopted throughout unless
explicitly stated.

\setcounter{chapter}{0}
\pagenumbering{arabic}
\chapter{Nonrelativistic Effective Theories\label{chap:nrgold}}
In this chapter we shall derive effective theories governing the low-energy,
small-momentum behavior of some nonrelativistic systems.  The systems
considered here have in common a spontaneously broken global symmetry.  On
account of Goldstone's theorem such a breakdown is always accompanied by
gapless modes.  If the original symmetry G is spontaneously broken to some
subgroup H, the Goldstone modes parameterize the coset space G/H.  Being
gapless, these modes are the dominant excitations at low energy and small
momentum and for this reason the relevant degrees of freedom to build
effective theories from.

In a Lorenz-invariant theory, massive elementary excitations have a spectrum
of the form $E^2({\bf k}) \sim {\bf k}^2 + m^2$, with ${\bf k}$ the momentum
and $m$ the mass of the excitation.  In this case, a gapless excitation is
easily described by taking the limit $m \rightarrow 0$ in the massive
theory.  In a Galilei-invariant theory, however, it is a priori not clear
how to describe a gapless mode.  Here, elementary excitations of a free
theory have a spectrum of the form $E({\bf k}) \sim {\bf k}^2/2 m$, so that
the limit $m \rightarrow 0$ cannot be taken.  Although elements of the
solution to this problem were present in the literature (see, e.g.,
\cite{Popov,Takahashi}), it was not until a paper by Greitner, Wilczek,
and Witten \cite{GWW} that the issue was settled.  They considered the
nonrelativistic Goldstone mode of a spontaneously broken global U(1)
symmetry generated by the total particle number.  Their general
considerations based on symmetry principles were powerful enough to fully
determine the effective theory.

In the following, we shall compute the effective theory describing the
nonrelativistic Goldstone modes as they arise in an ideal classical fluid
(Sec.\ \ref{sec:hydro}), in a BCS superconductor (Sec.\ \ref{sec:bcs}), in
superfluid $^3$He-a (Sec.\ \ref{sec:he3}), and in a weakly interacting Bose
gas (Sec.\ \ref{sec:bec}).  In addition, we investigate the
strong-coupling limit of the pairing theory (Sec.\ \ref{sec:comp}) and
the high-temperature limit of a weakly interacting Bose gas (Sec.\
\ref{sec:ftbec}) as well as that of the BCS theory (Sec.\ \ref{sec:ftbcs}).
\section{Effective Theory of Hydrodynamics}
\label{sec:hydro}
The hydrodynamics of an ideal, classical fluid was already well understood in
the 19th century.  The case of isentropic flow, for which the entropy per unit
mass is constant, is particularly simple.  The pressure $P$ is then a function
of the mass density $\rho$ only, and the flow is automatically a potential
flow.  A feature of such a fluid is that it supports unattenuated sound waves,
i.e., propagating density oscillations.  The waves are unattenuated because
viscosity and thermal conductivity, which usually serve to dissipate the
energy of a propagating mode, are absent.  More important to our present
considerations is that sound waves are gapless.  It would be gratifying to
identify them as the Goldstone mode of a broken continuous symmetry since that
would explain their gaplessness.  We will argue in this section that such an
identification is indeed possible. 

To describe the hydrodynamics of an isentropic fluid, we use Eckart's
variational principle \cite{Eckart} and start with the Lagrangian
\begin{equation} 
\label{hydro:L}
{\cal L} = \tfrac{1}{2} \rho {\bf v}^2 - \rho e + \phi [\partial_0
\rho + \nabla \cdot (\rho {\bf v})],
\end{equation}
where ${\bf v}$ is the velocity field, $\rho$ the mass density, and $e$ the
internal energy per unit mass.  For isentropic flow $e$ is a function of
$\rho$ alone.  The first and second term in (\ref{hydro:L}) represent the
kinetic and potential energy density, respectively.  The variable $\phi$ is
a Lagrange multiplier introduced to impose the conservation of mass:
\begin{equation}  \label{hydro:mass}
\partial_0 \rho + \nabla \cdot (\rho {\bf v}) = 0;
\end{equation} 
its dimension is $[\phi] = {\rm m}^2 {\rm s}^{-1}$.  
The variation of (\ref{hydro:L}) with respect to ${\bf v}$ yields the equation
\begin{equation} 
\label{hydro:v}
{\bf v} = \nabla \phi.
\end{equation}
It shows that, indeed, isentropic flow is automatically a potential flow and
it also identifies the Lagrange multiplier $\phi$ as the velocity
potential.  (How to incorporate vortices in this variational approach will
be discussed in the next section.)  With (\ref{hydro:v}), the Lagrangian
(\ref{hydro:L}) becomes
\begin{equation} 
\label{hydro:Ltheta}
{\cal L} = - \rho [\partial_0 \phi + \tfrac{1}{2} (\nabla \phi)^2 +
e],
\end{equation}
where we performed integrations by part.  A second field equation can be
obtained by varying the Lagrangian with respect to $\rho$.  This yields
the Bernoulli equation
\begin{equation} 
\label{hydro:rhovar}
\partial_0 \phi + \tfrac{1}{2} (\nabla \phi)^2 + h  = 0,
\end{equation} 
with $h = \partial (\rho e)/ \partial \rho$ the specific enthalpy.  From this
definition of $h$, one can easily derive the thermodynamic relation for
isentropic flow
\begin{equation} \label{hydro:thermo}
\nabla h = \frac{1}{\rho} \nabla P,
\end{equation}
with $P = \rho^2 \partial e/\partial \rho$ the pressure.  On taking the
gradient of (\ref{hydro:rhovar}) and using (\ref{hydro:thermo}) one obtains
Euler's equation
\begin{equation}
\label{hydro:Euler} 
\partial_0 {\bf v} + \tfrac{1}{2} \nabla {\bf v}^2 + \frac{1}{\rho}
\nabla P = 0
\end{equation}
governing the flow of the fluid.

We next wish to investigate the symmetry content of the theory.   Classical
hydrodynamics has the following invariances:
\begin{itemize}
\item[(i)] Invariance under spacetime translations, $x_\mu
\rightarrow x_\mu + \epsilon_\mu$, with $\epsilon_\mu$ a constant vector.
\item[(ii)] Invariance under global translations of the velocity
potential, $\phi \rightarrow \phi + \alpha$, with $\alpha$ a constant.
\item[(iii)] Invariance under Galilei boosts,
\begin{equation}   \label{hydro:boost}
\begin{array}{ll} 
t \rightarrow t' = t, & {\bf x} \rightarrow {\bf x}' = {\bf x} - {\bf u}
t;  \\
\partial_0 \rightarrow \partial_0' = \partial_0 + {\bf u} \cdot
\nabla, & \nabla \rightarrow \nabla' = \nabla,
\end{array}
\end{equation}      
with ${\bf u}$ a constant velocity.
\end{itemize}
According to Noether's theorem, symmetries imply conservation laws.
Associated with the above invariances we obtain the following conservation
laws:
\begin{itemize}
\item[(i)] $\tilde{\partial}_\mu t_{\mu \nu} = 0$, where $t_{\mu
\nu}$ $(\mu,\nu = 0,1,2,3)$ is the energy-momentum tensor, with $t_{0i}
= p_i$ the 
momentum density and $t_{00} = {\cal H}$ the Hamiltonian;
\item[(ii)] $\tilde{\partial}_\mu g_\mu = 0$, with $g_0$ the mass
density and $g_i$ the mass current;
\item[(iii)] $\tilde{\partial}_\mu g_{\mu j} = 0$, with $g_{0j} = -
g_0 x_j + t \, p_j$ and $g_{ij} = g_i x_j - t \, t_{ij}$ the
corresponding charge densities and currents.  Physically, the
conservation $d G_{0i} /dt = 0$ of the charges $G_{0i} = \int_{\bf x}
g_{0i}$ means that the center of mass of the fluid, ${\bf X} = \int_{\bf
x} {\bf x} \rho/ M$, with $M= \int_{\bf x}
\rho$ the total mass, moves with constant velocity,
\begin{equation} 
M \frac{d\bf X}{dt} = \int_{\bf x} {\bf p}.
\end{equation}   
Here, the right-hand side denotes the total momentum of the fluid.
\end{itemize}
From the Lagrangian (\ref{hydro:Ltheta}) we obtain as explicit form for the
various charge densities and currents \cite{Kronig}:
\begin{eqnarray} 
g_0 &=& -\frac{\partial {\cal L}}{\partial \partial_0 \phi} =
\rho \label{hydro:hydroa} \\
g_i &=& -\frac{\partial {\cal L}}{\partial \partial_i \phi} =
\rho v_i  \\
p_j (= t_{0 j}) &=& -\frac{\partial {\cal L}}{\partial \partial_0
\phi} \partial_j \phi = \rho v_j  \\
t_{i j} &=& {\cal L} \delta_{i j} - \frac{\partial {\cal L}}{\partial
\partial_i \phi} \partial_j \phi =
P \delta_{i j} + \rho v_i v_j \\
{\cal H} (= t_{0 0}) &=& \frac{\partial {\cal L}}{\partial \partial_0
\phi} \partial_0 \phi - {\cal L} =  \frac{\rho}{2} {\bf v}^2  +
\rho e \label{hydro:hydrof} \\
t_{i 0} &=& \frac{\partial {\cal L}}{\partial \partial_i \phi}
\partial_0 \phi = ({\cal H} + P) v_i  .
\end{eqnarray}
Time derivatives $\partial_0 \phi$ have been eliminated through the field
equation (\ref{hydro:rhovar}), so that, for example, ${\cal L}$ in the last
equation is replaced with
\begin{equation}  \label{hydro:LP}
{\cal L} \rightarrow \rho h - \rho e = \left(\rho \frac{\partial}{\partial
\rho}-1\right) (\rho e) = P.
\end{equation} 
A few remarks are in order.  First, the Hamiltonian ${\cal H}$ is the sum of
the kinetic and potential energy density, as required.  Second, the
equivalence of the mass current ${\bf g}$ and the momentum density ${\bf
p}$, which is a hallmark of Galilei invariance, is satisfied by the theory.
Finally, the set of equations (\ref{hydro:hydroa})--(\ref{hydro:hydrof})
constitutes all the equations of hydrodynamics.  This brings us to the
conclusion that the Lagrangian (\ref{hydro:Ltheta}) encodes all the relevant
information for the description of an isentropic fluid.

We next turn to the description of sound waves.  We restrict ourselves to
waves of small amplitude.  These generate only small deviations in the mass
density $\bar{\rho}$ and pressure $\bar{P}$ of the fluid at rest, so that we can
expand the Lagrangian (\ref{hydro:Ltheta}) in powers of $\rho - \bar{\rho} =
\tilde{\rho}$, with $|\tilde{\rho}| << \bar{\rho}$:
\begin{equation} 
{\cal L} = - (\partial_0 \phi + \tfrac{1}{2}
{\bf v}^2) (\bar{\rho} + \tilde{\rho}) -  \bar{e} \bar{\rho} - \bar{h}
\tilde{\rho} - \tfrac{1}{2} \bar{h}' \tilde{\rho}^2  + {\cal O}(\tilde{\rho}^3).
\end{equation}
The derivative $(')$ is with respect to $\rho$ and is to be evaluated at $\rho
= \bar{\rho}$.  Since for the system at rest $\phi$ is constant, it follows from
(\ref{hydro:rhovar}) that $\bar{h} = 0$.  If we denote the thermodynamic derivative
$\partial P/ \partial \rho$ by $c^2$, which has the dimension of a velocity
squared, the coefficient of the quadratic term in $\tilde{\rho}$ can be
written as
\begin{equation} 
\bar{h}' =  \frac{1}{\bar{\rho}} \bar{P}' = \frac{c_0^2}{\bar{\rho}}.
\end{equation}
Apart from an irrelevant constant term ($-\bar{e} \bar{\rho}$) the Lagrangian
becomes to this order
\begin{equation} \label{hydro:Lexpand}
{\cal L} = - (\partial_0 \phi + \tfrac{1}{2} {\bf v}^2) (\bar{\rho} +
\tilde{\rho}) - \frac{c_0^2}{2\bar{\rho}} \tilde{\rho}^2.
\end{equation}
We next eliminate $\tilde{\rho}$ by substituting
\begin{equation}
\label{hydro:rhotilde} 
\tilde{\rho} = -\frac{\bar{\rho}}{c_0^2} (\partial_0 \phi + \tfrac{1}{2}
{\bf v}^2),
\end{equation}
which follows from expanding the field equation (\ref{hydro:rhovar}).
Physically, this equation reflects Bernoulli's principle: in regions of
rapid flow, the mass density $\rho = \bar{\rho} + \tilde{\rho}$ and therefore
the pressure is low.  It also shows that the expansion in $\tilde{\rho}$ is
one in derivatives.  The higher-order terms that have been neglected
correspond to higher-order derivatives.  At low energy and small momentum,
these additional terms can be ignored.  After eliminating $\tilde{\rho}$, we
obtain a Lagrangian governing the velocity potential $\phi$ \cite{hydro}:
\begin{equation} \label{hydro:Leff}
{\cal L}_{\rm eff} = - \bar{\rho} (\partial_0 \phi + \tfrac{1}{2} {\bf v}^2) +
\frac{\bar{\rho}}{2 c_0^2} (\partial_0 \phi + \tfrac{1}{2} {\bf v}^2)^2.
\end{equation}
This is precisely the effective theory of a nonrelativistic (Abelian)
Goldstone mode obtained in Ref.~\cite{GWW} using general arguments.  Apart
from an irrelevant constant it is identical to a proposal by Takahashi
\cite{Takahashi} which was also based on symmetry principles.

The field equation one obtains for the velocity potential $\phi$ from
(\ref{hydro:Leff}) is nonlinear:
\begin{equation} \label{hydro:feq}
\bar{\rho} (\partial_0^2 \phi + \tfrac{1}{2} \partial_0 {\bf v}^2) - \rho c_0^2
\nabla \cdot {\bf v} + \tfrac{1}{2} \bar{\rho} ( \partial_0 {\bf v}^2 + {\bf v}
\cdot \nabla {\bf v}^2) =0.
\end{equation} 
The information contained in this equation cannot be more than the
conservation of mass because $\phi$ was initially introduced in
(\ref{hydro:L}) as a Lagrange multiplier precisely to enforce this
conservation law.  Indeed, remembering that the combination $\bar{\rho}(\partial_0
\phi + \tfrac{1}{2} {\bf v}^2)$ denotes $c_0^2$ times $\tilde{\rho}(x) =
\rho(x) -
\bar{\rho}$, with $\bar{\rho}$ the constant mass density of the fluid at rest, we
see that (\ref{hydro:feq}) reproduces (\ref{hydro:mass}) in this
approximation.  To simplify (\ref{hydro:feq}), we replace $\bar{\rho}$ in the
first and last term with the full mass density $\rho$ (which is justified to
this order) to arrive at the known \cite{Shivamoggi}, but unfamiliar field
equation
\begin{equation} \label{hydro:unfamiliar}
\partial_0^2 \phi - c_0^2 \nabla^2 \phi = - \partial_0 {\bf v}^2 -
\tfrac{1}{2} {\bf v} \cdot \nabla {\bf v}^2
\end{equation} 
of sound waves.  If we ignore the nonlinear terms, it becomes the more
familiar wave equation
\begin{equation}
\label{hydro:wave}
\partial_0^2 \phi - c_0^2 \nabla^2 \phi = 0,
\end{equation}
implying a gapless linear spectrum, and identifying $c$, which was
introduced via the thermodynamic derivative $\partial P/ \partial \rho
= c^2$, as the sound velocity.  

The combination $\partial_0 \phi + \tfrac{1}{2} (\nabla \phi)^2$ appearing
in the description of a nonrelativistic gapless field is dictated by Galilei
invariance.  To obtain the transformation property of the velocity potential
$\phi$ under a Galilei boost (\ref{hydro:boost}) we note that since $\nabla
\phi$ is a velocity field, $\nabla \phi(x) \rightarrow \nabla'
\phi'(x') = \nabla \phi(x) - {\bf u}$, with $x' = (t,{\bf x} - {\bf u}
t)$.  This gives as transformation rule for $\phi$
\begin{equation}
\label{hydro:transfotheta}
\phi(x) \rightarrow \phi'(x') = \phi(x) -
{\bf u} \cdot {\bf x} + f(t),
\end{equation}
with $f(t)$ a yet undetermined function of time.  To determine $f(t)$ we note
that the factor $-\partial_0 \phi$ in the Lagrangian (\ref{hydro:Ltheta}) is
the chemical potential per unit mass.  Indeed, using the standard definition
$\mu = \partial {\cal H} / \partial \rho$, we find
\begin{equation}
\mu = \tfrac{1}{2} {\bf v}^2 + h = -\partial_0 \phi,
\end{equation}
where in the second equality we used the field equation
(\ref{hydro:rhovar}).  This identification fixes the transformation rule of
$-\partial_0 \phi$:
\begin{equation}
\label{hydro:partialtheta}
-\partial_0 \phi(x) \rightarrow -\partial'_0 \phi'(x')
= -\partial_0 \phi (x) - {\bf u} \cdot {\bf v}(x) +
\tfrac{1}{2} {\bf u}^2
\end{equation}
and in combination with (\ref{hydro:transfotheta}), yields for $f(t)$
\begin{equation}
\label{hydro:f(t)}
\partial_0 f(t) = \tfrac{1}{2} {\bf u}^2, \;\;\; {\rm or}
\;\;\; f(t) = \tfrac{1}{2} {\bf u}^2 t
\end{equation}
up to an irrelevant constant.  It is easily checked that both the combination
$\partial_0 \phi + \tfrac{1}{2} (\nabla \phi)^2$ appearing in the
effective theory (\ref{hydro:Leff}) as well as the field equation
(\ref{hydro:unfamiliar}) are invariant under Galilei boosts.  So, contrary to
what is sometimes stated in the literature \cite{Jackson}, sound waves are
invariant under Galilei boosts.  The linearized wave equation
(\ref{hydro:wave}) is, of course, not invariant because essential nonlinear
terms are omitted.

Greitner, Wilczek, and Witten arrived at the effective Lagrangian
(\ref{hydro:Leff}) by requiring that the effective theory of sound
waves can only be constructed with the help of the Galilei-invariant
combination $\partial_0
\phi + \tfrac{1}{2} (\nabla \phi)^2$, and that it should give the
dispersion relation $E^2({\bf k}) = c^2 {\bf k}^2$.

From the effective Lagrangian (\ref{hydro:Leff}) one can again calculate the
various Noether charge densities and currents \cite{Takahashi}.  They are, as
might be expected, of the same form as the exact expressions
(\ref{hydro:hydroa})--(\ref{hydro:hydrof}), but now with the approximations
\begin{equation} 
\label{hydro:aprho}
\rho \simeq \bar{\rho} - \frac{\bar{\rho}}{c_0^2} (\partial_0 \phi +
\tfrac{1}{2} {\bf v}^2)
\end{equation}
as follows from Eq.\ (\ref{hydro:rhotilde}),
\begin{equation}
{\cal H} \simeq \frac{\rho}{2} {\bf v}^2 + \frac{c_0^2}{2 \bar{\rho}} (\rho -
\bar{\rho})^2,
\end{equation}
and
\begin{equation}
P \simeq -\bar{\rho} ( \partial_0 \phi + \tfrac{1}{2} {\bf v}^2) \simeq c_0^2
(\rho - \bar{\rho}).
\end{equation}
This last equation is consistent with the expression one obtains from
directly expanding the pressure: $P(\rho) = \bar{P} + \tilde{\rho} \bar{P}'$
since $\bar{P} = 0$ and $\bar{P}' = c_0^2$.    

To recapitulate, we have derived an effective theory describing a gapless
mode starting from the Lagrangian (\ref{hydro:Ltheta}) which entails the
complete hydrodynamics of an isentropic fluid.  We now wish to argue that
this gapless mode is a Goldstone mode associated with a spontaneously broken
symmetry.  A first indication in favor of such an interpretation follows
because the Hamiltonian ${\cal H}$ displays a property typical for a system
with broken symmetry, namely that it is a function not of the velocity
potential itself, but of the gradient of the field.  The energy is
minimal if $\phi$ is uniform throughout the sample, i.e., there is rigidity
\cite{PWA}.

Usually the broken symmetry can be identified by the general property that the
Goldstone mode is translated under the broken symmetry operations.  (There may
be other effects too, but the translation is always present.)  Here, this
general characteristic does not uniquely identify the broken symmetry because
$\phi$ is translated under two symmetry operations.  According to the
transformation rule (\ref{hydro:transfotheta}) with $f(t)$ given in
(\ref{hydro:f(t)}), we have that under a Galilei boost
\begin{equation}
\label{hydro:LeeG}
\delta^{G}_{\bf u} \phi (x) := \phi' (x) - \phi (x) = - {\bf u} \cdot
{\bf x} + t {\bf u} \cdot \nabla \phi (x),
\end{equation}
where we took the transformation parameter ${\bf u}$ infinitesimal small so
that quadratic and higher powers in ${\bf u}$ may be ignored.  The first
term at the right-hand side shows that the velocity potential is translated
under a Galilei boost.  The second symmetry under which $\phi$ is
translated is generated by the total particle number, or, equivalently, by
the total mass $M = \int_{\bf x} \rho$.  To see this we first compute from
(\ref{hydro:Ltheta}) the conjugate momentum $\pi_\phi$ of $\phi$,
\begin{equation}
\label{hydro:pitheta}
\pi_\phi = \frac{\partial {\cal L}}{\partial \partial_0 \phi} =
-\rho,
\end{equation} 
implying that $\phi$ and $\rho$ are canonically conjugate \cite{London}:
\begin{equation}
\label{hydro:can}
\{\phi(t,{\bf x}), \rho(t,{\bf x}') \} = -\delta ({\bf x} - {\bf x}'),
\end{equation}
where $\{\, ,\}$ denotes the Poisson bracket.  We then use a central result of
classical field theory stating that the charge $Q$ of a continuous symmetry is
the generator of the corresponding transformations of the fields, $\chi (x)
\rightarrow \chi' (x)$.  More specifically, for an infinitesimal
transformation $\delta_{\alpha}^Q \chi (x) = \chi' (x) -
\chi (x)$ one has
\begin{equation}
\delta_{\alpha}^Q \chi (x) = - \alpha \{\chi (x), Q\}, 
\end{equation}
with $\alpha$ the transformation parameter.  Equation (\ref{hydro:can}) thus
implies that under the global U(1) symmetry generated by the total mass,
$\phi$ is indeed translated
\begin{equation}
\delta_{\alpha}^M \phi (x) = - \alpha \{\phi (x), M\} =
\alpha.
\end{equation}
The point is that the generators of Galilei boosts $G_{0j} = \int_{\bf x} (-
x_j g_0 + t p_j)$ also contain the mass density $\rho = g_0$.  It is therefore
impossible to distinguish a broken Galilei invariance from a broken mass
symmetry by considering the algebra alone.

Let us at this point pause for a moment and consider the case of a
superfluid.  It is well established that in the normal-to-superfluid phase
transition, the global U(1) symmetry generated by the total mass is
spontaneously broken.  This comes about because in a superfluid, which is a
quantum system, many particles Bose-Einstein condense in a single quantum
state.  The coherence allows us to describe the system by a complex field
\begin{equation} \label{hydro:psiexpl}
\psi(x) = \sqrt{\rho(x)/m} \, {\rm e}^{i m \varphi(x)/\hbar}
\end{equation}  
normalized such that $|\psi|^2$ yields the particle number density $\rho/m$
of the condensate.  To underscore the quantum nature of a superfluid,
Planck's constant has been made explicit. The field $\varphi$, which turns
out to describe the Goldstone mode of the broken global U(1) symmetry, is a
phase field and therefore compact.  The field $\psi$ is frequently referred
to as the condensate wavefunction, even in the modern literature.  In our
view, this is somewhat misleading.  As has been stressed by Feynman
\cite{FeynmanSM}, $\psi$ is a classical field describing the coherent behavior
of many condensed particles in the same way as the classical field of
electrodynamics describes the behavior of many photons in a single state.
For these classical fields there is no probability interpretation as is
required for wavefunctions \cite{Fick}.

It has been argued by Feynman \cite{FeynmanSM} that the $\psi$-field is governed
by a nonrelativistic $|\psi|^4$-theory defined by the Lagrangian
\begin{equation} \label{hydro:Lpsi}
{\cal L}_\psi = i \hbar \psi^* \partial_0 \psi - \frac{\hbar^2}{2m} |\nabla
\psi|^2 - \frac{c_0^2}{2 \bar{\rho}} (m |\psi|^2  - \bar{\rho})^2.
\end{equation}
The potential energy has its minimum along a circle away from the origin at
$|\psi|^2 = \bar{\rho}/m$, implying a spontaneous breakdown of the global U(1)
symmetry.  The Lagrangian (\ref{hydro:Lpsi}) with (\ref{hydro:psiexpl})
reduces to the one given in (\ref{hydro:Lexpand}) when the term $-\hbar^2
(\nabla
\rho)^2/8m^2\rho$ is ignored.  Using the expression for the pressure, cf.\
(\ref{hydro:LP})
\begin{equation} 
P = \left[ \rho \left(\frac{\partial }{\partial \rho} -  \partial_i
\frac{\partial }{\partial \partial_i \rho}\right) -1 \right] (\rho e),
\end{equation} 
we find that it gives the contribution $-\hbar^2 (\nabla^2 \rho)/4m^2$
to the pressure---the so-called quantum pressure.  The reason for calling it
this way is that it is the only place where Planck's constant appears in the
equations.  To the order in which we are working, it is consistent to ignore
this term.  Because the field equation for $\psi$ derived from
(\ref{hydro:Lpsi}) has the {\em form} of a nonlinear Schr\"odinger equation,
we will refer to $\psi$ as a Schr\"odinger field.  We trust however that the
reader realizes that it is a classical field unrelated to a Schr\"odinger
wavefunction.

Let us compare the transformation properties of the Schr\"odinger field with
that of the velocity potential $\phi$ of an isentropic fluid.  Under a
Galilei boost, $\psi(x)$ transforms as \cite{Fick}
\begin{equation} \label{hydro:Schroeder} 
\psi(x) \rightarrow \psi'(x') = \exp[i (-{\bf u} \cdot {\bf
x} + \tfrac{1}{2} {\bf u}^2 t)m/\hbar] \, \psi (x).
\end{equation} 
With $m \varphi /\hbar$ denoting the phase of the Schr\"odinger field, we see
that $\varphi$ transforms in the same way as does the velocity potential.

Using that the canonical conjugate of the $\psi$-field is $\pi_\psi = i \hbar
\psi^*$, we easily derive the Poisson bracket 
\begin{equation} \label{hydro:Poisson}
\{ \psi (x), M \} = - i (m/\hbar)\psi (x).
\end{equation}   
This shows that the total mass $M$ generates phase transformations on the
$\psi$-field: $\psi (x) \rightarrow \psi'(x) = \exp(i
\alpha m /\hbar) \psi (x)$.  The phase $\varphi$ of the Schr\"odinger
field is consequently translated under the symmetry, just like the velocity
potential $\phi$.  This transformation property identifies $\varphi$ as
the Goldstone mode of the broken U(1) mass symmetry.  The Poisson bracket
(\ref{hydro:Poisson}) also implies that $\varphi$ and $\rho$ are canonical
conjugate
\cite{PWA}, cf.\ (\ref{hydro:can})
\begin{equation} 
\{\varphi (t,{\bf x}), \rho(t,{\bf x}') \} = - \delta ({\bf x} - {\bf
x}').
\end{equation} 
A similar relation holds for superconductors.  On quantizing, the Poisson
bracket is replaced by a commutator.  The Heisenberg uncertainty relation that
results for the conjugate pair has recently been demonstrated experimentally
\cite{Delft}.

As remarked above, a necessary condition for the spontaneous breakdown of
the global U(1) symmetry is the presence of a condensate.  Such an intrinsic
quantum phenomenon, requiring many particles in a single state, has no
analog in a classical setting.  Hence, the U(1) symmetry cannot be broken in
classical hydrodynamics.  This leaves us with the second possibility, namely
that of a spontaneously broken Galilei invariance.  The breakdown is a
result of the presence of a finite mass density.  This can be inferred
\cite{Takahashi} from considering the transformation of the momentum density
${\bf p}(x)$ under a Galilei boost:
\begin{equation}
\delta^G_{\bf u} p_i (x) = -u_j \{p_i(x), G_{0j}(t) \} =
-u_i \rho (x) + t {\bf u} \cdot \nabla p_i (x),
\end{equation}
or with $\delta^G_{\bf u} = u_j \delta^G_j$
\begin{equation}
\delta^G_j p_i (x) = -\rho (x) \delta_{j i} + t \partial_j
p_i (x).
\end{equation}
If the mass density $\rho$ is finite, the right-hand side is nonzero, which is
a symmetry-breaking condition.
\section{Including Vortices}
\label{sec:vortices}
There is an essential difference between the spontaneous breakdown of the
Galilei symmetry and that of the global U(1) symmetry.  Although both
symmetries are Abelian, the latter is a compact symmetry, whereas the
Galilei group is noncompact.  More specifically, the transformation
parameter $\alpha$ of the U(1) group has a finite domain ($0 \leq \alpha < 2
\pi$), while the domain of ${\bf u}$, the transformation parameter of the
Galilei group, is infinite.  As a result, the velocity potential of
classical hydrodynamics cannot be represented as the phase of a complex
field.  An immediate physical manifestation of the difference is that a
system with broken U(1) invariance supports topologically stable vortices,
whereas a system with broken Galilei invariance does not.  This is not to
say that vortices are absent in the latter case, it merely states that their
stability is not guaranteed by topological conservation laws.  Closely
connected to this is that the circulation is not quantized in classical
hydrodynamics, which is known to exist in superfluids.  Yet, the circulation
is conserved also in isentropic fluids.  This is again not for topological,
but for dynamical reasons, the conservation being proven by invoking Euler's
equation (\ref{hydro:Euler}) as was first done for an ideal, incompressible
fluid by Helmholtz \cite{Helmholtz} and generalized to a compressible fluid
by Thomson \cite{Thomson}.

The easiest way to observe vortices in a classical fluid is to punch a hole in
the bottom of the vessel containing the fluid.  As the fluid pours out a
vortex is formed in the remaining fluid---a phenomenon daily observed by
people unplugging a sinkhole.  Often, as happens in, for example, superfluid
$^4$He, the core of a vortex is in the normal state so that the symmetry is
restored there.  In the present context this would mean that inside the vortex
core, the fluid mass density $\rho$ is zero.  This is indeed what is observed:
the vortex core consists of air, therefore no fluid is present and $\rho = 0$
there.

In the eye of a tropical cyclone---another example of a vortex, nature
does its best to restore the Galilei symmetry, record low atmospheric
pressures being measured there.  (A complete restoration would imply
the absence of air corresponding to zero pressure.)

It is customary to incorporate vortices in a potential flow via the
introduction of so-called Clebsch potentials \cite{Clebsch}.  We will
not follow this route, but instead use the powerful principle of defect
gauge symmetry developed by Kleinert \cite{GFCM,KleinertPl,KleinertCam}.
In this approach, one introduces a so-called vortex gauge field
$\phi_\mu^{\rm P} = (\phi_0^{\rm P},
\bbox{\phi}^{\rm P})$ in the Lagrangian via minimally coupling to the
Goldstone field:
\begin{equation} \label{hydro:minimal}
\tilde{\partial}_\mu \phi \rightarrow \tilde{\partial}_\mu \phi +
\phi_\mu^{\rm P}, 
\end{equation}
with $\tilde{\partial}_\mu = (\partial_0,-\nabla)$ and
\begin{equation} \label{hydro:vorticity}
\nabla \times \bbox{\phi}^{\rm P} = -2 \bbox{\omega},
\end{equation}
so that $\nabla \times {\bf v} = 2 \bbox{\omega}$ yields (twice) the
vorticity $\bbox{\omega}$ of the vortex.  The combination
$\tilde{\partial}_\mu \phi + \phi_\mu^{\rm P}$ is invariant under
the local gauge transformation
\begin{equation} 
\phi(x) \rightarrow \phi(x) + \alpha(x); \;\;\;\;\;
\phi^{\rm P}_\mu \rightarrow \phi^{\rm P}_\mu - \tilde{\partial}_\mu
\alpha(x),
\end{equation} 
with $\phi^{\rm P}_\mu$ playing the role of a gauge field.  The left-hand
side of (\ref{hydro:vorticity}) may be thought of as defining the ``magnetic
field'' associated with the vortex gauge field ${\bf B}^{\rm P}= \nabla \times
\bbox{\phi}^{\rm P}$.

For illustrative purposes, let us consider an external, static vortex with
circulation $\Gamma$ located along a line $L$, which may be closed or
infinitely long \cite{Helmholtz}.  Then, $\bbox{\omega} = \tfrac{1}{2}
\Gamma \bbox{\delta } (L)$, where $\bbox{\delta} (L)$ is a delta function on
the line $L$,
\begin{equation} 
\delta_i (L) = \int_L \dd y_i \, \delta({\bf x} - {\bf y}).
\end{equation} 
This model with a static, external vortex may be thought of as describing the
steady flow in the presence of a vortex pinned to a fixed impurity.  The field
equation for $\phi$ obtained after the substitution (\ref{hydro:minimal})
reads
\begin{equation} \label{hydro:addeddefect}
\partial_0 \mu + c_0^2 \nabla \cdot {\bf v}
= \partial_0 {\bf v}^2 + \tfrac{1}{2} {\bf v} \cdot \nabla {\bf v}^2 - {\bf v}
\cdot {\bf E}^{\rm P}
\end{equation} 
with $\mu = -(\partial_0 \phi + \phi^{\rm P}_0)$ the chemical potential
and ${\bf v} = \nabla \phi - \bbox{\phi}^{\rm P}$ the velocity of the flow
in the presence of the vortex.  The last term gives a coupling of the velocity
field to the ``electric field'' associated with $\phi^{\rm P}_\mu$,
\begin{equation} 
{\bf E}^{\rm P} = -\nabla \phi_0^{\rm P} - \partial_0 \bbox{\phi}^{\rm P}.
\end{equation} 
Note that the field equation (\ref{hydro:addeddefect}) is invariant under
local vortex gauge transformations.  Ignoring the higher-order terms and
choosing the gauge $\phi_0^{\rm P} = 0$, we obtain as equation for the flow
in the presence of a static vortex:
\begin{equation}
\nabla \cdot {\bf v} = 0, \;\;\;\; {\rm or} \;\;\;\; \nabla \cdot (\nabla
\phi - \bbox{\phi}^{\rm P}) = 0,
\end{equation}
which is solved by
\begin{equation}
\label{hydro:solution}
\phi ({\bf x}) = - \int_{\bf y}  G({\bf x} - {\bf y}) \nabla \cdot
\bbox{\phi}^{\rm P}({\bf y}).
\end{equation}
Here, $G({\bf x})$ is the Green function of the Laplace operator
\begin{equation}
G({\bf x}) = \int_{\bf k} \frac{ {\rm e}^{i {\bf k}
\cdot {\bf x}}}{{\bf k}^2} = \frac{1}{4 \pi |{\bf x}|}.
\end{equation}
Straightforward manipulations then yield the well-known Biot-Savart
law for the velocity field in the presence of a static vortex
\cite{Helmholtz,GFCM} 
\begin{equation} \label{hydro:vv}
{\bf v}_{\rm v} ({\bf x}) = \frac{\Gamma}{4 \pi} \int_L \dd {\bf y} \times
\frac{{\bf x} -{\bf y}}{|{\bf x} - {\bf y}|^3},
\end{equation}
where the integration is along the vortex.  This exemplifies the viability of
the vortex gauge principle as an alternative to describe vortices in a
potential flow.

Let us continue to study the dynamics of vortices---a subject that has
recently received considerable attention in the literature \cite{Sonin}.
Because the vortex motion is determined by the flow itself, the vortex can no
longer be considered as external.  We shall see that the nonlinear part of
the field equation (\ref{hydro:addeddefect}) becomes relevant here.

In the absence of external forces, the vortex moves with a constant velocity,
${\bf v}_L$ say.  The flow in the presence of a moving vortex can be obtained
from the static solution (\ref{hydro:vv}) by replacing the coordinate ${\bf
x}$ with ${\bf x} - {\bf v}_L t$.  This implies that
\begin{equation} \label{hydro:vchange}
\partial_0 {\bf v}_{\rm v}({\bf x} -{\bf v}_Lt) = -{\bf v}_L \cdot \nabla
{\bf v}_{\rm v}({\bf x} -{\bf v}_Lt). 
\end{equation} 
Since the solution ${\bf v}_{\rm v}$ is curl-free outside the vortex core,
the right-hand side may there be written as $- \nabla ({\bf v}_L \cdot {\bf
v}_{\rm v})$.  To study sound waves in the presence of a moving vortex, we
write the velocity field as ${\bf v}(x) = {\bf v}_{\rm v}({\bf x} - {\bf
v}_L t) + \nabla \tilde{\phi}(x)$, with $\tilde{\phi}$ describing small
variations around the moving vortex solution.  Equation
(\ref{hydro:vchange}) then requires that we write for the chemical potential
in (\ref{hydro:addeddefect})
\begin{equation} 
\mu (x) = {\bf v}_L \cdot {\bf v}_{\rm v}({\bf x} - {\bf v}_L t) -
\partial_0 \tilde{\phi} (x).
\end{equation}
This leads to the linearized field equation \cite{Pitaevskii}
\begin{equation} \label{hydro:AB}
\partial^2_0 \tilde{\phi}(x) - c_0^2 \nabla^2 \tilde{\phi}(x) = - {\bf
v}_{\rm v}({\bf x}) \cdot \nabla \partial_0 [2 
\tilde{\phi}(t,{\bf x}) - \tilde{\phi}(t,0)],
\end{equation}  
describing sound waves in the presence of a moving vortex.  In deriving
(\ref{hydro:AB}) we again used the gauge $\phi^{\rm P}_0=0$, and
approximated ${\bf v}_{\rm v}({\bf x} - {\bf v}_L t)$ by ${\bf v}_{\rm
v}({\bf x})$.  To linear order in $\tilde{\phi}$ this is allowed since
the vortex, being driven by the sound wave, has a velocity
\begin{equation} 
{\bf v}_L(t) = \nabla \tilde{\varphi}(t,{\bf x} = {\bf v}_L t)
\approx \nabla \tilde{\varphi}(t,0).
\end{equation}   
We also neglected a term quadratic in ${\bf v}_{\rm v}$ which is justified
because this velocity is much smaller than the sound velocity outside the
vortex core \cite{Pitaevskii}.  The first term at the right-hand side in
(\ref{hydro:AB}) stems from the nonlinear term $\partial_0 {\bf v}^2$ in the
general field equation (\ref{hydro:addeddefect}).  Thus the nonlinearity of
sound waves becomes detectable.  Equation (\ref{hydro:AB}) can be used as a
basis to study the scattering of phonons of a free moving vortex
\cite{Sonin}.

So far we have contrasted the spontaneous breakdown of the Galilei
invariance (caused by a finite mass density) and that of the global U(1)
symmetry (caused by a nonzero condensate).  A superfluid, however, has a
finite mass density as well as a nonzero condensate.  Both symmetries are
therefore broken and we expect two different Goldstone modes to be present.
This is indeed what is observed in superfluid $^4$He.  The system supports
besides first sound, which are the usual density waves associated with the
spontaneously broken Galilei invariance, also second sound, or entropy
waves.  The latter mode depends crucially on the presence of the condensate
and is the Goldstone mode associated with the spontaneously broken global
U(1) symmetry generated by the total mass.  At the transition point, the
second sound velocity vanishes, whereas the first sound velocity remains
finite.  This is as expected since only the condensate vanishes at the
superfluid phase transition; the total mass density remains finite.
\section{Anderson-Bogoliubov Mode}
\label{sec:bcs}
In this section we study the so-called Anderson-Bogoliubov mode
\cite{Anderson,BTS} of a neutral superconductor.  This mode is known to be the
gapless Goldstone mode associated with the spontaneously broken global U(1)
symmetry generated by the total mass.  Given the general arguments of
Ref.~\cite{GWW} we expect that the effective theory of a superconductor is
exactly of the form (\ref{hydro:Leff}) we obtained for sound waves in
classical hydrodynamics, but with $\phi$ replaced by a compact field.

Our starting point is the famous microscopic model of Bardeen, Cooper, and
Schrieffer (BCS) defined by the Lagrangian \cite{BCS}
\begin{eqnarray}  \label{bcs:BCS}
     {\cal L} &=& \psi^{\ast}_{\uparrow} [i\partial_0 - \xi(-i \nabla)]
\psi_{\uparrow}  
     + \psi_{\downarrow}^{\ast} [i \partial_0 - \xi(-i \nabla)]\psi_{\downarrow}
     - \lambda_0 \psi_{\uparrow}^{\ast}\,\psi_{\downarrow}
     ^{\ast}\,\psi_{\downarrow}\,\psi_{\uparrow}          \nonumber  \\
     &:=& {\cal L}_{0} + {\cal L}_{\rm i},                  
\end{eqnarray} 
where ${\cal L}_{\rm i} = - \lambda_0 \psi_{\uparrow}^{\ast} \,
\psi_{\downarrow}^{\ast}\,\psi_{\downarrow}\,\psi_{\uparrow}$ is a
contact interaction term, representing the effective, phonon mediated,
attraction between electrons with coupling constant $\lambda_0 < 0$, and
${\cal L}_{0}$ is the remainder.  In (\ref{bcs:BCS}), the field
$\psi_{\uparrow (\downarrow )}$ is an anticommuting field describing the
electrons with mass $m$ and spin up (down); $\xi(-i \nabla) = \epsilon(-i
\nabla) - \mu_0$, with $\epsilon(-i \nabla) = - \nabla^2/2m$, is the kinetic
energy operator with the chemical potential $\mu_0$ subtracted.

The Lagrangian (\ref{bcs:BCS}) is invariant under global U(1)
transformations.  Under such a transformation, the electron fields pick up
an additional phase factor
\begin{equation} \label{bcs:3g}
\psi_{\sigma} \rightarrow \mbox{e}^{i \alpha }
\psi_{\sigma}                                
\end{equation}
with $\sigma = \uparrow, \downarrow$ and $\alpha$ a constant.
Notwithstanding its simple form, the microscopic model (\ref{bcs:BCS}) is a
good starting point to describe BCS superconductors.  The reason is that the
interaction term allows for the formation of Cooper pairs which below a
critical temperature condense.  This results in a nonzero expectation value
of the field $\Delta$ describing the Cooper pairs, and a spontaneous
breakdown of the global U(1) symmetry.  This in turn gives rise to the
gapless Anderson-Bogoliubov mode which---after incorporating the
electromagnetic field---lies at the root of most startling properties of
superconductors \cite{Weinberg}.

To obtain the effective theory of this mode, the fermionic degrees of freedom
have to be integrated out.  To this end we introduce Nambu's notation and
rewrite the Lagrangian (\ref{bcs:BCS}) in terms of a two-component field
\begin{equation} \label{bcs:32}
\psi = \left( \begin{array}{c} \psi_{\uparrow} \\ 
           \psi_{\downarrow}^{\ast}  \end{array} \right) \:\:\:\:\:\:
    \psi^{\dagger} = (\psi_{\uparrow}^{\ast},\psi_{\downarrow}).
\end{equation} 
In this notation, ${\cal L}_{0}$ becomes 
\begin{equation}    \label{bcs:33}
{\cal L}_{0} = \psi^{\dagger}\,
\left(\begin{array}{cc}
i \partial_0 - \xi(-i \nabla) & 0              \\
0           &  i \partial_0 + \xi(-i \nabla) 
    \end{array}\right) \, \psi,                        
\end{equation}
where we explicitly employed the anticommuting character of the electron
fields and neglected terms which are a total derivative.  The partition
function,
\begin{equation}     \label{bcs:34}
Z = \int \DD \psi^{\dagger} \DD \psi \exp \left( i \int_x
\,{\cal L} \right),                                            
\end{equation} 
must for our purpose be written in a form bilinear in the electron
fields.  This is achieved by rewriting the quartic interaction term as a
functional integral over auxiliary fields $\Delta$ and $\Delta^*$:
\begin{eqnarray}   \label{bcs:35} 
\lefteqn{
\exp \left( -i \lambda_0 \int_x \psi_{\uparrow}^{\ast}
\, \psi_{\downarrow}^{\ast} \, \psi_{\downarrow}\, \psi_{\uparrow} 
\right)  = }                                           \\
& & \!\!\!\!\!\! \int \DD \Delta^* \DD \Delta \exp \left[ -i
\int_x \left( \Delta^* \, \psi_{\downarrow}\,\psi_{\uparrow} +
\psi_{\uparrow}^{\ast} \, \psi_{\downarrow}^{\ast} \, \Delta -
\frac{1}{\lambda_0 } \Delta^* \Delta \right) \right], \nonumber 
\end{eqnarray} 
where, as always, an overall normalization factor is omitted.  Classically,
$\Delta$ merely abbreviates the product of two electron fields
\begin{equation}  \label{bcs:del}
\Delta = \lambda_0 \psi_{\downarrow} \psi_{\uparrow}.       
\end{equation} 
It would therefore be more appropriate to give $\Delta$ two spin labels
$\Delta_{\downarrow \uparrow}$. Since $\psi_{\uparrow}$ and
$\psi_{\downarrow}$ are anticommuting fields, $\Delta$ is antisymmetric in
these indices.  Physically, it describes the Cooper pairs of the
superconducting state.

By employing (\ref{bcs:35}), we can cast the partition function in the desired
bilinear form:
\begin{eqnarray}  \label{bcs:36}
Z = \int \DD \psi^{\dagger} \DD \psi \int \!\!\!\!\! && \!\!\!\!\! \DD
\Delta^* \DD \Delta   \; \exp\left(\frac{i}{\lambda_0} \int_x
\Delta^*  \Delta \right)       \\  &&
\!\!\!\!\!\!\!\!\!\!\!\!\! \times \exp  \left[  
    i \int_x \, \psi^{\dagger} \left( \begin{array}{cc} 
i \partial_{0} - \xi(-i \nabla)  & -\Delta \\
-\Delta^*  & i \partial_{0} + \xi(-i \nabla) 
\end{array} \right)   \psi \right] \nonumber .  
\end{eqnarray} 
Changing the order of integration and performing the Gaussian integral over
the Grassmann fields, we obtain
\begin{equation}   \label{bcs:37}
Z = \int \DD \Delta^* \DD \Delta \, \exp \left(i S_{\rm eff} [
\Delta^*, \Delta] + \frac{i}{\lambda_0}
\int_x \Delta^* \Delta \right),  
\end{equation}
where $S_{\rm eff}$ is the one-loop effective action which, using the
identity Det(A) = exp[Tr ln(A)], can be cast in the form
\begin{equation}  \label{bcs:312}
S_{\rm eff}[\Delta^*, \Delta] = -i \, {\rm Tr} \ln \left(
\begin{array}{cc} p_{0} - \xi ({\bf p}) & -\Delta \\ -\Delta^* &
p_{0} + \xi ({\bf p})
\end{array}\right),
\end{equation} 
where $p_0 = i \partial_0$ and $\xi({\bf p}) = \epsilon({\bf p}) - \mu_0$,
with $\epsilon({\bf p}) = {\bf p}^2/2m$.  The trace Tr appearing here needs
some explanation.  Explicitly, it is defined as \label{pag:derbegin}
\begin{equation}   \label{bcs:explicit}
S_{\rm eff} = -i {\rm Tr} \, \ln \left[K(p,x) \right] = -i {\rm tr} \ln\left[
K(p,x) \delta (x - y)\bigr|_{y = x} \right],
\end{equation}
where the trace tr is the usual one over discrete indices.  We
abbreviated the matrix appearing in (\ref{bcs:312}) by $K(p,x)$ so as to
cover the entire class of actions of the form
\begin{equation} 
S = \int_x \psi^\dagger(x) K(p,x) \psi(x).
\end{equation} 
The delta function in (\ref{bcs:explicit}) arises because $K(p,x)$ is
obtained as a second functional derivative of the action
\begin{equation}  
\frac{\delta^{2} S}{\delta \psi^\dagger(x) \, \delta \psi(x)} =
K(p,x)  \,  \delta (x - y) \bigr|_{y =  x},
\end{equation} 
each of which gives a delta function.  Since the action has only one
integral $\int_x$ over spacetime, one delta function remains.  Because it is
diagonal, it may be taken out of the logarithm and (\ref{bcs:explicit}) can
be written as
\begin{eqnarray}  \label{bcs:Trexplicit}
S_{\rm eff} &=& -i {\rm tr} \, \int_x 
\ln \left[ K(p,x) \right] 
\delta (x - y) \bigr|_{y = x}  \nonumber  \\ &=&
-i {\rm tr} \, \int_x \int_k {\rm e}^{i k \cdot x} \, \ln \left[ K(p,x)
\right]  {\rm e}^{-i k \cdot x}.
\end{eqnarray}
In the last step, we used the integral representation of the
delta function:
\begin{equation}
\delta (x) = \int_k {\rm e}^{-i k \cdot x},
\end{equation}
shifted the exponential function $\exp (i k \cdot y)$ to the left, which is
justified because the derivative $p_\mu$ does not operate on it, and,
finally, set $y_\mu$ equal to $x_\mu$.  We thus see that the trace Tr in
(\ref{bcs:explicit}) stands for the trace over discrete indices as well as
the integration over spacetime and over energy and momentum.  The integral
$\int_k$ arises because the effective action calculated here is a one-loop
result with $k_\mu$ the loop energy and momentum.

The integrals in (\ref{bcs:Trexplicit}) cannot in general be evaluated in
closed form because the logarithm contains energy-momentum operators and
spacetime-dependent functions in a mixed order.  To disentangle the
integrals resort has to be taken to a derivative expansion \cite{FAF} in
which the logarithm is expanded in a Taylor series.  Each term contains
powers of the energy-momentum operator $p_\mu$ which acts on every
spacetime-dependent function to its right.  All these operators are shifted
to the left by repeatedly applying the identity
\begin{equation} 
f(x) p_\mu g(x) = (p_\mu - i \tilde{\partial}_\mu) f(x) g(x),
\end{equation} 
where $f(x)$ and $g(x)$ are arbitrary functions of spacetime and the
derivative $\tilde{\partial}_\mu = (\partial_0,-\nabla)$ acts {\it only} on
the next object to the right.  One then integrates by parts, so that all the
$p_\mu$'s act to the left where only a factor $\exp(i k \cdot x)$ stands.
Ignoring total derivatives and taking into account the minus signs that
arise when integrating by parts, one sees that all occurrences of $p_\mu$
(an operator) are replaced with $k_\mu$ (an integration variable).  The
exponential function $\exp(-i k \cdot x)$ can at this stage be moved to the
left where it is annihilated by the function $\exp(i k \cdot x)$.  The
energy-momentum integration can now in principle be carried out and the
effective action be cast in the form of an integral over a local density
${\cal L}_{\rm eff}$:
\begin{equation}   
S_{\rm eff} = \int_x {\cal L}_{\rm eff}.
\end{equation} 
This is in a nutshell how the derivative expansion works. \label{pag:derend}

In the mean-field approximation, the functional integral (\ref{bcs:37}) is
approximated by the saddle point:
\begin{equation}   \label{bcs:38}
Z =  \exp \left(i S_{\rm eff}
[ \Delta^*_{\rm mf}, \Delta_{\rm mf} ]  + \frac{i}{\lambda_0} \int_x
\Delta^*_{\rm mf} \Delta_{\rm mf}  \right),                   
\end{equation} 
where $\Delta_{\rm mf}$ is the solution of mean-field equation
\begin{equation}     \label{bcs:gap}
\frac{\delta S_{\rm eff} }{\delta \Delta^*
(x) } = - \frac{1}{\lambda_0} \Delta. 
\end{equation}
If we assume the system to be spacetime independent so that $\Delta_{\rm
mf}(x) = \bar{\Delta}$, Eq.\ (\ref{bcs:gap}) yields the celebrated BCS gap
\cite{BCS} equation:
\begin{eqnarray}   \label{bcs:gape} 
\frac{1}{\lambda_0} &=& - i  \int_k \frac{1}{k_{0}^{2} - E^{2}(k) + i \eta}
\nonumber \\ &=& - \frac{1}{2} \int_{\bf k} \frac{1}{E({\bf k})},
\end{eqnarray} 
where $\eta$ is an infinitesimal positive constant that is to be set to
zero at the end of the calculation, and 
\begin{equation}  \label{bcs:spec}
E({\bf k}) = \sqrt{\xi^2({\bf k}) + |\bar{\Delta}|^2}
\end{equation}  
is the spectrum of the elementary fermionic excitations.  If this equation
yields a solution with $\bar{\Delta} \neq 0$, the global U(1) symmetry
(\ref{bcs:3g}) is spontaneously broken since
\begin{equation}
\bar{\Delta} \rightarrow \mbox{e}^{2i \alpha } \bar{\Delta} \neq  
\bar{\Delta}                             
\end{equation}
under this transformation.  The factor $2$ in the exponential function arises
because $\Delta$, describing the Cooper pairs, is built from two electron
fields.  It satisfies Landau's definition of an order parameter as its value
is zero in the symmetric, disordered state and nonzero in the state with
broken symmetry.  It directly measures whether the U(1) symmetry is
spontaneously broken.

In the case of a spacetime-independent system, the effective action
(\ref{bcs:312}) is readily evaluated.  Writing 
\begin{eqnarray} 
\left(
\begin{array}{cc} p_{0} - \xi ({\bf p}) & -\bar{\Delta} \\ -\bar{\Delta}^* &
p_{0} + \xi ({\bf p}) \end{array}\right) = && \!\!\!\!\!\!\!\!\!\!\!
\left(
\begin{array}{cc} p_{0} - \xi ({\bf p}) & 0 \\ 0 &
p_{0} + \xi ({\bf p}) \end{array}\right) \nonumber \\ &&
\!\!\!\!\!\!\!\!  - \left(
\begin{array}{cc} 0 & \bar{\Delta} \\ \bar{\Delta}^* & 0 \end{array}\right),
\end{eqnarray} 
and expanding the second logarithm in a Taylor series, we recognize the
form 
\begin{eqnarray}  
S_{\rm eff}[\bar{\Delta}^*, \bar{\Delta}] = && \!\!\!\!\!\!\!\!\!\!\!  -i \,
{\rm Tr} \ln \left(
\begin{array}{cc} p_{0} - \xi ({\bf p}) & 0 \\ 0 &
p_{0} + \xi ({\bf p}) \end{array}\right) \nonumber \\ && 
\!\!\!\!\!\!\!\!\!\!\! - i \, {\rm Tr}
\ln \left(1 - \frac{|\bar{\Delta}|^2}{p_0^2 - \xi^2({\bf p})} \right),
\end{eqnarray}  
up to an irrelevant constant.  The integral over the loop energy $k_0$ gives
for the corresponding effective Lagrangian
\begin{equation} 
{\cal L}_{\rm eff} = \int_{\bf k} \left[ E({\bf k}) - \xi({\bf k})
\right].
\end{equation} 
To this one-loop result we have to add the tree term $|\bar{\Delta}|^2/\lambda_0$.
Expanding $E({\bf k})$ in a Taylor series, we see that the effective
Lagrangian also contains a term quadratic in $\bar{\Delta}$.  This term amounts
to a renormalization of the coupling constant; we find to this order for the
renormalized coupling constant $\lambda$:
\begin{equation} \label{bcs:reng}
\frac{1}{\lambda} = \frac{1}{\lambda_0} + \frac{1}{2} \int_{\bf k}
\frac{1}{|\xi({\bf k})|}.
\end{equation} 
The integral at the right-hand side diverges, to regularize it we introduce
a momentum cutoff $\Lambda$.  In this way we obtain
\begin{equation} \label{bcs:ren}
\frac{1}{\lambda} = \frac{1}{\lambda_0} + \frac{m}{2 \pi^2}
\Lambda,
\end{equation} 
where we omitted the (irrelevant) finite part of the integral.  It should be
remembered that the bare coupling constant $\lambda_0$ is negative, so that
there is an attractive interaction between the fermions.  We can distinguish
two limits.  One, if the bare coupling constant is taken to zero, $\lambda_0
\rightarrow 0^-$, which is the famous weak-coupling BCS limit.  Second, the
limit where the bare coupling is taken to minus infinity $\lambda_0
\rightarrow - \infty$.  This is the strong-coupling limit, where the
attractive interaction is such that the fermions form tightly bound pairs
\cite{Leggett}.  These composite bosons have a weak repulsive interaction
and can undergo a Bose-Einstein condensation (see succeeding section).

To this order there is no renormalization of the chemical potential, so that
we can write $\mu = \mu_0$.

Since there are two unknowns contained in the theory, $\bar{\Delta}$ and
$\mu$, we need a second equation to determine these variables in the
mean-field approximation.  To find the second equation we note that the
average fermion number $N$, which is obtained by differentiating the
effective action (\ref{bcs:312}) with respect to $\mu$
\begin{equation} 
N = \frac{\partial S_{\rm eff}}{\partial \mu},
\end{equation} 
is fixed.  If the system is spacetime independent, this reduces to
\begin{equation} \label{bcs:n}
\bar{n} = - i\, {\rm tr} \int_k \, G_0(k) \tau_3,
\end{equation} 
where $\bar{n}=N/V$, with $V$ the volume of the system, is the constant
fermion number density, $\tau_3$ is the diagonal Pauli matrix in Nambu space,
\begin{equation} 
\tau_3 = \left(
\begin{array}{cr} 1 & 0 \\ 0 & -1
\end{array} \right),
\end{equation} 
and $G_0(k)$ is the Feynman propagator,
\begin{eqnarray}    \label{bcs:prop}
G_0(k) &=&
\left( \begin{array}{cc} k_0 - \xi  ({\bf k}) 
& -\bar{\Delta} \\ -\bar{\Delta}^*  & k_0 + \xi ({\bf k}) 
\end{array} \right)^{-1}  \\ &=& 
\frac{1}{k_0^2 - E^2({\bf k}) + i  \eta } 
\left( \begin{array}{cc} k_{0} \, {\rm e}^{i k_0 \eta } + \xi
({\bf k})  & 
\bar{\Delta} \\ \bar{\Delta}^* & k_{0} \, {\rm e}^{-i k_0 \eta}- \xi
({\bf k}) \end{array} \right). \nonumber 
\end{eqnarray}
Here, $\eta$ is an infinitesimal positive constant that is to be set to zero
at the end of the calculation.  The exponential functions in the diagonal
elements of the propagator are an additional convergence factor needed in
nonrelativistic theories \cite{Mattuck}.  If the integral over the loop
energy $k_0$ in the particle number equation (\ref{bcs:n}) is carried out,
it takes the familiar form
\begin{equation} \label{bcs:ne} 
\bar{n} = \int_{\bf k} \left(1 - \frac{\xi({\bf k})}{E({\bf k})} \right)
\end{equation} 
The two equations (\ref{bcs:gape}) and (\ref{bcs:n}) determine $\bar{\Delta}$
and $\mu$.  They are usually evaluated in the weak-coupling BCS limit.
However, as was first pointed out by Leggett \cite{Leggett}, they can also
be easily solved in the strong-coupling limit (see succeeding section),
where the fermions are tightly bound in pairs.  More recently, also the
crossover between the weak-coupling BCS limit and the strong-coupling
composite boson limit has been studied in detail
\cite{Haussmann,DrZw,MRE,MPS}.

We continue by writing the order parameter $\Delta_{\rm mf}$ as
\begin{equation} \label{bcs:London} 
\Delta_{\rm mf}(x) = \bar{\Delta} \, {\rm e}^{2i \varphi (x)}, 
\end{equation}   
where $\bar{\Delta}$ is a spacetime-independent solution of the mean-field
equation (\ref{bcs:gap}).  This approximation, where the phase of the order
parameter is allowed to vary in spacetime while the modulus is kept fixed,
is called the London limit.  The phase field $\varphi(x)$, which is a
background field, physically represents the Goldstone mode of the
spontaneously broken U(1) symmetry. Our objective is to derive the effective
theory describing this Goldstone mode. 

To this end we decompose the Grassmann field as, cf.\  \cite{ESA}
\begin{equation}   \label{bcs:decompose}
\psi_\sigma(x) = {\rm e}^{i \varphi(x)} \chi_\sigma(x)
\end{equation}
and substitute the specific form (\ref{bcs:London}) of the order
parameter in the partition function (\ref{bcs:36}).  Instead of the
effective action (\ref{bcs:312}) we now obtain
\begin{equation} 
S_{\rm eff} = -i {\rm Tr} \ln \left( 
\begin{array}{cc} p_{0} - \partial_0 \varphi - \xi ({\bf p} +
\nabla \varphi) & -\bar{\Delta} \\
-\bar{\Delta}^* &  p_{0} + \partial_0 \varphi + \xi  ({\bf p} -
\nabla \varphi) 
\end{array} \right),
\end{equation}
where the derivative $\tilde{\partial}_\mu \varphi$ of the Goldstone field
plays the role of an Abelian background gauge field.  We next write this
effective action in the equivalent form
\begin{equation}  \label{bcs:Seff}
S_{\rm eff} = -i {\rm Tr} \ln \left[ G_0^{-1} \left(1- G_0 \Lambda
\right) \right] = i {\rm Tr} \sum_{\ell=1}^\infty \frac{1}{\ell} \left(G_0
\Lambda  \right)^\ell, 
\end{equation} 
where in the last step we ignored an irrelevant constant and the matrix
$\Lambda$ is given by
\begin{equation}  \label{bcs:matrix}
\Lambda ({\bf p}) = U  \tau_3 +
\frac{1}{m}  {\bf p} \cdot \nabla \varphi  + \frac{i}{2m} \nabla^2 \varphi,
\end{equation} 
with $U$ the Galilei-invariant combination
\begin{equation}  \label{bec:Phi}
U = \partial_0 \varphi + \frac{1}{2m} (\nabla \varphi)^2.
\end{equation} 
We shall consider only the first two terms in the series at the right-hand
side of (\ref{bcs:Seff}), and ignore for the moment higher than first
derivatives on the Goldstone field $\varphi$.  The first term yields
\begin{equation}  \label{bcs:Seff1}
S_{\rm eff}^{(1)} = i\, {\rm Tr} \, G_0(p) \tau_3 \, \left[ \partial_0
\varphi + \frac{1}{2m} (\nabla
\varphi)^2\right].
\end{equation} 
On account of (\ref{bcs:n}), this can be written as
\begin{equation} \label{bcs:seff1}
S_{\rm eff}^{(1)} = -  \int_x \, \bar{n} \left[
\partial_0 \varphi + \frac{1}{2m} (\nabla \varphi)^2\right] ,
\end{equation} 
where $\bar{n} = k_{\rm F}^3/3 \pi^2$ is the fermion number density.  We see that
the phase of the order parameter appears only in the Galilei-invariant
combination (\ref{bec:Phi}).

We continue with the second term in the series (\ref{bcs:Seff}).  We
shall first restrict ourselves to the weak-coupling BCS limit, where the
chemical potential is well approximated by the Fermi energy, $\mu
\approx k_{\rm F}^2/2m$.  In this limit, we can make the approximation
\begin{equation} 
\int_{\bf k}  \approx \nu(0) \int
\frac{\dd^2 \hat{k}}{4\pi}  \int_0^\infty \dd \epsilon \approx \nu(0) \int
\frac{\dd^2 \hat{k}}{4\pi}  \int_{-\infty}^\infty \dd \xi ,     
\end{equation} 
where $\nu(0) = m k_{\rm F} / 2 \pi^{2}$ is the density of states per
spin degree of freedom at the Fermi level and $\int \dd^2 \hat{k}$
denotes the integral over the solid angle.  In the last step, we
extended the range of the $\xi = \epsilon -\mu$ integration from $[-\mu,
\infty)$ to $(-\infty , \infty)$ which is justified by the rapid
convergence of the integrals involved.  After some algebra, using the
integrals
\begin{equation} 
\int_{k_0} \frac{1}{(k_0^2 - E^2 + 
i \eta )^l} = i \, (-1)^l \frac{\Gamma(l-\tfrac{1}{2})}{\Gamma(l)}
\frac{1}{E^{2l - 1}},         
\end{equation} 
\begin{equation} 
\int_\xi  \frac{1}{E^{2l +1}} = \sqrt{\pi} 
\frac{\Gamma(l)}{\Gamma(l + \tfrac{1}{2})} \frac{1}{|\bar{\Delta}|^{2l}},
\end{equation} 
with $l=1,2, \ldots$, and $\Gamma(l)$ the gamma function, we obtain
\begin{equation} \label{bcs:seff2}
S_{\rm eff}^{(2)} = \nu(0) \int_x \left[
\partial_0 \varphi + \frac{1}{2m} (\nabla \varphi)^2 \right]^2,
\end{equation}  
which is again invariant under Galilei transformations.

Equations (\ref{bcs:seff1}) and (\ref{bcs:seff2}) give the effective theory
of the Goldstone mode associated with the spontaneous breakdown of the
global U(1) symmetry \cite{bcs},
\begin{equation} \label{bcs:LGold}
{\cal L}_{\rm eff} = - \bar{n} \left[\partial_0 \varphi + \frac{1}{2m} (\nabla
\varphi)^2\right]  + \nu(0) \left[
\partial_0 \varphi + \frac{1}{2m} (\nabla \varphi)^2 \right]^2.
\end{equation} 
As expected, it is of the same form as the effective theory we obtained in
Eq. (\ref{hydro:Leff}) of the previous section describing a hydrodynamic
sound wave.  After rescaling $\varphi \rightarrow m \varphi$, the two forms
become identical.  The only difference being that the present Goldstone
field $\varphi$ is a compact field.  For the velocity $c$ of the
Anderson-Bogoliubov mode we find
\begin{equation} \label{bcs:velo}
c^2 = \tfrac{1}{3} v_{\rm F}^2,
\end{equation} 
where $v_{\rm F}= k_{\rm F}/m$ is the Fermi velocity.  The effective theory
(\ref{bcs:LGold}) has been rederived by various other authors
\cite{AATZ,Stone,LoSh}.

\begin{figure}
\begin{center}
\epsfxsize=8.cm
\mbox{\epsfbox{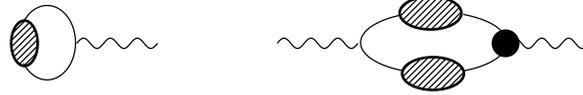}}
\end{center}
\caption{Graphical representation of the effective theory
(\protect\ref{bcs:LGold}). The symbols are explained in the
text. \label{fig:effective}}
\end{figure}
Graphically, the effective theory is represented by the diagrams depicted in
Fig.~\ref{fig:effective}, where a line with a shaded bubble inserted stands
for $i$ times the {\it full} Green function $G$ and the black bubble denotes
$i$ times the {\it full} interaction $\Gamma$ of the $\chi_\sigma$-fields,
introduced in (\ref{bcs:decompose}), with the background field $U$
which is denoted by a wiggly line.  Both $G$ and $\Gamma$ are $2 \times 2$
matrices.  The full interaction is obtained from the inverse Green function
by differentiation with respect to the chemical potential,
\begin{equation}  \label{bcs:defga}
\Gamma = - \frac{\partial G^{-1}}{\partial \mu}.		
\end{equation}
This follows because $U$, as defined in (\ref{bec:Phi}), appears in
the theory only in the combination $\mu - U$.  To lowest order, the
interaction is given by $-\tau_3$.  Employing Eq. (\ref{bcs:n}), with $G_0$
replaced by the full Green function, we conclude that the first diagram
indeed corresponds to the first part of the effective theory
(\ref{bcs:LGold}).  The second diagram without the wiggly lines denotes $i$
times the (0 0)-component of the {\it full} polarization tensor, $\Pi_{0
0}$, at zero energy transfer and low momentum ${\bf q}$,
\begin{equation} 
i \lim_{{\bf q} \rightarrow 0} \Pi_{0 0}(0,{\bf q}) =  \lim_{{\bf q}
\rightarrow 0} {\rm tr} \int_k \tau_3 G \,  \Gamma \, G \, (k_0,{\bf k}+
{\bf q}), 
\end{equation}
where the minus sign associated with the fermion loop is included.  To see
that this represents the second part of the effective theory we invoke an
argument due to Gavoret and Nozi\`eres \cite{GN}. By virtue of relation
(\ref{bcs:defga}) between the full Green function $G$ and the full
interaction $\Gamma$, the (0 0)-component of the polarization tensor can be
cast in the form
\begin{eqnarray}  \label{bcs:cruc}
\lim_{{\bf q} \rightarrow 0} \Pi_{0 0} (0,{\bf q}) &=& i  \lim_{{\bf q}
\rightarrow 0} {\rm tr} \int_k  \tau_3 G \, \frac{\partial G^{-1}}{\partial
\mu} \, G (k_0,{\bf k}+ {\bf q}) \nonumber \\ &=& - i \frac{\partial
}{\partial \mu} \lim_{{\bf q} 
\rightarrow 0} {\rm tr} \int_k \tau_3 G (k_0,{\bf k}+ {\bf q}) \nonumber \\
&=& \frac{\partial \bar{n}}{\partial \mu} = -\frac{1}{V} \frac{\partial^2
\Omega }{\partial \mu^2},
\end{eqnarray} 
where $\Omega$ is the thermodynamic potential and $V$ the volume of the
system.  The right-hand side of (\ref{bcs:cruc}) is $\bar{n}^2 \kappa$, with
$\kappa$ the compressibility.  Because it is related to the macroscopic sound
velocity $c$ via
\begin{equation}
\kappa = \frac{1}{m \bar{n} c^2},
\end{equation}
we conclude that the (0 0)-component of the full polarization tensor
satisfies the so-called compressibility sum rule of statistical
physics \cite{GN} 
\begin{equation}          \label{bec:rel}           
\lim_{{\bf q} \rightarrow 0} \Pi_{0 0} (0,{\bf q}) = - \frac{1}{V}
\frac{\partial ^2 \Omega }{\partial \mu^2} = \frac{\bar{n}}{m c^2}.         
\end{equation}
It now follows immediately that the second diagram in Fig.\
\ref{fig:effective} represents the second part of the effective theory
(\ref{bcs:LGold}).

Both the particle number density and the sound velocity can thus also be
obtained from the thermodynamic potential $\Omega$ by differentiating it
with respect to the chemical potential:
\begin{equation}  \label{bec:thermo}
\bar{n} = - \frac{1}{V} \frac{\partial \Omega }{\partial \mu}; \;\;\;\;\;\;
\frac{1}{c^2} = - \frac{1}{V} \frac{m}{\bar{n}} \frac{\partial^2 \Omega
}{\partial \mu^2}. 
\end{equation} 
In the weak-coupling BCS limit, the spectrum of the elementary fermionic
excitations does not differ significantly from the free-particle spectrum.
The thermodynamic potential may therefore be approximated by that of a free
fermion gas.  The partition function (\ref{bcs:34}) with ${\cal L}$ the free
Lagrangian (\ref{bcs:33}) reads
\begin{equation} \label{bcs:Z0} 
Z_{\rm free} = \exp\left( -i \int_x {\cal V}_{\rm free}\right),
\end{equation} 
where the spacetime integral over the potential ${\cal V}_{\rm free}$ is
given by minus the effective action (\ref{bcs:312}) with $\Delta$ set to
zero.  We used a potential rather than an action in (\ref{bcs:Z0}) because
the free fermion system is homogeneous in spacetime.  Carrying out the
integrals over the loop energy and momentum, we obtain for the potential
\begin{equation} 
{\cal V}_{\rm free} = -\frac{4 \sqrt{2}}{15 \pi^2} m^{3/2} \mu^{5/2}.
\end{equation}  
This physically represents the ground-state energy per unit volume.  The
thermodynamic potential (at the absolute zero of temperature) is
obtained by integrating ${\cal V}$ over space,
\begin{equation} 
\Omega = \int_{\bf x} {\cal V}.
\end{equation} 
With the help of (\ref{bec:thermo}), the velocity of the Anderson-Bogoliubov
mode can now be calculated; this yields again (\ref{bcs:velo}).

One can of course continue and calculate higher-order terms of the effective
theory.  Since these involve always higher-order derivatives, they are
irrelevant at low energy and small momentum.  To identify the expansion
parameter, let us compute the quadratic terms in the Goldstone fields,
involving fourth-order derivatives.  A somewhat tedious but
straightforward calculation yields in the weak-coupling BCS limit:
\begin{equation}   \label{bcs:L2pp}
{\cal L}_{\rm eff}'' = \frac{\nu(0)}{6 |\bar{\Delta}|^2}
\left[ (\partial_{0}^2 \varphi)^2 - \tfrac{2}{3} v_{\rm F}^2 (\partial_{0}^2
\varphi) (\nabla^2 \varphi) + \tfrac{1}{5} v_{\rm F}^4 (\nabla^2 \varphi)^2 
\right].
\end{equation} 
This equation shows us that the higher-order terms have either an extra
factor $\partial_0^2/|\bar{\Delta}|^2$ or $(v_{\rm F}
\nabla)^2/|\bar{\Delta}|^2$, where we recall that the BCS correlation
length $\xi_0$ at zero temperature is given by
\begin{equation} 
\xi_0 = \frac{v_{\rm F}}{\pi |\bar{\Delta}|}.
\end{equation}   
It sets the scale over which the modulus of the order parameter varies.  At
low energy and small momentum these additional terms can indeed be ignored.
In deriving (\ref{bcs:L2pp}) we integrated various times by parts and
neglected the ensuing total derivatives.  With the additional terms
included, the energy spectrum of the Anderson-Bogoliubov mode becomes
\begin{equation}              \label{bcs:spectrum}
E^2({\bf k}) = \tfrac{1}{3} v_{\rm F}^2 {\bf k}^2 \left( 1 - \tfrac{2}{45}
\pi^2 \xi_0^2{\bf k}^2 \right),
\end{equation} 
in accordance with Refs.\ \cite{KleinertFS,BrPo,Popov}.  The minus sign in
(\ref{bcs:spec}) shows the stability of this Goldstone mode against
decaying.

When one considers even higher energies and momenta, one also has to account
for variations in the modulus of the order parameter so that the London
limit, where the modulus is taken to be constant, is no longer applicable.
\section{Composite Boson Limit}
\label{sec:comp}
In this section we shall investigate the strong-coupling limit of the
pairing theory.  In this limit, the attractive interaction between the
fermions is such that they form tightly bound pairs of mass $2m$.  To
explicate this limit in arbitrary space dimension $d$, we swap the bare
coupling constant for a more convenient parameter---the binding energy
$\epsilon_a$ of a fermion pair in vacuum \cite{RDS}.  Both parameters
characterize the strength of the contact interaction.  To see the connection
between the two, let us consider the Schr\"odinger equation for the problem
at hand.  In reduced coordinates, it reads
\begin{equation} 
\left[- \frac{\nabla^2}{m} + \lambda_0 \, \delta({\bf x}) \right] \psi({\bf
x}) = - \epsilon_a,
\end{equation} 
where the reduced mass is $m/2$ and the delta-function potential, with
$\lambda_0 < 0$, represents the attractive contact interaction ${\cal
L}_{\rm i}$ in (\ref{bcs:BCS}).  We stress that this is a two-particle
problem in vacuum; it is not the famous Cooper problem of two interacting
fermions on top of a filled Fermi sea.  The equation is most easily solved
by Fourier transforming it.  This yields the bound-state equation
\begin{equation} 
\psi({\bf k}) = - \frac{\lambda_0}{{\bf k}^2/m + \epsilon_a} \psi(0),
\end{equation} 
or
\begin{equation} 
- \frac{1}{\lambda_0} = \int_{\bf k} \frac{1}{{\bf k}^2/m + \epsilon_a} .
\end{equation} 
This equation allows us to swap the coupling constant for the binding energy
$\epsilon_a$.  When substituted in the gap equation (\ref{bcs:gape}),  
the latter becomes
\begin{equation} \label{bcs:reggap}
\int_{\bf k} \frac{1}{{\bf k}^2/m + \epsilon_a} = \frac{1}{2}
\int_{\bf k} \frac{1}{E({\bf k})}.
\end{equation} 
By inspection, it is easily seen that this equation has a solution given
by \cite{Leggett}
\begin{equation} \label{comp:self}
\bar{\Delta} \rightarrow 0, \;\;\;\;\; \mu_0 \rightarrow - \tfrac{1}{2}
\epsilon_a,
\end{equation}   
where it should be noted that the chemical potential is negative here.  This
is the strong-coupling limit.  To appreciate the physical significance of
the specific value found for the chemical potential in this limit, we note
that the spectrum $E_{\rm b}({\bf q})$ of the two-fermion bound state
measured relative to the pair chemical potential $2\mu_0$ reads
\begin{equation} 
E_{\rm b}({\bf q}) = - \epsilon_a + \frac{{\bf q}^2}{4m} -2 \mu_0.
\end{equation} 
The negative value for $\mu_0$ found in (\ref{comp:self}) is precisely the
condition for a Bose-Einstein condensation of the composite bosons in the
${\bf q} = 0$ state.

To investigate this limit further, we consider the effective action
(\ref{bcs:312}) and expand $\Delta(x)$ around a constant value $\bar{\Delta}$
satisfying the gap equation (\ref{bcs:gape}),
\begin{equation} 
\Delta(x) = \bar{\Delta} + \tilde{\Delta}(x).
\end{equation} 
We obtain in this way,
\begin{equation} 
S_{\rm eff} = i \, {\rm Tr} \sum_{\ell =1}^\infty \frac{1}{\ell} \left[ G_0(p) 
\left( \begin{array}{cc} 0 & \tilde{\Delta} \\
\tilde{\Delta}^* & 0 \end{array} \right) \right]^\ell,
\end{equation} 
where $G_0$ is given in (\ref{bcs:prop}).  We are interested in terms
quadratic in $\tilde{\Delta}$.  Employing the derivative expansion
outlined on the pages \pageref{pag:derbegin}-\pageref{pag:derend}, we find
\begin{eqnarray}  \label{comp:Seff}
S_{\rm eff}^{(2)}(q) \!\!\!\! &=& \!\!\!\! \tfrac{1}{2}i \, {\rm Tr} \,
\frac{1}{p_0^2 - E^2({\bf p})} 
\frac{1}{(p_0 + q_0)^2 - E^2({\bf p} - {\bf q})}  \\ &&
\;\;\;\; \times 
\Bigr\{ \bar{\Delta}^2 \, \tilde{\Delta}^* \tilde{\Delta}^*  
+ [p_0 + \xi({\bf p})] [p_0 + q_0 - \xi({\bf p} - {\bf q})] \tilde{\Delta}
\tilde{\Delta}^* \nonumber \\ && \;\;\;\;\;\;\;\;\; +
\bar{\Delta}^{*^{\scriptstyle{2}}} 
\tilde{\Delta} \tilde{\Delta}  
+ [p_0 - \xi({\bf p})] [p_0 + q_0 + \xi({\bf p} - {\bf q})] \tilde{\Delta}^*
\tilde{\Delta} \Bigl\}, \nonumber 
\end{eqnarray} 
where $q_\mu = i\tilde{\partial}_\mu$.  It is to be recalled here that the
derivative $p_\mu$ operates on everything to its right, while
$\tilde{\partial}_\mu$ operates only on the first object to its right.  Let
us for a moment ignore the derivatives in this expression.  After carrying
out the integral over the loop energy $k_0$ and using the gap equation
(\ref{bcs:gape}), we then obtain
\begin{equation} \label{comp:Lag1}
{\cal L}^{(2)}(0) = -\frac{1}{8} \int_{\bf k} \frac{1}{E^3({\bf k})}
\left(\bar{\Delta}^2 \, \tilde{\Delta}^*{}^2 +
\bar{\Delta}^{*^{\scriptstyle{2}}}  \tilde{\Delta}^2 + 2 
|\bar{\Delta}|^2 |\tilde{\Delta}|^2 \right).
\end{equation} 
In the composite boson limit $\bar{\Delta} \rightarrow 0$, so that the spectrum
(\ref{bcs:spec}) of the elementary fermionic excitations can be
approximated by
\begin{equation} 
E({\bf k}) \approx  \epsilon({\bf k}) + \tfrac{1}{2} \epsilon_a.
\end{equation} 
The remaining integrals in (\ref{comp:Lag1}) become elementary in this
limit,
\begin{equation}
\int_{\bf k} \frac{1}{E^3({\bf k})} = \frac{4 \Gamma(3-d/2)}{(4 \pi)^{d/2}}
m^{d/2} \epsilon_a^{d/2-3}.
\end{equation} 

We next consider the terms involving derivatives in (\ref{comp:Seff}).
Following Ref.\ \cite{Haussmann} we set $\bar{\Delta}$ to zero here.  The
integral over the loop energy is easily carried out, with the result
\begin{eqnarray} 
{\cal L}^{(2)}(q) &=& - \frac{1}{2} \int_{\bf k}
\frac{1}{q_0 - {\bf k}^2/m + 
2 \mu_0 - {\bf q}^2/4m} \tilde{\Delta} \tilde{\Delta}^* \nonumber \\ && 
 - \frac{1}{2} \int_{\bf k} \frac{1}{-q_0 - {\bf k}^2/m +
2 \mu_0 - {\bf q}^2/4m} \tilde{\Delta}^* \tilde{\Delta}.
\end{eqnarray} 
The integral over the loop momentum ${\bf k}$ will be carried out using the
dimensional-regularized integral
\begin{equation} \label{dimreg}
\int_{\bf k} \frac{1}{({\bf k}^2 + M^2)} = \frac{\Gamma(1
-d/2)}{(4 \pi)^{d/2}} \frac{1}{\left(M^2\right)^{1-d/2}}
\end{equation} 
to suppress irrelevant ultraviolet divergences.  To illustrate the power of
dimensional regularization, let us consider the case $d=3$ in detail.
Introducing a momentum cutoff, we find in the large-$\Lambda$ limit
\begin{equation} 
\int_{\bf k} \frac{1}{({\bf k}^2 + M^2)} = \frac{\Lambda}{2 \pi^2} -
\frac{1}{4 \pi} M + {\cal O} \left(\frac{1}{\Lambda}\right).
\end{equation} 
From (\ref{dimreg}) however only the finite part emerges.  This exemplifies
that terms diverging with a strictly positive power of the momentum cutoff
are suppressed in dimensional regularization.  These contributions, which
come from the ultraviolet region, cannot physically be very relevant because
the simple BCS model (\ref{bcs:BCS}) stops being valid here and new theories
are required.  It is a virtue of dimensional regularization that these
irrelevant divergences are suppressed.  

We thus obtain  in the strong-coupling limit
\begin{eqnarray}    \label{comp:Lag2}
\lefteqn{\int_{\bf k} \frac{1}{q_0 - {\bf k}^2/m -\epsilon_a - {\bf q}^2/4m}
=} \\ &&- \frac{\Gamma(1-d/2)}{(4 \pi)^{d/2}} m^{d/2} (-q_0 +
\epsilon_a + {\bf q}^2/4m)^{d/2-1}, \nonumber 
\end{eqnarray} 
or expanded in derivatives
\begin{eqnarray} 
\lefteqn{\int_{\bf k} \frac{1}{q_0 - {\bf k}^2/m - \epsilon_a - {\bf
q}^2/4m} =} \\ && 
- \frac{ \Gamma(1-d/2)}{(4 \pi)^{d/2}} m^{d/2} \epsilon_a^{d/2-1} -
\frac{ \Gamma(2-d/2)}{(4 \pi)^{d/2}} m^{d/2}
\epsilon_a^{d/2-2} \left(q_0 - \frac{{\bf q}^2}{4m} \right).
\nonumber 
\end{eqnarray}  
The first term at the right-hand side yields as contribution to the
effective theory
\begin{equation} \label{bcs:con}
{\cal L}^{(2)}_\lambda = \frac{\Gamma(1-d/2)}{(4 \pi)^{d/2}} m^{d/2}
\epsilon_a^{d/2-1} |\tilde{\Delta}|^2.
\end{equation} 
To this we have to add the contribution $|\tilde{\Delta}|^2/\lambda_0$
coming from the tree potential, i.e., the last term in the partition
function (\ref{bcs:37}).  But this combination is no other than the one
needed to define the renormalized coupling constant via (\ref{bcs:reng}),
which in the strong-coupling limit reads using dimensional regularization
\begin{equation} 
\frac{1}{\lambda} = \frac{1}{\lambda_0} + 
\frac{\Gamma(1-d/2)}{(4 \pi)^{d/2}} m^{d/2} \epsilon_a^{d/2-1}.
\end{equation} 
In other words, the contribution (\ref{bcs:con}) can be combined with the
tree contribution to yield the term $|\tilde{\Delta}|^2/\lambda$.  Expanding
the square root in (\ref{comp:Lag2}) in powers of the derivative $q_\mu$
using the value (\ref{comp:self}) for the chemical potential, and pasting
the pieces together, we obtain for the terms quadratic in $\tilde{\Delta}$
\cite{Haussmann}, 
\begin{equation} 
{\cal L}^{(2)} = \frac{1}{2} \frac{\Gamma(2-d/2)}{(4 \pi)^{d/2}} m^{d/2}
\epsilon_a^{d/2-2}\, \tilde{\Psi}^\dagger \, 
M_0(q) \, \tilde{\Psi}, \;\;\;\;\;  \tilde{\Psi} = \left(\begin{array}{l}
\tilde{\Delta} \\ \tilde{\Delta}^* \end{array} \right),
\end{equation} 
where $M_0(q)$ is the $2 \times 2$ matrix,
\begin{eqnarray}    \label{comp:M} 
\lefteqn{M_0(q) =} \\ && \!\!\!\!\!\!\!\!
\left( \begin{array}{cc}
q_0 - {\bf q}^2/4m - (2-d/2) |\bar{\Delta}|^2/ \epsilon_a & 
\!\!\!\! - (2-d/2) \bar{\Delta}^2/ \epsilon_a   \\
- (2-d/2) \bar{\Delta}^{*^{\scriptstyle{2}}} / \epsilon_a
& \!\!\!\! -q_0 - {\bf q}^2/4m - (2-d/2) |\bar{\Delta}|^2/ \epsilon_a
\end{array} \right). \nonumber 
\end{eqnarray} 
As we will see shortly in Sec.~\ref{sec:bec}, this describes the superfluid
state of a weakly interacting composite boson system.  On comparing with
Eq.~(\ref{bec:M}) below, we conclude that the composite bosons have---as
expected---a mass $m_{\rm b}=2m$ twice the fermion mass $m$, and a small
chemical potential
\begin{equation} 
\mu_{0,{\rm b}} = (2-d/2) \frac{|\bar{\Delta}|^2}{\epsilon_a}.
\end{equation} 
From (\ref{comp:M}) one easily extracts the so-called Bogoliubov spectrum
and the velocity $c_0$ of the sound mode it describes (see
Sec.~\ref{sec:bec}),
\begin{equation} 
c_0^2 = \frac{\mu_{0,{\rm b}}}{m_{\rm b}} = (1-d/4)
\frac{|\bar{\Delta}|^2}{m \epsilon_a}. 
\end{equation} 
Also the number density $\bar{n}_{0,{\rm b}}$ of condensed composite bosons, 
\begin{equation} 
\bar{n}_{0,{\rm b}} = \frac{\Gamma(2-d/2)}{(4 \pi)^{d/2}} m^{d/2}
\epsilon_a^{d/2-2} |\bar{\Delta}|^2
\end{equation} 
as well as the weak repulsive interaction $\lambda_{0,{\rm b}}$ between the
composite bosons,
\begin{equation} \label{comp:lambda}
\lambda_{0,{\rm b}} = (4 \pi)^{d/2} \frac{1-d/4}{\Gamma(2-d/2)}
\frac{\epsilon_a^{1-d/2}}{m^{d/2}}
\end{equation}   
follow immediately.  We in this way have explicitly demonstrated that
the BCS theory in the composite boson limit maps onto the Bogoliubov
theory.

In concluding this section, we remark that in $d=2$ various integrals we
encountered become elementary for arbitrary values of $\bar{\Delta}$.  For
example, the gap equation (\ref{bcs:reggap}) reads explicitly in $d=2$
\begin{equation} 
\epsilon_a = \sqrt{\mu^2 + |\bar{\Delta}|^2} - \mu,
\end{equation} 
while the particle number equation (\ref{bcs:ne}) becomes
\begin{equation} 
\bar{n} = \frac{m}{2 \pi} \left(\sqrt{\mu^2 + |\bar{\Delta}|^2} + \mu \right).
\end{equation} 
Since in two dimensions, 
\begin{equation} 
\bar{n} = \frac{k_{\rm F}^2}{2 \pi} = \frac{m}{\pi} \epsilon_{\rm F},
\end{equation} 
with $k_{\rm F}$ and $\epsilon_{\rm F} = k_{\rm F}^2/2m$ the Fermi momentum
and energy, the two equations can be combined to yield \cite{RDS}
\begin{equation}
\frac{\epsilon_a}{\epsilon_{\rm F}} = 2 \frac{\sqrt{\mu^2 +
|\bar{\Delta}|^2} - \mu}{\sqrt{\mu^2 + |\bar{\Delta}|^2} + \mu}.
\end{equation}  
The composite boson limit we have been discussing in this section is
easily retrieved from these more general equations.  Also note that in
this limit, $\bar{n} = 2 \bar{n}_{0,{\rm b}}$.
\section{Superfluid $^3$He-a}
\label{sec:he3}
In this section we extend the previous analysis to a more complicated
system, namely that of superfluid $^3$He \cite{VoWo}.  Whereas the
Cooper-pairing in a BCS superconductor is in a spin-singlet state, the
pairing in $^3$He is in a triplet state.  This means that besides the global
U(1)$^N$ group generated by the particle number $N$ also the SO(3)$^S$ spin
rotation group has to be considered.  Moreover, since on account of the
exclusion principle the pairs must have an odd angular momentum, also the
space rotation group must be included.  In $^3$He, it is generally excepted
that the pairing is in the $L=1$ state.  The relevant symmetry group is now
much larger than that of a BCS system which leads to a much richer structure
of possible superfluid phases \cite{BrVo,ScBa}.  Since in most phases the
three symmetries become intertwined, superfluid $^3$He displays surprising
physical properties \cite{Volovik}.

We will consider a $^3$He film.  The specific superfluid state we will study
is the two-dimensional analog of the three-dimensional $^3$He-A phase, the
so-called $^3$He-a phase, first considered by Stein and Cross \cite{SC} (see
also Ref.~\cite{BrPo}). This state is characterized by the magnetic quantum
numbers $m_S=0, m_L = -1$, where the orbital magnetic quantum number $m_L$
may be considered as the projection of the orbital angular momentum on the
nonexisting $z$-axis.  The corresponding symmetry breakdown is
\begin{equation}             \label{he3:1}
{\rm SO}(3)^S \times {\rm SO}(2)^L \times {\rm U}(1)^{N} \supset
{\rm SO}(2)^S \times {\rm U}(1)^{L+N},         
\end{equation} 
where the generator of the residual symmetry group U(1)$^{L+N}$ is the sum
of the generators of the group SO(2)$^L$ of space rotations and the group
U(1)$^{N}$ of phase transformations.  We see that in this state, the spin
rotation group SO(3)$^S$ is spontaneously broken to SO(2)$^S$, as in
ferro- and antiferromagnets.  Since $m_S=0$, the superfluid $^3$He-a phase
is, in addition, an antiferromagnet.  We choose to represent the order
parameter by the matrix $A^\alpha_i$ with spin index $\alpha =1,2,3$
referring to the $x,\,y$ or $z$ component in spin space, and with orbital
index $i=1,2$ referring to the $x$ or $y$ component in real space.  (We will
always denote vector spin indices---such as $\alpha$---as superscript to
distinguish them from ordinary space indices.)  In this basis, the $^3$He-a
order parameter is given by \cite{CaSc}
\begin{equation}          \label{he3:2}  
A^\alpha_i = \bar{\Delta} \left(
\begin{array}{lr}
0  & 0     \\
0  & 0    \\
1  & -i     \\
\end{array}
\right) ,                                         
\end{equation} 
where $\bar{\Delta}$ is a constant whose physical significance will become clear
when we proceed.  Without loss of generality, we have chosen $\bar{\Delta}$ to
be real.

It follows from (\ref{he3:1}) that the coset space is three dimensional,
implying three Goldstone modes.  To make these degrees of freedom explicit, we
write the order parameter in the more general form
\begin{equation}   \label{he3:3} 
A^\alpha_i = \bar{\Delta} d^\alpha \,({\bf e}^1 - i {\bf e}^2)_i \, {\rm
e}^{2i\varphi}, 
\end{equation}                                             
where ${\bf e}^1$ and ${\bf e}^2$ are two orthonormal vectors in real space,
$d^\alpha$ is a unit vector in spin space, parameterizing the coset
SO(3)$^S$/SO(2)$^S \, \simeq \, {\rm S}^2$, and $2\varphi$ is the phase of
the order parameter.  The residual U(1)$^{L+N}$ symmetry is reflected in the
invariance of the order parameter (\ref{he3:3}) under the combined action of
a U(1) transformation $\varphi
\rightarrow \varphi +\tfrac{1}{2} \alpha$ and a rotation of the ${\bf
e}$-frame through $\alpha$ about the $z$ axis: $({\bf e}^1 - i {\bf e}^2)
\rightarrow ({\bf e}^1 - i {\bf e}^2)\exp({-i\alpha})$.  The transformation
parameter $\alpha$ may be taken to be spacetime dependent.  The specific form
(\ref{he3:2}) is recovered by choosing $d^\alpha$ in the $z$ direction of
spin space, and ${\bf e}^1$ and ${\bf e}^2$ in the $x$ and $y$ direction of
real space, respectively, and by taking $\varphi = 0$.

Superfluid $^3$He can be modeled by the Lagrangian \cite{KleinertFS}
\begin{equation}    \label{he3:4}
{\cal L} = \psi^{\ast}_{\sigma} [i \partial_0 - \xi(-i \nabla)]
\psi_{\sigma} -\tfrac{1}{2} \lambda_0 [\psi_{\sigma}^{\ast}(-i
\hat{\partial}_i)
\psi_{\tau}^{\ast}]\,[\psi_{\tau}(-i \hat{\partial}_i) \psi_{\sigma}], 
\end{equation} 
where the last term, with $\lambda_0$ a positive coupling constant, is an
attractive interaction term appropriate for P-wave pairing.  In
(\ref{he3:4}), $\psi_{\sigma}$ are Grassmann fields describing the $^3$He
atoms with spin $\sigma = \uparrow, \downarrow$, and $\xi(-i \nabla)$ is as
defined below (\ref{bcs:BCS}).  Finally, $\hat{\partial}_i = (\partial_i -
\stackrel{\leftarrow}{\partial}_i)/2k_{\rm F}$, with $i=1,2$.  We again
linearize the theory by introducing auxiliary fields $\Delta$ and $\Delta^*$
using the functional identity
\begin{eqnarray}       \label{he3:5}
\lefteqn{ 
\exp \left\{- i \frac{\lambda_0}{2} \int_x [\psi_{\sigma}^{\ast} 
(-i \hat{\partial}_i) \psi_{\tau}^{\ast}] \, [\psi_{\tau} (-i\hat{\partial}_i) 
\psi_{\sigma}]\right\}  =}   \nonumber \\   
& & \int \DD \Delta^* \DD \Delta  \exp \Biggl( -\frac{i}{2} \int_x  \Bigl\{
\Delta^{* \, i}_{\sigma \tau} \, [\psi_{\tau} (-i \hat{\partial}_i)
\psi_{\sigma}] + 
[\psi_{\sigma}^{\ast} (-i \hat{\partial}_i) \psi_{\tau}^{\ast}] \,
\Delta^{i}_{\tau
\sigma} \nonumber \\ && 
\;\;\;\;\;\;\;\;\;\;\;\;\;\;\;\;\;\;\;\;\;\;\;\;\;\;\;\;\;\;\;\;\;\;\;\;\;\;\;
- \frac{1}{\lambda_0} \Delta^{* \, i}_{\sigma \tau}
\Delta^{i}_{\tau  \sigma} \Bigr\} \Biggr).   
\end{eqnarray}  
The Euler-Lagrange equation for $\Delta$ shows that classically it
merely stands for a product of two Grassmann fields:
\begin{equation}       \label{he3:6}
\Delta^{i}_{\sigma \tau} = \lambda_0 \psi_{\sigma} (-i \hat{\partial}_i) 
\psi_{\tau}.                                            
\end{equation} 
As may be inferred from this equation, $\Delta$ is symmetric in its spin 
indices, i.e., it is a spin triplet.  Moreover, the index $i$ indicates 
that the field is a vector under spatial rotations.  The partition 
function $Z$ of the theory can now be represented by the functional 
integral:
\begin{eqnarray}     \label{he3:7}
Z = && \!\!\!\!\!\!\!\!\! \int \DD \Psi^{\dagger} \DD \Psi \! \int \DD 
\Delta^* \DD \Delta \,
\exp \left( \frac{i}{2 \lambda_0} \int_x  \Delta^{* \,
i}_{\sigma \tau} \Delta^{i}_{\tau  \sigma}  \right)
\\ && \!\!\!\!\!\!\!\!\! \times \exp \!
\left[\frac{i}{2} 
\int_x \Psi^{\dagger} \! \left( \begin{array}{cc} 
p_0 - \xi ({\bf p}) & -\tfrac{1}{2} \{p_i/k_{\rm F}, \Delta^{i}\}
g^{\dagger} \\ -\tfrac{1}{2} g \{p_i/k_{\rm F}, \Delta^{* \, i}\} & p_{0}
+ \xi ({\bf p}) \end{array} \right) \! \Psi \right] , \nonumber
\end{eqnarray} 
where $\{\, , \, \}$ denotes the anticommutator.  We recall that the
operator $p_\mu = i\tilde{\partial}_\mu$ acts on everything to its right.
The derivatives $\partial_i \Delta^i$ and $\partial_i \Delta^{* \, i}$
contained in (\ref{he3:7}) arise after a partial integration in the
corresponding terms in (\ref{he3:5}).  The tensor $g$ in (\ref{he3:7}) is
the metric spinor
\begin{equation}             \label{he3:8}
g = \left( \begin{array}{rr}
0   & 1     \\
-1  & 0              
\end{array} \right),                       
\end{equation} 
and $\Psi$ stands for the Nambu multiplet
\begin{equation}          \label{he3:9}
\Psi = \left( \begin{array}{c}
\psi \\ g \psi^{*}
\end{array} \right), \;\;\;\;\; \psi = \left( \begin{array}{c}
\psi_\uparrow \\ g \psi_\downarrow \end{array} \right).
\end{equation} 
The reason for introducing the spinor $g$ in this multiplet is that the two
symmetric matrices $\Delta^i$ $(i=1,2)$ may be expressed as a linear
combination of the symmetric matrices $\sigma^{\alpha}g$, with
$\sigma^{\alpha} \, (\alpha =1,2,3)$ the Pauli matrices:
\begin{equation}            \label{he3:10}
\Delta^{i}_{\sigma \tau} = A^\alpha_i (\sigma^{\alpha} g)_{\sigma \tau},
\end{equation} 
where $A^\alpha_i$ are the expansion coefficients, so that
\begin{equation}         \label{he3:11}
\Delta^i g^{\dagger} = A^\alpha_i \sigma^{\alpha}.       
\end{equation} 
In this way, the metric spinor $g$ disappears from the argument of the
exponential function in (\ref{he3:7}).  This will simplify the calculation.
With the value (\ref{he3:3}) for the matrix $A^\alpha_i$, the Lagrangian
appearing in the partition function (\ref{he3:7}) can be cast in the concise
form
\begin{equation}  \label{he3:consice}
{\cal L} = \tfrac{1}{2} \Psi^\dagger \left(p_{0} - \xi ({\bf p})\tau_3 -
\Lambda({\bf p}) -\frac{\bar{\Delta}}{2k_{\rm F}} \tau_i \{p_i, d^\alpha\}
\sigma^{\alpha}\right) \Psi,
\end{equation} 
where $\tau_i$ $(i=1,2)$ are Pauli matrices in Nambu space which should not
be confused with the Pauli matrices $\sigma^\alpha$ in spin space.  The
vertex $\Lambda({\bf p})$ is the same as the one we encountered before in
the context of the BCS theory [see Eq.\ (\ref{bcs:matrix})].  In deriving
(\ref{he3:consice}) we have given the ${\bf e}$-frame the standard
orientation.

We proceed in the same way as in the previous section, and decompose the
Grassmann fields as in Eq.~(\ref{bcs:decompose}).  The partition function
(\ref{he3:7}) then becomes after carrying out the Gaussian integration over
the fermionic degrees of freedom
\begin{equation}      \label{he3:12} 
Z = \int \DD \Delta^* \DD \Delta \, \exp \left(i S_{\rm eff} [
\Delta^*, \Delta] + \frac{i}{2 \lambda_0} 
\Delta^{* \, i}_{\sigma \tau} \Delta^{i}_{\tau \sigma} \right),
\end{equation} 
with $S_{\rm eff}$ the effective action,  
\begin{equation}       \label{he3:13}
S_{\rm eff} = -\tfrac{1}{2}i \, {\rm Tr} \log
\left(p_{0} - \xi ({\bf p})\tau_3 - \Lambda({\bf p})
-\frac{\bar{\Delta}}{2k_{\rm F}} \tau_i \{p_i, d^\alpha\}
\sigma^{\alpha}\right).
\end{equation} 
A general orientation of the ${\bf e}$-frame, which may vary in spacetime,
is accounted for by requiring the action to be invariant under the residual
U(1)$^{L+N}$ symmetry.  This is achieved by the replacement
\begin{equation}        \label{he3:35}
\tilde{\partial}_{\mu} \varphi \rightarrow \tilde{\partial}_{\mu} \varphi -
{\bf e}^1 \cdot  \tilde{\partial}_{\mu} {\bf e}^2 .                      
\end{equation} 
Since in this section we are not interested in the Goldstone modes of the
spontaneously broken spin rotation group, we keep the spin vector $d^\alpha$
fixed in the $z$ direction.

We continue as in the previous section and expand the effective action in a
Taylor series:
\begin{equation}  \label{he3:Seff}
S_{\rm eff} = \tfrac{1}{2}i {\rm Tr} \sum_{\ell=1}^\infty
\frac{1}{\ell}\left[G_0(p) \Lambda({\bf p}) \right]^\ell,
\end{equation} 
where the Feynman propagator now reads
\begin{eqnarray}          \label{he3:20}
G_0(k) &=& \left[k_0 - \xi({\bf k}) \tau ^3 -
\frac{\bar{\Delta}}{k_{\rm F}} {\bf k} \cdot \bbox{\tau} \sigma^3 \right]^{-1}
\\ &=&
\frac{1}{k_0^2 - E^2({\bf k}) + i \eta} \left[k_0 {\rm e}^{i k_0 \eta
\tau_3} + \xi({\bf k})
\tau_3 + \frac{\bar{\Delta}}{k_{\rm F}} {\bf k}
\cdot \bbox{\tau} \sigma^3 \right].  \nonumber  
\end{eqnarray} 
Here, $E({\bf k})$ is the spectrum of the elementary fermionic excitations
\begin{equation}   \label{he3:21}
E^2({\bf k}) = \xi^2({\bf k}) + \left(\frac{\bar{\Delta}}{k_{\rm F}}\right)^2.
\end{equation}
The constant $\bar{\Delta}$ introduced in (\ref{he3:3}) is---apart from a factor
$k_{\rm F}$---seen to be the energy gap of these excitations at the Fermi
circle.

The expression (\ref{he3:Seff}) can be evaluated in the weak-coupling
limit along the lines of Sec.\ \ref{sec:bcs} to obtain the effective
theory of the Abelian Goldstone mode \cite{he-a}.  We arrive in this way
again at the theory (\ref{bcs:LGold}) we calculated from the BCS theory
with $\nu(0) = m/2 \pi$ now the two-dimensional density of states per
spin degree of freedom and $\bar{n}=k_{\rm F}^2/2 \pi$ the two-dimensional
fermion number density.  It describes a sound wave traveling with the
speed $v_{\rm F}/\sqrt{2}$.  It is remarkable that to this order the
additional contributions arising from the derivatives in the interaction
term of the Lagrangian (\ref{he3:4}) cancel so that the same effective
theory is obtained as in the previous section where these derivatives
were absent.

In the next chapter we shall consider the effective theory describing the
antiferromagnetic spin waves associated with the spontaneous breakdown of the
spin rotation group SO(3)$^S \supset $SO(2)$^S$ in this particular superfluid
$^3$He phase.  Since this symmetry is an internal symmetry unrelated to Galilei
invariance, there is no reason to expect a theory that is invariant under
Galilei transformations.
\section{Weakly Interacting Bose Gas}
\label{sec:bec}
The model commonly used to describe a weakly interacting Bose gas (for a
general introduction, see the textbooks \cite{FW,AGD,Huang}), is defined
by the Lagrangian \cite{GrPi}
\begin{equation} \label{bec:Lagr}
{\cal L} = \phi^* \bigl[i \partial_0 - \epsilon(-i \nabla) + \mu_0\bigr] \phi
- \lambda_0 |\phi|^4,
\end{equation} 
where $\epsilon(-i \nabla) = - \nabla^2/2m$ and $\mu_0$ is the chemical
potential.  The self-coupling is taken to be positive, $\lambda_0 > 0$, so
that the contact interaction is repulsive.  We will treat this model
perturbatively in a loop expansion.  For this to be applicable the
(renormalized) coupling constant must be small.  Superfluid $^4$He is a
strongly interacting system where this premise does not hold.  Fortunately,
after almost two decades of experimental efforts by various groups in the US
as well as in Europe, Bose-Einstein condensation (BEC) has recently been
found in weakly interacting Bose gases.  A group in Boulder, Colorado were
the first to produce such a BEC in a cloud of $^{87}$Rb atoms trapped in a
magnetic field
\cite{Boulder}.  The condensate formed at a temperature around 170 nK and
was comprised of up to 2000 atoms; it was stable for about 15 s.  A second
unequivocal observation of BEC was subsequently established in a system of
Na atoms, with a condensate containing two orders of magnitude more
particles than in the Colorado experiments \cite{MITBEC}.  Various other
systems of bosonic alkali atoms are presently under study.

\begin{figure}
\begin{center}
\epsfxsize=8.cm
\mbox{\epsfbox{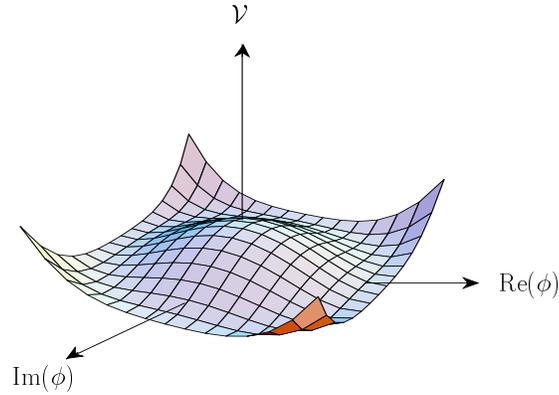}}
\end{center}
\caption{Graphical representation of the potential energy 
(\protect\ref{bec:V}). \label{fig:potential}}
\end{figure}

At zero temperature, the global U(1) symmetry is spontaneously broken by a
nontrivial ground state.  This can be easily seen by considering the shape
of the potential energy
\begin{equation} \label{bec:V}
{\cal V} = - \mu_0 |\phi|^2 + \lambda_0 |\phi|^4,
\end{equation} 
depicted in Fig.~\ref{fig:potential} which  is seen to have a minimum away
from the origin $\phi = 0$.
To account for this, we shift $\phi$ by a (complex) constant $\bar{\phi}$ and
write
\begin{equation}  \label{bec:newfields}
\phi(x) = {\rm e}^{i \varphi(x)} \, [\bar{\phi} + \tilde{\phi}(x)].
\end{equation}
The scalar field $\varphi(x)$ is a background field representing the
Goldstone mode of the spontaneously broken global U(1) symmetry.  At
zero temperature, the constant value
\begin{equation}  \label{bec:min}
|\bar{\phi}|^2 = \frac{1}{2} \frac{\mu_0}{\lambda_0 }                     
\end{equation}
minimizes the potential energy.  It physically represents the number density
of particles contained in the condensate for the total particle number
density is given by
\begin{equation} 
n(x)  = |\phi(x)|^2.
\end{equation} 
Because $\bar{\phi}$ is a constant, the condensate is a uniform, zero-momentum
state.  That is, the particles residing in the ground state are in the ${\bf
k}=0$ mode.  In terms of the new variables, the quadratic terms of the
Lagrangian (\ref{bec:Lagr}) may be cast in the matrix form
\begin{equation}  \label{bec:L0}
{\cal L}_0 = \tfrac{1}{2} \tilde{\Phi}^{\dagger} M_0(p)
\tilde{\Phi}, \;\;\;\;\;\;  \tilde{\Phi} = \left(\begin{array}{l} \tilde{\phi} \\
\tilde{\phi}^* \end{array} \right),
\end{equation}
with
\begin{eqnarray}  \label{bec:M} 
\lefteqn{M_0(p) =} \\ \nonumber && \!\!\!\!\!\!\!
\left( \begin{array}{cc}
p_0 - \epsilon({\bf p}) + \mu_0 - U - 4 \lambda_0 |\bar{\phi}|^2 & 
\!\!\!\! - 2 \lambda_0 \bar{\phi}^2  \\
- 2 \lambda_0 \bar{\phi}^{*^{\scriptstyle{2}}}     & \!\!\!\! -p_0 -
\epsilon ({\bf p})  
+ \mu_0 - U - 4 \lambda_0 |\bar{\phi}|^2
\end{array} \right),
\end{eqnarray} 
where $U$ is the Galilei-invariant combination (\ref{bec:Phi}).  In
writing this we have omitted a term $\nabla^2 \varphi$ containing two
derivatives which is irrelevant at low momentum and also a term of the form
$\nabla \varphi \cdot {\bf j}$, where ${\bf j}$ is the Noether current
associated with the global U(1) symmetry,
\begin{equation} 
{\bf j} =  \frac{1}{2 i m}  \phi^*
\stackrel{\leftrightarrow}{\nabla} \phi. 
\end{equation}  
This term, which after a partial integration becomes $- \varphi \nabla
\cdot {\bf j}$, is also irrelevant at low energy and small momentum because in a
first approximation the particle number density is constant, so that the
classical current satisfies the condition
\begin{equation} 
\nabla \cdot {\bf j} =0.
\end{equation}
The Feynman propagator of the theory is easily extracted by inverting
the matrix $M_0$ with the background field $U$ set to zero.
This yields upon using the mean-field value (\ref{bec:min}) for $\bar{\phi}$
\begin{eqnarray}  \label{bec:prop}
\lefteqn{G_0(k) = \frac{1}{k_0^2 - E^2({\bf k}) + i \eta }} \\ &&
\times  \left(
\begin{array}{cc} 
k_0 \, {\rm e}^{i k_0 \eta } + \epsilon({\bf k}) + 2 \lambda_0
|\bar{\phi}|^2 & - 2 \lambda_0 \bar{\phi}^2 \\ - 2 \lambda_0
\bar{\phi}^{*^{\scriptstyle{2}}}  
& \;\; - k_0 \, {\rm e}^{-i k_0 \eta } + 
\epsilon({\bf k}) + 2 \lambda_0  |\bar{\phi}|^2
\end{array} \right) \nonumber ,  					
\end{eqnarray} 
with $\eta $ a small positive constant that is to be set to zero after
the $k_0$ integration has been performed, and $E({\bf k})$ the
single-particle Bogoliubov spectrum \cite{Bogoliubov},
\begin{eqnarray}  \label{bec:bogo}
E({\bf k}) &=& \sqrt{ \epsilon ^2({\bf k}) + 2 \mu_0 \epsilon({\bf k}) }
\nonumber \\ &=& \sqrt{ \epsilon ^2({\bf k}) + 4 \lambda_0 |\bar{\phi}|^2
\epsilon({\bf k}) }. 
\end{eqnarray} 
The most notable feature of this spectrum is that it is gapless,
behaving for small momentum as
\begin{equation} \label{bec:micror}
E({\bf k}) \sim u_0 \, |{\bf k}|, 				
\end{equation} 
with $u_0 = \sqrt{\mu_0/m}$ a velocity which is sometimes referred to as the
microscopic sound velocity.  It was first shown by Beliaev \cite{Beliaev}
that the gaplessness of the single-particle spectrum persists at the
one-loop order.  This was subsequently proven to hold to all orders in
perturbation theory by Hugenholtz and Pines \cite{HP}.  For large momentum,
the Bogoliubov spectrum takes a form
\begin{equation} \label{bec:med}
E({\bf k}) \sim \epsilon({\bf k}) + 2 \lambda_0 |\bar{\phi}|^2
\end{equation} 
typical for a nonrelativistic particle with mass $m$ moving in a medium.
To highlight the condensate we have chosen here the second form in
(\ref{bec:bogo}) where $\mu_0$ is replaced with $2 \lambda_0 |\bar{\phi}|^2$.

Since gapless modes in general require a justification for there existence,
we expect the gaplessness of the single-particle spectrum to be a result of
Goldstone's theorem.  This is corroborated by the relativistic version of
the theory.  There, one finds two spectra, one corresponding to a massive
Higgs particle which in the nonrelativistic limit becomes too heavy and
decouples from the theory, and one corresponding to the Goldstone mode of
the spontaneously broken global U(1) symmetry \cite{BBD}.  The latter
reduces in the nonrelativistic limit to the Bogoliubov spectrum.  Also, when
the theory is coupled to an electromagnetic field, one finds that the
single-particle spectrum acquires an energy gap.  This is what one expects
to happen with the spectrum of a Goldstone mode when the Higgs mechanism is
operating.  The equivalence of the single-particle excitation and the
collective density fluctuation has been proven to all orders in perturbation
by Gavoret and Nozi\`eres \cite{GN}.

Given this observation we immediately infer that at low energy and small
momentum the superfluid phase is described by the theory we encountered
before
\begin{equation}  \label{bec:Lef}  
{\cal L}_{\rm eff} = - \bar{n} \left[ \partial _0 \varphi + \frac{1}{2 m}
(\nabla \varphi )^2 \right] + \frac{\bar{n}}{2 m c^2} \left[ \partial_0
\varphi + \frac{1}{2 m} (\nabla \varphi )^2 \right]^2,
\end{equation}
of a nonrelativistic sound wave, with the real scalar field $\varphi$
representing the Goldstone mode of the spontaneously broken global U(1)
symmetry.  The alert reader might be worrying about an apparent mismatch in
the number of degrees of freedom in the normal and the superfluid phase.
Whereas the normal phase is described by a complex $\phi$-field, the
superfluid phase is described by a real scalar field $\varphi$.  The
resolution of this paradox lies in the spectrum of the modes
\cite{Leutwyler}.  In the normal phase, the spectrum $E({\bf k}) = {\bf
k}^2/2m$ is linear in the energy, so that only positive energies appear
in the Fourier decomposition, and one needs---as is well know from
standard quantum mechanics---a complex field to describe a single
particle.  In the superfluid phase, where the spectrum is $E^2({\bf k}) = c^2
{\bf k}^2$, the counting goes differently.  The Fourier decomposition
now contains positive as well as negative energies and a single real
field suffice to describe this mode.  In other words, although the
number of fields is different, the number of degrees of freedom is the
same in both phases.

The effective theory is graphically again represented by Fig.\
\ref{fig:effective}.  To lowest order, the inverse propagator is given by
the matrix $M_0$ in (\ref{bec:M}), so that the vertex describing the
interaction between the $U$ and the $\tilde{\Phi}$-fields is minus the
unit matrix.  In terms of the full Green function $G$, the particle number
density now reads
\begin{equation}
\bar{n} = \frac{i}{2} \, {\rm tr} \int_k G (k).
\end{equation}
The (0 0)-component of the {\it full} polarization tensor, $\Pi_{0 0}$, at
zero energy transfer and low momentum ${\bf q}$,
\begin{equation} \label{bec:Pi}
i \lim_{{\bf q} \rightarrow 0} \Pi_{0 0}(0,{\bf q}) = -\frac{1}{2}
\lim_{{\bf q} \rightarrow 0} {\rm tr} \int_k G \, \Gamma \, G \, (k_0,{\bf
k}+ {\bf q})
\end{equation} 
contains a symmetry factor $\tfrac{1}{2}$ which is absent in fermionic
theories.  Following the same steps as in the case of superconductors, we
again arrive at the compressibility sum rule (\ref{bec:rel}).  

The diagrams of Fig.~\ref{fig:effective} can also be evaluated in a loop
expansion (thereby integrating out the complex scalar field
$\tilde{\phi}$) to obtain explicit expressions for the particle number
density $\bar{n}$ and the sound velocity $c$ to any given order
\cite{effbos}.  In doing so, one encounters---apart from ultraviolet
divergences which can be renormalized away---also infrared divergences
because the Bogoliubov spectrum is gapless.  When however all on-loop
contributions are added together, these divergences are seen to cancel
\cite{effbos}.  We will in this report not proceed in this way, but instead
compute the thermodynamic potential $\Omega$ from which both $\bar{n}$ and
$c$ are obtained by differentiating with respect to the chemical potential
[see (\ref{bec:thermo})].

In the approximation (\ref{bec:L0}) of ignoring higher than second order in
the fields, the integration over the $\tilde{\Phi}$ field is Gaussian.
Carrying out this integral, we obtain for the partition function
\begin{equation}  \label{bec:Z}
Z = {\rm e}^{-i \int_x {\cal V}_0} \int \DD \tilde{\Phi} \exp \left(i \int_x
{\cal L}_0 \right) = {\rm e}^{- i \int_x {\cal V}_0} \, {\rm Det}^{-1/2}
[M_0(p)],
\end{equation} 
where ${\cal V}_0$ is the potential (\ref{bec:V}) with $\phi$ replaced by
$\bar{\phi}$,
\begin{equation}
{\cal V}_0 = - \frac{1}{4} \frac{\mu_0^2}{\lambda_0} .
\end{equation} 
Writing
\begin{equation} 
Z = \exp\left[-i \int_x ({\cal V}_0 +{\cal V}_{\rm eff}) \right],
\end{equation} 
we find from (\ref{bec:Z}) that the effective potential is given to this
order by
\begin{equation} \label{bec:Veff}
{\cal V}_{\rm eff} = - \frac{i}{2} {\rm tr} \int_k \ln [M_0(k)].
\end{equation} 
The use of a potential rather than an action here is to indicate that we are
working with a spacetime-independent condensate, so that also $U=0$.
The easiest way to evaluate the loop integral over $k_\mu$ is to
assume---without loss of generality---that $\bar{\phi}$ is real and to first
differentiate the expression with respect to $\mu_0$
\begin{eqnarray} 
\lefteqn{\!\!\!\!\!\!\!\!\!\!\!\!\!\!\!\!\!\!\!\!\!\!\!
\frac{\partial}{\partial \mu_0}  {\rm tr}
\, \int_k \ln \left( \begin{array}{cc}
k_0 - \epsilon({\bf k}) - \mu_0 & 
- \mu_0 \\ - \mu_0 & -k_0 - \epsilon ({\bf k}) - \mu_0
\end{array} 
\right) =} \nonumber  \\ &&
\;\;\;\;\;\;\;\;\;\;\;\;\;\;\;\;\;\;\;\;\;\;\;\;\;\;\;\;\;\; 
-2 \, \int_k 
\frac{\epsilon({\bf k})}{k_0^2 - E^2({\bf k}) + i \eta },
\end{eqnarray} 
where $E({\bf k})$ is the Bogoliubov spectrum (\ref{bec:bogo}).  The integral
over $k_0$ can be carried out with the help of a contour integration,
yielding
\begin{equation}
\int_k \frac{\epsilon({\bf k})}{k_0^2 - E^2({\bf k}) + i \eta } =
- \frac{i}{2} \, \int_{\bf k} \frac{\epsilon({\bf k})}{E({\bf k})}.
\end{equation} 
This in turn is easily integrated with respect to $\mu_0$.  Putting the
pieces together, we obtain for the effective potential
\begin{equation} \label{bec:formalV}
{\cal V}_{\rm eff} = \frac{1}{2} \, \int_{\bf k} E ({\bf k}). 
\end{equation} 
The remaining integral over the loop momentum ${\bf k}$ is diverging in the
ultraviolet.  We regularize it by introducing a momentum cutoff $\Lambda$.
The model (\ref{bec:Lagr}) is only valid far below the cutoff.  Above it, the
model breaks down and new physics starts.  In the large-$\Lambda$ limit, we
arrive at
\begin{equation} \label{bec:Vnon}
{\cal V}_{\rm eff} =  \frac{1}{12
\pi^2} \Lambda^3 \mu_0 - \frac{1}{4 \pi^2} \Lambda m \mu_0^2 + \frac{8}{15
\pi^2} m^{3/2} \mu_0^{5/2}, 
\end{equation}
ignoring an irrelevant term ($\propto \Lambda^5$) independent of the
chemical potential.  Equation (\ref{bec:Vnon}) contains two
ultraviolet-diverging terms.  It should be realized that they arise from a
region where the model (\ref{bec:Lagr}) is not applicable.  They cannot
therefore be of deep significance.  As a consequence of the uncertainty
principle, stating that large momenta correspond to small distances, terms
arising from the ultraviolet region are always local and can be absorbed by
redefining the parameters appearing in the Lagrangian
\cite{Donoghue}.  Since $\mu_0 = 2 \lambda_0 |\bar{\phi}|^2$, the two
ultraviolet-diverging terms in (\ref{bec:Vnon}) can be absorbed by
introducing the renormalized parameters
\begin{eqnarray} 
\mu &=& \mu_0 - \frac{1}{6\pi^2} \lambda_0 \Lambda^3 \label{bec:renmu} \\
\lambda &=& \lambda_0 - \frac{1}{\pi^2} m \lambda_0^2
\Lambda. \label{bec:renla} 
\end{eqnarray} 
Because the diverging terms are---at least to this order---of a form already
present in the original Lagrangian, the theory is called ``renormalizable''.
The renormalized parameters are the physical ones that are to be identified
with those measured in experiment.  In this way, we see that the
contributions to the loop integral stemming from the ultraviolet region are
of no importance.  What remains is the finite part
\begin{equation} \label{bec:finite}
{\cal V}_{\rm eff} = \frac{8}{15 \pi^2} m^{3/2} \mu^{5/2}.
\end{equation}

This result could have been obtained directly without renormalization if we,
instead of introducing a momentum cutoff to regularize the integrals, had
employed analytic regularization.  In such a regularization scheme, where
for example the integrals are analytically continued to arbitrary real
values of the space dimension $d$, divergences proportional to powers of the
cutoff never show up as was demonstrated in Sec.\ \ref{sec:comp}.  Only
logarithmic divergences appear as $1/\epsilon$ poles, where $\epsilon$ tends
to zero when the parameter in which the analytic continuation is carried out
is given its physical value.  These logarithmic divergences
$\ln(\Lambda/E)$, with $E$ an energy scale, are relevant also in the
infrared because for fixed cutoff $\ln(\Lambda/E)
\rightarrow -\infty$ when $E$ is taken to zero.

In so-called ``nonrenormalizable'' theories, the terms which are diverging
in the ultraviolet are still local but not of a form present in the original
Lagrangian.  Whereas in former days such theories were rejected
because there supposed lack of predictive power, the modern view is that
there are no fundamental theories and that there is no basic difference
between renormalizable and nonrenormalizable theories \cite{CaoSc}.  Even a
renormalizable theory like (\ref{bec:Lagr}) should be extended to include
all higher-order terms such as a $|\phi|^6$-term which are allowed by
symmetry.  These additional terms render the theory ``nonrenormalizable''.
This does not however change the predictive power of the theory.  The point
is that when describing the physics at an energy scale $E$ far below the
cutoff, the higher-order terms are suppressed by powers of $E/\Lambda$, as
follows from dimensional analysis.  Therefore, far below the cutoff, the
nonrenormalizable terms are negligible.

The thermodynamic potential $\Omega$ becomes to the order in which we are
working
\begin{equation} \label{bec:omega-V}
\Omega = \int_{\bf x} \left( {\cal V}_0 + {\cal V}_{\rm eff} \right).
\end{equation}
We are now in a position to determine the particle number density and the
sound velocity using (\ref{bec:thermo}).  We find
\begin{equation} \label{bec:n}
\bar{n} =  \frac{1}{2} \frac{\mu}{\lambda} \left(1 - \frac{8\lambda
}{3 \pi^2} m^{3/2} \mu^{1/2} \right)
\end{equation} 
and
\begin{equation} \label{bec:c}
c^2 =  \frac{\mu}{m} \left(1 + \frac{4 \lambda}{3 \pi^2} m^{3/2}
\mu^{1/2} \right),
\end{equation} 
where in the last equation we made an expansion in the (renormalized)
coupling constant $\lambda$ which was assumed to be small.  These equations
reveal that the expansion is more precisely one in terms of the
dimensionless parameter $\lambda m^{3/2} \mu^{1/2}$, or reintroducing
Planck's constant $\lambda m^{3/2} \mu^{1/2}/\hbar^3$.  

Up to this point, we have considered the chemical potential to be the
independent parameter, thereby assuming the presence of a reservoir that can
freely exchange particles with the system under study.  It can thus contain
any number of particles, only the average number is fixed by external
conditions.  From the experimental point of view it is, however, often more
realistic to consider the particle number fixed.  If this is the case, the
particle number density $\bar{n}$ should be considered as independent
variable and the chemical potential should be expressed in terms of it. This
can be achieved by inverting relation (\ref{bec:n}):
\begin{eqnarray} \label{bec:na}
\mu &=& 2 \bar{n} \lambda \left[1 + \frac{8}{3 \pi^2} (2 \bar{n} m^3
\lambda^3)^{1/2} \right] \nonumber \\ &=& \frac{4 \pi \bar{n} a}{m} \left[1 +
\frac{32}{3} \left(\frac{\bar{n} a^3}{\pi}\right)^{1/2} \right],
\end{eqnarray} 
where in the last step we employed the relation between the (renormalized)
coupling constant $\lambda$ and the S-wave scattering length $a$
\cite{Hugenholtz,AGD},
\begin{equation} \label{bec:a}
\lambda = \frac{2 \pi a}{m}.
\end{equation} 
This relation follows from comparing the differential cross sections for
scattering of a slowly moving boson from a hard-core sphere of radius $a$
and from a delta function potential with strength $\lambda$ in the Born
approximation.  For the sound velocity (\ref{bec:c}) expressed in terms of
the particle number density we find
\begin{eqnarray} \label{bec:ca}
c^2 &=&  \frac{2 \bar{n} \lambda}{m} \left[1 + \frac{4}{\pi^2}
\left(2 \bar{n} m^3 \lambda^3 \right)^{1/2}  \right] 
\nonumber \\ &=& \frac{4 \pi \bar{n} a}{m^2} \left[1 + 16
\left(\frac{\bar{n} a^3}{\pi}\right)^{1/2} \right].
\end{eqnarray}
It is important to note that $c^2$ is linear in the coupling constant.
Without the interparticle interaction characterized by $\lambda$, the sound
velocity would be zero.  Moreover, the interaction must be repulsive for the
system to support sound waves.  Let us in this connection mention that a
third experimental group has reported evidence of BEC in a system of $^7$Li
atoms \cite{Rice}.  This would be a surprising result as $^7$Li atoms have---in
contrast to $^{87}$Rb and Na atoms---an attractive interaction and thus a
negative scattering length.

The effective theory (\ref{bec:Lef}) can also be put in a equivalent form
\begin{equation} \label{eff:quick}
{\cal L}_{\rm eff} = -  \bar{n} U(x) + \frac{1}{4} U(x) \frac{1}{\lambda_0}
U(x),
\end{equation} 
which can be easily generalized to systems with long-ranged interactions.  A
case of particular interest to us is the Coulomb potential
\begin{equation} 
V({\bf x}) = \frac{e_0^2}{|{\bf x}|},
\end{equation} 
whose Fourier transform in $d$ space dimensions reads
\begin{equation} 
V({\bf k}) = 2^{d-1} \pi^{(d-1)/2} \Gamma\left[\tfrac{1}{2}(d-1)\right]
\frac{e_0^2}{|{\bf k}|^{d-1}}.
\end{equation} 
The simple contact interaction $L_{\rm i} = - \lambda_0 \int_{\bf x} 
|\phi(x)|^4$ in (\ref{bec:Lagr}) gets now replaced by
\begin{equation}  
L_{\rm i} = - \frac{1}{2} \int_{{\bf x}, {\bf y}} |\phi(t,{\bf x})|^2
V({\bf x} - {\bf y}) |\phi(t,{\bf y})|^2.
\end{equation} 
The rationale for using the three-dimensional Coulomb potential even when
considering charges confined to move in a lower dimensional space is that
the electromagnetic interaction remains three-dimensional.  The effective
theory (\ref{eff:quick}) now becomes in the  Fourier representation
\begin{equation}  \label{effCoul}
{\cal L}_{\rm eff} = - \bar{n} U(k)  + \frac{1}{2} U(k_0,{\bf k})
\frac{1}{V({\bf k})} U(k_0,-{\bf k})
\end{equation}
and leads to the dispersion relation
\begin{equation}
E^2({\bf k}) =  2^d \pi^{(d-1)/2} \Gamma\left[\tfrac{1}{2}(d-1)\right]
\frac{\bar{n} e_0^2}{m} |{\bf k}|^{3-d}.
\end{equation}
For $d=3$, this yields the famous plasma mode with an energy gap given by
the plasma frequency $\omega_{\rm p}^2 = 4 \pi \bar{n} e_0^2/m$.

To appreciate under which circumstances the Coulomb interaction becomes
important, we note that for electronic systems $1/|{\bf x}| \sim k_{\rm
F}$ for dimensional reasons and the fermion number density $\bar{n} \sim
k_{\rm F}^d$, where $k_{\rm F}$ is the Fermi momentum.  The ratio of the
Coulomb interaction energy $\epsilon_{\rm C}$ to the Fermi energy
$\epsilon_{\rm F} = k_{\rm F}^2/2m$ is therefore proportional to
$\bar{n}^{-1/d}$.  This means that the lower the electron number
density, the more important the Coulomb interaction becomes.

For later reference, we close this section by calculating the fraction of
particles residing in the condensate.  In deriving the Bogoliubov spectrum
(\ref{bec:bogo}), we set $|\bar{\phi}|^2 = \mu_0/2 \lambda_0$.  For our
present consideration we have to keep $\bar{\phi}$ as independent variable.
The spectrum of the elementary excitation expressed in terms of $\bar{\phi}$
is
\begin{equation} \label{bec:bogog}
E({\bf k}) = \sqrt{\bigl[ \epsilon({\bf k})  - \mu_0 + 4 \lambda_0
|\bar{\phi}|^2 \bigr]^2 - 4 \lambda_0^2 |\bar{\phi}|^4 } \, .  	
\end{equation}
It reduces to the Bogoliubov spectrum when the mean-field value
(\ref{bec:min}) for $\bar{\phi}$ is inserted.  Equation (\ref{bec:formalV})
for the effective potential is still valid, and so is
(\ref{bec:omega-V}).  We thus obtain for the particle number density
\begin{equation} \label{bec:intn} 
\bar{n} = \left.  \frac{\mu_0}{2 \lambda_0} - \frac{1}{2}
\frac{\partial}{\partial \mu_0}  
\int_{\bf k} E ({\bf k})  \right|_{|\bar{\phi}|^2 = \mu_0/2 \lambda_0},
\end{equation} 
where the mean-field value for $\bar{\phi}$ is to be substituted after the
differentiation with respect to the chemical potential has been carried out.
The integral in (\ref{bec:intn}) gives rise to an ultraviolet-diverging term
which can can be cancelled by going over to the renormalized parameters
(\ref{bec:renla}) and (\ref{bec:renmu}) in the first term of
(\ref{bec:intn}).  In the second term, being a one-loop result, we may to
this order simply replace the bare by the (one-loop) renormalized parameters.
We find in this way \cite{Weichman} 
\begin{equation}
\bar{n} = \frac{\mu}{2 \lambda} + \frac{1}{3 \pi^2} (2 m \lambda
|\bar{\phi}|^2)^{3/2}, 
\end{equation} 
or for the so-called depletion of the condensate \cite{LHY,TN}
\begin{equation} \label{bec:depl}
\frac{\bar{n}}{\bar{n}_0} -1 \approx \frac{8}{3} \left(\frac{\bar{n}
a^3}{\pi}\right)^{1/2}, 
\end{equation} 
where 
\begin{equation}   \label{bec:n0}
\bar{n}_0 = \frac{\mu}{2 \lambda}
\end{equation}
is the number density of particles in the condensate.  Equation
(\ref{bec:depl}) shows that even at zero temperature not all the particles
reside in the condensate.  Due to the interparticle repulsion, particles are
removed from the ground state and put in states of finite momentum.  It has
been estimated that in superfluid $^4$He---a strongly interacting
system---only about 8\% of the particles condense in the zero-temperature
state
\cite{PeOn}.  However, it is well known (see next section) that at zero
temperature all the particles nevertheless participate in the superfluid
motion \cite{NoPi}.  Apparently, the condensate drags the normal fluid along
with it.
\section{BEC at Finite Temperature}
\label{sec:ftbec}
We continue to consider BEC in a weakly interacting Bose gas at finite
temperature.  Some finite-temperature considerations can be found in a
textbook by Popov \cite{PopovD}.  At the absolute zero of temperature, we
saw that the Bogoliubov spectrum (\ref{bec:bogo}) vanishes linearly as the
momentum goes to zero.  Since this property is a direct consequence of the
spontaneously broken global U(1) symmetry, we expect it to hold at any
temperature below the critical temperature $T_{\rm c}$.  We shall compute
the sound velocity in the vicinity of $T_{\rm c}$.

To calculate the thermodynamic potential at finite temperature, we have to
generalize the zero-temperature quantum field theory used up to this point
to finite temperature.  For our purposes, this is achieved simply by going
over to imaginary time $t \rightarrow - i \tau$, with $0 \leq \tau
\leq 1/T$, and by replacing integrals over energy $k_0$ with summations over
frequencies \cite{Kapusta},
\begin{equation} \label{bcs:ftft}
\int_{k_0} g(k_{0})\rightarrow  i\, T
\sum_{n} \, g( i\,\omega_{n}),   
\end{equation} 
where $g$ is an arbitrary function, and $\omega_n$ are the so-called
Matsubara frequencies,
\begin{equation} \label{bcs:Mats}
\omega_n = \left\{ \begin{array}{ll}
2n \pi T      & {\rm for \;\; bosons}    \\
(2n + 1) \pi T & {\rm for \;\; fermions}.
\end{array}        \right.                                       
\end{equation} 
These rules allow us to calculate the partition function $Z$ at finite
temperature.  The thermodynamic potential is then obtained using the relation
\begin{equation} 
\Omega = - T \ln(Z).
\end{equation} 

As in the previous section, we evaluate the thermodynamic potential for the
case where the condensate is spacetime independent.  That is, $\bar{\phi}$
introduced in (\ref{bec:newfields}) is assumed to depend only on temperature
and the scalar field $\varphi$ representing the Goldstone mode is set to
zero.  It was shown in (\ref{bec:omega-V}) that in the Bogoliubov
approximation of ignoring terms higher than second order in the complex
field $\tilde{\phi}$, the thermodynamic potential is given by
the effective potential.  Applying the rules outlined above to the
zero-temperature expression (\ref{bec:Veff}), we obtain
\begin{equation} \label{bec:VeffT}
{\cal V}_{\rm eff}(T) = \frac{T}{2} \, \sum_n
\int_{\bf k}  {\rm tr} \, \ln [M_0(i \omega_n,{\bf k})],
\end{equation} 
with the matrix $M_0$ given by (\ref{bec:M}) with $U = 0$.  The
mean-field value of the potential energy is given by
\begin{equation} \label{bec:V0T}
{\cal V}_0(T) = - \mu_0 |\bar{\phi}(T)|^2 + \lambda_0 |\bar{\phi}(T)|^4,
\end{equation}
with $|\bar{\phi}(T)|^2$ denoting the number density of particle residing in
the condensate.

The summation over the Matsubara frequencies can be carried out using
standard methods.  In this way, one obtains for the effective
potential
\begin{equation} \label{bec:Vfinal}
{\cal V}_{\rm eff}(T) = \frac{1}{2} \int_{\bf k} E({\bf k}) + T
\int_{\bf k} \ln \left(1 - {\rm e}^{-E({\bf k})/T} \right).
\end{equation} 
The first term at the right-hand side has been analyzed in the previous
section and was shown to lead to a renormalization of the parameters
$\lambda_0$ and $\mu_0$ in addition to the contribution (\ref{bec:finite})
which we will ignore here.  We shall study the second term in the vicinity
of the critical point by expanding it in a high-temperature series.  The
expansion is justified only if $T_{\rm c}$ is in the high-temperature
regime.  It will turn out that this is indeed the case for the weak-coupling
theory we are discussing.  We obtain for the thermodynamic potential
\begin{eqnarray}     \label{bec:expan}
\frac{\Omega}{V} = \!\!\!\!\! && \!\!\!\!\! {\cal
V}(T)  + \frac{(2 m)^{3/2}}{2 \pi^2} T^{5/2}
\int_0^{\infty} \dd q  \nonumber \\  &&  \!\!\!\!\! \times \Biggl\{ 
q^2 \, \ln\left(1 - {\rm e}^{-q^2}\right) + \frac{1}{T} \bigl(4 \lambda
|\bar{\phi}|^2 - \mu\bigr) \,
\frac{q^2}{{\rm e}^{q^2} - 1} \nonumber \\ && \;\;\; -
\frac{1}{2 T^2} \left[4 \lambda^2 |\bar{\phi}|^4 \, \frac{1}{{\rm
e}^{q^2} - 1} + \bigl(4 \lambda |\bar{\phi}|^2 - \mu \bigr)^2 \, \frac{q^2 {\rm
e}^{q^2}}{\left({\rm e}^{q^2} - 1\right)^2} \right] \nonumber \\ && \;\;\; 
+ {\cal O}\left(\frac{1}{T^3}\right) 
\Biggr\},                                             
\end{eqnarray}
where the integration variable is $q = |{\bf k}|/\sqrt{2 m T}$.  The
potential is denoted by ${\cal V}(T)$ to indicate that we included the
zero-temperature renormalization of the parameters in (\ref{bec:V0T}).  The
first integral appearing here is finite and yields
\begin{equation} 
\int_0^{\infty} \dd y y^2 \, \ln \left(1 - {\rm e}^{-y^2} \right) = -
\tfrac{1}{2} \Gamma(\tfrac{3}{2}) \zeta (\tfrac{5}{2}),
\end{equation} 
with $\Gamma$ the gamma function and $\zeta$ the Riemann zeta function.  The
last two integrals, however, are infrared divergent.  We regularize these by
analytically continue the following equations to arbitrary values of $\alpha$
\begin{equation} 
\int_0^{\infty} \frac{\dd x}{x} \frac{x^{\alpha} }{{\rm e}^x - 1} =
\Gamma(\alpha ) \zeta(\alpha ),
\end{equation} 
\begin{equation} 
\int_0^{\infty} \frac{\dd x}{x} \frac{x^{\alpha} }{({\rm e}^x - 1)^2}
= \Gamma (\alpha ) [ \zeta (\alpha -1) - \zeta (\alpha )],
\end{equation} 
with the result
\begin{eqnarray}    
\frac{\Omega}{V} = {\cal V}(T) + \left( \frac{m}{2 \pi} \right)^{3/2}
\Bigl\{ \!\!\!\!\!\! && \!\!\!\!\!\!  -\zeta (\tfrac{5}{2}) T^{5/2} +
\bigl(4 \lambda |\bar{\phi}|^2 - \mu 
\bigr) \zeta (\tfrac{3}{2}) T^{3/2} \nonumber \\ \!\!\!\!\!\! &&
\!\!\!\!\!\! - \left[4 \lambda^2 |\bar{\phi}|^4 + \tfrac{1}{2} \bigl(4
\lambda |\bar{\phi}|^2 - \mu \bigr)^2 \right] \zeta (\tfrac{1}{2}) T^{1/2}
\Bigr\}. \nonumber \\
\end{eqnarray} 
To this order in $1/T$ the thermodynamic potential is of the Landau
form, involving terms up to fourth order in the order parameter
$\bar{\phi}$:
\begin{equation} \label{bec:lanform}
\frac{\Omega}{V} = \alpha_0 - \alpha_1 |\bar{\phi}|^2 + \alpha_2
|\bar{\phi}|^4,
\end{equation} 
with
\begin{equation}
\alpha_0 =\left( \frac{m}{2 \pi} \right)^{3/2} \left[ -\zeta (\tfrac{5}{2})
T^{5/2} - \mu  \zeta (\tfrac{3}{2}) T^{3/2} -\tfrac{1}{2} \mu^2 \zeta
(\tfrac{1}{2} ) T^{1/2}\right]
\end{equation}
a $\bar{\phi}$-independent term, and
\begin{eqnarray} \label{bec:c1}
\alpha_1 &=& \mu - 4 \lambda \left( \frac{m}{2 \pi} \right)^{3/2} \left[\zeta
(\tfrac{3}{2}) T^{3/2} + \mu \zeta (\tfrac{1}{2} ) T^{1/2} \right]
\\
\alpha_2 &=& \lambda \left[1 - 12 \lambda \left( \frac{m}{2 \pi}
\right)^{3/2} \zeta (\tfrac{1}{2} ) T^{1/2} \right].
\end{eqnarray} 
We note, however, that our expansion is not one in terms of $\bar{\phi}$ since the
higher-order terms in $1/T$ contain, besides higher-order terms in $\bar{\phi}$,
also $|\bar{\phi}|^2$- and $|\bar{\phi}|^4$-terms.  

The critical temperature is determined by setting the coefficient $\alpha_1$
to zero.  An approximate solution to this equation is given by
\begin{equation}   \label{bec:jus}
T_{\rm c} = \frac{\pi}{\left[\sqrt{2}\,
\zeta(\tfrac{3}{2})\right]^{2/3}} 
\frac{1}{m} \left( \frac{\mu}{\lambda } \right)^{2/3} - \tfrac{2}{3}
\frac{ \zeta (\tfrac{1}{2})}{\zeta (\tfrac{3}{2})} \mu + {\cal
O}(\lambda^{2/3}),                                    
\end{equation}
where we expanded in the coupling constant $\lambda$.  This justifies
the high-temperature expansion we have been using because the leading
term is of the order $\lambda ^{-2/3}$ which is large for a
weak-coupling theory.

Equation (\ref{bec:jus}) expresses the critical temperature in terms of
$\mu$, the chemical potential.  As we mentioned in the previous section,
from the experimental point of view it is sometimes more realistic to
consider the particle number density as independent variable.  To
express $T_{\rm c}$ in terms of the particle number density, we invert
the equation for $n$ obtained from (\ref{bec:lanform}) with $\bar{\phi}=0$,
\begin{equation}
\bar{n}(T_{\rm c}) = - \frac{1}{V} \frac{\partial \Omega}{\partial \mu}
\Biggr|_{T=T_{\rm c}} =
\zeta(\tfrac{3}{2}) \left(\frac{m T_{\rm c}}{2 \pi} \right)^{3/2} + 
\mu \left(\frac{m}{2 \pi} \right)^{3/2} \zeta(\tfrac{1}{2}) T_{\rm c}^{1/2},
\end{equation}
to obtain $\mu = \mu(\bar{n})$.  When this is substituted in the condition
$\alpha_1=0$, the latter can be manipulated in the form
\begin{equation}
\bar{n}(T_{\rm c}) - 4  \bar{n}(T_{\rm c}) \lambda 
\left(\frac{m}{2 \pi} \right)^{3/2}  \zeta (\tfrac{1}{2}) T_{\rm c}^{1/2} -
\left(\frac{m}{2 \pi} \right)^{3/2}  \zeta(\tfrac{3}{2})T_{\rm c}^{3/2} = 0, 
\end{equation}
from which the critical temperature as function of the particle number density
can be determined.  Setting $\lambda$ equal to zero, we obtain the critical
temperature 
\begin{equation}
T_0 = \frac{2 \pi}{m} \left( \frac{\bar{n}}{\zeta(\tfrac{3}{2})}
\right)^{2/3}
\end{equation}
of a free Bose gas.  To first nontrivial order in an expansion in $\lambda$,
we find
\begin{eqnarray} 
T_{\rm c} &=& T_0 \left[ 1 - \frac{4}{3 \pi} \frac{\zeta
(\tfrac{1}{2})}{\zeta^{1/3} (\tfrac{3}{2})} m \lambda \bar{n}^{1/3}(T_{\rm c})
\right] \nonumber  \\ &=& T_0 \left[ 1 - \frac{8}{3} \zeta (\tfrac{1}{2})
\left(\frac{\bar{n}(T_{\rm c}) a^3}{\zeta (\tfrac{3}{2})}
\right)^{1/3}\right],
\end{eqnarray} 
where in the last equation we replaced the coupling constant $\lambda$ with
the scattering length $a$ using relation (\ref{bec:a}).  This equation
implies a (slight) increase of the critical temperature due to the weak
interaction since $\zeta(\tfrac{1}{2}) <0$.  This is qualitatively different
from the strongly interacting $^4$He system.  A free gas with $^4$He
parameters at vapor pressure would have a critical temperature of about 3
K, whereas liquid $^4$He becomes superfluid at the {\it lower} temperature
of 2.17 K.

The value of the order parameter $\bar{\phi}(T)$ obtained from
Eq.\ (\ref{bec:lanform}) reads
\begin{equation} \label{bec:mini}
|\bar{\phi}(T)|^2 = \frac{1}{2} \frac{\alpha_1}{\alpha_2}.		
\end{equation}
Near the critical temperature  this can be cast in the form
\begin{eqnarray} 	\label{bec:chi0T}
|\bar{\phi}(T)|^2 &\approx& 3 \zeta (\tfrac{3}{2}) \left( \frac{ m T_{\rm c}}{2
\pi}\right)^{3/2}  \left( 1 - \frac{T}{T_{\rm c}}\right) \nonumber \\ &\approx& 
3 \bar{n}(T_{\rm c}) \left( 1 - \frac{T}{T_{\rm c}}\right).
\end{eqnarray} 
Recalling that $|\bar{\phi} (T)|^2$ represents the particle number density
$\bar{n}_0(T)$ of the condensate, we see that when the critical temperature
is approached from below, the Bose-Einstein condensate is drained of
particles, and that at $T_{\rm c}$ it vanishes altogether.

We are now in a position to investigate the spectrum of the elementary
excitations at finite temperature.  If one simply substitutes the value
(\ref{bec:mini}) for $\bar{\phi}(T)$ in the Bogoliubov spectrum
(\ref{bec:bogog}) one finds, contrary to what is expected, that it has an
energy gap.  The solution to this paradox, which also exists in the
relativistic $|\phi|^4$-theory \cite{Kapusta}, lays in the observation that,
as at zero temperature \cite{Hugenholtz}, perturbation theory should be
carried out consistently.  This has not been done up to this point.  Whereas
Eq.\ (\ref{bec:mini}) included thermal fluctuations, the spectrum
(\ref{bec:bogog}) did not.  To fix this, we go back to Eq.\ (\ref{bec:c1})
and note that it reflects a change in the chemical potential due to thermal
fluctuations.  With this modification the finite-temperature spectrum
becomes gapless---in accordance with a theorem due to Hohenberg and Martin
\cite{HM}.

Let us work this out in some detail near the critical temperature, where
$|\bar{\phi}(T)|^2$ is given by (\ref{bec:chi0T}).  Retaining only the
leading term in the $1/T$ expansion, we infer from (\ref{bec:c1}) the
following change in the chemical potential
\begin{equation} 
\mu \rightarrow \tilde{\mu} = \mu - 4 \zeta(\tfrac{3}{2}) \lambda
\left(\frac{m T}{2 \pi} \right)^{3/2},
\end{equation} 
and $2 \lambda |\bar{\phi}|^2 = \tilde{\mu}$, where $\mu$ is given a tilde to
indicate that it is dressed by thermal fluctuations.  In this
way, the finite-temperature spectrum becomes
\begin{equation}
E({\bf k}) = \sqrt{\epsilon^2({\bf k})  + 4 \lambda  |\bar{\phi}(T)|^2
\epsilon({\bf k})},
\end{equation}
which is indeed gapless.  Since we included the zero-temperature
renormalization of the parameters stemming from the first term at the
right-hand side of (\ref{bec:Vfinal}), the renormalized parameters $\lambda$
and $\mu$ feature in these equations.  The gaplessness of the
finite-temperature spectrum was shown to be true in all orders of
perturbation theory by Hohenberg and Martin \cite{HM}.  The result we just
obtained is completely analogous to the zero-temperature Bogoliubov spectrum
(\ref{bec:bogo}).  The expression for the sound velocity at finite
temperature we extract from the spectrum reads
\begin{eqnarray} 
c^2(T) &=& \frac{2 \lambda}{m} |\bar{\phi}(T)|^2 \nonumber \\
&=& 6 \frac{\lambda}{m} \bar{n}(T_{\rm c}) \left(1 - \frac{T}{T_{\rm c}} \right).
\end{eqnarray} 
It vanishes when the temperature approaches $T_{\rm c}$.  This is in
accord with the observation that the gapless Goldstone mode of the
spontaneously broken global U(1) symmetry vanishes at the critical
temperature.  We note that also at finite temperature, the sound velocity
squared is linear in the coupling constant---without the interparticle
repulsion the sound velocity would be zero.

At the end of the preceding section we mentioned the well-known fact that
despite only a fraction of the particles at zero temperature reside in the
zero-momentum state, all particles participate in the superfluid motion.  To
show this, let us assume the entire system moves with a velocity ${\bf u}$
relative to the laboratory system.  As is known from standard hydrodynamics
the time derivate in the frame following the motion of the fluid is
$\partial_0 + {\bf u} \cdot \nabla$ [see Eq.\ (\ref{hydro:boost})]. If we
insert this in the Lagrangian (\ref{bec:Lagr}) of the near-ideal Bose gas,
it becomes
\begin{equation}   \label{bec:Lagu}
{\cal L} = \phi^* [i \partial_0 - \epsilon(-i \nabla) + \mu_0 - {\bf u}
\cdot (-i \nabla)] \phi - \lambda_0 |\phi|^4,      				
\end{equation} 
where the extra term features the total momentum $\int_{\bf x}
\phi^* (-i \nabla) \phi$ of the system.  The velocity ${\bf u}$ multiplying this
is on the same footing as the chemical potential which multiplies the
particle number $\int_{\bf x} |\phi|^2$.  Whereas $\mu_0$ is associated
with particle number conservation, ${\bf u}$ is related to the conservation
of momentum.

In the two-fluid picture, the condensate can move with a different velocity
${\bf v}_{\rm s}$ as the rest of the system.  To bring this out we introduce
new fields, cf.\ (\ref{bec:newfields}) 
\begin{equation} 
\phi (x) \rightarrow \phi'(x) = {\rm e}^{im {\bf v}_{\rm s} \cdot {\bf x}}
\phi (x) 
\end{equation} 
in terms of which the Lagrangian becomes \cite{Brown}
\begin{equation}   \label{bec:Lagus}
{\cal L} = \phi^* \bigl[i\partial_0 - \epsilon(-i \nabla)  + \mu_0 -
\tfrac{1}{2} m {\bf v}_{\rm s} \cdot ({\bf v}_{\rm s} - 2 {\bf u}) - ({\bf
u} - {\bf v}_{\rm s}) \cdot (-i \nabla) \bigr] \phi - \lambda_0 |\phi|^4,
\end{equation} 
where we dropped the primes on $\phi$ again.  Both velocities appear in this
expression.  Apart from the change ${\bf u} \rightarrow {\bf u} - {\bf
v}_{\rm s}$ in the second last term, the field transformation resulted in a
change of the chemical potential
\begin{equation} \label{bec:mureplacement}
\mu_0 \rightarrow \mu_{\rm eff} :=
\mu_0 - \tfrac{1}{2} m {\bf v}_{\rm s} \cdot ({\bf v}_{\rm s} -  2 {\bf u}) 
\end{equation} 
where $\mu_{\rm eff}$ may be considered as an effective chemical potential.

The equations for the Bogoliubov spectrum and the thermodynamic potential
are readily written down for the present case when we keep these two changes
in mind.  In particular, the effective potential (\ref{bec:Vfinal}) reads
\begin{equation} \label{bec:Vus}
{\cal V}_{\rm eff}(T) = \frac{1}{2} \int_{\bf k}
E({\bf k}) + T \int_{\bf k} \ln \left(1 - {\rm e}^{-[E({\bf k})
+ ({\bf u} - {\bf v}_{\rm s}) \cdot {\bf k}]/T} \right),
\end{equation} 
where $E({\bf k})$ is the Bogoliubov spectrum (\ref{bec:bogog}) with the
replacement Eq.\ (\ref{bec:mureplacement}).  The mean-field potential ${\cal
V}_0(T)$ is given by (\ref{bec:V0T}) with the same replacement.
The momentum density, or equivalently, the mass current ${\bf g}$ of the
system is obtained in this approximation by differentiating the potential
${\cal V}_0(T) + {\cal V}_{\rm eff}(T)$ with respect to $-{\bf u}$.  We find,
using the equation
\begin{equation}
\frac{\partial \mu_{\rm eff}}{\partial {\bf u}} = m {\bf v}_{\rm s}
\end{equation} 
that it is given by
\begin{equation} \label{bec:j}
{\bf g} = m  n  {\bf v}_{\rm s} - \int_{\bf k}
\frac{{\bf k}}{\exp\{[E({\bf k}) + ({\bf u} - {\bf v}_{\rm s}) \cdot {\bf
k}]/T\} - 1}.
\end{equation}
The last term is the contribution stemming from the elementary excitations.
In the zero-temperature limit, this term vanishes and ${\bf g} = m n
{\bf v}_{\rm s}$.  This equation, comprising the total particle number
density $n$, shows that at zero temperature indeed all the particles
are involved in the superflow, despite the fact that only a fraction of them
resides in the condensate \cite{NoPi}.  When the condensate moves with the
same velocity as the rest (${\bf v}_{\rm s} = {\bf u}$), the last term in
(\ref{bec:j}) vanishes again, now by symmetry.

Assuming that the difference between the normal and superfluid velocity
is small, we may expand the last term in (\ref{bec:j}) to linear
order in this difference to find
\begin{equation} \label{bec:lr}
{\bf g}  = \rho {\bf v}_{\rm s} + 
\frac{1}{3 T}  \int_{\bf k} {\bf k}^2 \, \frac{{\rm e}^{E({\bf k})/T}}{({\rm
e}^{E({\bf k})/T} - 1)^2} \, ({\bf u} - {\bf v}_{\rm s}),
\end{equation}
where $\rho =m n$ is the total mass density of the fluid, and we used
the general result
\begin{equation}
\int_{\bf k} k^i k^j f(|{\bf k}|) = \frac{1}{3} \delta^{i j} \int_{\bf k} {\bf
k}^2 f(|{\bf k}|),
\end{equation} 
with $f(|{\bf k}|)$ an arbitrary function depending only on the length of
${\bf k}$.  The last term stemming from the elementary excitations may be used
to define the normal mass density $\rho_{\rm n}$,
\begin{equation}
\rho_{\rm n} = \frac{1}{3T} \int_{\bf k} {\bf k}^2 \, \frac{{\rm
e}^{E({\bf k})/T}}{({\rm e}^{E({\bf k})/T} - 1)^2}.
\end{equation} 
Writing 
\begin{equation} 
\rho = \rho_{\rm s} + \rho_{\rm n}
\end{equation} 
for the total mass density, we may cast (\ref{bec:lr}) in the form
\begin{equation} \label{bec:md}
{\bf g} = \rho_{\rm s} {\bf v}_{\rm s} +\rho_{\rm n} {\bf u}_{\rm n}.
\end{equation} 
These last two equations are the basic equations of the two-fluid model
\cite{NoPi}.  The model was introduced by Tisza \cite{Tisza} using idea's of
London to give a phenomenological description of superfluid $^4$He.  It not
only successfully explained various startling experimental properties of
the strongly interacting system, but also predicted new phenomena which
were later confirmed by experiment
\section{BCS at Finite Temperature}
\label{sec:ftbcs}

We in this section apply the high-temperature expansion to the BCS
theory to show that it yields the usual Ginzburg-Landau theory \cite{GL}.

The one-loop effective potential for a uniform system at finite temperature
is readily seen to take the form
\begin{equation}  \label{omegabcs} 
{\cal V}_{\rm eff} = - \int_{\bf k} E({\bf k}) - 2 \, T \int_{\bf k} \ln
\left(1 + {\rm e}^{-E({\bf k})/T} \right),
\end{equation}
with $E({\bf k})$ the BCS spectrum (\ref{bcs:spec}) of the elementary
fermionic excitations.  The factor $-2$ in (\ref{omegabcs}) arises
because there are two fermion species, with spin $\sigma = \uparrow$ and
$\downarrow$, respectively.  To the one-loop potential we have to add
the tree contribution
\begin{equation}  \label{ep0}
{\cal V}_0 = - \frac{|\bar{\Delta} |^2}{\lambda_0 }.
\end{equation}
The form (\ref{omegabcs}) for the effective potential is usually not taken
as starting point to derive the Ginzburg-Landau theory.  Normally, one first
carries out the integration over the loop energy $\xi$ and then performs the
sum over the Matsubara frequencies \cite{Gorkov}.  But here we follow---in
analogy with the previous calculation---the opposite route and have carried
out the summation first and will now perform the integration over $\xi$ (in
the weak-coupling limit),
\begin{equation} \label{epint}
{\cal V}_{\rm eff} = -2 \, T \nu(0) \int_\xi \ln \left( 1 + {\rm e}^{-E/T}
\right),
\end{equation}
where we ignored the first term in (\ref{omegabcs}) corresponding to the
one-loop quantum contribution to the effective potential which we studied in
Sec.\ \ref{sec:bcs}, save for the renormalization of $\lambda_0$ it leads
to.  Our problem is seen to reduce to one in a single dimension.  To cope
with divergences that will appear we dimensional regularize the integral and
consider the problem in $1 - \epsilon$ dimensions.  Equation (\ref{epint})
then becomes
\begin{equation}  \label{yint} 
{\cal V}_{\rm eff} = -4 \, T^2 \nu(0) \int_0^{\infty} \dd y \, y^{-\epsilon }
\ln \left( 1 + {\rm e}^{- \sqrt{y ^2 + |\bar{\Delta} |^2/T^2 }} \right),
\end{equation}
with $y$ the dimensionless variable $y = \xi/T$.  We kept the
one-dimensional integration measure $K_1$, where
\begin{equation}  \label{Kd}
K_d = \frac{2}{(4\pi)^{d/2} \Gamma(\tfrac{d}{2})}, 
\end{equation}  
is the area of a unit sphere in $d$ spatial dimensions divided by
$(2\pi)^d$.  Because $\bar{\Delta}$ appears only in the combination
$|\bar{\Delta}|^2/T^2$, it follows that an expansion in this parameter is
tantamount to one in $1/T$.  Carrying out this expansion, we arrive at
\begin{eqnarray}   \label{hight} 
{\cal V}_{\rm eff} = 2 \nu(0) \int_0^{\infty} \dd y \, y^{-\epsilon}
\Biggl\{\!\!\!\!\!\! && \!\!\!\!\!\! -2 \, T^2 \ln \left( 1 + {\rm e}^{- y}
\right) +   |\bar{\Delta}|^2 \frac{1}{y} \frac{1}{{\rm e}^{y} + 1} 
\nonumber \\ \!\!\!\!\!\! && \!\!\!\!\!\! 
- \frac{1}{4} \frac{|\bar{\Delta}|^4}{T^2}
\left[\frac{1}{y^3} \frac{1}{{\rm e}^y + 1} + \frac{1}{y^2}
\frac{{\rm e}^y}{({\rm e}^{y} + 1)^2 } \right] \nonumber \\ \!\!\!\!\!\! &&
\!\!\!\!\!\! + {\cal O} \left(\frac{1}{T^4} \right)  \Biggr\}. 
\end{eqnarray} 
In analogy with what we did in the case of a weakly interacting Bose gas,
we regularize the infrared-divergent integrals by analytically continue the
following equations to arbitrary values of $\alpha$
\begin{equation} 
\int_0^{\infty} \dd x \, \ln \left(1 + {\rm e}^{-x} \right) =
\frac{\pi^2}{12}, 
\end{equation} 
\begin{equation} 
\int_0^{\infty} \frac{\dd x}{x} \frac{x^{\alpha} }{{\rm e}^x + 1} =
\Gamma(\alpha ) (1 - 2^{1 - \alpha }) \zeta(\alpha ), 
\end{equation} 
\begin{equation} 
\int_0^{\infty} \frac{\dd x}{x} \frac{x^{\alpha} }{({\rm e}^x + 1)^2}
= \Gamma (\alpha ) \left[ (1-2^{1 - \alpha }) \zeta (\alpha) - (1 -
2^{2 - \alpha } )\zeta (\alpha - 1) \right].
\end{equation} 
In this way we obtain in the limit $\epsilon \rightarrow 0$
\begin{eqnarray}   \label{inter}
{\cal V}_{\rm eff} = \!\!\!\!\! && \!\!\!\!\!  -\frac{\pi^2}{3} \nu(0)
T^2 - \nu(0) |\bar{\Delta} |^2 \left[\frac{1}{\epsilon } + \gamma + 2 \ln(2) + 2
\zeta '(0)
\right] \nonumber \\ \!\!\!\!\!\! && \!\!\!\!\!\! - \frac{7}{4} \zeta'(-2)
\nu(0) \frac{|\bar{\Delta} |^4}{T^2}.  
\end{eqnarray} 
To make contact with the standard approach we use the identities
\begin{equation}
\zeta '(0) = - \tfrac{1}{2} \ln (2 \pi), \hspace{1.cm}
\zeta '(-2) = - \frac{1}{4 \pi^2} \zeta (3)
\end{equation}
and substitute
\begin{equation}
\frac{1}{\epsilon } \rightarrow \ln \left(\frac{\omega _{\rm D}}{T}
\right), 
\end{equation}
where the Debeye energy $\omega_{\rm D}$, being a measure of the inverse
lattice spacing, is the physical ultraviolet cutoff and the temperature $T$
is the relevant infrared scale. [This correspondence between the pole
$1/\epsilon$ of dimensional regularization and the logarithm $\ln(\Lambda)$
appearing in the regularization with a cutoff is commonly used in the
context of quantum field theory.]  The critical temperature is again
determined by the condition that the coefficient of the term quadratic in
the order parameter be zero, i.e.,
\begin{equation}
\nu(0) \left[ \ln \left(\frac{\omega _{\rm D}}{T} \right) + \gamma - \ln
\left( \tfrac{1}{2} \pi \right) \right] + \frac{1}{\lambda} = 0,
\end{equation}
where we included the tree contribution (\ref{ep0}) with the bare coupling
$\lambda_0$ replaced with the renormalized one (\ref{bcs:reng}).  This yields
the standard result for the critical temperature $T_{\rm c}$
\begin{equation}
T_{\rm c} = \frac{2}{\pi} {\rm e}^{\gamma } \omega_{\rm D} {\rm
e}^{1/\nu(0) \lambda} ,
\end{equation}
where it should be kept in mind that the coupling constant $\lambda$ is
negative in the weak-coupling limit.  Employing this expression for the
critical temperature, we can manipulate the effective potential
(\ref{inter}) with the tree contribution (\ref{ep0}) added in the canonical
form
\begin{equation}
{\cal V} = -\frac{\pi^2}{3} \nu(0) T^2 + \nu(0) \ln \!\left( \frac{T}{T_{\rm
c}} \right) |\bar{\Delta} |^2 + \frac{7 \zeta(3)}{16 \pi^2} 
\nu(0)\frac{|\bar{\Delta} |^4}{T^2},
\end{equation}
valid close to the critical temperature.  The minimization of this potential
with respect to $\bar{\Delta}$ yields the well-known temperature dependence of
the order parameter near the critical temperature
\begin{equation} 
|\bar{\Delta} (T)|^2 \approx \frac{8 \pi^2 }{7 \zeta (3)}
T^2_{\rm c} \left( 1 - \frac{T}{T_{\rm c}} \right),
\end{equation} 
which should be compared with Eq.\ (\ref{bec:chi0T}) we obtained for a
superfluid.

\chapter{Induced Quantum Numbers \label{chap:indq}}
Solitons play an important role in condensed matter physics (for a
general introduction see Ref.\ \cite{Rajaraman,ChLu}).  One of the most
famous solitons is the magnetic vortex in a superconductor first discussed
by Abrikosov \cite{Abrikosov}.  In various models, solitons are found to
have peculiar quantum numbers associated with them.  We, in this
chapter, exclusively study solitons arising in fermionic systems.  The
unusual quantum numbers induced by the solitons can then be computed by
integrating out the fermionic degrees of freedom.  The effective theory
thus obtained contains the same information as the original fermionic
model at low energy and small momentum.  In particular, approximate
expressions for the fermionic currents---so-called Goldstone-Wilczek
currents \cite{GoWi}---can be extracted from it.  To see how this is
connected to the quantum numbers of solitons, it should be realized that
the effective theory and the ensuing Goldstone-Wilczek currents are
built from background fields, which in our definition include possible
Goldstone fields.  But solitons are precisely specified by these fields.
Hence, when a specific field configuration describing a soliton is
substituted in the Goldstone-Wilczek currents, the induced quantum
numbers localized on the soliton can be evaluated.  These quantum
numbers arising in purely bosonic effective theories can be peculiar in
that they can have fermionic, or even anyonic characteristics.  We will
encounter various examples of this along the way.

We will also consider phase transitions where the induced quantum numbers
localized on the solitons change from fermionic to bosonic.  Such a
transition, first considered by Wen and Zee \cite{WenZee}, is called a
statistics-changing phase transition.
\section{Skyrmion Lattice}
\label{sec:af}
In this section we investigate the effective theory describing the spin
degrees of freedom of superfluid $^3$He-a.  In this superfluid state, as
we remarked in Sec.~\ref{sec:he3}, the SO$^S$(3) spin rotation group is
spontaneously broken to the group SO$^S$(2) of spin rotations about the
preferred spin axis.  Since the state is characterized by the magnetic
quantum number $m_S=0$, we expect it to display besides superfluid also
antiferromagnetic behavior.  The spin waves of an ordinary uniaxial
antiferromagnet, which are the Goldstone modes of the spontaneously
broken SO$^S$(3) symmetry, are known to be described by the O(3)
nonlinear sigma model \cite{Fradkin}.  In 2+1 dimensions, the model
allows for a topological term in the action---the so-called Hopf term
\cite{WenZee}.

It has been suggested by Dzyaloshinski, Polyakov, and Wiegmann
\cite{DPW} that the Hopf term with $\theta = \pi$ should be included in
the effective theory describing a spin-$\tfrac{1}{2}$ Heisenberg
antiferromagnet in 2+1 dimensions.  But subsequent microscopic calculations
showed that the topological term cannot be derived from the Heisenberg model
\cite{HaFS}.  It will turn out that in $^3$He-a the situation is different
in that here the quantum-induced effective action does accommodate such a
topological term.

To derive the effective theory, we use the well-known fact that the unit
spin vector $d^\alpha(x)$ in (\ref{he3:3}) can always be rotated in the
third direction by introducing a spacetime-dependent $2\times2$ SU(2)-matrix
$s(x)$:
\begin{equation}  \label{af:rot}
d^\alpha(x)  \sigma^\alpha = s(x) \sigma^3 s^{\dagger}(x).           
\end{equation} 
We next introduce the decomposition
\begin{equation}   \label{af:decom}
\Psi = \left( \begin{array}{c} \psi \\ g \psi^{\ast} \end{array} \right) 
= \left( \begin{array}{c} s\chi \\ s g \chi^{\ast} \end{array} \right)
\end{equation} 
in the Lagrangian (\ref{he3:consice}).  The theory then takes the form of an
SU(2) gauge theory:
\begin{equation}  \label{af:consice}
{\cal L} = \frac{1}{2} X^{\dagger} \left(p_0 - B_0 - \xi(p_i -
B_i) \tau_3 -\frac{\bar{\Delta}}{2k_{\rm F}} \{(p_i - B_i)
\tau_i, \sigma^3 \} \right) X,
\end{equation}   
where $X=(\chi, g \chi^{\ast})^{\rm T}$, with the superscript T indicating the
transpose, and the $2 \times 2$ matrix-valued field
\begin{equation}       \label{af:B}
B_{\mu} = -i s^{\dagger}\tilde{\partial}_{\mu} s =  B_{\mu}^{\alpha} 
\sigma^{\alpha}
\end{equation} 
plays the role of a gauge field.  Remember that $\tilde{\partial}_\mu =
(\partial_0, - \nabla)$.  Also recall that the Pauli matrices $\tau_3$
and $\tau_i$ $(i=1,2)$ operate in Nambu space, whereas the Pauli matrices
$\sigma^\alpha$ $(\alpha =1,2,3)$ operate in spin space.  After integrating
out the fermionic degrees of freedom, we now obtain instead of
(\ref{he3:13}) the effective one-loop action
\begin{equation}   \label{af:Seff}
S_{\rm eff} = -\frac{i}{2} \, {\rm Tr} \log
\left(p_{0} - B_0 - \xi({\bf p} - {\bf B})\tau_3 -\frac{\bar{\Delta}}{2k_{\rm
F}} \{p_i - B_i, \sigma^3\} \tau_i \right).
\end{equation} 
Expanded in a Taylor series this becomes, apart from an irrelevant constant,
\begin{equation}   \label{af:Taylor}
S_{{\rm eff}} = \frac{i}{2} \, {\rm Tr} \sum_{\ell=1}^{\infty} \frac{1}{\ell}
[ G_0(p) \Lambda({\bf p})]^{\ell},
\end{equation} 
where the vertex $\Lambda({\bf p})$ is now given by
\begin{equation}  \label{af:vertex}
\Lambda ({\bf p} ) = B_0 + \frac{1}{2m}[({\bf B})^2 - 2 {\bf p} \cdot {\bf B}
- i \nabla \cdot {\bf B}] \tau_3 -
\frac{\bar{\Delta}}{k_{\rm F}} B_i^3 \tau_i ,        
\end{equation} 
and $G_0(p)$ is the propagator (\ref{he3:20}).  We have written ${\bf B}
\cdot {\bf B}$ as $({\bf B})^2$ so as not to confuse it with the second spin
component ($\alpha=2$) of the spatial vector ${\bf B}^\alpha$.

The first term $S_{\rm eff}^{(1)}$ in the expansion (\ref{af:Taylor}) is
readily shown to give
\begin{equation}   \label{af:S1}
S_{\rm eff}^{(1)} = - \frac{\bar{n}}{4m} \, {\rm tr} \int_x ({\bf B})^2 .
\end{equation} 
The trace tr stands for the trace over the remaining sigma matrix
indices, where one should bear in mind that the $B_{\mu}$'s defined by
(\ref{af:B}) are elements of the SU(2) algebra.  For the second term
$S_{\rm eff}^{(2)}$ in the expansion (\ref{af:Taylor}) we obtain to
lowest order in derivatives:
\begin{equation}    \label{af:S2}
S_{\rm eff}^{(2)} =  \frac{1}{4} \nu(0) \, {\rm tr}  \int_x
\left\{(B_0)^2 - (\sigma^3 B_0)^2 + \frac{v_{\rm F}^2}{2} \left[ ({\bf
B})^2 + \gamma (\sigma^3 {\bf B})^2\right] \right\},
\end{equation} 
where $\gamma$ stands for
\begin{equation} 
\gamma = 1 + \left(\frac{\bar{\Delta}}{\mu}\right)^2 \!
\ln{\left(\frac{c \Lambda}{\bar{\Delta}} \right)},
\end{equation} 
with $\Lambda$ an ultraviolet energy cutoff and $c$ an irrelevant
numerical constant. The term proportional to $(\bar{\Delta}/\mu)^2$ stems
from the last term in the vertex (\ref{af:vertex}).  Since $\bar{\Delta} <<
\mu \approx \epsilon_{\rm F}$ in the weak-coupling limit we are
considering, the factor $(\bar{\Delta}/\mu)^2 \ln{(c \Lambda/\bar{\Delta})}$ is
negligible compared to 1, and may be set to zero.  Adding the
contributions (\ref{af:S1}) and (\ref{af:S2}), and using the identity
\begin{equation} 
{\rm tr}[(B_{\mu})^2 - (\sigma^3 B_{\mu})^2] = (\partial_{\mu}
d^\alpha)^2,
\end{equation} 
which is easily derived from the definitions (\ref{af:rot}) and
(\ref{af:B}), we conclude that, apart from a possible topological term,
the effective theory describing the antiferromagnetic spin waves is the
O(3) nonlinear sigma model \cite{he-a}
\begin{equation}   \label{af:sigma}
{\cal L}_{\rm eff} = \tfrac{1}{4} \nu(0) \left[(\partial_0 d^\alpha)^2 -
\tfrac{1}{2} v_{\rm F}^2 (\partial_i d^\alpha)^2 \right].
\end{equation} 
It is gratifying to see that the ${\bf B}$-terms obtained from the first and
second term in the expansion of the effective action conspire to precisely
generate the right combination ${\rm tr}[({\bf B})^2 - (\sigma^3{\bf B})^2]$.

The O(3) nonlinear sigma model can be equivalently represented by the
CP$^1$ model
\begin{equation}    \label{af:cp}
{\cal L}_{\rm eff} = \nu(0) (\partial_\mu z^\dagger \partial_\mu
z - B_\mu^3 B_\mu^3 ),
\end{equation} 
adopting units such that the spin-wave velocity is unity $v_{\rm
F}/\sqrt{2} = 1$. Here, the complex scalar $z = (z_1, z_2)^{\rm T}$,
which is subject to the constraint $z^\dagger z=1$, is defined by the
equation
\begin{equation}   \label{af:z}
d^\alpha = z^{\dagger} \sigma^\alpha z,    
\end{equation} 
and is related to the $2 \times 2$ matrix $s$ in the following way
\begin{equation}  
s = \left( \begin{array}{lr} z_1 & -z_2^* \\
z_2 &  z_1^* \end{array} \right).
\end{equation} 
In terms of $z$, the field $B_\mu^3$ reads 
\begin{equation} 
B_\mu^3 = -i z^{\dagger} \tilde{\partial}_{\mu} z.
\end{equation} 
Since $d^\alpha$ defines $z$ only up to a phase, the Lagrangian
(\ref{af:cp}) has a {\it local} gauge symmetry:
\begin{equation}    \label{af:gauge}
z(x) \rightarrow {\rm e}^{i \alpha(x)} z(x).
\end{equation} 
In the original variables, such a transformation corresponds to a spin
rotation about the preferred spin axis.  That is, the unbroken component of
the global SO$^S$(3) spin rotation group becomes a {\it local} gauge symmetry
in the effective theory.  We will refer to this symmetry as local spin gauge
symmetry.  The reader is referred to Ref.~\cite{BKY} for a general discussion
on this phenomenon.

We next turn to the topological term.  When written in terms of the spin gauge
field $B_{\mu}^3$, the Hopf term takes the form of a Chern-Simons term:
\begin{equation}     \label{af:CS}
\theta  S_{\rm CS} =   \frac{\theta}{4\pi^2} \int_x \epsilon_{\mu \nu
\lambda } B^3_\mu \tilde{\partial}_\nu B^3_\lambda,
\end{equation} 
where $\theta$ is a real parameter.  

The question whether a Chern-Simons term is induced at the quantum level
can be addressed in great generality.  To the author's knowledge, So
\cite{So}; Ishikawa and Matsuyama \cite{IMa} were the first to reveal the
underlying structure, showing that $\theta$ is determined by the Feynman
propagator $G(k)$ of the underlying fermionic theory
\begin{equation}      \label{af:theta}
\theta = \frac{1}{24\pi} \epsilon_{\mu \nu \lambda } \int_k
\, {\rm tr} \left( G \frac{\partial  G^{-1}}{\partial  k_\mu} G
\frac{\partial G^{-1}}{\partial k_\nu} G \frac{\partial  G^{-1}}{\partial
k_\lambda} \right).
\end{equation} 
They also showed that the right-hand side of (\ref{af:theta}) can acquire a
topological meaning.  This work appears to have been largely ignored in the
literature, and the results have later been rederived by others
\cite{Volovik89,theta}.

For $^3$He-a, where according to Eq.~(\ref{he3:20}) the inverse
propagator reads
\begin{equation} 
G_0^{-1}(k) = k_0 - \xi({\bf k}) \tau_3 -
\frac{\bar{\Delta}}{k_{\rm F}} {\bf k} \cdot \bbox{\tau} \sigma^3,
\end{equation} 
the $\theta$-parameter takes the value \cite{VSY}
\begin{equation} 
\theta = {\rm sgn}({\bf e}^1 \times {\bf e}^2) \pi.
\end{equation} 
Here, sgn denotes the signum, and it should be realized that the cross
product of two vectors is a scalar in two space dimensions.  The dependence
of $\theta$ on the orientation of the ${\bf e}$-frame is crucial because it
changes sign under both a parity transformation and time inversion.  Under a
parity transformation in 2+1 dimensions, one spatial coordinate is
reflected:
\begin{equation}  \label{af:parx}
(t,x_1,x_2) \rightarrow (t,-x_1,x_2);
\end{equation}
and an arbitrary vector $V_{\mu}$ transforms as:
\begin{equation}     \label{af:parV}
V_{0,2}(t,x_1,x_2) \rightarrow V_{0,2}(t,-x_1,x_2), \; V_1(t,x_1,x_2)
\rightarrow -V_1(t,-x_1,x_2).
\end{equation}   
The sign change in $\theta$ offsets the one in the Chern-Simons term ${\cal
L}_{\rm CS}$ under these transformations, so that the combination $\theta
{\cal L}_{\rm CS}$ is invariant \cite{Volovik89,VSY}.

The appearance of a Chern-Simons term in the effective theory of a
nonrelativistic model might seem surprising.  However, the two possible
terms $\epsilon_{i j} B^3_0 \partial_i B^3_j$ and $\epsilon_{i j} B^3_i
\partial_0 B^3_j$ one can write down are related by the requirement of
invariance under the spin gauge transformation $B^3_{\mu} \rightarrow
B^3_\mu + \tilde{\partial}_\mu \alpha$, in such a way that the two can
be combined into a single Chern-Simons term.  The presence of both terms
can be checked explicitly by invoking the derivative expansion we used
before.  The specific form of the interaction, reflected in the last
term of the vertex (\ref{af:vertex}), as well as of the propagator
(\ref{he3:20}), are crucial in obtaining them.

It was pointed out by Wilczek and Zee \cite{WiZe} that the Hopf term
imparts spin to the solitons of the (2+1)-dimensional O(3) nonlinear
sigma model.  These solitons, first discussed by Belavin and Polyakov
\cite{BeP}, belong to the class of topological defects that are regular
throughout coordinate space.  That is, they do not possess a singular core
as is often the case.  They are characterized by the topological charge
\cite{Rajaraman}
\begin{equation}   \label{af:topcharge}
Q = \frac{1}{8\pi} \int_{\bf x} \epsilon_{i j} \epsilon^{\alpha \beta
\gamma} d^\alpha \partial_i d^\beta \partial_j d^\gamma ,
\end{equation} 
where the spin indices $\alpha, \beta, \gamma$ run over $1,2,3$, and the
antisymmetric Levi-Civita symbol is defined such that $\epsilon^{123}
=1$.  We repeat that vector spin indices are always represented by
superscripts.  

The natural language for the description and classification of defects
in ordered systems is provided by homotopy theory.  (For general
accounts on homotopy theory see, for example, \cite{Mermin},
\cite{Trebin} or \cite{Mineev}.)  The way to characterize a general
defect of dimension $\epsilon_{\rm d}$ is to surround it by a spherical
surface of dimension $r$ such that \cite{ToKl}
\begin{equation} \label{af:homotopy}
r = d - \epsilon_{\rm d} - 1,
\end{equation} 
with $d$ the dimension of the medium under consideration.  The minus one
at the right-hand side represents the distance from the defect to the
surrounding hypersphere.  In each point of this surface, the Goldstone
fields define a map of the $r$-sphere $S^r$ onto the coset space $G/H$.
Topologically stable defects arise when the contour in the coset space
is not contractable. They are therefore classified by the homotopy
groups $\pi_r(G/H)$.  

For the case at hand, the topological charge (\ref{af:topcharge}) is the
winding number of the map
\begin{equation}  \label{af:winding} 
d^\alpha({\bf x}): {\rm S}^2_{\bf x} \rightarrow {\rm S}^2
\end{equation} 
of compactified space ${\rm S}_{\bf x}^2$ into the internal two-sphere
${\rm S}^2$ parameterized by $d^\alpha$, where it is to be noted that
the second homotopy group $\pi_2({\rm S}^2) = Z$.  The corresponding
topological current, which is conserved independently of the field
equations, is
\begin{equation}   \label{af:J}
J_\mu = \frac{1}{8\pi} \epsilon_{\mu \nu \lambda} \epsilon^{\alpha \beta
\gamma} d^\alpha \tilde{\partial}_\nu d^\beta \tilde{\partial}_\lambda
d^\gamma . 
\end{equation} 
This current may alternatively be written in terms of the field
$B^3_\mu$ as:
\begin{equation}   \label{af:top}
J_\mu = \frac{1}{2\pi} {\tilde G}_\mu,         
\end{equation} 
where the dual field ${\tilde G}_\mu$ is defined by 
\begin{equation}    \label{af:Ftilde}
{\tilde G}_\mu = \epsilon_{\mu \nu \lambda} \tilde{\partial}_\nu
B^3_\lambda.
\end{equation}  

The simplest soliton has $Q=1$ and is called a skyrmion.  Due to the
Chern-Simons term, skyrmions acquire fractional spin $\theta/2\pi$ and
statistics $\theta$ \cite{WiZe,BKW}.  The values $\theta = \pm \pi$
found in superfluid $^3$He-a imply that here the skyrmions have spin
$\tfrac{1}{2}$ and are fermions.  The fractional statistics (for an
introductory review, see Ref.\ \cite{KleinertPath}) imparted by the
topological term to the skyrmion can be easily understood by rewriting
the Chern-Simons term (\ref{af:CS}) as
\begin{equation}   \label{af:charge}
\theta {\cal L}_{\rm CS} = \frac{\theta}{2\pi} J_\mu B_\mu^3,  
\end{equation} 
where $J_\mu$ is the topological current (\ref{af:J}).  This shows that the
effect of the Chern-Simons term is to couple the topological current to the
spin gauge field $B^3_\mu$, with a charge $\theta/2\pi$.  In other words,
besides carrying flux, skyrmions also carry charge.  Now, if a skyrmion
winds around another skyrmion, its wavefunction will acquire a phase $\exp(i
\theta)$ through the Aharonov-Bohm effect, thus turning skyrmions into anyons
with statistics parameter $\theta$. 

\begin{figure}
\begin{center}
\epsfxsize=5.cm
\mbox{\epsfbox{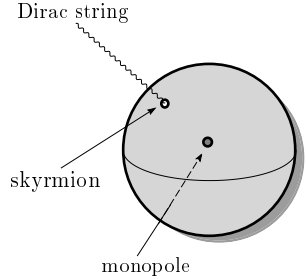}}
\end{center}
\caption{Compactified space represented by a two-sphere with a monopole at its
origin.  The Dirac string pierces the plane at the center of the skyrmion.
\label{fig:dirac}}
\end{figure}

A skyrmion configuration is given by
\begin{equation} 
d^\alpha({\bf x}) = \left( \begin{array}{c}    \cos \phi \, \sin u(r) \\
\sin \phi \, \sin u(r) \\ \cos u(r) \end{array} \right),  
\end{equation} 
where $(r, \phi)$ are circle coordinates in the spatial plane, while
$u(r)$ is a function with $u(r) = 0$ at the origin of the skyrmion
$(r=0)$ and $u(r) \rightarrow \pi$ for $r \rightarrow \infty$.  The
corresponding spin gauge-field configuration $B^3_{\mu}$, which can be
found by solving (\ref{af:rot}) for $s$, reads
\begin{equation} 
B^3_\mu = \tilde{\partial}_\mu \alpha + \tfrac{1}{2}(1 - \cos u)
\tilde{\partial}_\mu \phi,
\end{equation} 
where $\alpha$ is an arbitrary gauge parameter originating from the
circumstance that $s$ is determined by $d^\alpha$ only up to a phase factor
$\exp(i \alpha \sigma^3)$.  We fix the gauge by demanding that the
``magnetic'' flux 
\begin{equation} 
\Phi = \int_{\bf x} {\tilde G}_0
\end{equation}  
is regular everywhere.  Since $\epsilon_{i j} \partial_i \partial_j
\phi$ is singular at the origin, we choose $\alpha = 0$.  This leads to
\begin{equation}  \label{af:G0} 
{\tilde G}_0 = \frac{1}{2r} \frac{\dd u}{\dd r} \sin u.      
\end{equation} 
The corresponding magnetic flux $\Phi (R)$ through a disk of 
radius $R$ is 
\begin{equation}      \label{af:flux}
\Phi (R) = \pi [1 - \cos u(R)].                        
\end{equation} 
With $u(\infty) = \pi$, it follows that the flux piercing the plane is
$2\pi$.  That is, each skyrmion carries one unit of flux.  As pointed out by
Huang, Koike, and Polonyi \cite{HKP}, these conclusions may be nicely
visualized by placing a Dirac monopole with unit magnetic charge outside of
the system at the origin of a two-sphere representing the compactified
spatial plane, and letting the Dirac string pierce the plane at infinity.
Alternatively, one may extract a unit flux tube so that the Dirac string
pierces the plane at the origin where the center of the skyrmion resides
(see Fig.~\ref{fig:dirac}).

Let us mention that we are working in the ordered phase of the O(3)
nonlinear sigma model.  The (2+1)-dimensional model ${\cal L} =
(\partial_\mu d^\alpha)^2/2g^2$, with $g$ the coupling constant, has a
phase transition into a disordered, strong-coupling phase at \cite{AA}
\begin{equation} 
g_{\rm cr}^{-2} = \frac{3}{2\pi^2} \Lambda,    
\end{equation} 
where $\Lambda$ is an ultraviolet energy cutoff.  Since in our case,
according to (\ref{af:sigma}), $g^{-2} = \nu(0)/2 = m/4\pi$, with $m
\sim {\rm GeV} >> \Lambda$, we are indeed in the ordered, weak-coupling
phase.  The two phases differ in the way the so-called flux symmetry
generated by the topological current (\ref{af:top}) is realized
\cite{KRCP}.  In the weak-coupling phase the flux symmetry is unbroken,
while in the strong-coupling phase it is realized in the so-called
Kosterlitz-Thouless mode.  In the latter phase, there is---at least in
perturbation theory---algebraic long-range order due to the presence of
a gapless Kosterlitz-Thouless boson.  It has been pointed out by Huang,
Koike, and Polonyi \cite{HKP}, that a condensation of the Dirac
monopoles would invalidate this picture.  If this happens, they argued, the
flux symmetry becomes anomalous and the Kosterlitz-Thouless boson
acquires an energy gap so that the algebraic long-range order of the
ordered, strong-coupling phase is lost.  These authors found numerical
evidence supporting this scenario, but Bitar and Manousakis
\cite{BiMa} criticized their numerical analysis and concluded that the
monopoles do not spoil the perturbative picture.

We next study the connection between the microscopic fermionic model and the
effective theory, i.e., the nonlinear sigma model with a Chern-Simons term
added.  In particular, we will be interested in how the spin currents
$j^{\alpha}_\mu$ of the original model
\begin{eqnarray}   \label{af:jspin}
j^{\alpha}_0 &=& \psi^{\dagger} \tfrac{1}{2} \sigma^\alpha \psi \nonumber \\
j^{\alpha}_i &=& \frac{1}{2m}	\psi^{\dagger} \tfrac{1}{2}
\sigma^\alpha 
(-i\stackrel{\leftrightarrow}{\partial_i}) \psi,
\end{eqnarray} 
with $j^\alpha_0$ the spin density, are represented in the effective
model.  In (\ref{af:jspin}), $\stackrel{\leftrightarrow}{\partial_i} \,
= \partial_i -\! \stackrel{\leftarrow}{\partial}_i$, is the gradient
operator acting to the right and left.  Since the global SO$^S$(3) spin
rotation group is spontaneously broken to SO$^S$(2) in $^3$He-a, the two
components $j^{a}_{\mu}$, with $a = 1,2$ are supercurrents, meaning that
the corresponding charges are transported by the spin waves without
dissipation.  We shall demonstrate that the induced spin currents are
represented by the isospin currents $I^\alpha_\mu$ of the nonlinear
sigma model,
\begin{equation}  \label{af:iso}
I_\mu^\alpha = \epsilon^{\alpha \beta \gamma} d^{\beta}
\frac{\partial {\cal L}}{\partial \!\left(\partial_\mu
d^\gamma\right)} = \frac{1}{2} \nu(0) \epsilon^{\alpha \beta \gamma }
d^\beta \tilde{\partial}_\mu d^\gamma;
\end{equation}  
and that due to the presence of the Chern-Simons term, the spin currents
acquire an additional contribution.

Let us rewrite the spin currents (\ref{af:jspin}) in terms of $\chi$ and
the fields $B_{\mu} = -i s^{\dagger} \tilde{\partial}_\mu s$, where
$\psi$ and $\chi$ are related via (\ref{af:decom}).  We find
\begin{eqnarray}  \label{af:curr}
j^\alpha_0 &=& R^{\alpha \beta} \, \chi^{\dagger} \tfrac{1}{2}
\sigma^\beta
\chi \nonumber \\ j^\alpha_i &=& R^{\alpha \beta} \, \frac{1}{2m} \left[
\chi^\dagger \tfrac{1}{2} \sigma^\beta
(-i\stackrel{\leftrightarrow}{\partial_i}) 
\chi - \chi^{\dagger} B_i^{\beta} \chi \right],
\end{eqnarray} 
where $R^{\alpha \beta}$ is an orthogonal rotation matrix defined via
\begin{equation}       \label{af:R}
s^{\dagger} \sigma^\alpha s  = R^{\alpha \beta} \sigma^\beta.
\end{equation}  
If we differentiate the Lagrangian (\ref{af:consice}) with respect to the
gauge fields, we obtain
\begin{eqnarray}   \label{af:deri}
-\frac{1}{2} \frac{\partial {\cal L}}{\partial B_0^\alpha } &=& \chi^{\dagger}
\tfrac{1}{2} \sigma^\alpha \chi       \nonumber  \\
\frac{1}{2} \frac{\partial {\cal L}}{\partial B_i^\alpha} &=& 
\frac{1}{2m} \left[ \chi^\dagger \tfrac{1}{2} \sigma^\alpha 
(-i\stackrel{\leftrightarrow}{\partial_i}) \chi -
\chi^\dagger B_i^\alpha  \chi \right],              
\end{eqnarray} 
where in the last equation with $\alpha = 3$ we neglected the contribution
arising from the interaction term. This is justified because
$\bar{\Delta}/k_{\rm F} << 1$.  It follows from (\ref{af:curr}) and
(\ref{af:deri}) that
\begin{equation}  \label{af:source}
j_0^\alpha = -\frac{1}{2}  R^{\alpha \beta} \frac{\partial {\cal
L}}{\partial B^\beta_0}, \;\;\;
j_i^\alpha = \frac{1}{2}  R^{\alpha \beta} \frac{\partial {\cal
L}}{\partial B^\beta_i}
\end{equation} 
If we now wish to calculate $j_\mu^\alpha$ in the effective theory, we have to
replace ${\cal L}$ in this equation with the effective Lagrangian.  Omitting
the Chern-Simons term for the moment, we obtain in this way
\begin{equation} 
j_\mu^\alpha = - \nu(0) R^{\alpha  b} B^b_{\mu}, \;\;\;\;  (b=1,2).
\end{equation} 
We proceed to show that the right-hand side of this equation is the isospin
current (\ref{af:iso}).  To this end we write:
\begin{eqnarray} 
I^\alpha_\mu &=& \tfrac{1}{4} i \nu(0)   \tilde{\partial}_\mu
d^\beta d^\gamma {\rm tr}\, \sigma^\alpha \sigma^\beta \sigma^\gamma
\nonumber  \\ &=& \tfrac{1}{4}i \nu(0)  {\rm tr}\, s^{\dagger}
\sigma^\alpha \tilde{\partial}_\mu s + \tfrac{1}{4}\nu(0)  {\rm tr}\,
s^{\dagger} \sigma^\alpha s \sigma^3 B_\mu \sigma^3,
\end{eqnarray} 
where we used (\ref{af:rot}) and the definition of $B_{\mu}$.  Then,
employing (\ref{af:R}) and performing the trace, we see that the induced
spin currents are indeed given by the isospin currents of the nonlinear
sigma model \cite{a2}
\begin{equation}    \label{af:equi}
j_\mu^\alpha = I_\mu^\alpha.
\end{equation} 

We next turn our attention to the anomalous contribution to the spin 
currents arising from the Chern-Simons term:
\begin{equation} 
j_{\theta,\mu}^\alpha = - \frac{\theta}{4\pi^2} R^{\alpha 3} \,
\tilde{G}_\mu.
\end{equation} 
In particular, we are interested in the spin carried by a skyrmion.  The
zero-component of the dual field describing a skyrmion configuration is
given by Eq.~(\ref{af:G0}).  To find the spin, we calculate the projection
of the spin density $j_{\theta,0}^\alpha$ onto the spin-quantization axis.
Usually this axis is fixed in spacetime, but for a skyrmion configuration the
spin-quantization axis $d^\alpha({\bf x})$ varies in space.  The spin
$\sigma$ of a skyrmion is therefore given by
\begin{equation} 
\sigma = \int_{\bf x} d^\alpha j_{\theta,0}^\alpha.
\end{equation} 
This quantity is most easily evaluated  when use is made of the identity
\begin{equation}  \label{af:id}
d^\alpha R^{\alpha \beta} = \delta^{3 \beta},
\end{equation} 
which is readily derived by multiplying the definition of the $R$
matrix, Eq.~(\ref{af:R}), with $d^\alpha$ and using (\ref{af:rot}).  We
find in this way that the spin of a skyrmion is related to its
topological charge (\ref{af:top}) via
\begin{equation}   \label{af:an} 
\sigma =  \frac{\theta}{2 \pi} Q.       
\end{equation} 
More precisely, a skyrmion with topological charge $Q=1$ has a spin
$\theta/2 \pi$ in accord with its statistics.

This result can also be derived by directly employing (\ref{af:source}) with
(\ref{af:id}):
\begin{equation}  \label{af:source?}
d^\alpha j_0^\alpha = -\frac{1}{2} \frac{\partial {\cal L}}{\partial
B^3_0},
\end{equation} 
and replacing ${\cal L}$ with the Chern-Simons term.

\begin{figure}
\begin{center}
\epsfxsize=6.5cm
\mbox{\epsfbox{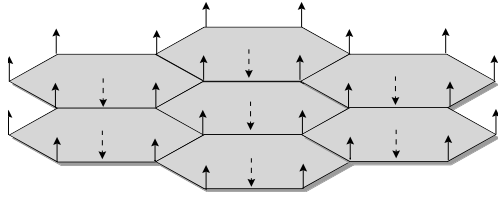}}
\end{center}
\caption{Hexagonal spin-$\tfrac{1}{2}$ skyrmion lattice.  The arrows denote
the direction of the spin vector $d^\alpha({\bf x})$.
\label{fig:lattice}}
\end{figure}
Equation (\ref{af:an}) is the anomalous contribution to the total spin
of the system.  Hence, the difference in the number
$N_{\uparrow,\downarrow}$ of spin-$\uparrow$ and spin-$\downarrow$
$^3$He atoms is given by the skyrmion number \cite{a2}
\begin{equation} 
N_{\uparrow} - N_{\downarrow} = \pm Q.                
\end{equation} 
It follows that a finite number of skyrmions can be created by taking
$N_{\uparrow} \neq N_{\downarrow}$.  If the difference $|N_{\uparrow} -
N_{\downarrow}|$ is large enough the skyrmions, which are fermions due to
the Chern-Simons term, may form a hexagonal lattice (see
Fig.~\ref{fig:lattice}) analogous to the one formed by so-called
Anderson-Toulouse-Chechetkin vortices \cite{ATC} in rotating superfluid
$^3$He-A \cite{SV}.  Experimentally, however, such a skyrmion lattice may be
difficult to realize because an external magnetic field necessary to spin
polarize the system locks the spin-quantization axis $d^\alpha$ to a plane
normal to the field.
\section{Peierls Instability}
\label{sec:rp}
In this section we shall discuss a model describing quasi
one-dimensional metals.  The effective low-energy, small-momentum
continuum theory is shown to have two phases both of which possess a
rich soliton structure.  The two phases differ in a topological term
which is induced in one phase, but not in the other.  This so-called
$\theta$-term is shown to change the statistics of certain solitons.

One-dimensional metals are frequently discussed in terms of the Hubbard model
defined by the lattice Hamiltonian \cite{Fradkin}
\begin{equation}    \label{rp:hu}
H_{\rm H} = -t  \sum_j (c^{\dagger}_{j \sigma} c_{j+1, \sigma} + {\rm 
h.c.}) + U \sum_j n_{j \uparrow} n_{j \downarrow}.   
\end{equation} 
It contains two parameters, $U$ and $t$, representing the Coulomb interaction
between a pair of valence electrons on the same atom, and the matrix element
for the hopping of an electron to a neighboring atom, respectively
\cite{EmIA}.  When $U$ is chosen positive, the Coulomb interaction is
repulsive.  In (\ref{rp:hu}), the operator $c^{\dagger}_{j \sigma}$
creates an electron of spin $\sigma (=\uparrow, \downarrow)$ at the
location of the $j$th atom, and $n_{j \sigma} = c^{\dagger}_{j \sigma}
c_{j \sigma}$ is the electron-number operator at site $j$.  The sum
$\sum_j$ is over all lattice sites. It is assumed that the Coulomb
interaction is highly screened, with a screening length of the order of
a lattice spacing $a$, so that we can restrict ourselves to the on-site
Coulomb repulsion.  In addition, it is assumed that the electrons are
well localized, so that the tight-binding approximation is valid.

The Hubbard Hamiltonian (\ref{rp:hu}) may be rewritten in the equivalent 
form
\begin{equation}    \label{rp:hu'}
H_{\rm H} = -t \sum_j(c^{\dagger}_{j \sigma} c_{j+1, \sigma} + {\rm
h.c.}) - \frac{2}{3} U \sum_j \left(S^\alpha_j\right)^2,
\end{equation}   
where we dropped an irrelevant constant.  Here, $S_j^\alpha =
\tfrac{1}{2} c^{\dagger}_{j \sigma} \sigma_{\sigma \tau}^\alpha c_{j
\tau}$, with $\sigma^\alpha$ $(\alpha=1,2,3)$ the Pauli matrices, is the
electron spin operator at site $j$.  The representation (\ref{rp:hu'})
which explicitly exhibits the spin operator is most convenient to derive
the effective theory.

In the large-$U$ limit, the Hubbard model at half-filling, where the
number of electrons equals the number of sites, can be mapped onto the
spin-$\tfrac{1}{2}$ Heisenberg model \cite{Fradkin}
\begin{equation}      \label{rp:hei}
H = J \sum_j S^\alpha_j S^\alpha_{j+1},       
\end{equation} 
with coupling constant $J=4t^2/U >0$.  Because $J$ is positive, spins on
neighboring sites favor antiferromagnetic coupling.  This model, solved by
Bethe, is known to be gapless \cite{Bethe}.  In the opposite limit of weak
coupling $U << t$, the Coulomb repulsion is a small perturbation and we expect
the model to behave qualitatively the same as a free-electron gas.  These two
arguments taken together make it plausible that the one-dimensional Hubbard
model at half-filling is gapless for all values of the coupling constant,
implying that it exhibits metallic behavior.

Although the Hubbard chain is meant to describe a one-dimensional metal, it
completely ignores the electron-phonon interaction.  This coupling,
particularly in one dimension, can however be very important.  A famous
argument due to Peierls \cite{Peierls} shows that a one-dimensional metal
with a partially filled conduction band is unstable towards a periodic
distortion of the linear lattice which opens an energy gap at the Fermi
points.  Because of this, Peierls concluded that such a one-dimensional
system at the absolute zero of temperature would probably never have
metallic properties and would instead be an insulator.  Fr\"ohlich 
\cite{Frohlich} showed that a periodic lattice distortion is accompanied by a
modulation of the electron number density with the same periodicity---a
so-called charge density wave  \cite{CDW}.

Being a fundamental property of one-dimensional metals, we wish to
extend the Hubbard model so as to account for the Peierls instability.
At half-filling, the periodic distortion has a period of twice the
lattice spacing.  The modulation of the electron-number density
resulting from this instability is given by
\begin{equation}        \label{rp:pei}
H_{\rm P} = \Delta \sum_{j \sigma} (-1)^j n_{j \sigma},    
\end{equation} 
where $\Delta$ is the so-called Peierls energy gap, it is typical of the order
of 100 K.  By adding this term to the Hamiltonian (\ref{rp:hu'}) we account for
a crucial aspect of the one-dimensional electron-phonon interaction in the
Hubbard model.  

We examine the model in the limit of low energy and small momentum by going to
the continuum.  This is facilitated by introducing the operators $a_{\rm e},
\, a_{\rm o}$ with engineering dimension $\tfrac{1}{2}$ via
\begin{eqnarray}      \label{rp:cc}
c_{2j} &\rightarrow& i^{2j} \sqrt{2a} \, a_{\rm e}(2j) \nonumber \\
c_{2j+1} &\rightarrow& i^{2j+1} \sqrt{2a} \, a_{\rm o}(2j+1),
\end{eqnarray} 
for even and odd sites, respectively.  Here, as well as in the following, we
suppress spin indices.  Summation over these hidden indices is always
implied.  The powers of $i$ in (\ref{rp:cc}) represent the factor $\exp{(i
k_{\rm F} x_1)}$, with $x_1= 2ja$ for even sites and $x_1=(2j+1)a$ for odd
sites, respectively, where $k_{\rm F} = \pi/2a$ is the Fermi momentum at
half-filling.  In terms of the new operators, the hopping term of the
Hubbard model becomes
\begin{eqnarray}     
-t \sum_{j} (c^{\dagger}_{j} c_{j+1} + {\rm h.c.}) \! \rightarrow
&& \!\!\!\!\!\!\!\!\!\! -2ita 
\sum_j \Bigl\{ \left[ a^{\dagger}_{\rm e}(2j) - a^{\dagger}_{\rm 
e}(2j+2) \right] a_{\rm o} (2j+1) \nonumber \\ && \;\;\;\;\;\;\;\; +
a_{\rm o}^{\dagger} (2j+1) \left[ a_{\rm e}(2j+2) - a_{\rm e}(2j) \right]
\Bigr\} .\nonumber \\
\end{eqnarray} 
With the derivative of the operator $a_{\rm o}$ defined by
\begin{equation}
\partial_1 a_{\rm o} = \lim_{a \rightarrow 0} \frac{a_{\rm o}(2j+1) - a_{\rm 
o}(2j-1)}{2a},
\end{equation}  
and a similar definition for $\partial_1 a_{\rm e}$, we obtain for this term
in the continuum limit
\begin{equation}           \label{rp:chop}
-t \sum_{j} (c^{\dagger}_{j} c_{j+1} + {\rm h.c.}) \rightarrow 2ta
\int_{x_1} \psi^\dagger \alpha (-i \partial_1) \psi.
\end{equation} 
Here, we introduced the multiplet $\psi$ containing the even-site and
odd-site species of fermions,
\begin{equation}
\psi = \left( \begin{array}{c} a_{\rm e} \\ a_{\rm o} \end{array} 
\right), 
\end{equation} 
with $\alpha = \tau_1$ the first Pauli matrix which should not be
confused with the spin matrices $\sigma^\alpha$.  In deriving (\ref{rp:chop})
we replaced the summation over the lattice sites with an integral: $2a
\sum_j \rightarrow \int \dd x_1$.  The right-hand side of (\ref{rp:chop}) is
of the form of a massless Dirac Hamiltonian.  It describes Bloch electrons
with a gapless dispersion relation that is linear in the crystal momentum
$k_1$
\begin{equation}       \label{rp:dis}
E(k_1) = v_{\rm F} |k_1|,                           
\end{equation} 
where $v_{\rm F} =2ta$ is the Fermi velocity.  A one-dimensional model
described by this dispersion relation would exhibit metallic properties.  As
expected, the term (\ref{rp:pei}) representing the Peierls instability
drastically changes this behavior.  In the continuum limit, this term becomes
a mass term in the Dirac theory
\begin{eqnarray}            \label{rp:cpei}
\Delta \sum_{j} (-1)^j n_{j} \!\!\!\!\!\!\!\! && \rightarrow 2a \Delta
\sum_j \left[ a^{\dagger}_{\rm e}(2j) a_{\rm e}(2j) - a^{\dagger}_{\rm o}(2j+1)
a_{\rm o}(2j+1) \right] \nonumber \\
\!\!\!\!\!\!\!\! && \rightarrow  \Delta \int_{x_1}  \psi^\dagger \beta \psi,
\end{eqnarray}   
with $\beta = \tau_3$ being the diagonal Pauli matrix.  As a result, the
gapless spectrum (\ref{rp:dis}) changes into a spectrum with an energy gap
\begin{equation}                    \label{rp:dis'}
E^2(k_1) = v_{\rm F}^2 k_1^2 + \Delta^2,         
\end{equation} 
and the metallic properties of the free model are lost.

We treat the Coulomb interaction of the Hubbard model (\ref{rp:hu'}) by
replacing it with
\begin{equation}         \label{rp:rep}
- \frac{2}{3} U \sum_j \left(S^\alpha_j\right)^2 \rightarrow \frac{4}{3} U
\sum_j M_j \left[ \frac{M_j}{2} \left(d^\alpha_j\right)^2 - (-1)^j S^\alpha_j 
d^\alpha_j \right],
\end{equation} 
where the operator field $d^\alpha_j$ satisfies the constraint equation
\begin{equation}
M_j d^\alpha_j = (-1)^j S^\alpha_j
\end{equation} 
and $\left(d^\alpha_j\right)^2 = 1$ at every site $j$.  In the low-energy,
small-momentum limit, the modulus $M_j$ can be taken as a constant
$\bar{M}$. The factor $(-1)^j$, alternating sign from site to site, is included
in (\ref{rp:rep}) because we are interested in the antiferromagnetic
properties of the model.  In the mean-field approximation, the operator
$d^\alpha_j$ is considered to be a classical field.  It then describes the
staggered magnetization and thus becomes the order parameter of the N\'eel
state.  In this approximation, the continuum limit of the relevant term of
the Coulomb interaction becomes
\begin{equation}      \label{rp:ccou} 
- \frac{4}{3} \bar{M} U \sum_j (-1)^j S^\alpha_j d^\alpha_j \rightarrow \Sigma
\int_{x_1} \psi^\dagger \beta d^\alpha \sigma^\alpha \psi, 
\end{equation} 
where we introduced the effective coupling constant $\Sigma = -\tfrac{2}{3}
\bar{M} U$.  The term (\ref{rp:ccou}) together with the two other terms
(\ref{rp:chop}) and (\ref{rp:cpei}) describe the N\'eel state of the
extended Hubbard model in the mean-field approximation.  The corresponding
Lagrangian reads
\begin{equation}       \label{rp:lag}
{\cal L} = {\bar \psi}  (i \overlay{/}{\tilde \partial} - \Delta - \Sigma
\, d^\alpha \sigma^\alpha)\psi, \;\;\;\; \;\;\; {\bar \psi} = \psi^{\dagger}
\beta ,                                       
\end{equation} 
where the Fermi velocity $v_{\rm F}$ is set to unity, and
\begin{equation} \label{rp:Dirac} 
\gamma_0 = \beta = \tau_3, \;\;\;\; \gamma_1 = \beta \alpha = i \tau_2
\end{equation} 
are the Dirac matrices.  Moreover, $\overlay{/}{\tilde \partial} =
\tilde{\partial}_\mu \gamma_\mu$.  Although mean-field theory should in
general not be trusted in lower dimensions, it in this case seems to capture
the essential physics.

The effective theory we are seeking is obtained by integrating out the
electron fields.  (We tacitly switched here from the operator formalism to
the functional-integral approach.)  In our discussion, $\Delta$ is
considered to be a constant.  If one is interested in the dynamics of the
charge density wave, which we are not, $\Delta$ is assumed to have a
spacetime-dependent phase, $\Delta (x) = \bar{\Delta} \exp[i\varphi (x)]$.

Since the Lagrangian (\ref{rp:lag}) is quadratic in the electron fields, these
are readily integrated out.  The ensuing effective action $S_{\rm eff}$ reads
\begin{equation}                   \label{rp:effs}
S_{\rm eff} = -i {\rm Tr}\, \ln ( \overlay{/}{p} - \Delta -
\Sigma \, d^\alpha  \sigma^\alpha),
\end{equation} 
which we again evaluate in a derivative expansion \cite{FAF}.  On our way, we
encounter integrals which diverge in the ultraviolet.  To handle these, we
use dimensional regularization and generalize the integrals to arbitrary
spacetime dimensions $D$.

The lowest-order term is of the form $r (d^\alpha)^2$, where $r$ is a
constant that tends to infinity when the dimensional-regularization
parameter $\epsilon = 1 - \tfrac{1}{2} D$ is taken to zero.  This term
merely renormalizes the first term at the right-hand side of (\ref{rp:rep}).
Neglecting for the moment a possible topological term, we obtain as next
order in derivatives the contribution \cite{peierls}
\begin{equation}   \label{rp:o3}
{\cal L}_{\rm eff} = \frac{1}{2g^2} \left(\partial_{\mu} d^\alpha\right)^2,
\end{equation} 
where $1/g^2 = \Sigma^2/6 \pi \Delta^2$.  This is a kinetic term for the
staggered magnetization induced by quantum effects.  Note that $1/g^2 > 0$,
which is required for stability.  The O(3) nonlinear sigma model
(\ref{rp:o3}) in 1+1 spacetime dimensions is known to have only one phase,
viz.\ the so-called quantum disordered phase, with a finite correlation
length \cite{Polyakov}.  Due to strong infrared interactions of the
``Goldstone modes'', they acquire an energy gap for all values of the
coupling constant $g$.  For this reason one frequently refers to the
(1+1)-dimensional state described by the effective theory (\ref{rp:o3}),
with the constraint $\left(d^\alpha\right)^2=1$, as a {\it short-ranged}
N\'eel state.  In higher spacetime dimensions the model has a phase
transition between an ordered, weak-coupling phase $(g < g_{\rm cr})$ and a
disordered, strong-coupling $(g > g_{\rm cr})$ phase, where $g_{\rm c}$ is
the critical coupling.

The form (\ref{rp:effs}) of the action is not convenient to derive the
topological term.  To obtain this term we instead follow Jaroszewics
\cite{Jaroszewics} and introduce, as we have done in the previous section, the
decomposition $\psi = s \chi$, with $s(x)$ the spacetime-dependent SU(2)
spin rotation matrix defined in (\ref{af:rot}).  The effect of this unitary
transformation is to rotate the staggered magnetization in every point of
spacetime into a fixed direction.  With the help of this decomposition, the
Lagrangian (\ref{rp:lag}) can be cast in the equivalent form
\begin{equation}               \label{rp:lag'}
{\cal L} = \bar{\chi} ( i \overlay{/}{\tilde \partial} - \Delta - \Sigma
\sigma^3 - \overlay{/}{B} \, ) \chi,
\end{equation} 
where $B_{\mu} = -i s^{\dagger} \tilde{\partial}_{\mu} s = B_{\mu}^{\alpha}
\sigma^\alpha$ is the $2 \times 2$ matrix-valued field we encountered
before in the context of superfluid $^3$He-a.  As was the case there,
this Lagrangian is invariant under the {\it local} spin gauge
transformation
\begin{equation}  \label{rp:local}
\chi(x) \rightarrow {\rm e}^{-i \alpha(x) \sigma^3} \chi(x), \;\;\;
B_{\mu}^3(x) \rightarrow B_{\mu}^3(x) + \tilde{\partial}_{\mu} \alpha(x),
\end{equation} 
with spin gauge field $B_{\mu}^3$.  This local gauge freedom derives from the
circumstance that the transformed rotation matrix $s' = s \exp(i\alpha
\sigma^3)$ also satisfies (\ref{af:rot}), i.e., $s'^{\dagger} \sigma^\alpha
d^\alpha s' = \sigma^3$.  That is to say, $d^\alpha$ determines the spin
rotation matrix $s$ only up to a multiplicative U(1) factor $\exp(i\alpha
\sigma^3)$.  In terms of $B_\mu$, the kinetic term $\left(\partial_\mu
d^\alpha\right )^2$ is given by
\begin{equation}      \label{rp:corr}
\left(\partial_\mu d^\alpha \right)^2 = 4 \left(B_\mu^a \right)^2, 
\end{equation} 
where the summation over the spin index only involves the components
$a=1,2$.  The kinetic term is thus seen to be independent of the spin gauge
field $B_\mu^3$.  The term $\left(B_\mu^3\right)^2$, which would give the
gauge field a mass, is not generated as it would violate local spin gauge
symmetry.  On the other hand, the topological term that in principle can
arise in this context depends only on $B_\mu^3$
\begin{equation}             \label{rp:top}
{\cal L}_{\theta} = \frac{\theta}{2 \pi} \epsilon_{\mu \nu}
\tilde{\partial}_\mu B_\nu^3.
\end{equation} 
It contains, as is frequently the case with topological terms, the
antisymmetric Levi-Civita symbol and it is linear in derivatives.  It
can alternatively be written in terms of the O(3) nonlinear
sigma field as \cite{Fradkin}:
\begin{equation}           
{\cal L}_{\theta} = \frac{\theta}{8\pi} \epsilon_{\mu \nu} \epsilon^{\alpha
\beta \gamma} d^\alpha \tilde{\partial}_\mu d^\beta \tilde{\partial}_\nu
d^\gamma,  
\end{equation} 
where we recognize the winding number of the map
\begin{equation} 
d^\alpha(x): {\rm S}^2_x \rightarrow {\rm S}^2
\end{equation} 
of compactified spacetime ${\rm S}_x^2$ into the internal two-sphere
${\rm S}^2$ parameterized by $d^\alpha$, cf.\ (\ref{af:winding}).

Since it is independent of $B_\mu^1$ and $B_\mu^2$, these components may, as
far as deriving the topological term is concerned, be neglected in the
Lagrangian (\ref{rp:lag'}).  This leads us to consider the following
Lagrangian
\begin{eqnarray}               \label{rp:lag''}
{\cal L} &=& \bar{\chi}_{\rm e} \left[ i \overlay{/}{\tilde \partial} -
(\Delta + \Sigma) - \overlay{/}{B}^{\, 3} \right]
\chi_{\rm e} \nonumber \\ && \!\!\!\! + \bar{\chi}_{\rm o} \left[
i \overlay{/}{\tilde \partial} - (\Delta - \Sigma) + \overlay{/}{B}^{\, 3}
\right] \chi_{\rm o},
\end{eqnarray}   
where we made explicit the two fermion species $\chi_{\rm e}$ and $\chi_{\rm
o}$, which are seen to have different masses and to couple to the local spin
gauge field in an opposite way.  This is because the N\'eel state is
characterized by a staggered magnetization, flipping sign when we go from an
even to an odd site.

To investigate whether the topological term (\ref{rp:top}) is induced by
the fermions, we could proceed as before and apply the derivative
expansion.  But in 1+1 dimensions we have another tool available, namely
that of bosonization, which we shall use instead.  Since
(\ref{rp:lag''}) contains only a U(1) symmetry, we can employ Abelian
bosonization.  One could also bosonize the original theory
(\ref{rp:lag}), as was first done by Wen and Zee \cite{WenZee}, but this
involves the more intricate non-Abelian bosonization.  The
Abelian bosonization rules---on which we shall comment in the next
chapter---read
\begin{eqnarray}    \label{rp:bosrules}
\bar{\chi}_{\rm e} i \overlay{/}{\tilde \partial} \chi_{\rm e} &\rightarrow&
\tfrac{1}{2}  (\partial_{\mu} \phi_{\rm e})^2 \nonumber \\
\bar{\chi}_{\rm e} \chi_{\rm e} &\rightarrow&  K \cos(\sqrt{4\pi} \phi_{\rm
e}) \\ \bar{\chi}_{\rm e} \gamma_\mu \chi_{\rm e} &\rightarrow&
\frac{1}{\sqrt{\pi}}
\epsilon_{\mu \nu} \tilde{\partial}_{\nu} \phi_{\rm e},   \nonumber
\end{eqnarray} 
with $K$ a for our purposes irrelevant positive constant and $\phi_{\rm e}$ a
real Bose field.  With these and similar rules for $\chi_{\rm o}$, the
Lagrangian (\ref{rp:lag''}) can be represented in bosonized form as
\begin{equation}        \label{rp:bos}
{\cal L}_{\rm bos} = \frac{1}{2} (\partial_\mu \phi_{+})^2 + \frac{1}{2}
(\partial_\mu \phi_{-})^2 - \sqrt{\frac{2}{\pi}} \epsilon_{\mu \nu} \phi_{-}
\, \tilde{\partial}_\mu  B_\nu^3 - {\cal V},
\end{equation} 
where we introduced the fields 
\begin{equation} 
\phi_{\pm} = (\phi_{\rm e} \pm \phi_{\rm o})/\sqrt{2}, 
\end{equation} 
and where the potential ${\cal V}$ is given by
\begin{equation}      \label{rp:pot}
{\cal V} = 2 \Delta K \cos (\sqrt{2\pi} \phi_{+}) \, \cos (\sqrt{2\pi}
\phi_{-}) - 2 \Sigma K \sin (\sqrt{2\pi} \phi_{+}) \, \sin (\sqrt{2\pi} 
\phi_{-}).                                          
\end{equation} 
Note that in (\ref{rp:bos}) only the field $\phi_{-}$ couples to the local
spin gauge field.  The potential ${\cal V}$ has a minimum
\begin{equation}   \label{rp:potmin}
{\cal V}_{\rm min} = \Bigl\{ \begin{array}{l} -2 |\Delta| K \;\;\; {\rm for}
\;\;\; |\Delta| > |\Sigma| \\ -2 |\Sigma| K \;\;\; {\rm for} \;\;\; |\Delta|
< |\Sigma|.  \end{array}                                     
\end{equation} 
at
\begin{equation}
\phi_{\pm} = 0, \; |\phi_{\mp}| =  \sqrt{\frac{\pi}{2}} \;\;\; {\rm for}
\;\; \Delta > \Sigma >0;                                        
\end{equation} 
\begin{equation}      \label{rp:min}
\phi_{+} = \phi_{-} = \pm \frac{1}{2} \sqrt{\frac{\pi}{2}} \;\;\; {\rm 
for} \;\; 0 < \Delta < \Sigma,
\end{equation} 
and similar values for $\Delta$ and $\Sigma$ negative.  With these
constant $\phi$-values, the third term in the Lagrangian (\ref{rp:bos})
takes the form of the topological term (\ref{rp:top}).  We find that the
$\theta$-parameter, which is an angle, is given by \cite{peierls}
\begin{equation}           \label{rp:th}
\theta = \pi [{\rm sgn}(\Delta + \Sigma) - {\rm sgn}(\Delta - \Sigma)],
\end{equation} 
valid for arbitrary sign of $\Delta$ and $\Sigma$.  

More specific, $\theta = 0 \, ({\rm mod} \, 2\pi)$ when $|\Delta| >
|\Sigma|$, and $\theta = \pi \, ({\rm mod} \, 2\pi)$ when $|\Delta| <
|\Sigma|$.  That is, for $|\Delta| > |\Sigma|$ the effective theory is
simply the O(3) nonlinear sigma model (\ref{rp:o3}), with coupling
constant squared $g^2 > 6
\pi$.  As we remarked before, the state it describes is the short-ranged N\'eel
state with finite correlation length. 

\begin{figure}
\begin{center}
\epsfxsize=5.5cm
\mbox{\epsfbox{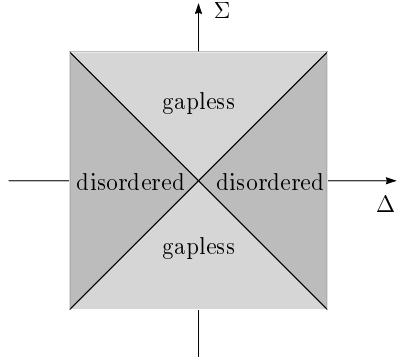}}
\end{center}
\caption{Phase diagram of the extended Hubbard model (\protect\ref{rp:lag}),
where $\Sigma$ represents the on-site Coulomb interaction and $\Delta$ is the
Peierls energy gap. \label{fig:phasedia}}
\end{figure}

For $|\Delta| < |\Sigma|$, on the other hand, (\ref{rp:o3}) is to be
augmented by the topological term (\ref{rp:top}).  In terms of the field
$d^\alpha$, the effective theory now reads
\begin{equation}     \label{rp:o3+}
{\cal L}_{\rm eff} = \frac{1}{2g^2} \left(\partial_{\mu}
d^\alpha\right)^2 + \frac{\theta}{8\pi} \epsilon_{\mu \nu}
\epsilon^{\alpha \beta \gamma} d^\alpha
\tilde{\partial}_\mu d^\beta \tilde{\partial}_\nu d^\gamma, 
\end{equation} 
with $\theta = \pm \pi$.  The $\theta$-term changes the physical content
of the O(3) nonlinear sigma model in a dramatic way.  This term has
been extensively studied in the context of the antiferromagnetic
Heisenberg spin chain which can also be mapped onto the nonlinear
sigma model \cite{HaA}.  Whereas the half-integer spin chain is known
to be gapless, the integer spin chain has a gap in the excitation
spectrum---the so-called Haldane gap.  The two cases differ in the
topological term which is induced with a coefficient $\theta = 2
\pi \sigma$ $({\rm mod} \, 2 \pi)$, where $\sigma$ is the spin.  For
integer spins the coefficient is zero and the excitation spectrum has an
energy gap.  For half-integer spins, on the other hand, the coefficient is
nonzero and it can be shown that as a consequence the spectrum becomes
gapless \cite{SRA}.  Using these results, we obtain the phase diagram
depicted in Fig.~\ref{fig:phasedia}.  The standard Hubbard chain is
recovered by setting $\Delta = 0$.  It is therefore effectively described by
(\ref{rp:o3}), implying that the model has a gapless spectrum and thus
metallic properties for all values of the Coulomb interaction $U$.  The
disordered phase of the effective theory, describing the insulating state of
the extended Hubbard model, becomes accessible when the Peierls energy gap
$\Delta$ is taken sufficiently large compared to $\Sigma$, measuring the
strength of the Coulomb interaction.

The fermions of the original theory appear in the bosonized theory as
solitons to which we next turn.  An example of a soliton in the regime $0 <
\Delta < \Sigma$, where the groundstates are given by (\ref{rp:min}), is the
configuration $\phi_{+}(x_1)=\phi_{-}(x_1)=f(x_1)$, where the function
$f(x_1)$ interpolates between the two different ground states
\begin{equation}
f(-\infty) = \frac{1}{2} \sqrt{\frac{\pi}{2}}, \;\;\; f(\infty) = -\frac{1}{2}
\sqrt{\frac{\pi}{2}}.
\end{equation} 
This soliton carries one unit of fermion-number charge $N$.  This follows
because according to the bosonization rules (\ref{rp:bosrules}), the
fermion-number current $j_\mu$ is represented as
\begin{equation}
j_\mu = \sqrt{\frac{2}{\pi}} \epsilon_{\mu \nu} \tilde{\partial}_\nu \phi_{+},
\end{equation} 
implying that
\begin{equation}
N = \int_{x_1} j_0 = -\sqrt{\frac{2}{\pi}} \left[\phi_{+}(\infty) - 
\phi_{+}(-\infty)\right] = 1.
\end{equation} 
Similar solitons of unit fermion-number charge can be constructed in the
other sectors of the phase diagram.

It can be easily demonstrated that these solitons carry in addition a spin
$\tfrac{1}{2}$, so that they have the same quantum numbers as the original
fermions \cite{WenZee}.  Using arguments similar to those of the previous
section, one can show that the spin current, given by
\begin{equation} 
j_\mu^{\alpha} = {\bar \psi} \tfrac{1}{2} \sigma^{\alpha} \gamma_{\mu}\psi,
\end{equation} 
is again represented in the effective theory by the isospin current
$I_{\mu}^{\alpha}$ of the O(3) nonlinear sigma model:
\begin{equation}
I_\mu^\alpha = \epsilon^{\alpha \beta \gamma} d^{\beta} \frac{\partial {\cal
L}}{\partial (\partial_\mu d^\gamma)}.
\end{equation} 
Also Eq.~(\ref{af:source}) is valid, so that
\begin{equation} 
d^\alpha j_0^\alpha = -\frac{1}{2} \frac{\partial {\cal L}}{\partial
B^3_0}= -\frac{1}{2} \sqrt{\frac{2}{\pi}} \partial_1 \phi_{-}.
\end{equation}
This gives as spin for the soliton under consideration
\begin{equation} 
\sigma = \int_{x_1}  d^\alpha j_0^\alpha = \pm \frac{1}{2}.
\end{equation} 
It may therefore be thought of as representing the fundamental fermion of
the original theory.  Similar conclusions hold for the solitons in the other
sectors of the phase diagram. 

Up to this point we have only considered the part of the bosonized theory
involving the scalar fields $\phi_\pm$.  We now turn to the nonlinear
sigma field $d^\alpha$ and will argue that twists in this field may give
certain solitons an anomalous spin.  For definiteness we shall assume in the
following that the space dimension is compactified into a circle $-L \leq
x_1 \leq L$, with $x_1 = -L$ and $x_1 = L$ labeling the same point.  The
model (\ref{rp:lag}) admits two possible boundary conditions for $d^\alpha$,
namely periodic $d^\alpha (-L) =d^\alpha (L)$ and antiperiodic $d^\alpha
(-L) = - d^\alpha(L)$ ones \cite{WenZee}.  That is, the two phases split up
into an even and odd sector.

We are interested in the contribution $j_{\theta, \mu}^\alpha$ to the spin
current stemming from the topological term ${\cal L}_{\theta}$:
\begin{equation}
j_{\theta, \mu}^{\alpha} = \epsilon^{\alpha \beta \gamma} d^\beta
\frac{\partial {\cal L}_{\theta}}{\partial  (\partial_\mu
d^\gamma )} =
\frac{\theta}{4\pi} \epsilon_{\mu \nu} \tilde{\partial}_\nu d^\alpha.
\end{equation} 
The corresponding charge $S_\theta^{\alpha}$ is zero in the even sector,
and
\begin{equation}
S_\theta^{\alpha} = \int_{x_1} j_{\theta, 0}^{\alpha} =
\frac{\theta}{4\pi}
\left[ d^\alpha(L) - d^\alpha(-L) \right] = \frac{\theta}{2\pi}
d^\alpha(-L),
\end{equation} 
in the odd sector of the theory.  It follows that $S_\theta^\alpha$ is
nonzero only in the gapless phase where $\theta = \pm \pi$, and with
antiperiodic boundary conditions imposed on $d^\alpha$.  A soliton in this
part of the theory acquires consequently an extra contribution
$\sigma_{\theta}$ to the spin given by
\begin{equation}
\sigma_\theta = \pm \tfrac{1}{2}.
\end{equation} 
Since the solitons have a canonical spin of $\tfrac{1}{2}$, the topological
term in the odd sector of the gapless phase transmutes these fermions into
bosons \cite{WenZee}.  This ability of ${\cal L}_\theta$ to change
statistics it shares with the Chern-Simons term in (2+1)-dimensional
theories.
         
We next consider the critical lines $\Delta = \pm \Sigma$.  At these lines,
either the component $\chi_{\rm e}$ or $\chi_{\rm o}$ in (\ref{rp:lag''}) is
gapless, and the above analysis does not apply.  Let us first focus on the
critical line $\Delta = \Sigma$.  The bosonized form of (\ref{rp:lag''})
then reads
\begin{eqnarray}        \label{rp:bos'}
{\cal L} &=& \frac{1}{2} (\partial_\mu \phi_{\rm e})^2 + \frac{1}{2}
(\partial_\mu \phi_{\rm o})^2 - (\Delta + \Sigma) K
\cos(\sqrt{4\pi} \phi_{\rm e}) \nonumber \\ && - \sqrt{\frac{1}{\pi}}   
\epsilon_{\mu \nu} (\phi_{\rm e} - \phi_{\rm o})  \tilde{\partial}_{\mu}
B_\nu^3 , 
\end{eqnarray} 
where the gaplessness of the field $\chi_{\rm o}$ is reflected in the
absence of a cosine term for $\phi_{\rm o}$.  As a result, the
integration over this field is a simple Gaussian which can easily be
carried out to give a gauge-invariant mass term for the local spin gauge
field
\begin{equation}
{\cal L}_{\rm mass} = \frac{1}{2\pi} B^3_\mu \left( g_{\mu \nu} - 
\frac{\tilde{\partial}_\mu \tilde{\partial}_\nu}{\partial^2} \right) B^3_\nu.
\end{equation} 
This mass generation by gapless fermions in two spacetime dimensions is the
famous Schwinger mechanism \cite{Schwingermodel}.  The contribution of
$\chi_{\rm e}$ to the effective theory may be approximated by substituting
the value $\phi_{\rm e} = 0$, or $\sqrt{\pi}/2$ in (\ref{rp:bos'}),
corresponding to the minimum of the potential $(\Delta + \Sigma) K
\cos(\sqrt{4\pi}\phi_{\rm e})$ for $\Delta = \Sigma < 0$, or $\Delta =
\Sigma > 0$.  It follows that only in the latter case a topological term
with $\theta = \pi$ is generated.  The Schwinger mechanism operates
similarly at the critical line $\Delta = -\Sigma$.  Here, the topological
term (with $\theta = \pi$) persists only at the lower half of the line,
defined by $\Delta = -\Sigma>0$.
\section{Statistics-Changing Phase Transition}
\label{sec:pt}
The (1+1)-dimensional model (\ref{rp:lag'}) we discussed in the previous
section has similar characteristics in 2+1 dimensions
\cite{DoMa,KoRocat,prd,neuchatel}.  While the Dirac Hamiltonian naturally
appears in (1+1)-dimensional systems where it describes the fermionic
excitations around the two Fermi points \cite{Fradkin}, this is not the case
in 2+1 dimensions.  However, there exist two-dimensional systems---so-called
semimetals---that have point-like Fermi surfaces \cite{semimetals}.  In
these materials, the valence and conduction band intersect in discrete
points.  Near such a degeneracy of two energy levels, the Hamiltonian
describing the full theory may be approximated by a $2 \times 2$ Dirac
Hamiltonian describing just the two-level subsystem in the vicinity of the
diabolic points \cite{GWS84,ll1,gordon}.  In this section, we briefly
discuss some of the salient features of the (2+1)-dimensional model.

For our purposes, it suffice to consider the simplified, Abelian version of
the model given by (\ref{rp:lag''}).  We have chosen a two-dimensional
representation of the Dirac algebra
\begin{equation}   \label{pt:gamma}
\gamma_0 = \beta = \tau_3, \;\; \gamma_1 = i \tau_2, \;\; \gamma_2 = i
\tau_1,   
\end{equation} 
\begin{equation}   \label{pt:dirac}
\{\gamma_\mu, \gamma_\nu\} = 2 g_{\mu \nu}, \;\;  g_{\mu \nu} = {\rm 
diag}(1,-1,-1),                                       
\end{equation} 
with $\tau$ the Pauli matrices.  With the help of the derivative
expansion \cite{FAF}, one can easily compute the induced fermion-number
current $j_\mu$.  It is given by \cite{prd}
\begin{equation}  \label{pt:j}
j_\mu = \frac{\vartheta}{2\pi} \tilde{G}_\mu,
\end{equation} 
where $\tilde{G}_\mu$ is the dual field $\tilde{G}_\mu = \epsilon_{\mu \nu
\lambda} \tilde{\partial}_\nu B^3_\lambda$ and
\begin{equation} 
\vartheta = \tfrac{1}{2} [{\rm sgn}(\Delta + \Sigma) - {\rm
sgn}(\Delta -\Sigma)].
\end{equation} 
The first term here is the contribution from the even-site spinor $\chi_{\rm
e}$, while the second term is the contribution from the odd-site spinor
$\chi_{\rm o}$.  The relative minus sign reflects the opposite charge of
$\chi_{\rm e}$ and $\chi_{\rm o}$ with respect to the local spin gauge
symmetry.  We encountered the dual field $\tilde{G}_\mu$ previously in the
context of superfluid $^3$He-a [see (\ref{af:Ftilde})], where it was shown
that it is proportional to the topological current $J_\mu$ of the O(3)
nonlinear sigma model.  The corresponding charge $Q$ is the winding number
of the map (\ref{af:winding}).

We thus arrive at the conclusion that the induced fermion-number current
$j_\mu$ is proportional to the topological current $J_\mu$ of the effective
theory,
\begin{equation}   \label{pt:link}
j_\mu = \vartheta J_\mu.
\end{equation} 
The proportionality constant $\vartheta$ is nonzero for $|\Sigma| >
|\Delta|$, where ${\rm sgn}(\Delta + \Sigma) - {\rm sgn}(\Delta -\Sigma) = 2
{\rm sgn}(\Sigma)$, and zero for $|\Sigma| < |\Delta|$.

Let us first discuss the case $|\Sigma| > |\Delta|$.  The situation of
two different conserved currents which become proportional to each other
is typical for ordered states where the residual symmetry links up
different groups of the original symmetry.  For the model under
consideration, (\ref{pt:link}) implies that the fermion-number symmetry
U(1)$^N$ generated by the charge $N = \int_{\bf x} j_0$, and the flux
symmetry U(1)$^\Phi$ generated by the flux
\begin{equation} \label{pt:flux}
\Phi = \int_{\bf x} \tilde{G}_0,
\end{equation} 
are spontaneously broken in the following manner:
\begin{equation}   \label{pt:sb}
{\rm U(1)}^N \times {\rm U(1)}^{\Phi} \supset {\rm U(1)}^{N - \vartheta
\Phi}.
\end{equation} 
To identify the Abelian Goldstone mode associated with this spontaneous
symmetry breakdown, we also calculate the induced ``Maxwell'' term,
\begin{equation} 
{\cal L} = - \tfrac{1}{4} \Pi G^2_{\mu \nu},
\end{equation} 
with $G_{\mu \nu} = \tilde{\partial}_\mu B^3_\nu - \tilde{\partial}_\nu
B^3_\mu$.  To the one-loop order we find for the coefficient $\Pi$ appearing
here \cite{prd}
\begin{equation} 
\Pi = \frac{1}{8\pi} \left( \frac{1}{|\Delta + \Sigma|} + 
\frac{1}{|\Delta-\Sigma|}\right).
\end{equation} 
If we imagine integrating over the local spin gauge field, the partition
function can be written as
\begin{equation}  \label{pt:Z} 
Z = \int \DD B^3_\mu \exp\left[ i \int_x \left( -\tfrac{1}{4} \Pi G_{\mu
\nu}^2  - \vartheta \epsilon_{\mu \nu \lambda} A_\mu \tilde{\partial}_\nu
B^3_\lambda \right)\right],
\end{equation} 
where we omitted a gauge-fixing factor for the spin gauge field and coupled
the current $j_\mu$ to a background field $A_\mu$.  This coupling allows us,
by differentiating with respect to $A_\mu$, to compute the induced
fermion-number current (\ref{pt:link}).

To make the Goldstone mode explicit we employ a procedure frequently used to
derive dual theories and introduce the change of variables $B^3_\mu
\rightarrow \tilde{G}_\mu$ in the functional integral.  Since the dual field
$\tilde{G}_\mu$ fulfills the Bianchi identity $\tilde{\partial}_\mu
\tilde{G}_\mu = 0$, we introduce a Lagrange multiplier $\varphi$ in the
functional integral to impose this constraint:
\begin{equation}   \label{pt:Max}
Z = \int \DD \tilde{G}_\mu \DD \varphi \exp \left[ i \int_x \left( -
\tfrac{1}{2} \Pi  \tilde{G}^2_\mu - \vartheta A_\mu \tilde{G}_\mu + \varphi 
\tilde{\partial}_\mu \tilde{G}_\mu \right) \right].   
\end{equation} 
Performing the Gaussian integral over the dual field $\tilde{G}_\mu$, we
obtain an expression for the partition function in terms of a gapless scalar
field:
\begin{equation} 
Z = \int \DD \varphi \, \exp \left[\frac{i}{2 \Pi}  \int_x 
(\tilde{\partial}_\mu \varphi + \vartheta A_\mu)^2 \right].
\end{equation} 
This is the Goldstone field we were seeking.  It should be noted that in 2+1
dimensions, both a massless vector field and a real scalar field represent
one degree of freedom.  This can be easily understood by noting that a
photon has only one transverse degree of freedom in two space dimensions.
In terms of the Goldstone field $\varphi$, the fermion-number current
becomes
\begin{equation} \label{pt:current}
j_\mu = -\frac{\vartheta}{\Pi}(\tilde{\partial}_\mu \varphi + \vartheta
A_\mu)
\end{equation} 
as in BCS theory.  From this we conclude that the (2+1)-dimensional model
(\ref{rp:lag'}) exhibits superconductivity.  It should be noted that the
mechanism leading to this, with the spin gauge field playing a decisive
role, is entirely different from that in classic superconductors.

On comparing the expression (\ref{pt:current}) for the fermion-number
current with the one obtained directly from (\ref{pt:Z}), we conclude that
the dual field $\tilde{G}_\mu$ is related to the Goldstone field via
\begin{equation}  \label{pt:photon} 
\tilde{G}_\mu = - \frac{1}{\Pi} \tilde{\partial}_\mu \phi.
\end{equation} 

We next turn to the case $|\Sigma| < |\Delta|$.  When one of the two spinors
$\chi_{\rm e}$ or $\chi_{\rm o}$ become gapless, i.e., $\Delta + \Sigma =
0$ or $\Delta -\Sigma = 0$, the induced fermion-number current
discontinuously drops to zero, as was first noticed by Chen and Wilczek
\cite{ChWi}.  So, for $|\Sigma| < |\Delta|$ the argument that the
fermion-number current and the topological current are proportional, which
led to the conclusion that we have the spontaneous symmetry breaking
(\ref{pt:sb}), is invalid and the symmetries U(1)$^N$ and U(1)$^{\Phi}$
remain unbroken here.  The discontinuities in the induced charges at the
lines $|\Sigma| = |\Delta|$ merely reflect that these are critical lines
\cite{prd}.

The restoration of symmetry when one passes from the broken phase to the
symmetric phase should be accompanied by a loss of the gapless mode.  To see
that this is indeed the case we note that besides the Maxwell term, also a
Chern-Simons terms is generated here,
\begin{equation}   \label{pt:CS}
{\cal L}_\theta = \frac{\theta}{4\pi^2} \epsilon_{\mu \nu \lambda}
B^3_\mu \tilde{\partial}_\nu B^3_\lambda,
\end{equation} 
with $\theta = - \tfrac{1}{2} \pi [{\rm sgn}(\Delta + \Sigma) + {\rm
sgn}(\Delta -\Sigma)]$.  The first and second term here arise from the
even-site and odd-site spinors, $\chi_{\rm e}$ and $\chi_{\rm o}$,
respectively.  Observe that in the ordered, gapless phase where $|\Sigma| >
|\Delta|$, the Chern-Simons term is zero.  The Euler-Lagrange equation for
$B^3_\mu$ in the presence of a Chern-Simons term and with the background
field $A_\mu$ set to zero yields $\tilde{\partial}_\mu G_{\mu \nu} \propto
\tilde{G}_\nu$, or in dual form $\tilde{\partial}_\mu
\tilde{G}_\nu - \tilde{\partial}_\nu \tilde{G}_\mu \propto G_{\mu \nu}$.
Upon taking the divergence of this last equation and using the previous one
as well as the Bianchi identity $\tilde{\partial}_\mu \tilde{G}_\mu = 0$,
one finds
\cite{DJT}
\begin{equation} 
\partial^2 \tilde{G}_{\mu} \propto \tilde{G}_{\mu}.   
\end{equation} 
This equation together with (\ref{pt:photon}) shows that $\varphi$ has
become massive.  From this we conclude that due to the presence of the
Chern-Simons term in the symmetric phase, the gapless mode of the ordered
phase vanishes, as it should be.

A last point we like to mention is that the phase transition we have been
discussing here is a statistics-changing phase transition.  This is easily
understood using the result \cite{WiZe} that a Chern-Simons term imparts
spin to a skyrmion.  The presence of this term (\ref{pt:CS}) in the unbroken
phase turns a skyrmion, which is an ordinary boson in the broken phase where
no Chern-Simons term is generated, into a fermion.  Whence, on crossing the
phase boundary $|\Sigma|=|\Delta|$, skyrmions undergo a spin transmutation
and change their statistics from fermionic to bosonic or {\it vice versa}.
\section{Fluxons}
\label{sec:fl}
In this section, we examine a gas of electrons confined to a plane.  We
present a simple method to calculate the quantum numbers induced by a
uniform magnetic background field \cite{ll1,ll2}.  We will in particular
focus on the quantum numbers carried by a fluxon---a point-like object
carrying one unit of magnetic flux that can be pictured as being obtained by
piercing the spatial plane with a magnetic vortex.  It is found that it
carries both the charge and spin of a fermion.  Unlike what one expects, the
spin of the fluxon does not originate from a Chern-Simons term, but from a
so-called mixed Chern-Simons term involving two different gauge fields,
viz.\ the electromagnetic gauge field $A_\mu$ and spin gauge field $B_\mu^3$
which describes the spin degrees of freedom.

Let us consider the Lagrangian
\begin{equation}  \label{fl:Lag}
\mbox{$\cal L$} =\psi ^{\dagger}(i \partial_0 + \mu  -H_{{\rm
P}})\psi +  
b \psi ^{\dagger}\tfrac{1}{2} \sigma^3 \psi, \;\;\;\;\; \psi =
\left( \begin{array}{c} \psi_\uparrow \\ \psi_\downarrow \end{array} \right),
\end{equation} 
governing the dynamics of the Pauli spinor field $\psi$, with Grassmann
components $\psi_{\uparrow}$ and $\psi_{\downarrow}$ describing the
electrons of spin-$\uparrow$ and $\downarrow$ and chemical potential $\mu
\approx \epsilon_{\rm F}= k_{\rm F}^2/2m$ which is well approximated by the
Fermi energy.  We introduced an external source $b$ coupled to the spin
density, to compute the induced spin.  The Pauli Hamiltonian
\begin{equation}       \label{fl:Pauli} 
H_{\rm P} = \frac{1}{2m}({\bf p} - e {\bf A})^2- g_0 \mu_{\rm B}
\tfrac{1}{2} \sigma^3 H + eA_{0},                          
\end{equation} 
with $\mu_{\rm B}=e/2m$ the Bohr magneton and $g_0=2$ the free electron
$g$-factor, contains a Zeeman term which couples the electron spins to the
background magnetic field $H$.  Usually this term is omitted.  The reason is
that in realistic systems the $g$-factor is much larger two---the value for
a free electron.  In strong magnetic fields relevant to, say, the QHE, the
energy levels of
spin-$\downarrow$ electrons are too high to be occupied so that the system
is spin polarized, and the electron's spin is irrelevant to the problem.
Since we consider a free electron gas, we include the Zeeman term.  We
describe the uniform magnetic field $H$ by the vector potential $A_0=A_1=0;
A_2= H x_1$.  The energy eigenvalues of the Pauli Hamiltonian are the famous
Landau levels 
\begin{equation}    \label{fl:nrll}
E_l^\pm = \frac{|eH|}{m} \left(l+ \tfrac{1}{2} \right) - \frac{eH}{m}
\sigma_\pm,
\end{equation} 
with $\sigma_\pm = \pm \tfrac{1}{2}$ for spin-$\uparrow$ and spin-$\downarrow$
electrons, respectively.  

Integrating out the fermionic degrees of freedom, one finds the one-loop
effective action:
\begin{equation} 
S_{\rm eff}= \int_x {\cal L}_{\rm eff} = -i{\rm Tr}\ln\left(p_0 - H_{\rm P}+
\mu + \tfrac{1}{2}b \sigma^3\right).
\end{equation} 
The key observation to evaluate this expression is that with our gauge
choice, the theory is translational invariant.  Although a translation
by a distance $\Delta_1$ in the $x_1$-direction changes the vector
potential, $A_2 \rightarrow A_2 + H \Delta_1$, this change can be
canceled by a gauge transformation $A_\mu \rightarrow A_\mu +
\tilde{\partial}_\mu \alpha$, with gauge parameter $\alpha = H \Delta_1
x_2$, so that the theory is invariant under the combined symmetry.  As a
result, the theory effectively reduces to a (0+1)-dimensional theory, and
each Landau level is infinitely degenerate.  The number of degenerate states
per unit area is given by $|eH|/2 \pi$ for each level.  With these
observations, we can write ${\cal L}_{\rm eff}$ as
\begin{eqnarray}      \label{fl:nreff}
{\cal L}_{\rm eff} = -i \frac{|eH |}{2\pi } \sum_{l=0}^{\infty } 
\int_{k_0} \bigl[ \!\!\!\!\!\! && \!\!\!\!\!\! \ln \left(k_0-E_{l,+} +
\mu + \tfrac{1}{2}b\right) \nonumber \\ && \!\!\!\!\!\! +  \ln 
\left(k_0-E_{l,-} + \mu - \tfrac{1}{2}b \right)\bigr].
\end{eqnarray} 
The presence of only an integral over the loop energy and no momentum
integrals reflects that this is effectively a theory in zero space
dimensions.

The induced fermion number is obtained by differentiating ${\cal L}_{\rm
eff}$ with respect to the chemical potential, as can be inferred from the
original Lagrangian (\ref{fl:Lag}), and setting the external source $b$ to
zero,
\begin{equation} 
j_0 = \frac{\partial {\cal L}_{\rm eff}}
{\partial \mu}\Bigr|_{b=0}.
\end{equation} 
To evaluate the resulting energy integral, we employ the integral
\begin{equation}              
\int \frac{\dd k_0}{2\pi i}  \frac{{\rm e}^{i k_0 \eta }}
{k_0 + \xi - i \xi \eta } = \theta (\xi ) ,
\end{equation} 
containing, as usual in nonrelativistic calculations \cite{Mattuck}, an
additional convergence factor $\exp (i k_0 \eta)$.  The function
$\theta(\xi)$ at the right-hand side is the Heaviside unit step function.
The value for the induced fermion-number density thus obtained is
\begin{equation}    \label{fl:j0} 
j_0 = \frac{|eH|}{2\pi} (l_+ + l_-),                    
\end{equation} 
where $l_\pm$ is the number of filled Landau levels for spin-$\uparrow$ and 
spin-$\downarrow$ electrons, 
\begin{equation}     \label{fl:npm}
l_\pm = \Xi \left( \frac{m \mu_\pm }{|eH|} + \frac{1}{2}\right) ,
\end{equation}                     
with $\Xi(x)$ the integer-part function denoting the largest integer less
than $x$, and
\begin{equation}     \label{fl:chemical}
\mu_\pm = \mu + \frac{eH}{m} \sigma_\pm     
\end{equation}
their effective chemical potentials.  Implicit in this framework is the
assumption that the Fermi energies of the spin-$\uparrow$ and the
spin-$\downarrow$ electrons lie between two Landau levels, so that the
integer-part function is well defined.  We see from (\ref{fl:j0}) that the
so-called filling factor $\nu_H$, defined as
\begin{equation} 
\nu_H = \frac{j_0}{|eH|/2\pi}= l_+ + l_-,
\end{equation}  
takes on integer values only.  This was to be expected for an ideal electron
gas at zero temperature; given a value of the Fermi energy $\epsilon_{\rm
F}$, a Landau level below the Fermi surface is filled, while a level above
it is empty.  (When the Fermi energy and that of a Landau level coincide,
the value of the integer-part function $\Xi$ is ambiguous.)  

If, in addition to a magnetic field, there is also an electric field ${\bf
E}$ in the plane, a Hall current is induced perpendicular to the two fields.
The current carried by the filled Landau levels is obtained by multiplying
the induced density (\ref{fl:j0}) with the drift velocity $E/H$.  In this
way one finds:
\begin{equation}  \label{fl:j}
j_2 = \frac{{\rm sgn}(eH)}{2\pi} e (l_+ + l_-) E,                    
\end{equation} 
where, without loss of generality, the electric field is chosen in the $x_1$
direction: $A_0 = -E x_1$.  The induced fermion-number density (\ref{fl:j0})
and Hall current (\ref{fl:j}) correspond to a Chern-Simons term,
\begin{equation}          \label{fl:cs}
{\cal L}_\theta = \tfrac{1}{2} \theta e^2 
\epsilon_{\mu \nu \lambda} A_\mu \tilde{\partial}_\nu A_\lambda
\end{equation}
in the effective theory, with
\begin{equation}    \label{fl:nrth}
\theta = \frac{{\rm sgn}(eH)}{2\pi} (l_+ + l_-).  
\end{equation}
Because of the factor ${\rm sgn}(eH)$, which changes sign under a parity
transformation, this Chern-Simons term is invariant under such
transformations.

We next turn to the magnetic properties of the two-dimensional electron gas
in a constant magnetic field.  The induced spin density calculated from the
effective Lagrangian via
\begin{equation} 
s = \frac{\partial {\cal L}_{\rm eff}}
{\partial b}\Bigr|_{b=0},
\end{equation} 
turns out to be independent of the filled Landau levels $l_\pm$, viz.\
\begin{equation}   \label{fl:spin}
s=\frac{eH }{4\pi}.               
\end{equation} 
This follows from the symmetry in the spectrum $E_{l+1}^+ = E_l^-$ if
$eH>0$, and $E_l^+ = E_{l+1}^-$ if $eH<0$.  The spin magnetic moment, or
magnetization $M$, is obtained from (\ref{fl:spin}) by multiplying $s$ with
twice the Bohr magneton $\mu_{\rm B}$,
\begin{equation}  \label{ft:mag}
M = g_0 \mu_{\rm B} s = \frac{e^2}{4 \pi m} H.
\end{equation}
This leads to the text-book result for the magnetic spin susceptibility
$\chi_{\rm P}$
\begin{equation}              \label{fl:textb}
\chi_{\rm P} = \frac{\partial M}{\partial H} = \frac{e^2}{4 \pi m} = 2 \mu_{\rm 
B}^2 \, \nu(0),                    
\end{equation}
with $\nu(0) = m/2\pi$ the density of states per spin degree of freedom in
two space dimensions.  To see how the spin contribution (\ref{fl:textb}) to
the magnetic susceptibility compares to the orbital contribution, we
evaluate the $k_0$-integral in the Lagrangian (\ref{fl:nreff}) with $b=0$
to obtain
\begin{equation}
{\cal L}_{\rm eff} =  \frac{|eH|}{2 \pi} \sum_{l=0}^{\infty} 
\sum_{\varsigma =\pm} (\mu - E_l^\varsigma)
\theta(\mu - E_l^\varsigma).
\end{equation}
The summation over the Landau levels $l$ is easily carried out with the
result for small fields
\begin{equation}                   \label{ft:nr}
{\cal L}_{\rm eff} = \frac{1}{4 \pi} \sum_{\varsigma =\pm} \left[ 
\mu_\varsigma^2 m - \frac{(eH)^2}{4 m} \right]
= \frac{\mu^2 m}{2 \pi} + \frac{(eH)^2}{8 \pi m} [(2\sigma)^2
-1],
\end{equation}
where $\sigma=\tfrac{1}{2}$ and $\mu_\pm$ is given by
(\ref{fl:chemical}).  The first term at the right-hand side of
(\ref{ft:nr}), which is independent of the magnetic field, is minus the
energy density of a free electron gas particle contribution,
\begin{equation}
-\frac{\mu^2 m}{2 \pi} = 2 \int_{\bf k} \left( 
\frac{k^2}{2m} - \mu \right) \theta \left(\mu -
\frac{k^2}{2m} \right). 
\end{equation}
The factor $2$ at the right-hand side accounts for the spin-$\uparrow$ and
spin-$\downarrow$ fermions, while the step function shows that only energy
levels below the Fermi level contribute.  The energy density is negative
because the levels are measured relative to the Fermi energy.  The second
term in (\ref{ft:nr}) yields the low-field susceptibility
\begin{equation}   \label{ft:nrchi}
\chi = (-1)^{2 \sigma +1}  2 \mu_{\rm B}^2 \, \nu(0) \left[(2\sigma)^2 - 
1)\right], \;\;\;\;\;\;\;\;\; (d=2).
\end{equation}
We have cast it in a form valid for $\sigma=0,\tfrac{1}{2},1$.  The term
involving the factor $(2 \sigma)^2$ is the spin contribution which reduces
to (\ref{fl:textb}) for $\sigma = \tfrac{1}{2}$.  Let us compare
(\ref{ft:nrchi}) with the three-dimensional expression
\begin{equation}       \label{ft:chi3D}
\chi = (-1)^{2\sigma+1}  2 \mu_{\rm B}^2 \, \nu(0) \left[(2\sigma)^2 - 
\tfrac{1}{3}\right],    \;\;\;\;\;\;\;\;\;  (d=3).            
\end{equation}
where $\nu(0) = m k_F/2\pi^2$ now denotes the three-dimensional density of
states per spin degree of freedom at the Fermi sphere.  We see that the
ratio of orbital to spin contribution to $\chi$ is different in the two
cases.  In addition, whereas a three-dimensional electron gas is
paramagnetic $(\chi>0)$ because of the dominance of the spin contribution,
the two-dimensional gas is not magnetic $(\chi=0)$ at small fields since the
diamagnetic orbital and paramagnetic spin contributions to $\chi$ cancel.

The induced currents we just calculated may be used to compute the induced
quantum numbers carried by certain scalar field configurations.  The case of
interest to us is the fluxon.  This is a point-like object carrying one unit
of magnetic flux $2 \pi/e$ which can be described by a magnetic field
\begin{equation} 
H_\otimes = \frac{2 \pi}{e} \delta ({\bf x}).
\end{equation} 
According to (\ref{fl:spin}), a single fluxon carries a spin $\sigma_\otimes
= \tfrac{1}{2}$ and, since for small fields
\begin{equation}  \label{fl:jf}
j_0 \rightarrow \frac{\mu m}{\pi } + \frac{|eH|}{2 \pi},
\end{equation}
where the first term at the right-hand side is the two-dimensional
fermion-number density $\mu m/\pi = k_{\rm F}^2/2\pi$, also one
unit of fermion-number charge.  That is, in a nonrelativistic electron gas,
a fluxon carries the quantum numbers of a fermion.  However, the close
connection between the spin of a fluxon and induced Chern-Simons term for
arbitrary large fields that exists in a relativistic context, is lost.  This
can be traced back to the fact that in nonrelativistic theories, the
electron spin is an independent degree of freedom.  Below, we point out that
the spin of the fluxon does not derive from the ordinary Chern-Simons term,
but from a so-called mixed Chern-Simons term. Such a term is absent in a
relativistic context.

It is interesting to note that whereas the spin of an antifluxon, which can
be described by the magnetic field
\begin{equation} 
H_\odot = -\frac{2 \pi}{e} \delta ({\bf x}),
\end{equation} 
is $-\tfrac{1}{2}$, its fermion-number charge is the same as that for a
fluxon because of the absolute values appearing in (\ref{fl:jf}).  This is
what one expects in a nonrelativistic theory.

Because fluxons carry the quantum numbers of fermions, the exclusion
principle forbids two fluxons to be in the same state.  This is important
when calculating, for example, the orbital angular momentum $L$ of a state
with $N_\otimes$ fluxons since they have to be put in successive orbital
angular momentum states.  We find for $L$ \cite{Forte}
\begin{equation}                    
L = 2 \sigma_\otimes \sum_{\ell=1}^{N_\otimes} (\ell-1) = \sigma_\otimes
N_\otimes (N_\otimes -1).
\end{equation}
In this way, the total angular momentum $J=S+L$, with $S=\sigma_\otimes
N_\otimes$ the total spin carried by the fluxons, becomes
\begin{equation}
J = S+L = \sigma_\otimes N^2_\otimes = \tfrac{1}{2} N^2_\otimes.
\end{equation}

We next investigate the origin of the induced spin density (\ref{fl:spin})
we found in the nonrelativistic electron gas.  To this end we slightly
generalize the theory (\ref{fl:Lag}) and consider the Lagrangian
\begin{equation}         \label{fl:Lagext} 
{\cal L} =\psi ^{\dagger}\left[p_0 - e A_0 + \mu
-\frac{1}{2m}({\bf p} - e {\bf A})^{2}  
\right] \psi +g_0 \mu_{\rm B} H^\alpha\psi^{\dagger }\tfrac{1}{2}
\sigma^\alpha \psi.  
\end{equation} 
It differs from (\ref{fl:Lag}) in that the spin source term is omitted and
the magnetic field in the Zeeman term is allowed to point in any direction
in spin space labeled by the index $\alpha=1,2,3$.  It is convenient to
consider a magnetic field with fixed magnitude, but a spacetime-dependent
orientation
\begin{equation}                   
H^\alpha (x)= H d^\alpha(x),                  
\end{equation} 
where $H$ is constant, and $d^\alpha$ is a unit vector in spin space.  The
gauge field $A_\mu$ appearing in the extended Lagrangian (\ref{fl:Lagext})
still gives the magnetic field perpendicular to the plane $\nabla \times
{\bf A} = H$.  As we did various times before, we make the decomposition 
\begin{equation} 
\psi (x)=s(x)\chi (x),
\end{equation} 
with $s(x)$ the spacetime-dependent SU(2) matrix first introduced in
(\ref{af:rot}).  In terms of these new variables the Lagrangian
(\ref{fl:Lagext}) becomes
\begin{equation} 
{\cal L} = \chi ^{\dagger }\left[p_0 - eA_0 -B_0 + \mu
- \frac{1}{2m} ({\bf p} - e {\bf A} - {\bf B})^{2}\right]\chi + \frac{eH
}{2m}\chi ^{\dagger}\sigma^3\chi,
\end{equation} 
where $B_\mu = -is^{\dagger}\tilde{\partial}_\mu s$ is the field
(\ref{af:B}).  The theory takes again the form of a gauge theory with gauge
field $B_\mu$.  The spin-density operator
\begin{equation} 
j_{0}^\alpha = \psi^{\dagger} \tfrac{1}{2} \sigma^\alpha  \psi,      
\end{equation} 
becomes in these new variables [see (\ref{af:source})] 
\begin{equation} 
j_{0}^\alpha = R^{\alpha \beta} \chi^{\dagger} \tfrac{1}{2} \sigma^\beta
\chi = -\frac{1}{2} R^{\alpha \beta} \frac{\partial {\cal L} }{\partial
B_0^\beta}. 
\end{equation} 
In deriving the first equation we employed the identity (\ref{af:R}),
relating the SU(2) matrices in the $j=\tfrac{1}{2}$ representation, 
\begin{equation} 
s = \exp\left(i \tfrac{1}{2} \theta^\alpha \sigma^\alpha \right),
\end{equation} 
to those in the adjoint representation $(j=1)$, 
\begin{equation} 
R = \exp \left(i \theta^\alpha J_{\rm adj}^\alpha\right).
\end{equation}   
The matrix elements of the generators in the latter representation are
\begin{equation} 
\left(J^\alpha_{\rm adj} \right)^{\beta \gamma} = -i\epsilon^{\alpha \beta
\gamma}.
\end{equation} 

The projection of the spin density $j_0^\alpha$ onto the spin-quantization
axis, i.e., the direction $d^\alpha$ of the applied magnetic field [see
Eq.~(\ref{af:source?})],
\begin{equation} 
s = d^\alpha  j_0^\alpha =  -\frac{1}{2} \frac{\partial {\cal L}}
{\partial B_{0}^3},                                       
\end{equation} 
only involves the spin gauge field $B_\mu^3$.  So, when calculating the
induced spin density we may set the fields $B^1_\mu$ and $B^2_\mu$ to zero
and consider the simpler theory
\begin{equation}
{\cal L} = \sum_{\varsigma = \pm} \chi_{\varsigma}^{\dagger }
\left[p_0 - eA_0^{\varsigma} + \mu_{\varsigma} -  
\frac{1}{2m} ({\bf p} - e {\bf A}^{\varsigma})^2\right]\chi_{\varsigma},
\end{equation}
where the effective Fermi energies for the spin-$\uparrow$ and
spin-$\downarrow$ electrons are given in (\ref{fl:chemical}) and $eA_\mu^\pm
= eA_\mu \pm B_\mu^3$.  Both components $\chi_\uparrow$ and
$\chi_\downarrow$ induce a Chern-Simons term, so that in total we have
\begin{eqnarray}           \label{fl:total}
{\cal L}_\theta &=& \tfrac{1}{2} e^2 \epsilon_{\mu \nu \lambda} (\theta_+
A^+_\mu \tilde{\partial}_\nu A^+_\lambda + \theta_- A^-_\mu
\tilde{\partial}_\nu A^-_\lambda) \nonumber \\ &=& \tfrac{1}{2} (\theta_+ + 
\theta_-) \epsilon_{\mu \nu \lambda} (e^2 A_\mu \tilde{\partial}_\nu
A_\lambda + B^3_\mu \tilde{\partial}_\nu B^3_\lambda) \\ && + e (\theta_+ -
\theta_-) \epsilon_{\mu \nu \lambda} B^3_\mu \tilde{\partial}_\nu
A_\lambda, \nonumber
\end{eqnarray}
where the last term involving two different vector potentials is a mixed
Chern-Simons term.  The coefficients are given by, cf.~(\ref{fl:nrth})
\begin{equation}
\theta_\pm = \frac{{\rm sgn}(e H)}{2 \pi} \,   l_\pm ,
\end{equation}
assuming that $|eH|> \tfrac{1}{2} |\epsilon_{i j} \partial_i B_j^3|$, so
that the sign of $eH$ is not changed by spin gauge contributions.  The
integers $l_\pm$ are the number of filled Landau levels for spin-$\uparrow$
and spin-$\downarrow$ electrons given by (\ref{fl:npm}).  Since $l_+ -l_- =
{\rm sgn} (e H)$, we obtain for the induced spin density $s$ precisely the
result (\ref{fl:spin}) we found in the preceding section,
\begin{equation}
s = -\frac{1}{2} 
\frac{\partial \mbox{$\cal L$}_{\rm eff}}
{\partial B_0^3} \Bigr|_{B_\mu^3 = 0} = \frac{e H}{4 \pi}.
\end{equation}
The present derivation shows that the induced spin in the nonrelativistic
electron gas originates not from the standard Chern-Simons term
(\ref{fl:cs}), but from a mixed Chern-Simons term involving the
electromagnetic and spin gauge potential.

The first term in (\ref{fl:total}) is the standard Chern-Simons term, the
combination $\theta_+ + \theta_-$ precisely reproduces the result
(\ref{fl:nrth}) and is related to the induced fermion-number density
(\ref{fl:j0}).

\chapter{Dual Theories \label{chap:dual}}
In this chapter we shall discuss dual theories of a superfluid $^4$He film
and of a BCS superconductor in two and three space dimensions.  The first
use of a duality transformation in contemporary physics is generally
attributed to Kramers and Wannier who applied such a transformation to the
Ising model on a square lattice \cite{KrWa}.  The models we consider in this
chapter all possess vortex solutions.  The stability of these defects, which
appear in the original formulations as singular objects, is guaranteed by a
nontrivial topology.  The surrounding hyper sphere of a vortex is a circle
[$r=1$ in Eq.\ (\ref{af:homotopy})], so that the relevant homotopy group is
the fundamental homotopy group.  In $d=2$, a vortex is a point object
($\epsilon_d=0$), while in $d=3$ it is a line defect ($\epsilon_d=1$).  A
loop circling any of these vortices cannot be deformed to a point without
encountering a singularity.

The duality transformations of the models discussed here are aimed at
obtaining a field theoretic description of the vortices they contain.
Whereas these excitations are of topological nature in the original
formulations, they become the elementary excitations of the dual theory.
\section{Superfluid $^4$He Film}
\label{sec:kt}
In this section, we shall derive the dual theory describing a $^4$He film.
It is well known that this system undergoes a Kosterlitz-Thouless phase
transition at a temperature well below the bulk transition temperature.  The
superfluid low-temperature state is characterized by tightly bound
vortex-antivortex pairs which at the Kosterlitz-Thouless temperature unbind
and thereby disorder the superfluid state.  The disordered state, at
temperatures still below the bulk transition temperature, consists of a plasma
of unbound vortices.  We shall see in what way the dual theory, which gives
a field theoretic description of the vortices, accounts for these phenomena.

\begin{figure}
\begin{center}
\epsfxsize=5cm
\mbox{\epsfbox{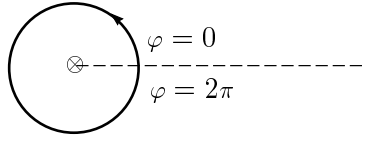}}
\end{center}
\caption{Vortex of unit winding number.  The dashed line denotes the cut
in the spatial plane along which $\varphi$ makes a jump of $2 \pi$.
\label{fig:cut}}   
\end{figure}

The phase transition is an equilibrium transition, we can accordingly ignore
any time dependence.  The important fluctuations here, at temperatures below
the bulk transition temperature, are phase fluctuations so that we can
consider the London limit and take as Hamiltonian
\begin{equation} \label{kt:HHe}
{\cal H} = \tfrac{1}{2} \bar{\rho}_{\rm s} {\bf v}^2_{\rm s},
\end{equation} 
where $\bar{\rho}_{\rm s}$ is the superfluid mass density which we assume to be
constant and ${\bf v}_{\rm s}$ is the superfluid velocity
\begin{equation} \label{kt:vs}
{\bf v}_{\rm s} = \frac{1}{m} (\nabla \varphi - \bbox{\varphi}^{\rm P}).   
\end{equation}
Following Kleinert \cite{GFCM}, we included a vortex gauge field
$\bbox{\varphi}^{\rm P}$ to account for possible vortices in the system.  A
vortex in two space dimensions is, as we mentioned in the introduction, a
point-like object.  It is characterized by the winding number of the map
\begin{equation} 
\varphi({\bf x}) : {\rm S}^1_{\bf x} \rightarrow {\rm S}^1
\end{equation} 
of a circle S$^1_{\bf x}$ around the vortex into the internal circle S$^1$
parameterized by $\varphi$.  When we circle an elementary vortex once,
$\varphi$ changes by $2 \pi$ (see Fig.~\ref{fig:cut}).
In itself this is harmless because $\varphi$ is a compact field with
periodicity $2\pi$, i.e., $\varphi$ and $\varphi +2\pi$ are identified.
However, the jumps taking place along a cut in the spatial plane would
turn the superfluid velocity into a discontinuous field.  This is
physically unacceptable.  The field $\bbox{\varphi}^{\rm P}$ is
introduced with the purpose to compensate for these jumps.  Its curl
yields the vortex density $n_{\rm v}$ consisting of delta functions
at the vortex positions ${\bf x}^\alpha$
\begin{equation} \label{kt:vdens}
\nabla \times \bbox{\varphi}^{\rm P}({\bf x}) = - 2\pi \, n_{\rm v} ({\bf
x}) = -2\pi \sum_{\alpha} w^{\alpha} \delta({\bf x} - {\bf x}^{\alpha }),
\end{equation} 
leading to the vorticity
\begin{equation} \label{kt:vorticity}
\nabla \times {\bf v}_{\rm s} = 2 \pi \frac{n_{\rm v}}{m}.
\end{equation}
The integer $w_\alpha$ is the winding number of the vortex located at ${\bf
x}^\alpha$.  We shall restrict ourselves to vortices of unit winding number,
so that $w_\alpha = \pm 1$ for a vortex and antivortex, respectively.

The canonical partition function describing the equilibrium configuration of
$N_+$ vortices and $N_-$ antivortices in a superfluid $^4$He film is given by
\begin{equation}  \label{kt:Zorig}
Z_N = \frac{1}{N_+! N_-!} \prod_\alpha \int_{{\bf
x}^\alpha} \int \DD \varphi \, \exp\left(-\beta \int_{\bf x} {\cal
H}\right),  
\end{equation} 
with ${\cal H}$ the Hamiltonian (\ref{kt:HHe}) and $N = N_+ + N_-$ the total
number of vortices and antivortices.  The factors $N_+!$ and $N_-!$ arise
because the vortices and antivortices are indistinguishable, and
$\prod_\alpha
\int_{{\bf x}^\alpha}$ denotes the integration over the positions of the
vortices.  The functional integral over $\varphi$ is Gaussian and therefore
easily carried out, with the result
\begin{equation}   \label{kt:Zx}
Z_N = \frac{1}{N_+! N_-!} \prod_\alpha \int_{{\bf x}^\alpha}
\exp\left[\pi \frac{\beta\bar{\rho}_{\rm s}}{m^2} \sum_{\alpha, \beta}
w_\alpha w_\beta \ln\left(\frac{|{\bf x}^\alpha - {\bf
x}^\beta|}{a}\right) \right].
\end{equation} 
The constant $a$, with the dimension of a length, in the argument of the
logarithm is included for dimensional reasons.  Physically, it represents
the vortex core diameter.  Apart from an irrelevant normalization factor,
Eq.~(\ref{kt:Zx}) is the canonical partition function of a two-dimensional
Coulomb gas with charges $q_\alpha = q w_\alpha = \pm q$, where
\begin{equation} 
q = \sqrt{2 \pi \bar{\rho}_{\rm s} }/m.
\end{equation}
Let us rewrite the sum in the exponent appearing in (\ref{kt:Zx}) as
\begin{eqnarray} 
\sum_{\alpha, \beta} q_\alpha
q_\beta \ln\left(\frac{|{\bf x}^\alpha - {\bf x}^\beta|}{a}\right) &=&
\sum_{\alpha, \beta} q_\alpha q_\beta \left[ \ln\left(\frac{|{\bf
x}^\alpha - {\bf x}^\beta|}{a}\right) - \ln(0) \right] \nonumber \\ && +
\ln(0) \left(\sum_\alpha q_\alpha \right)^2,
\end{eqnarray} 
where we isolated the self-interaction in the last term at the right-hand
side.  Since $\ln(0) = -\infty$, the charges must add up to zero so as to
obtain a nonzero partition function.  From now on we will therefore assume
overall charge neutrality, $\sum_\alpha q_\alpha = 0$, so that $N_+ = N_- =
N/2$, where $N$ must be an even integer.  To regularize the remaining
divergence, we replace $\ln(0)$ with an undetermined, negative constant $-c$.
The exponent of (\ref{kt:Zx}) thus becomes
\begin{equation} \label{kt:reg}
\frac{\beta}{2} \sum_{\alpha, \beta} q_\alpha q_\beta \ln\left(\frac{|{\bf
x}^\alpha - {\bf x}^\beta|}{a} \right) = \frac{\beta}{2} \sum_{\alpha \neq
\beta} q_\alpha q_\beta \ln\left(\frac{|{\bf x}^\alpha - {\bf x}^\beta|}{a}
\right) -  \beta \epsilon_{\rm c} N,
\end{equation} 
where $\epsilon_{\rm c}= c q^2/2$ physically represents the core energy,
i.e., the energy required to create a single vortex.  In deriving this we
used the identity $\sum_{\alpha \neq \beta} q_\alpha q_\beta = -
\sum_\alpha q_\alpha^2 = -N q^2$ which follows from charge neutrality.
Having dealt with the self-interaction, we limit the integrations
$\prod_\alpha \int_{{\bf x}^\alpha}$ in (\ref{kt:Zx}) over the location of
the vortices to those regions where they are more than a distance $a$ apart,
$|{\bf x}^\alpha - {\bf x}^\beta| >a$.  The grand-canonical partition
function of the system can now be cast in the form
\begin{equation} \label{kt:coul}
Z = \sum_{N=0}^\infty \frac{z^{N}}{[(N/2)!]^2}
\prod_{\alpha} \int_{{\bf x}^\alpha} \exp\left[ \frac{\beta}{2}\sum_{\alpha
\neq \beta} q_\alpha q_\beta \ln\left(|{\bf x}^\alpha - {\bf x}^\beta|\right)
\right], 
\end{equation}
where $z = \exp(-\beta \epsilon_{\rm c})$ is the fugacity.  We suppressed an
irrelevant dimensionful factor $a^{N(\beta q^2/2 - 1)}$.  The system is known to
undergo a phase transition at the Kosterlitz-Thouless temperature
\cite{Berezinskii,KT73}
\begin{equation}  \label{jump} 
T_{\rm KT} = \frac{1}{4} q^2 =  \frac{\pi}{2} \frac{\bar{\rho}_{\rm s}}{m^2},
\end{equation} 
triggered by the unbinding of vortex-antivortex pairs.  It follows from this
equation that the two-dimensional superfluid mass density $\bar{\rho}_{\rm
s}(T)$, which varies from sample to sample, terminates on a line with
universal slope as $T$ approaches the Kosterlitz-Thouless temperature from
below \cite{NeKo}.

To derive the dual theory we note that $\ln(|{\bf x}|)$ is the
inverse of the Laplace operator $\nabla^2$, 
\begin{equation} 
\frac{1}{2 \pi} \nabla^2 \ln(|{\bf x}|) = \delta({\bf x}).
\end{equation} 
This allows us to represent the exponential function in (\ref{kt:coul})
as a functional integral over an auxiliary field $\phi$:
\begin{eqnarray}  \label{kt:aux}
\lefteqn{\exp\left[ \frac{\beta}{2} \sum_{\alpha \neq \beta} q_\alpha q_\beta
\ln\left(|{\bf x}^\alpha - {\bf x}^\beta|\right) \right] =} \nonumber \\ &&
\int \DD \phi \exp\left\{ - \int_{\bf x} \left[ \frac{1}{4 \pi \beta} (\nabla
\phi)^2 + i \rho \phi \right] \right\}, 
\end{eqnarray} 
where $\rho({\bf x}) = \sum_\alpha q_\alpha \delta({\bf x} - {\bf
x}^\alpha)$ is the charge density.  In this way, the partition function
becomes
\begin{equation} \label{kt:phi}
Z = \sum_{N=0}^\infty \frac{z^{N}}{[(N/2)!]^2}
\prod_{\alpha=1}^N \int_{{\bf x}^\alpha} 
\int \DD \phi \exp\left\{ - \int_{\bf x} \left[ \frac{1}{4 \pi \beta} (\nabla
\phi)^2 + i \rho \phi \right] \right\}.
\end{equation}
In a mean-field treatment, the functional integral over the auxiliary field
introduced in (\ref{kt:aux}) is approximated by the saddle point determined
by the field equation
\begin{equation} \label{kt:feq}
i T \nabla^2 \phi = - 2 \pi \rho.
\end{equation} 
When we introduce the scalar variable $\Phi := i T \phi$, this equation
becomes formally Gauss' law, with $\Phi$ the electrostatic scalar
potential.  The auxiliary field introduces in (\ref{kt:aux}) may therefore
be thought of as representing the scalar potential of the equivalent Coulomb
gas \cite{GFCM}.

On account of charge neutrality, we have the identity
\begin{equation} 
\left[ \int_{\bf x} \left( {\rm e}^{iq \phi({\bf x})} + {\rm e}^{-iq \phi({\bf
x})} \right) \right]^N = \frac{N!}{[(N/2)!]^2} \prod_{\alpha=1}^{N}
\int_{{\bf x}^\alpha} {\rm e}^{-i \sum_{\alpha} q_\alpha \phi({\bf
x}^\alpha)},
\end{equation} 
where we recall that $N$ is an even number.  The factor $N!/[(N/2)!]^2$
is the number of charge-neutral terms contained in the binomial
expansion of the left-hand side.  The partition function (\ref{kt:phi})
may thus be written as \cite{GFCM}
\begin{eqnarray} \label{kt:sG} 
Z &=& \sum_{N=0}^\infty \frac{(2z)^{N}}{N!}
\int \DD \phi \exp\left[ - \int_{\bf x} \frac{1}{4 \pi \beta} (\nabla 
\phi)^2 \right] \left[\cos\left(\int_{\bf x} q \phi \right)
\right]^N \nonumber \\ &=& 
\int \DD \phi \exp\left\{ - \int_{\bf x} \left[ \frac{1}{4 \pi \beta}
(\nabla \phi)^2 - 2z \cos(q \phi) \right] \right\},
\end{eqnarray} 
where in the final form we recognize the sine-Gordon model.  This is the
dual theory we were seeking.  Contrary to the original formulation
(\ref{kt:Zorig}), which contains the vortices as singular objects, the dual
formulation has no singularities.  To see how the vortices and the
Kosterlitz-Thouless phase transition are represented in the dual theory we
note that the field equation of the auxiliary field now reads
\begin{equation}  \label{kt:gauss}
i T \nabla^2 \phi = 2 \pi z q \left({\rm e}^{iq \phi} - {\rm e}^{-iq
\phi} \right).
\end{equation} 
On comparison with the previous field equation (\ref{kt:feq}), it follows
that the right-hand side represents the charge density of the Coulomb gas.
In terms of the scalar potential $\Phi$, Eq.~(\ref{kt:gauss}) becomes the
Poisson-Boltzmann equation
\begin{equation} \label{kt:PB} 
\nabla^2 \Phi = - 2 \pi q \left(z \, {\rm e}^{- \beta q \Phi} - z
\, {\rm e}^{\beta q \Phi} \right),
\end{equation}  
describing, at least for temperatures above the Kosterlitz-Thouless
temperature, a plas\-ma of positive and negative charges with
density $n_\pm$,
\begin{equation} \label{kt:spatiald} 
n_\pm = z \, {\rm e}^{\mp \beta q \Phi},
\end{equation} 
respectively.  The fugacity $z$ is the density at zero scalar potential.
(It is to recalled that we suppress factors of $a$ denoting the diameter
of the vortex cores.)  Equation (\ref{kt:PB}) is a self-consistent equation
for the scalar potential $\Phi$ giving the spatial distribution of
the charges via (\ref{kt:spatiald}).  It follows from this argument that
the interaction term $2z \cos(q \phi)$ of the sine-Gordon model
represents a plasma of vortices. 

The renormalization group applied to the sine-Gordon model reveals that at
the Kosterlitz-Thouless temperature $T_{\rm KT} = \tfrac{1}{4}q^2$ there is
a phase transition between a low-temperature phase of tightly bound neutral
pairs and a high-temperature plasma phase of unbound vortices
\cite{Schenker}.  In the low-temperature phase, the (renormalized) fugacity
scales to zero in the large-scale limit so that the interaction term,
representing the plasma of unbound vortices, is suppressed.  The
long-distance behavior of the low-temperature phase is therefore well
described by the free theory $(\nabla \phi)^2/4 \pi\beta$, representing the
gapless Kosterlitz-Thouless mode.  This is the superfluid state.  The
expectation value of a single vortex vanishes because in this gapless state
its energy diverges in the infrared.

An important characteristic of a charged plasma is that it has no gapless
excitations, the photon being transmuted into a massive plasmon.  To see
this we assume that $q \Phi << T$, so that $\sinh(\beta q \Phi)
\approx \beta q \Phi$.  In this approximation, the Poisson-Boltzmann equation
(\ref{kt:PB}) can be linearized to give
\begin{equation} \label{kt:mpoi}
(\nabla^2 - m_{\rm D}^2) \Phi = 0, \;\;\; m_{\rm D}^2 = 4 \pi \beta
z q^2.
\end{equation} 
This shows us that, in contradistinction to the low-temperature phase, in
the high-temperature phase, the scalar potential describes a massive
mode---the plasmon.  In other words, the Kosterlitz-Thouless mode acquires
an energy gap $m_{\rm D}$.  Since it provides the high-temperature phase with
an infrared cutoff, isolated vortices have a finite energy now and
accordingly a finite probability to be created.  This Debeye mechanism of
mass generation for the photon should be distinguished from the Higgs
mechanism which operates in superconductors (see below) and also generates a
photon mass.

Another property of a charged plasma is that it screens charges.  This
so-called Debeye screening may be illustrated by adding an external
charge to the system.  The linearized Poisson-Boltzmann equation
(\ref{kt:mpoi}) then becomes
\begin{equation} \label{kt:pois}
(\nabla^2 - m_{\rm D}^2) \Phi({\bf x}) = - 2 \pi q_0 \delta ({\bf x}), 
\end{equation} 
with $q_0$ the external charge which we have placed at the origin.  The
solution of this equation is given by $\Phi ({\bf x}) = q_0
K_0(m_{\rm D}|{\bf x}|)$ with $K_0$ a modified Bessel function.  The mass
term in (\ref{kt:pois}) is ($2 \pi$ times) the charge density induced by the
external charge, i.e.,
\begin{equation} 
\rho_{\rm ind}({\bf x}) = - \frac{1}{2 \pi} q_0 m_{\rm D}^2
K_0(m_{\rm D}|{\bf x}|).  
\end{equation} 
By integrating this density over the entire system, we see that the total
induced charge $\int_{\bf x} \rho_{\rm ind} = -q_0$ completely screens the
external charge---at least in the linear approximation we are using here.
The inverse of the plasmon mass is the screening length---the so-called
Debeye screening length.

To see that the sine-Gordon model gives a dual description of a $^4$He
film we cast the field equation (\ref{kt:feq}) in the form
\begin{equation} 
i T \nabla^2 \phi = - m q \nabla \times {\bf v}_{\rm s},
\end{equation} 
where we employed Eq.~(\ref{kt:vorticity}).  On integrating this
equation, we obtain up to an irrelevant integration constant
\begin{equation} 
i T \partial_i \phi = - q \epsilon_{i j} (\partial_j \varphi - \varphi_j^{\rm
P}).
\end{equation} 
This relation, involving the antisymmetric Levi-Civita symbol, is a typical
one between dual variables.  It also nicely illustrates that although the
dual variable $\phi$ is a regular field, it nevertheless contains the
information about the vortices which in the original formulation are
described via the singular vortex gauge field $\bbox{\varphi}^{\rm P}$.

Given this observation it is straightforward to calculate the
current-current correlation function $\langle g_i ({\bf k}) g_j(-{\bf
k}) \rangle$, with
\begin{equation} 
{\bf g} = \bar{\rho}_{\rm s} {\bf v}_{\rm s}
\end{equation} 
the mass current.  We find
\begin{equation} 
\langle g_i ({\bf k}) g_j(-{\bf k}) \rangle = - \frac{1}{2 \pi \beta^2} 
\bar{\rho}_{\rm s} \epsilon_{ik} \epsilon_{jl} k_k k_l \langle \phi({\bf k})
\phi(-{\bf k}) \rangle,
\end{equation} 
where the average is to be taken with respect to the partition function
\begin{equation} 
Z_0 = \int \DD \phi \exp\left[ - \frac{1}{4 \pi \beta} \int_{\bf x} (\nabla
\phi)^2 \right],
\end{equation} 
which is obtained from (\ref{kt:sG}) by setting the interaction term to
zero.  We obtain in this way the standard expression for a superfluid
\begin{equation} \label{kt:jj} 
\langle g_i ({\bf k}) g_j(-{\bf k}) \rangle = - \frac{\bar{\rho}_{\rm s}}{\beta} 
 \frac{1}{{\bf k}^2} \left( \delta_{ij} {\bf k}^2 - k_i k_j \right).
\end{equation}  
The $1/{\bf k}^2$ reflects the gaplessness of the $\phi$-field in the
low-temperature phase, while the combination $\delta_{ij} {\bf k}^2 - k_i
k_j$ arises because the current is divergent free, $\nabla \cdot {\bf
g}({\bf x}) = 0$, or ${\bf k} \cdot {\bf g}({\bf k}) = 0$.
\section{Two-dimensional Superconductor}
\label{sec:2sc}
We now turn to the dual description of a superconducting film.  We thereto
minimally couple the model of the preceding section to a magnetic field
described by the magnetic vector potential ${\bf A}$.  For the time being we
ignore vortices by setting the vortex gauge field $\bbox{\varphi}^{\rm P}$ to
zero.  The partition function of the system then reads
\begin{equation}  \label{2sc:znovor}  
Z = \int \DD\varphi \int \DD {\bf A} \, \Xi ({\bf A}) 
\, \exp\left( -\beta \int_{\bf x}{\cal H} \right),
\end{equation}
where $\Xi({\bf A})$ is a gauge-fixing factor for the gauge field ${\bf A}$,
and ${\cal H}$ is the Hamiltonian
\begin{equation} \label{2sc:H}
{\cal H} = \tfrac{1}{2} \bar{\rho}_{\rm s} {\bf v}_{\rm s}^2 + \tfrac{1}{2}
(\nabla \times {\bf A})^2 
\end{equation} 
with
\begin{equation} \label{2sc:vs}
{\bf v}_{\rm s} = \frac{1}{m} (\nabla \varphi - 2e {\bf A}).   
\end{equation}
The double charge $2e$ stands for the charge of the Cooper pairs which are
formed at the bulk transition temperature.  The functional integral over
$\varphi$ in (\ref{2sc:znovor}) is easily carried out with the result
\begin{eqnarray}   \label{2sc:schwlike} 
\lefteqn{Z =} \\ && \!\!\!\!\!\!\!\!\! \int \DD {\bf A} \, \Xi({\bf A}) \, {\rm
exp}\left\{-\frac{\beta}{2} \int_{\bf x} \left[(\nabla \times {\bf A})^2 +
m_A^2 A_i \left( \delta_{i j} - \frac{\partial _i
\partial_j}{\nabla^2} \right) A_j \right] \right\}, \nonumber 
\end{eqnarray} 
where the last term, with $m^2_A = 4 e^2 \bar{\rho}_{\rm s}/m^2$, is a
gauge-invariant, albeit nonlocal mass term for the gauge field
generated by the Higgs mechanism.  The number of degrees of freedom does
not change in the process.  This can be seen by noting that a gapless
gauge field in two dimensions represents no physical degrees of freedom.
(In Minkowski spacetime, this is easily understood by recognizing that
in $1+1$ dimensions there is no transverse direction.)  Before the Higgs
mechanism took place, the system therefore contains only a single
physical degree of freedom described by $\varphi$.  This equals the
number of degrees of freedom contained in (\ref{2sc:schwlike}).

We next introduce an auxiliary field $\tilde{h}$ to linearize the first term
in (\ref{2sc:schwlike}),
\begin{equation}     \label{2sc:efield}
\exp \left[-\frac{\beta}{2} \int_{\bf x}  (\nabla \times {\bf A})^2
\right] = \int \DD \tilde{h} \, {\rm exp}\left[-\frac{1}{2 \beta} \int_{\bf x}
\tilde{h}^2 + i \int_{\bf x} \tilde{h} (\nabla
\times {\bf A}) \right],
\end{equation} 
and integrate out the gauge-field fluctuations [with a gauge-fixing term
$(1/2\alpha)(\nabla \cdot {\bf A})^2$].  The result is a manifestly
gauge-invariant expression for the partition function in terms of a massive
scalar field $\tilde{h}$, representing the single degree of freedom
contained in the theory:
\begin{equation}   \label{2sc:massivescalar}
Z = \int \DD \tilde{h} \, {\rm exp}\left\{-\frac{1}{2 \beta} \int_{\bf x}
\left[ \frac{1}{m_A^2} (\nabla \tilde{h})^2 + \tilde{h}^2 \right] \right\}.
\end{equation} 
To understand the physical significance of this field, we note from
(\ref{2sc:efield}) that it satisfies the field equation
\begin{equation}  \label{2sc:id} 
\tilde{h} = i \beta \nabla \times {\bf A}.
\end{equation} 
That is, the fluctuating field $\tilde{h}$ represents the local magnetic
induction, which is a scalar in two space dimensions.  Equation
(\ref{2sc:massivescalar}) shows that the magnetic field has a finite
penetration depth $\lambda_{\rm L} = 1/m_A$.  In contrast to the original
description where the functional integral runs over the gauge potential, the
integration variable in (\ref{2sc:massivescalar}) is the physical field.

We next include vortices.  The penetration depth $\lambda_{\rm L}$ provides
the system with an infrared cutoff so that a single magnetic vortex in the
charged theory has a finite energy.  Vortices can therefore be thermally
activated.  This is different from the superfluid phase of the neutral
model, where the absence of an infrared cutoff permits only tightly bound
vortex-antivortex pairs to exist.  We expect, accordingly, the
superconducting phase to describe a plasma of vortices, each carrying one
magnetic flux quantum $\pm \pi/e$.  The partition function now reads
\begin{equation}     \label{2sc:vincluded}
Z = \sum_{N_{+},N_{-}=0}^{\infty} \frac{z^{N_{+}+N_{-}}}{N_{+}!\, N_{-}!}
\prod_{\alpha} \int_{{\bf x}^\alpha} \, \int \DD \varphi
\int \DD {\bf A} \,  \Xi({\bf A}) \, \exp \left(- \beta \int_{\bf
x} {\cal H} \right)
\end{equation} 
where $z$ is the fugacity, i.e., the Boltzmann factor associated with the
vortex core energy.  The velocity appearing in the Hamiltonian (\ref{2sc:H})
now includes the vortex gauge field
\begin{equation} 
{\bf v}_{\rm s} = \frac{1}{m} (\nabla \varphi - 2e {\bf A} -
\bbox{\varphi}^{\rm P}).    
\end{equation}

The vortex gauge field $\bbox{\varphi}^{\rm P}$ can be shifted from the
first to the second term in the Hamiltonian (\ref{2sc:H}) by applying
the transformation ${\bf A} \rightarrow {\bf A} - \bbox{\varphi}^{\rm
P}/2e$.  This results in the shift
\begin{equation} 
\nabla \times {\bf A} \rightarrow \nabla \times {\bf A} - B^{\rm P},
\end{equation} 
with the plastic field 
\begin{equation}  \label{2sc:BP}
B^{\rm P} = -\Phi_0 \sum_{\alpha} w_{\alpha} \, \delta({\bf x} - {\bf
x}^{\alpha })
\end{equation} 
representing the magnetic flux density.  Here, $\Phi_0 = \pi/e$ is the
elementary flux quantum.  Repeating the steps of the previous paragraph
we now obtain instead of (\ref{2sc:massivescalar})
\begin{eqnarray}  \label{2sc:vortexsum}
\lefteqn{Z = \sum_{N_\pm=0}^\infty \frac{z^{N_{+}+N_{-}}}{N_{+}!\,
N_{-}! }  \prod_{\alpha} \int_{{\bf x}^\alpha} \int \DD \tilde{h}} \\ &&
\times {\rm exp}\left\{-\frac{1}{2 \beta} \int_{\bf x} \left[
\frac{1}{m_A^2} (\nabla \tilde{h})^2 + \tilde{h}^2 \right] + i \int_{\bf x}
B^{\rm P} \tilde{h} \right\}, \nonumber
\end{eqnarray} 
where $\tilde{h}$ represents the physical local magnetic induction $h$
\begin{equation} 
\tilde{h} = i \beta (\nabla \times {\bf A} - B^{\rm P}) = i \beta h.
\end{equation} 
The field equation for $\tilde{h}$ obtained from (\ref{2sc:vortexsum})
yields for the magnetic induction:
\begin{equation} \label{2sc:fam}
- \nabla^2 h + m_A^2 h = m_A^2 B^{\rm P},
\end{equation} 
which is the familiar equation in the presence of magnetic vortices.

The last term in (\ref{2sc:vortexsum}) shows that the charge $g$ with which
a magnetic vortex couples to the fluctuating $\tilde{h}$-field is the
product of an elementary flux quantum (contained in the definition of
$B^{\rm P}$) and the inverse penetration depth $m_A = 1/\lambda_{\rm L}$,
\begin{equation} \label{2sc:g}
g = \Phi_0 m_A.
\end{equation} 
For small fugacities the summation indices $N_{+}$ and $N_{-}$ can be
restricted to the values $0,1$ and we arrive at the partition function
of the massive sine-Gordon model \cite{Schaposnik}
\begin{equation}   \label{2sc:sineGordon}
Z = \int \DD \tilde{h} \, {\rm exp} \left( - \int_{\bf x}
\left\{\frac{1}{2 \beta} \left[\frac{1}{m_A^2} (\nabla \tilde{h})^2 + 
\tilde{h}^2\right]- 2z \cos \left( \Phi_0 \tilde{h} \right) \right\}
\right).
\end{equation} 
This is the dual formulation of a two-dimensional superconductor.  The
magnetic vortices of unit winding number $w_\alpha = \pm 1$ turned the
otherwise free theory (\ref{2sc:massivescalar}) into an interacting one.

The final form (\ref{2sc:sineGordon}) demonstrates the rationales for going
over to a dual theory.  First, it is a formulation directly in terms of a
physical field representing the local magnetic induction.  There is no
redundancy in this description and therefore no gauge invariance.  Second,
the magnetic vortices are accounted for in a nonsingular fashion.  This is
different from the original formulation of the two-dimensional
superconductor where the local magnetic induction is the curl of an
unphysical gauge potential ${\bf A}$, and where the magnetic vortices appear
as singular objects.

Up to this point we have discussed a genuine two-dimensional superconductor.
As a model to describe superconducting films this is, however, not adequate.
The reason is that the magnetic interaction between the vortices takes place
mostly not through the film but through free space surrounding the film
where the photon is gapless.  This situation is markedly different from a
superfluid film.  The interaction between the vortices there is mediated by
the Kosterlitz-Thouless mode which is confined to the film.  A genuine
two-dimensional theory therefore gives a satisfactory description of a
superfluid film.

To account for the fact that the magnetic induction is not confined to the
film and can roam in outer space, the field equation (\ref{2sc:fam}) is
modified in the following way \cite{Pearl,deGennes}
\begin{equation}  \label{2sc:mod}
- \nabla^2 h({\bf x}_\perp,x_3) + \frac{1}{\lambda_\perp} \delta_d(x_3) h({\bf
x}_\perp,x_3) = \frac{1}{\lambda_\perp}  \delta_d(x_3) B^{\rm P}({\bf x}).
\end{equation}  
Here, $1/\lambda_\perp = d m_A^2 $, with $d$ denoting the thickness of the
superconducting film, is an inverse length scale, ${\bf x}_\perp$ denotes
the coordinates in the plane, $h$ the component of the induction field
perpendicular to the film, and $\delta_d(x_3)$ is a smeared delta function
of thickness $d$ along the $x_3$-axis
\begin{equation}  
\delta_d(x_3) \left\{ \begin{array}{cc} = 0 & {\rm for} \;\;\;\; |x_3| > d/2 \\
\neq 0 &  {\rm for} \;\;\;\; |x_3| \leq d/2 \end{array} \right. .
\end{equation} 
The reason for including the smeared delta function at the right-hand
side of (\ref{2sc:mod}) is that the vortices are confined to the film.  The
delta function in the second-term at the left-hand side is included
because this term is generated by screening currents which are also confined
to the film.

To be definite, we consider a single magnetic vortex located at the origin.
The induction field found from (\ref{2sc:mod}) reads
\begin{equation} 
h({\bf x}_\perp,0) = \frac{\Phi_0}{2 \pi} \int_0^\infty \dd q \frac{q}{1+ 2
\lambda_\perp q} J_0(q |{\bf x}_\perp|),
\end{equation} 
with $J_0$ the 0th Bessel function of the first kind.  At small distances
from the vortex core ($\lambda_\perp q >> 1$)
\begin{equation} \label{2sc:vincin}
h({\bf x}_\perp,0) \sim \frac{\Phi_0}{4 \pi \lambda_\perp |{\bf x}_\perp|},
\end{equation} 
while far away  ($\lambda_\perp q << 1$)
\begin{equation} 
h({\bf x}_\perp,0) \sim \frac{\Phi_0 \lambda_\perp}{\pi |{\bf x}_\perp|^3}.
\end{equation} 
This last equation shows that the field does not exponentially decay as
would be the case in a genuine two-dimensional system.  The reason for the
long range is that most of the magnetic interaction takes place in free
space outside the film where the photon is gapless.  If, as is often the
case, the length $\lambda_\perp =1/d m_A^2$ is much larger than the sample
size, it can be effectively set to infinity.  In this limit, the effect of
the magnetic interaction, as can be seen from (\ref{2sc:vincin}), diminishes
and the vortices behave as in a superfluid film.  One therefore expects a
superconducting film to also undergo a Kosterlitz-Thouless transition at
some temperature $T_{\rm KT}$ characterized by an unbinding of
vortex-antivortex pairs.  The first experiment to study this possibility was
carried out in Ref.\ \cite{BMO}.  Because the transition temperature $T_{\rm
KT}$ is well below the bulk temperature $T_{\rm c}$ where the Cooper pairs
form, the energy gap of the fermions remains finite at the critical point
\cite{CFGWY}.  This prediction has been corroborated by experiments
performed by Hebard and Palaanen on superconducting films \cite{HPsu2}.  For
temperatures $T_{\rm KT} \leq T \leq T_{\rm c}$, there is a plasma of
magnetic vortices which disorder the superconducting state.  At $T_{\rm KT}$
vortices and antivortices bind into pairs and algebraic long-range order
sets in.
\section{Bosonization}
\label{sec:bos}
The two dual models (\ref{2sc:massivescalar}) and (\ref{2sc:sineGordon}) we
encountered in the preceding section describing a superconducting film
without and with vortices included are reminiscent of the bosonized massless
and massive Schwinger model, respectively.  The massless Schwinger model
\cite{Schwingermodel} describes gapless fermions interacting with an
electromagnetic field in one space and one time dimension.  It is defined by
the Lagrangian
\begin{equation} 
{\cal L} = \bar{\psi}(i  \overlay{/}{\tilde \partial} - e \overlay{/}{A})
\psi - \tfrac{1}{4} F_{\mu \nu}^2,
\end{equation} 
where $\overlay{/}{A} = A_\mu \gamma_\mu$ with $\gamma_\mu$ the Dirac
matrices (\ref{rp:Dirac}), $\psi$ is a two-component Grassmann field
describing the fermions, and $\bar{\psi} =
\psi^\dagger \beta$.  Since the theory is bilinear in $\psi$, the
functional integral over the Grassmann fields in the partition function
\begin{equation}  \label{bos:Z}
Z = \int \DD \bar{\psi} \DD \psi \int \DD A_\mu \Xi(A_\mu) \, \exp \left(i
\int_x {\cal L} \right),
\end{equation} 
with $\Xi(A_\mu)$ a gauge-fixing factor, can be easily carried out.  It
yields a functional determinant 
\begin{equation}   
\int \DD \bar{\psi} \DD \psi \, {\rm exp}\left[ i \int_x
\bar{\psi}( i \overlay{/}{\tilde \partial} - e \overlay{/}{A}) \psi \right]
= {\rm Det}\, (\overlay{/}{p} - e \overlay{/}{A})
\end{equation} 
which can be evaluated in closed form using, for example, the derivative
expansion \cite{DaKa}.  One finds
\begin{equation} \label{bos:exact}
{\rm Det}\, (\overlay{/}{p} - e \overlay{/}{A}) = \exp \left[ \frac{ie^2}{2
\pi} \int_x A_{\mu} \left( g_{\mu \nu} - \frac{\tilde \partial_\mu \tilde
\partial_\nu}{\partial^2} \right) A_\nu
\right]. 
\end{equation} 
The partition function (\ref{bos:Z}) can now be written in a form,
\begin{eqnarray}     \label{bos:fuja}
\lefteqn{Z = \int \DD A_\mu \Xi(A_\mu) } \\ &&  \times  \exp \left\{ i \int_x
\left[-\frac{1}{4} F_{\mu \nu}^2 + \frac{1}{2} m_A^2 A_\mu \left(
g_{\mu \nu} - \frac{\tilde \partial_\mu \tilde \partial_\nu}{\partial^2}
\right) A_\nu \right] \right\}, \nonumber 
\end{eqnarray} 
which is the Minkowski analog of the partition function (\ref{2sc:schwlike})
describing a superconducting film without vortices.  Equation
(\ref{bos:fuja}) shows us that the photon has acquired a mass $m_A$, given by
$m_A^2 = e^2/\pi$, where it should be borne in mind that in one spatial
dimension the electric charge $e$ has mass dimension one.  This mass
generation due to gapless fermions is called the Schwinger mechanism.

We have seen that the partition function (\ref{2sc:schwlike}) could be
equivalently represented by the massive scalar field theory
(\ref{2sc:massivescalar}).  The same holds for the Schwinger model.  Rather
than linearizing the Maxwell term in (\ref{bos:fuja}), which would be the
analog of what we did in (\ref{2sc:efield}), we represent the term obtained
in (\ref{bos:exact}) as a functional integral over a Bose field
\cite{RRlect}:
\begin{eqnarray} 
\lefteqn{\int \DD \bar{\psi} \DD \psi \, {\rm exp}\left[ i \int_x
\bar{\psi}( i \overlay{/}{\tilde \partial} - e \overlay{/}{A}) \psi \right]
=} \nonumber \\ &&
\int \DD \phi \exp \left\{ i \int_x \left[ \frac{1}{2}
(\partial_\mu\phi)^2 - \frac{e}{\sqrt{\pi}} \epsilon_{\mu \nu} A_\mu
\tilde \partial_\nu \phi \right] \right\}.                                   
\end{eqnarray} 
We recognize here two of the bosonization rules (\ref{rp:bosrules}) we
employed before, viz.:
\begin{eqnarray}   \label{bos:bos}
\bar{\psi} i \overlay{/}{\tilde \partial} \psi &\rightarrow& \tfrac{1}{2} 
(\partial_\mu \phi)^2 \nonumber \\ j_\mu &\rightarrow& \frac{1}{\sqrt{\pi}}
\epsilon_{\mu \nu} \tilde \partial_\nu \phi,
\end{eqnarray} 
where $j_\mu = \bar{\psi} \gamma_\mu \psi$ is the electromagnetic current.
The integral over the gauge field can now be easily carried out, say in the
Lorentz gauge $\tilde{\partial}_\mu A_\mu=0$, to yield the bosonized form of
the massless Schwinger model
\begin{equation}  \label{bos:Zphi}
Z = \int \DD \phi \exp \left\{ \frac{i}{2} \int_x \left[ 
(\partial_\mu\phi)^2 -  m_A^2 \phi^2 \right] \right\}.
\end{equation} 
This shows that the model is equivalent to a massive scalar theory.  It
is the Minkowski analog of the dual theory (\ref{2sc:massivescalar})
of a superconducting film without taking into account vortices.

To understand the physical origin of the scalar field $\phi$ and why it
is massive, it should be noted that the massless Schwinger model has
besides the local U(1) gauge symmetry
\begin{equation}
\psi(x) \rightarrow {\rm e}^{i e \alpha(x) } \psi(x); \;\;\;\; A_\mu(x)
\rightarrow A_\mu(x) + \tilde \partial_\mu \alpha(x)
\end{equation} 
also a global U(1) chiral symmetry.  Under chiral transformations
\begin{equation} 
\psi \rightarrow {\rm e}^{i \lambda \gamma_5} \psi,
\end{equation} 
where $\lambda$ is the transformation parameter and $\gamma_5$ denotes the
matrix $\gamma_5 = i \gamma_0 \gamma_1$.  The corresponding Noether current
is the axial current
\begin{equation} 
j^5_\mu  = i \bar{\psi} \gamma_\mu \gamma_5 \psi.
\end{equation} 
At the quantum level, the chiral U(1) symmetry is spontaneously broken by a
fermion condensate \cite{Smilga}
\begin{equation}
\langle \bar{\psi} \psi \rangle =  - \frac{{\rm e}^\gamma}{2 \pi} m_A,
\end{equation} 
with $\gamma \approx 0.577216$ Euler's constant.  By virtue of the identity
\begin{equation} 
i \gamma_\mu \gamma_5 = \epsilon_{\mu \nu} \gamma_\nu,
\end{equation} 
the axial and  electromagnetic current are related via
\begin{equation}
j^5_\mu = \epsilon_{\mu \nu} j_\nu.
\end{equation}  
With the bosonization rule (\ref{bos:bos}) we then obtain the correspondence
\begin{equation} 
j^5_\mu \rightarrow - \frac{1}{\sqrt{\pi}} \tilde \partial_\mu \phi,
\end{equation} 
identifying $\phi$ as the Goldstone field of the broken U(1) chiral
symmetry.  This can, however, not be the end of the story \cite{NiScchiral}.
Goldstone modes are by definition gapless, but Eq.~(\ref{bos:Zphi}) shows
that $\phi$ has acquired a mass, meaning that the chiral symmetry is no
longer an exact symmetry of the bosonized theory.  The axial current, which
is classically conserved, has an anomaly at the quantum level as can be seen
by invoking the field equation obtained from (\ref{bos:Zphi}),
\begin{equation}  \label{bos:anomaly}
\tilde{\partial}_\mu j^5_\mu = \frac{m_A^2}{\sqrt{\pi}} \phi.
\end{equation} 
The destruction of the chiral symmetry at the quantum level can also be seen
directly in the fermionic formulation, where it can be attributed to the
effect of one-loop graphs.  The anomaly (\ref{bos:anomaly}) follows there
from integrating out the fermionic degrees of freedom.  Using, for example,
the derivative expansion \cite{DaKa}, one finds
\begin{equation} 
\tilde{\partial}_\mu j^5_\mu = -\frac{m_A}{\sqrt{\pi}} \epsilon_{\mu \nu}
\tilde \partial_\mu A_\nu.
\end{equation} 
On comparing the two expressions for the anomaly, we see that the Bose field
$\phi$ represents the dual field strength $\tilde{F} := \epsilon_{\mu \nu}
\tilde \partial_\mu A_\nu$,
\begin{equation} 
\phi  = -\frac{1}{m_A} \tilde{F}.
\end{equation} 
This is the Minkowski analog of the relation (\ref{2sc:id}) we found in the
context of a superconducting film.  

We next consider possible vortex contributions.  Let us go back to the
functional determinant appearing in (\ref{bos:exact}) and evaluate it in the
presence of a vortex.  To this end we go over to Euclidean spacetime where a
vortex becomes an instanton.  There exists a powerful index theorem
\cite{NiScindex} relating the zero eigenvalues of the massless Dirac operator
$i \overlay{/}{\tilde \partial} - e \overlay{/}{A}$ to the so-called
Pontryagin index
\begin{equation} 
Q = \frac{e}{4\pi} \int_x \epsilon_{\mu \nu} F_{\mu \nu},
\end{equation} 
which is the winding number of the vortex.  The theorem states that the
index of the Dirac operator, defined by the number $n_+$ of zero-modes with
positive chirality minus the number $n_-$ of zero-modes with negative
chirality, is given by the Pontryagin index,
\begin{equation} 
n_+ - n_- = Q.
\end{equation}
In the presence of vortices $Q \neq 0$, so that there are always zero-modes.
Since the determinant of an operator is the product of its eigenvalues, the
determinant appearing in (\ref{bos:exact}) is zero when vortices are
included.  Hence, in the massless Schwinger model, vortices do not
contribute to the partition function.  The suppression of instanton
contributions by gapless fermions is a common feature \cite{Rajaraman}.

The situation is different in the massive model, defined by the Lagrangian
\begin{equation} 
{\cal L} = \bar{\psi}(i \overlay{/}{\tilde \partial} - m - e
\overlay{/}{A}) \psi - \tfrac{1}{4} F_{\mu \nu}^2.
\end{equation} 
First of all, the mass term $-m \bar{\psi} \psi$ in the Lagrangian
explicitly breaks the chiral symmetry.  So, even if there were no anomaly,
we expect a massive bosonized theory.  Second, because the massive Dirac
operator contains no zero-modes, the index theorem is of no relevance
here and vortices should be included in (\ref{bos:Zphi}).  The bosonized
partition function then becomes, cf.~(\ref{2sc:vortexsum}),
\begin{eqnarray} 
\lefteqn{Z = \!\!\sum_{N_\pm=0}^{\infty} \frac{z^{N_{+}+N_{-}}}{N_{+}!\,
N_{-}!}\prod_{\alpha} \int_{x^\alpha} \int \DD \phi }  \\ && \times 
{\rm exp}\left\{\frac{i}{2} \int_x \left[ (\partial_\mu \phi)^2 - m_A^2
\phi^2
\right] + i \int_x \tilde{F}^{\rm P} \phi \right\}, \nonumber 
\end{eqnarray} 
where $z = \exp(i S_{\rm c})$, with $S_{\rm c}$ the vortex core action.
The plastic field $\tilde{F}^{\rm P}$ describes the vortices,
\begin{equation} 
\tilde{F}^{\rm P} = -\Phi_0 \sum_{\alpha}
w_{\alpha} \, \delta(x - x^{\alpha }),
\end{equation}
where the sum is over all vortex locations $x^\alpha$ and $w_\alpha$ is the
winding number of the $\alpha$th vortex.  The elementary flux quantum in the
Schwinger model is $\Phi_0 = 2 \pi/e$, which is twice the value in a
superconductor because the electric charge of a Cooper pair is twice that of
an electron.  We shall consider only vortices of unit winding number
$w_\alpha = \pm 1$.  For $z$ small we can restrict ourselves to
configurations having zero or one vortex.  The partition function then
becomes the Minkowski analog of (\ref{2sc:sineGordon})
\begin{equation} \label{bos:sineGordon}
Z = \int \DD \phi \, {\rm exp} \left( i \int_x \left\{\frac{1}{2}
\left[(\partial_\mu \phi)^2 - m_A^2 \phi^2\right]- 2z \cos \left(
g \phi\right) \right\} \right), 
\end{equation}
with $g= \Phi_0 m_A$ as in (\ref{2sc:g}).  From this we see that the
inclusion of the mass term in the Schwinger model resulted in the cosine
term in the bosonized theory.  That is, we have the correspondence
\begin{equation}   \label{bos:m}
m \bar{\psi} \psi \rightarrow 2z \cos \left(\sqrt{4 \pi} \phi\right),
\end{equation} 
where we used that
\begin{equation} 
g = \Phi_0 m_A =  \sqrt{4 \pi}
\end{equation} 
in the Schwinger model.  Surprisingly, (\ref{bos:m}) is the remaining
bosonization rule of (\ref{rp:bosrules}).  Instanton contributions, which
were suppressed in the massless model, generate the interaction term in the
bosonized form of the massive Schwinger model.  We note that even in the
absence of the anomaly, implying that the mass term in
(\ref{bos:sineGordon}) would be absent, the bosonized theory would still be
massive.  This follows from noting that the expansion of the cosine term
contains a term quadratic in $\phi$.
\section{Ginzburg-Landau Model}
\label{sec:gl}
In this section, we give arguments as to why the Ginzburg-Landau model of a
three-dimensional superconductor is not well suited to establish the
critical properties of a type-II superconductor at zero external field.

The model is defined by the Hamiltonian
\begin{equation}  \label{gl:H}   
{\cal H} = \left|(\nabla -  2 i e {\bf A})\phi\right|^2 +
m_\phi^2 |\phi|^2 + \lambda |\phi|^4 +
\frac{1}{2} (\nabla \times {\bf A})^2 + \frac{1}{2 \alpha} (\nabla \cdot
{\bf A})^2 ,
\end{equation}	
where the coefficients $2e$ and $m_\phi$ are the electric charge and mass of
the complex $\phi$-field, while $\lambda$ is a coupling constant.  The mass
term changes sign at the critical temperature $T_{\rm c}$.  We have added a
gauge-fixing term with parameter $\alpha$.  To acquire a physical
understanding of what the Hamiltonian (\ref{gl:H}) describes, we bring to
the attention the well-know fact \cite{GFCM} that a $|\phi|^4$-theory gives
a field theoretic description of a gas of closed loops with contact
repulsion.  This equivalence is based on Feynman's observation
\cite{Feynman50} that the Green function
\begin{equation} \label{gl:start}
G({\bf x}) = \int_{\bf k} \frac{{\rm e}^{i {\bf k}
\cdot {\bf x}}}{{\bf k}^2 + m_\phi^2}
\end{equation}
of the free theory with positive mass term,
\begin{equation} \label{gl:Hamilton}
{\cal H}_0 = |\nabla \phi|^2 + m_\phi^2 |\phi|^2 ,
\end{equation}
can be expressed as a path integral.  (In contrast to the previous two
chapters, we in this and the following chapter denote the free Green function
by $G$ without an index 0.)  To this end, the Schwinger proper-time method is
invoked to write the right-hand side of (\ref{gl:start}) as an integral over
the proper time $\tau$ \cite{proptime}:
\begin{eqnarray} \label{gl:green}
G({\bf x}) &=& \int_0^{\infty} \dd \tau \, {\rm e}^{-\tau m_\phi^2}
\int_{\rm k} {\rm e}^{i {\bf k} \cdot {\bf x} } {\rm e}^{-\tau{\bf k}^2}
\nonumber \\ &=&
\int_0^{\infty} \dd \tau \, {\rm e}^{-\tau m_\phi^2}
\left( \frac{1}{4 \pi \tau} \right)^{3/2} {\rm e}^{-\frac{1}{4} {\bf x}^2/\tau},
\end{eqnarray} 
where we used the identity
\begin{equation} \label{gl:gamma}
\frac{1}{a} = \int_0^\infty \dd \tau \, {\rm e}^{- \tau a}.
\end{equation} 
According to the path-integral formulation of quantum mechanics
\cite{Feynman48}, the right-hand side of (\ref{gl:green}) can be represented
as a sum over all paths of a relativistic particle with mass $m_\phi$
running from $0$ at imaginary time 0 to ${\bf x}$ at time $\tau$:
\begin{equation} \label{gl:feynrep}
G({\bf x}) = \int_0^{\infty} \dd \tau \int_{{\bf x}(0)=0}^{{\bf x}(\tau)={\bf x}}
{\cal D} {\bf x}(\tau') {\rm e}^{-S_0}.
\end{equation} 
The (Euclidean) action
\begin{equation} \label{gl:world}
S_0 = \int_0^\tau \dd \tau' \left[\tfrac{1}{4} \dot{\bf x}^2 (\tau') +
m_\phi^2 \right],
\end{equation} 
with $\dot{\bf x}(\tau) = \dd {\bf x}(\tau)/\dd \tau$, is an elusive
representation of ($m_\phi$ times) the length $L$ of the path
\begin{equation}  \label{gl:length}
m_\phi L = m_\phi \int_0^\tau \dd \tau' \sqrt{\dot{\bf x}^2 (\tau')} ,
\end{equation} 
with $m_\phi$ now being interpreted as the line tension.  To establish this
connection we consider the canonical momentum obtained from
(\ref{gl:length})
\begin{equation} 
{\bf p} = m_\phi \frac{\dot{\bf x}}{\sqrt{\dot{\bf x}^2}}
\end{equation} 
which is seen to satisfy the constraint 
\begin{equation} 
{\bf p}^2 = m_\phi^2.
\end{equation} 
To incorporate this constraint we go over to the Hamilton formalism and
write instead of (\ref{gl:length})
\begin{equation} 
m_\phi L \rightarrow \int_0^\tau \dd \tau' \left[{\bf p} \cdot \dot{\bf x} -
\alpha ({\bf p}^2 - m_\phi^2)\right]
\end{equation} 
where the constraint is implemented with the help of the Lagrange multiplier
$\alpha$.  The path integral is now over phase space.  Because the action is
invariant under reparameterization, i.e., under the transformation $\tau
\rightarrow \bar{\tau}(\tau)$, this multiplier can be given a fixed value;
$\alpha = 1$, for example.  The integral over ${\bf p}$ is Gaussian and
immediately yields the action (\ref{gl:world}).

The partition function (with fields and coupling constants rescaled such
that no explicit temperature dependence appears in the Boltzmann factor)
\begin{equation} 
Z_0 = \int \DD \phi^* \DD \phi \, \exp\left(- \int_{\bf x} {\cal H}_0
\right),
\end{equation} 
involves only closed paths:
\begin{eqnarray} 
\ln(Z_0) &=& -\ln [ {\rm Det} ({\bf p}^2 + m_\phi^2)] = - {\rm Tr} \ln (
{\bf p}^2 + m_\phi^2)  \\
&=& \int_0^{\infty}
\frac{d\tau}{\tau} {\rm e}^{-\tau m^2_\phi} \int_{\bf k} {\rm
e}^{-\tau {\bf k}^2} = \int_0^\infty \frac{d \tau}{\tau} \oint {\cal D} {\bf
x} (\tau') {\rm e}^{-S_0}, \nonumber
\end{eqnarray} 
with $S_0$ the action (\ref{gl:world}).  

The $|\phi|^4$-interaction in the Ginzburg-Landau model results in an
additional term 
\begin{equation} 
S_\lambda  = - \lambda \int_0^{\tau} \dd \tau' \dd \tau'' \, \delta \left[
{\bf x} (\tau') - {\bf x} (\tau'') \right]
\end{equation} 
in the action, which gives an extra weight each time two loops---one
parameterized by $\tau'$ and one by $\tau''$---intersect.  Physically, it
represents a repulsive contact interaction between loops.  Finally, the
coupling of the field $\phi$ to the magnetic vector potential ${\bf A}$ via
the electric current 
\begin{equation} 
{\bf j}_e = - 2 e i\phi^* \stackrel{\leftrightarrow}{\nabla} \phi -2
(2e)^2 {\bf A} |\phi|^2
\end{equation} 
with a charge $2e$, results in the additional term
\begin{equation} 
S_e = 2ie \int_0^{\tau} \dd \tau' \,  \dot{\bf x} \cdot {\bf A}[{\bf x}(\tau')], 
\end{equation} 
showing that the loops described by the Ginzburg-Landau model are electric
current loops.  

On entering the superconducting phase, characterized by a negative mass term
($m_\phi^2 < 0$), the electric current loops proliferate and $|\phi|$
develops a vacuum expectation value.  In the London limit, the fluctuations
in the modulus of $\phi$ are neglected and the field is written as
\begin{equation} \label{gl:London}
\phi ({\bf x}) = \frac{\bar{\phi}}{\sqrt{2}} {\rm e}^{i\varphi({\bf x})},
\end{equation} 
with $\bar{\phi}$ a constant.  Formally, this corresponds to taking the
limit $m_\phi^2 \rightarrow - \infty$ and $\lambda
\rightarrow \infty$ in the Hamiltonian such that
\begin{equation}  \label{gl:phi0}
|\bar{\phi}|^2 = - \frac{m_\phi^2}{\lambda}
\end{equation}
is finite.  Physically, $|\bar{\phi}|^2$ denotes the mass density $\bar{\rho}_{\rm
s}$ of the superconducting electrons.

Let us first define an order parameter describing the superconducting state.
The field $\phi$ as it stands can not play this role because it is not gauge
invariant.  But it can be used to construct a gauge-invariant
Mandelstam-like operator in the following way
\begin{equation} \label{scop}
W(L_{\bf z}) = \exp\left\{i \left[\varphi ({\bf z}) + \int_{\bf x} {\bf
A} ({\bf x}) \cdot {\bf J} ({\bf x})\right] \right\},
\end{equation} 
where $\varphi$ is the phase of $\phi$ and ${\bf J}$ describes an
external current line originating in ${\bf z}$,
\begin{equation}  \label{Epl}
\nabla \cdot {\bf J} ({\bf x}) = 2e \delta({\bf x} - {\bf z}),
\end{equation} 
or
\begin{equation}  \label{Jline}
J_i({\bf x}) = 2e \int_{L_{\bf z}} \dd y_i \delta ({\bf x} -
{\bf y}).
\end{equation} 
Here, $L_{\bf z}$ denotes a path running from ${\bf z}$ to infinity.  The
second term in (\ref{scop}) is incorporated to render the operator gauge
invariant. Indeed, under a gauge transformation
\begin{equation} 
{\bf A} ({\bf x}) \rightarrow {\bf A} ({\bf x}) + \nabla \alpha ({\bf
x} ), \; \; \;  \varphi ({\bf z}) \rightarrow \varphi ({\bf z}) + 2e
\alpha ({\bf z}),
\end{equation} 
the operator $W(L_{\bf z})$ is invariant
\begin{equation} 
W(L_{\bf z}) \rightarrow W(L_{\bf z}) \exp \left[ i \alpha ({\bf
z}) + i \int_{\bf x} \nabla \alpha ({\bf x}) \cdot {\bf J}
({\bf x}) \right] = W(L_{\bf z}),
\end{equation} 
where in the last step we performed an integration by parts.  The gauge
invariance of $W(L_{\bf z})$ can be made more explicit by writing it in the
equivalent form
\begin{equation} 
W(L_{\bf z}) = \exp \left[ -\frac{i}{2e} \int_{\bf x} ( \nabla \varphi -
2 e {\bf A}) \cdot {\bf J} \right].
\end{equation} 

To show that this operator indeed is the order parameter of the
superconducting state, we calculate the correlation function
\begin{equation} 
\langle W(L_{\bf z}) W^*(L_{\bar {\bf z}}) \rangle = \int \DD {\bf A}  \DD
\varphi \, W(L_{\bf z}) W^*(L_{\bar {\bf z}}) \, \exp\left(-\int_{\bf x}
{\cal H} \right), 
\end{equation} 
obtained by introducing besides an external current line
originating in ${\bf z}$ also one terminating at ${\bar {\bf z}}$.  The
Hamiltonian ${\cal H}$ is the Ginzburg-Landau Hamiltonian (\ref{gl:H}) in
the London limit.  Since both integrations are Gaussian, they are easily
carried out with the result \cite{MA}
\begin{eqnarray}  \label{expO} 
\lefteqn{\langle W(L_{\bf z}) W^*(L_{\bar {\bf z}})\rangle =}
\\ && \!\!\!\!\!\!\!\! \exp \biggl\{ - \frac{1}{2} \int_{{\bf x}, {\bf y}}
\left[ \frac{1}{m_A^2} \rho ({\bf x})\, G({\bf x} - {\bf y}) \, \rho ({\bf y})
+ J_i ({\bf x}) \, G({\bf x} - {\bf y}) \, J_i ({\bf y})
\right] \biggr\} \nonumber ,
\end{eqnarray} 
where $\rho ({\bf x}) = 2e[\delta ({\bf x} - {\bf z}) - \delta ({\bf x} -
{\bar {\bf z}} )]$ denotes the electric current source and sink, and
\begin{equation}  \label{Yuka}
G ( {\bf x} ) = \int_{\bf k} \frac{ {\rm e}^{i {\bf k} \cdot {\bf
x}}}{{\bf k}^2+m^2_A} = \frac{1}{4 \pi} \frac{{\rm e}^{-m_A |{\bf x}|}}{
|{\bf x}|}
\end{equation} 
is the scalar Green  function, with
\begin{equation} \label{photonmass}
m_A = 2e |\bar{\phi}|
\end{equation}  
the photon mass.

In the superconducting phase, the electric current lines proliferate and
carry no energy; they are of no physical significance, only the endpoints
are.  The external current ${\bf J}$ can then be written as a
gradient of a potential $U$,
\begin{equation}  \label{pot}
{\bf J} = - \nabla U,
\end{equation} 
with $\nabla^2 U ({\bf x}) = - \rho({\bf x})$.  In this way,
the correlation function becomes
\begin{equation} 
\langle W({\bf z}) W^* ({\bar {\bf z}}) \rangle = \exp \left[ -
\frac{1}{2 m_A^2} \int_{{\bf x},{\bf y}}  \rho({\bf x}) G_0
({\bf x} - {\bf y}) \rho ({\bf y}) \right],
\end{equation} 
where $G_0 ({\bf x} - {\bf y})$ is the gapless scalar Green function
obtained from the massive one (\ref{Yuka}) by taking the limit $m_A
\rightarrow 0$.  For ${\bf x} = {\bf y}$ we have a diverging
self-interaction which is irrelevant and can be eliminated by defining a
renormalized operator $W_{\rm r}$ via
\begin{equation}  \label{renopW}
W_{\rm r} ({\bf z}) = W({\bf z}) \exp\left[ \frac{1}{2|\bar{\phi}|^2}
G(0) \right].
\end{equation} 
We then find for this operator
\begin{equation} \label{opsc}
\langle W_{\rm r}({\bf z}) W_{\rm r}^* ({\bar {\bf z}}) \rangle =
\exp \left( \frac{1}{4 \pi |\bar{\phi}|^2} \frac{1}{|L_{ {\bf z} {\bar {\bf
z}}}|} \right),
\end{equation} 
where $L_{{\bf z}{\bar {\bf z}}}$ is a path connecting the source at
${\bf z}$ with the sink at ${\bar {\bf z}}$, and $|L_{{\bf z}{\bar {\bf
z}}}|$ is the length of the path.  For large separation this correlation
function tends to one, implying that the operator $W_{\rm r}({\bf z})$
develops an expectation value in the superconducting phase.

It is important to realize that this operator is gauge invariant.
Frequently, the superconducting phase is referred to as a phase of
spontaneously broken gauge symmetry \cite{Weinbergbook}.  This should,
however, not be taken too literally.  A celebrated theorem due to Elitzur
\cite{Elitzur} states that a local gauge symmetry can never be spontaneously
broken.  An operator that by developing a nonzero vacuum expectation value
would spontaneously break the gauge symmetry does not exist. To break the
symmetry, the operator must transform under the gauge group, but then it
cannot develop an expectation value because only gauge-invariant objects
can.  

In the normal phase, electric current lines carry energy and therefore
become physical.  The form (\ref{pot}) is then no longer applicable and
we have to use Eq.~(\ref{Jline}) instead.  We shall assume that the
current lines have a certain width of the order $1/|m_\psi|$, where the
mass $m_\psi$ will be identified later on in this chapter (see page
\pageref{pag:relevance}).  To prevent infrared divergences, we give the
magnetic vector potential ${\bf A}$ a small mass $\mu$.  The correlation
function (\ref{expO}) then becomes after renormalization
\begin{equation}  \label{opnp}  
\langle W_{\rm r}(L_{\bf z}) W_{\rm r}^*(L_{\bar {\bf z}})\rangle =
\exp(-M_W |L_{{\bf 
z}{\bar {\bf z}}}|) \exp\left[\frac{1}{4 \pi}\left(\frac{2e}{\mu}\right)^2
\frac{{\rm e}^{- \mu |L_{{\bf z}{\bar {\bf z}}}|}}{|L_{{\bf z}{\bar {\bf
z}}}|} \right],
\end{equation}
where now $L_{{\bf z}{\bar {\bf z}}}$ denotes the shortest path connecting
the source at ${\bf z}$ with the sink at ${\bar {\bf z}}$, while $|L_{{\bf
z}{\bar {\bf z}}}|$ denotes the length and $M_W$ the line tension of the
current line,
\begin{equation} \label{MW}
M_W = \frac{(2e)^2}{4 \pi} \ln\left(\frac{|m_\psi|}{\mu}\right).
\end{equation} 
We see that in the limit $|L_{{\bf z}{\bar {\bf z}}}| \rightarrow
\infty$, the expectation value vanishes.  Although the operator $W$
distinguishes the superconducting phase from the normal phase, having a zero
expectation value in the normal phase and a nonzero one in the
superconducting phase, it is not an order parameter in the strict sense of
Landau in that it makes no symmetry statement.

\begin{figure}
\begin{center}
\epsfysize=2cm
\mbox{\epsfbox{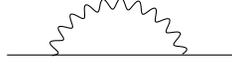}}
\end{center}
\vspace{-1.8cm}
\caption{One-loop mass correction.  The straight and wiggly lines represent
the $\phi$- and ${\bf A}$-field Green functions, respectively.
\label{fig:mass} }
\end{figure}

A few remarks are in order.  First, the line tension $M_W$ diverges when the
artificially introduced photon mass $\mu$ is taken to zero.  This means
that the current line connecting the source and sink becomes infinitely
heavy.  Isolated sources and sinks can therefore not exist in the normal
phase.  They are confined in neutral configurations of tightly bound pairs.
In Minkowski spacetime, where an electric current line becomes the worldline
of a charged particle, the energy (\ref{MW}) denotes the infrared-diverging
selfenergy \cite{Sen} of that particle obtained from evaluating the Feynman
graph depicted in Fig.~\ref{fig:mass}.
Second, the alert reader might wonder how the result (\ref{opsc}) obtained
in the normal phase is connected with Eq.\ (\ref{opnp}) obtained in the
superconducting phase.  In particular, the line tension $M_W$ of the current
line seems not to vanish at the critical temperature, implying that the
first exponential function in (\ref{opnp}) would survive the transition to
the superconducting phase, and that the operator $W$ would vanish there.
This is in apparent contradiction
\label{pag:contr} with previous statements that in the superconducting phase,
current lines carry no energy and proliferate.  This issue will be addressed
below (see page \pageref{pag:resol}) after we applied a duality
transformation to the Ginzburg-Landau model.

\begin{figure}
\epsfxsize=6.cm
\begin{center}
\mbox{\epsfbox{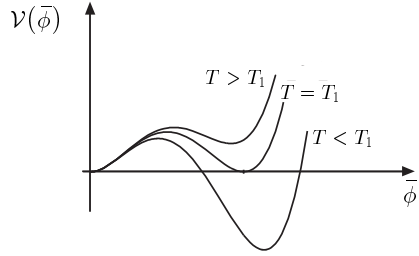}}
\end{center}
\caption{Effective potential up to one-loop order.}
\label{fig:potentialGL}
\end{figure}

We continue to investigate the phase transition by taking into account
fluctuations.  The one-loop effective potential is readily calculated in the
London limit (\ref{gl:London}).  In the gauge $\nabla \cdot {\bf A} = 0$,
one obtains
\begin{equation} \label{Veff}
{\cal V}_{\rm eff}(\bar{\phi}) = \int_{k<\Lambda } \ln(k^2 + m_A^2) = -
\frac{4}{3 \pi} e^3 |\bar{\phi}|^3.
\end{equation}
We regularized the integral by introducing an ultraviolet cutoff $\Lambda$.
An irrelevant ultraviolet divergence of the type $\Lambda^3\log \Lambda$ is
ignored, while a divergence of the type $\Lambda |\bar{\phi}|^2$ is absorbed
into a renormalization of the mass parameter, as usual.  The effective
potential ${\cal V}_{\rm eff}(\bar{\phi})$ has to be added to the tree
potential ${\cal V}_0 = \frac{1}{2}m_\phi^2 |\bar{\phi}|^2 +
\frac{1}{4}\lambda |\bar{\phi}|^4$.  It results in a 1st-order phase
transition taking place at a temperature $T_1$ {\it above} the temperature
$T_{\rm c}$ where the mass term of the Ginzburg-Landau model changes sign
(see Fig.\ \ref{fig:potentialGL}).
The (positive) value of $m_\phi^2$ at the 1st-order transition point,
obtained by equating the sum of the tree and the one-loop potential and also
its first derivative with respect to $\bar{\phi}$ to zero, is given by
\cite{alushta}
\begin{equation} 
m_\phi^2 = \frac{32}{9 \pi^2} \frac{e^6}{\lambda}.
\end{equation} 
At small values of the so-called Ginzburg-Landau parameter $\kappa_{\rm GL}$,
which is defined as the ratio of the two mass scales in the theory,
\begin{equation} \label{gl:gl}
\kappa_{\rm GL}= \frac{|m_\phi|}{m_A} = \frac{\lambda_{\rm L}}{\xi} =
\frac{\lambda}{e^2}, 
\end{equation} 
the reliability of this perturbative result has been verified numerically
\cite{DFK}.  It has been argued by Kleinert \cite{Kleinerttri} that for
larger values this result is, however, not meaningful anymore.  At some
critical value of the Ginzburg-Landau parameter $\kappa_{\rm GL}$, the
1st-order transition goes over into a 2nd-order transition.

Let us investigate the superconducting-to-normal transition further by
applying renormalization-group analysis to the Ginzburg-Landau model.  This
was first done by Halperin, Lubensky, and Ma \cite{HLM} in an $\epsilon( =
4-d)$-expansion around the upper critical dimension $d=4$.  We shall instead
use the fixed-dimension approach of Parisi \cite{Plect,Parisi}.  The model
(\ref{gl:H}) involves only a single complex field.  However, for reasons
that will become clear when we proceed, it is often generalized to contain
$2n$ real components:
\begin{equation} \label{gl:phi}
\phi \rightarrow \bbox{\phi} = \frac{1}{\sqrt{2}} \left( \begin{array}{c}
\phi_1 + i \phi_2 \\ \vdots \\ \phi_{2n-1} + i \phi_{2n} \end{array} \right).
\end{equation} 
The quantities appearing in (\ref{gl:H}) should be given an index $0$ to
indicate that they are bare quantities which will change in the process of
renormalization.  The renormalized fields and parameters are related to the
bare ones via
\begin{equation}   
{\bf A} = Z_{A}^{-1/2} {\bf A}^0 , \;\;\; e = Z_{A}^{1/2} e_0 , \;\;\;
\bbox{\phi} = Z_{\phi}^{-1/2} \bbox{\phi}^0 , \;\;\; \lambda =
Z_{\lambda}^{-1} Z_{\phi}^2 \, \lambda _0.
\end{equation} 
We shall work at criticality and to that end set the mass $m_\phi$ to zero.
To avoid infrared divergences, diagrams are evaluated at finite external
momentum $\kappa$.  The renormalization factors can be used to define the
following renormalization-group functions
\begin{equation}  
\gamma_A = \kappa \frac{\partial}{\partial \kappa} \ln(Z_A) \biggr|_0, \;\;\;
\gamma_\phi = \kappa \frac{\partial}{\partial \kappa} \ln(Z_\phi) \biggr|_0,
\end{equation} 
where the subscript 0 is to indicate that the bare coupling constants $e_0,
\lambda_0$ are kept fixed.  The dimensional parameter $\kappa$ plays the
role of renormalization group scale parameter.  The critical points
$\hat{e}^2_*$ and $\hat{\lambda}_*$ are determined by the zeros of the
renormalization-group beta functions,
\begin{equation} 
\beta_{e^2} = \kappa \frac{\partial}{\partial \kappa} \hat{e}^2 \biggr|_0,
\;\;\; \beta_{\lambda} = \kappa \frac{\partial}{\partial \kappa}
\hat{\lambda}\biggr|_0,
\end{equation} 
where $\hat{e}^2 = e^2 \kappa^{d-4}$ and $\hat{\lambda}= \lambda
\kappa^{d-4}$ are the dimensionless coupling constants.  When evaluated at
the critical point, the renormalization-group functions $\gamma_A$ and
$\gamma_\phi$ yield the anomalous dimension $\eta_A$ and $\eta_\phi$ of the
${\bf A}$- and $\phi$-field, respectively.

The various renormalization factors are readily evaluated in arbitrary
dimension $2 < d < 4$, using the fixed-dimension approach of Parisi
\cite{Plect,Parisi}.  To the one-loop order we find
\begin{eqnarray}   \label{gl:laZ} 
\!\!\!\!\!\!\!\!\!\!\!\!\!\! Z_A \!\!\! &=& \!\!\! 1 - n \frac{c(d)}{d-1}
\hat{e}^2 \label{gl:ZA} \\ \!\!\!\!\!\!\!\!\!\!\!\!\!\! Z_\phi \!\!\! &=&
\!\!\! 1 + c(d)[(d-1) - \alpha (d-3)] \hat{e}^2 \label{gl:phiZ} \\
\!\!\!\!\!\!\!\!\!\!\!\!\!\! Z_\lambda \!\!\! &=& 
\!\!\! 1 + c(d) \left[ (2n+8) \hat{\lambda} - 2 \alpha (d-3) \hat{e}^2
+\tfrac{1}{4} d(d-1) \hat{e}^4 /\hat{\lambda} \right],
\end{eqnarray} 
where $c(d)$ stands for the one-loop integral
\begin{equation} 
c(d) = \int_{\bf k} \frac{1}{{\bf k}^2 ({\bf k} + {\bf q})^2}
\biggr|_{{\bf q}^2=1} =
\frac{\Gamma(2-d/2) \Gamma^2(d/2-1)}{(4 \pi)^{d/2} \Gamma(d-2)}.
\end{equation}
In deriving this we used analytic regularization to handle the ultraviolet
divergences.  With the one-loop expressions for the renormalization factors,
we obtain for the beta functions
\begin{eqnarray}  
\beta_{e^2} \!\!\!\! &=& \!\!\!\! (d-4) \left[ \hat{e}^2 - n \frac{c(d)}{d-1}
\hat{e}^4 \right] \label{betae} \\ \beta_\lambda \!\!\!\! &=& \!\!\!\!
(d-4)\left\{ \hat{\lambda} - c(d) \left[(2n+8) \hat{\lambda}^2 - 2 (d-1)
\hat{\lambda} \hat{e}^2 + \tfrac{1}{4} d(d-1) \hat{e}^4
\right]\right\}. \nonumber \\ 
\label{betal} 
\end{eqnarray} 
Note that while $Z_\lambda$ and $Z_\phi$ depend on the gauge-fixing
parameter $\alpha$, the function $\beta_\lambda$ extracted from them is
independent thereof.  We expect this cancelation of gauge-dependence in 
$\beta_\lambda$ to persist in all orders of perturbation theory.
Although the beta functions are gauge independent, the locations of the
fixed points are not universal.  They depend on the regularization chosen as
well as on the renormalization prescription.  In the limit $d \rightarrow 4
-\epsilon$ our formulas reduce to the know form \cite{HLM,Kang}.

\begin{figure}
\vspace{-1.cm}
\begin{center}
\epsfxsize=4.cm
\mbox{\epsfbox{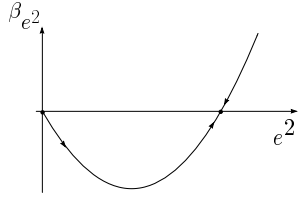}}
\end{center}
\vspace{-1.5cm}
\caption{Schematic representation of the
$\beta_{e^2}$-function. \label{fig:betae2}} 
\end{figure}

Remarkably, $\beta_{e^2}$ is independent of the self-coupling
$\lambda$.  By itself it gives rise to two fixed points (see
Fig.~\ref{fig:betae2}):
\begin{equation} 
\hat{e}^2_*= \left\{ \begin{array}{c} 0 \\
(d-1)/n c(d). \end{array} \right.
\end{equation} 
The neutral fixed point is ultraviolet stable, while the charged fixed point
is infrared stable in the $e^2$-direction.  Since also $Z_A$ as given in
Eq.~(\ref{gl:ZA}) and thus $\gamma_A$ are independent of $\lambda$, the
anomalous dimension $\eta_A$ of the gauge field can be calculated without
any knowledge of the location of the fixed points along the
$\lambda$-direction.  For the infrared-stable (IR) fixed point one finds at
the one-loop order \cite{HLM}
\begin{equation} \label{gl:etaA}
\eta_A = \gamma_A \bigr|_{\hat{e}_*^2} = 4-d.
\end{equation} 
It was pointed out by Herbut and Te\v{s}anovi\'{c} \cite{HeTe} that this is
an exact result which follows directly from the definition of the 
$\beta_{e^2}$-function in terms of the renormalization factor $Z_A$
\begin{equation}  
\beta_{e^2} = \kappa \frac{\partial}{\partial \kappa} \hat{e}^2 \bigr|_0 =
\hat{e}^2 [ (d-4) + \gamma_A].
\end{equation} 
That is to say, there are no higher-order corrections to the one-loop result
(\ref{gl:etaA}).  We have checked this explicitly to the two-loop order in
the $\epsilon (=4-d)$-expansion.  The anomalous dimension is such that the
dimension $d_A$ of the gauge field, which by naive power counting
equals $\tfrac{1}{2}(d-2)$, is exactly unity,
\begin{equation} 
d_A = 1 \;\;\;\;\;\;\; {\rm for} \;\;\;\;\; 2 < d <4.
\end{equation} 
It is as if we were in the upper critical dimension $d=4$.

Since the result is independent of the number $2n$ of field components
contained in the theory, this conclusion can be confirmed in the
$1/n$-expansion.  To leading order in $1/n$, the inverse gauge field
propagator reads in the gauge $\nabla \cdot {\bf A} =0$ and for $2 < d <4$
\begin{equation}  \label{gl:P1/n}     
\Pi_{i j}({\bf q}) = -\frac{c(d)}{d-1} P_{i j}({\bf q}) |{\bf q}|^{d-2}, 
\end{equation} 
with $P_{i j}({\bf q})$ the transverse projection operator
\begin{equation} \label{gl:Pi}
P_{i j}({\bf q}) = \delta_{i j} - \frac{q_i q_j}{{\bf q}^2}.
\end{equation} 
The exponent $d-2$ of the momentum in (\ref{gl:P1/n}) equals $d - 2 d_A$, so
that we again arrive at the conclusion that $d_A=1$ in $2 < d <4$.

It implies that near the charged fixed point, the electric charge scales
with the correlation length $\xi$ in a way one expects from naive power
counting
\begin{equation} 
e^2 = Z_A e_0^2 \sim \xi^{d-4}
\end{equation} 
since $Z_A$ scales as $\xi^{-\eta_A}$.  

The exact result $d_A=1$ has far reaching consequences as it implies
that the term $(\nabla \times {\bf A})^2$ in the Hamiltonian, which by naive
power counting is marginal, has dimension 4.  It is therefore an irrelevant
operator for all $d$ below the upper critical dimension and can be omitted.
Formally, this corresponds to taking the limit $e\rightarrow
\infty$ in the Ginzburg-Landau model as can be seen after a rescaling of the
gauge field ${\bf A} \rightarrow {\bf A}/e$.  Without the term $(\nabla
\times {\bf A})^2$, the Ginzburg-Landau model resembles the CP$^{n-1}$
model,
\begin{equation} \label{gl:CPN}
{\cal L}_{\rm CP} = \left|(\nabla - i {\bf A})\bbox{\phi}\right|^2,
\end{equation}	
which has the additional constraint 
\begin{equation} \label{gl:constraint}
|\bbox{\phi}|^2=1.
\end{equation}   
As regards to critical exponents this difference is however irrelevant.
Indeed, the well-studied O($n$) $\bbox{\phi}^4$-model and its nonlinear
version, the O($n$) nonlinear sigma model where the fields satisfy a
condition like (\ref{gl:constraint}), are generally accepted to be in the
same universality class \cite{HiBr}.  Whereas the linear model is usually
investigated in an $\epsilon$-expansion around the upper critical dimension
($d=4$), the nonlinear model is often treated in an expansion around the
lower critical dimension ($d=2$), so a direct comparison is not possible.
But the $1/n$-expansion, being the same for both theories and applicable in
arbitrary dimension $2 < d <4$, can be used to bridge these two regions.
The results obtained in this way for $d$ close to the upper and lower
critical dimension are identical to those obtained in the respective
$\epsilon$-expansions for large $n$.

The $1/n$-expansion has also been employed to argue that the
Ginzburg-Landau model with $2n$-com\-po\-nents and the CP$^{n-1}$ model,
too, are in the same universality class \cite{Hikami,VaNa}.  Our
conclusion that below the upper critical dimension, the term $(\nabla
\times {\bf A})^2$ in the Ginzburg-Landau model is irrelevant gives
additional support to this conjecture.

For a superconductor, having one complex field ($n=1$), this implies that
the relevant model is the CP$^0$ model.  Although the $\epsilon
(=d-2)$-expansion \cite{Hikami} predicts a nontrivial 2nd-order transition,
this model has no degrees of freedom.  Physically, this means that the
charged degrees of freedom are irrelevant at the phase transition \cite{KKR}.
In itself this is no reason to reject the Ginzburg-Landau model as a basis
for studying the phase transition---the ``unphysical'' limit $n
\rightarrow 0$ of the O($n$) model describes the long-distance behavior
of a flexible polymer.  But what is more worrying is that vortices,
which we saw to be of paramount importance in two dimensions, are not
accounted for in the discussion so far.  In addition, there are
technical problems with using the Ginzburg-Landau model to study the
critical properties of the phase transition.

\begin{figure}
\vspace{-1.0cm}
\begin{center}
\epsfxsize=7.cm
\mbox{\epsfbox{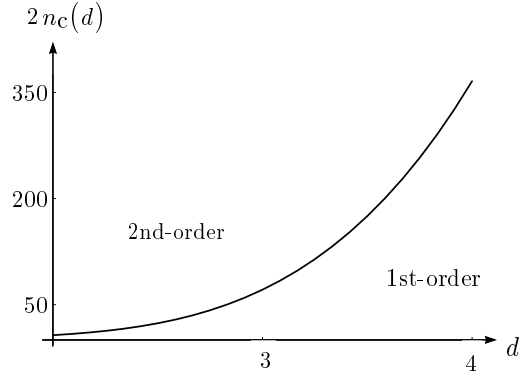}}
\end{center}
\vspace{-2.5cm}
\caption{The critical number of components $2n_c$ as a function of the
dimensionality $d$ below which the one-loop calculation predicts a 1st-order
transition. \label{fig:nc} }
\end{figure}

Namely, the two beta functions yield no IR fixed point
unless the number of field components is taken large enough.  Below a
critical number $n_c(d)$, determined by the condition that
\begin{equation} \label{deltad}
\Delta_d = 2 \sqrt{n^2 - 2 (d-2) (d-1)^2 (d+1) n - 4 (d-1)^3 (d+1)}
\end{equation} 
be real, the system has no IR fixed point and the one-loop
calculation predicts a 1st-order transition.  In Fig.~\ref{fig:nc} this
number is plotted as function of the dimensionality $d$.
We see that the number decreases when the dimensionality is reduced.  More
precisely,
\begin{eqnarray}  
2n_{\rm c}(4) &=&  12(15 + 4 \sqrt{15}) \approx 365.9 \\ 
2n_{\rm c}(3) &=&  16(2 + \sqrt{6}) \approx 71.2 \\  
2n_{\rm c} (2) &=&  4\sqrt{3} \approx 6.9,
\end{eqnarray} 
where the $d=4$-result is due to Halperin, Lubensky, and Ma \cite{HLM}.  The
figure also shows that for a fixed number of field components, 2nd-order
behavior is favored when $d \rightarrow 2$, while in the opposite limit, $d
\rightarrow 4$, 1st-order behavior is favored.  However, even close to the
lower critical dimension, where the CP$^{n-1}$ model predicts a 2nd-order
phase transition all the way down to $n=1$ \cite{Hikami}, the
Ginzburg-Landau model still requires more than $2\sqrt{3} \approx 3.5$
complex fields in order to have an IR fixed point.  The status of the
results obtained from the Ginzburg-Landau model in low-order perturbation
theory is therefore not clear.

Given the two-dimensional results, showing the importance of vortices, it is
natural to consider the dual theory of the three-dimensional Ginzburg-Landau
model as an alternative to study the phase transition.  As before, the dual
formulation focuses on the vortices in the system, which in three dimensions
are line defects.  It will again turn out that the dual formulation is one
directly in terms of physical variables and for that reason involves only a
global, not a local, symmetry.  We will see that the phase transition of a
type-II superconductor at zero external field becomes more tractable in the
dual formulation than in the original Ginzburg-Landau formulation.
\section{Dual Ginzburg-Landau Model}
\label{sec:dgl}
In this section, we apply a duality transformation to the Ginzburg-Landau
model.  The basic idea of this approach originates from
three-dimensional lattice studies carried out two decades ago
\cite{BMK,Peskin,TS,Savit}.  These studies were instigated by the
success of the Kosterlitz-Thouless theory of the phase transition in a
superfluid film \cite{Berezinskii,KT73}, which we discussed in Sec.\
\ref{sec:kt}.  The three-dimensional lattice studies of the neutral 
\cite{BMK} and charged $xy$-model \cite{Peskin,TS} were aimed at
obtaining a dual description in terms of vortices.  Following a suggestion
by Helfrich and M\"uller \cite{HeMu}, Dasgupta and Halperin \cite{DaHa}
carried out a lattice simulation of the superconductor-to-normal phase
transition in the dual formulation.  Their study revealed that for small
values of the electric charge $e$, which implies large values for the
Ginzburg-Landau parameter (\ref{gl:gl}), the transition was 2nd-order with
$xy$-exponents.  A detailed account of these matters as well as an extensive
list of references to the literature can be found in Ref.\ \cite{GFCM}.

Another development underscoring the importance of the dual approach to the
Ginzburg-Landau model was initiated in Ref.\ \cite{KRE}.  The basic
observation was that since a local gauge symmetry can never be broken
\cite{Elitzur}, a local gauge description of a phase transition is not
feasible.  It was argued that the three-dimensional Ginzburg-Landau theory
contains, in addition to the local gauge symmetry, another {\it global} U(1)
symmetry.  When considered in $2+1$ space-time dimensions with a Minkowski
metric, this symmetry is generated by the magnetic flux operator.  It was
demonstrated that this symmetry is broken in the normal phase, while it is
unbroken in the superconducting phase.  A genuine order parameter in the
sense of Landau was given, and it was shown that the massless photon of the
normal phase is the Goldstone mode associated with the broken flux symmetry.

To see in which regime magnetic vortices or Abrikosov flux tubes are
important, we introduce a magnetic monopole at some point ${\bf z}$ inside
the system.  A monopole is a source of magnetic flux.  Due to the Meissner
effect, the flux lines emanating from the monopole are squeezed into a flux
tube.  In this way we have created a magnetic vortex at zero external field.
Electrodynamics in the presence of a monopole was first described by Dirac
\cite{Dirac} who argued that the combination
\begin{equation}   \label{combi}
\nabla \times {\bf A}({\bf x}) - {\bf B}^{\rm P}({\bf x})
\end{equation} 
is the physical local magnetic induction ${\bf h}$.  The subtracted
plastic field
\begin{equation} \label{subtr}
B_i^{\rm P} ({\bf x}) = \Phi_0 \int_{L_{\bf z}} \dd y_i \, \delta ({\bf
x} - {\bf y}),
\end{equation}
with $\Phi_0 = \pi/e$ the magnetic flux quantum, removes the field of the
so-called Dirac string running along some path $L_{\bf z}$ from the location
${\bf z}$ of the monopole to infinity.  On account of Stokes' theorem, the
plastic field satisfies the equation
\begin{equation} 
\nabla \cdot {\bf B}^{\rm P} ({\bf x}) = \Phi_0 \, \delta
({\bf x} - {\bf z}).
\end{equation}  

In the presence of the monopole, the Ginzburg-Landau Hamiltonian becomes
in the London limit and after integrating out the phase field $\varphi$
\begin{equation}   \label{HP}
{\cal H}^{\rm P} = \frac{1}{2} (\nabla \times {\bf A} - {\bf B}^{\rm P})^2 +
\frac{1}{2} m_A^2 A_i \left( \delta_{i j} - \frac{\partial_i
\partial_j }{\nabla^2} \right) A_j + \frac{1}{2\alpha}(\nabla \cdot
{\bf A})^2,
\end{equation} 
where we gave ${\cal H}$ the superscript ${\rm P}$ to indicate the
presence of the monopole.  Let us first verify Dirac's assertion that
the combination (\ref{combi}) describes a point monopole with its Dirac
string removed and consider the field equation for the gauge field,
\begin{equation}  \label{clas}
A_i ({\bf x}) = \int_{\bf y} G_{i j} ({\bf x} - {\bf y})
\left[\nabla \times {\bf B}^{\rm P} ({\bf y})\right]_j.
\end{equation}  
The gauge-field Green function $G_{i j}$ appearing here is
\begin{equation} \label{gfprop} 
G{i j} ({\bf x}) = \int_{\bf k} \left(
\frac{\delta_{i j} - (k_i k_j)/{\bf k}^2}{{\bf k}^2+m_A^2} + \alpha
\frac{k_i k_j}{{\bf k}^4} \right) {\rm e}^{i {\bf k} \cdot {\bf x}}.
\end{equation}
The local magnetic induction corresponding to the classical solution
given in (\ref{clas}) is
\begin{equation} \label{bclas}
\nabla \times {\bf A}({\bf x}) - {\bf B}^{\rm P}({\bf x}) =
\Phi_0 \nabla G({\bf x} - {\bf z}) - m_A^2 \int_{\bf y}
G({\bf x}- {\bf y}) {\bf B}^{\rm P} ({\bf y}),
\end{equation} 
where $G$ is the Green function (\ref{Yuka}).  The first term at the
right-hand side corresponds to a screened Coulomb force generated by the
monopole.  The last term, which is only present in the superconducting phase
where the Meissner effect is operating and $m_A\neq 0$, describes the
magnetic vortex.  If we calculate from the right-hand side of (\ref{bclas})
the magnetic flux through a plane perpendicular to the Dirac string, we find
that precisely one flux quantum pierces the surface in the negative
direction
\begin{equation} \label{Diracisright}
\int \dd^2 x_i \left[ \Phi_0 \partial_i G({\bf x} - {\bf z})
- m_A^2 \int_{\bf y} G({\bf x}- {\bf y}) B^{\rm P}_i ({\bf y})
\right] = - \Phi_0.
\end{equation} 
Here, $\dd^2x_i$ denotes an element of the surface orthogonal to the Dirac
string.  Equation (\ref{Diracisright}) confirms Dirac's assertion that the
magnetic flux emanating from a monopole must be supplied by an
infinitesimally thin string of magnetic dipoles and that in order to obtain
the true local magnetic induction of a genuine point monopole, this string
has to be subtracted.  Whence, the ${\bf B}^{\rm P}$-term at the left-hand
side of (\ref{bclas}).  While the Dirac string is immaterial in the normal
phase, the last term at the right-hand side of (\ref{bclas}) shows that it
acquires physical relevance in the superconducting phase where it serves as
the core of the Abrikosov flux tube
\cite{Nambu}.

The energy $E_V$ of this configuration is obtained by substituting the
solution (\ref{clas}) back into the Hamiltonian (\ref{HP}).  It is
divergent in the ultraviolet because in the London limit, where the mass
$|m_\phi|$ of the $\phi$-field is taken to be infinite, the vortices are
considered to be ideal lines.  For a finite mass, a vortex core has a
typical width of the order of the coherence length $\xi =1/|m_\phi|$.
This mass therefore provides a natural ultraviolet cutoff to the theory.
Omitting the (diverging) monopole self-interaction, one finds
\cite{Nambu}
\begin{equation}   \label{moncon}
E_V = \frac{1}{2} g^2 \int_{L_{\bf z}} \dd x_i \int_{L_{\bf z}} \dd
y_i\, G ({\bf x} - {\bf y}) 
= M_V \left| L_{\bf z} \right|,
\end{equation}
with $g = \Phi_0 m_A$ a combination we also encountered in the
two-dimensional charged models, $|L_{\bf z}|$ the (infinite) length of the
flux tube, and
\cite{Abrikosov}
\begin{equation}  \label{M} 
M_V= \frac{1}{8\pi} g^2 \ln\left(
\frac{|m_\phi|^2}{m_A^2} \right) = \frac{1}{4\pi} g^2 \ln (
\kappa_{\rm GL} )  
\end{equation} 
the line tension.  The value $\kappa_{\rm GL} = 1/\sqrt{2}$ for the
Ginzburg-Landau parameter (\ref{gl:gl}) separates the type-II regime
$(\kappa_{\rm GL} > 1/\sqrt{2})$, where isolated vortices can exist, from
the type-I regime $(\kappa_{\rm GL} < 1/\sqrt{2})$, where a partial
penetration of an external field is impossible.  Remembering that $L_{\bf
z}$ was the Dirac string, we see from (\ref{moncon}) that it indeed becomes
the core of a vortex in the superconducting phase.

To identify the regime where vortices are important, we convert the line
tension $M_V$ into a magnetic field via
\begin{equation} 
H_{{\rm c}_1} = \frac{M_V}{\Phi_0},
\end{equation} 
and compare it with the critical field 
\begin{equation} 
H_{\rm c} = \frac{1}{2\sqrt{2} \pi} g  |m_\phi|,
\end{equation} 
obtained by equating the tree potential 
\begin{equation} 
{\cal V}_0 = \tfrac{1}{2} m_\phi^2 |\bar{\phi}|^2 + \tfrac{1}{4}\lambda
|\bar{\phi}|^4
\end{equation} 
to $-\tfrac{1}{2} H_{\rm c}^2$.  The critical field $H_{{\rm c}_1}$
physically denotes the smallest value of an external field needed to excite
vortices.  The field $H_{\rm c}$, on the other hand, is a measure of the
condensation energy which in turn sets the energy scale.  It physically
denotes the value of an external field at which a type-I superconductor can
no longer resist the magnetic pressure and reverts to the normal phase
characterized by a perfect penetration of the field.  The ratio of the two
fields
\begin{equation} 
\frac{H_{{\rm c}_1}}{H_{\rm c}} = \frac{1}{\sqrt{2}}\frac{\ln(
\kappa_{\rm GL})}{\kappa_{\rm GL}} 
\end{equation} 
shows that for increasing $\kappa_{\rm GL}$, vortices become easier to
excite.  In other words, the deeper one enters the type-II regime, the more
important vortex excitations become.  Since these have been ignored in the
calculation of the effective potential (\ref{Veff}), the prediction of a
1st-order transition is only reliable at small $\kappa_{\rm GL}$ and breaks
down at larger values of the Ginzburg-Landau parameter.

The above construct of inserting a monopole into the system can be used to
define a so-called disorder parameter \cite{GFCM}.  In contrast to the order
parameter (\ref{scop}), a disorder parameter should develop an expectation
value in the normal, not in the superconducting phase.  The operator
$V(L_{\bf z})$ describing the monopole with its emerging flux tube is easily
obtained by noting that in the functional-integral approach, a given field
configuration is weighted with a Boltzmann factor $\exp \left(-\int_{\bf x}
{\cal H}^{\rm P}\right)$, where the Hamiltonian is given by (\ref{HP}).
From this we infer that the explicit form of the operator is 
\begin{equation} \label{dop} 
V(L_{\bf z}) = \exp \left\{ \int_{\bf x} \left[(\nabla \times {\bf A})
\cdot {\bf B}^{\rm P} - \tfrac{1}{2} \left({\bf B}^{\rm P}
\right)^2  \right] \right\}.
\end{equation} 
We are interested in the correlation function $\langle V( L_{{\bf z}} ) V^*(
L_{\bar {\bf z}}) \rangle$, where $V^*(L_{\bar {\bf z}})$
describes an additional antimonopole brought into the system at ${\bar {\bf
z}}$, with $L_{\bar {\bf z}}$ being the accompanying Dirac string running
from infinity to ${\bar {\bf z}}$.  Since all the integrals involved are
Gaussian, this expectation value can be evaluated directly.  However, we
proceed in an indirect way to reveal some aspects of the nature of the dual
theory and first linearize the functional integral over the gauge field by
introducing an auxiliary field $\tilde{\bf h}$.  In the gauge $\nabla
\cdot {\bf A} = 0$, which corresponds to setting $\alpha=0$, we find
\begin{eqnarray}    \label{Vdu}
\lefteqn{\langle V( L_{\bf z} ) V^*(L_{\bar {\bf z}}) \rangle = }
\\ && \int \DD
{\bf A} \DD \tilde{\bf h} \exp \left\{ \int_{\bf x}  \left[ -\frac{1}{2}
\tilde{\bf h}^2 + i \tilde{\bf h} \cdot (\nabla \times {\bf A} -
{\bf B}^{\rm P}) - \frac{m_A^2}{2}  {\bf A}^2 \right] \right\},
\nonumber 
\end{eqnarray} 
where now $\nabla \cdot {\bf B}^{\rm P}({\bf x}) = \Phi_0 [ \delta ({\bf
x}-{\bf z})- \delta({\bf x} - {\bar {\bf z}})]$.  To appreciate the
physical relevance of the auxiliary field, let us consider its field
equation 
\begin{equation}
\label{phys} 
\tilde{\bf h} = i (\nabla \times {\bf A} - {\bf B}^{\rm P}) = i {\bf h}.
\end{equation}
It tells us that apart from a factor $i$, $\tilde{\bf h}$ can be thought of as
representing the local magnetic induction ${\bf h}$.

The integral over the vector potential is easily carried out by
substituting the field equation for ${\bf A}$,
\begin{equation}
\label{fieldeq} 
{\bf A} = \frac{i}{m_A^2}  \nabla \times \tilde{\bf h},
\end{equation} 
back into (\ref{Vdu}), with the result 
\begin{eqnarray}   \label{Vdua}
\lefteqn{\langle V( L_{\bf z} ) V^*(L_{\bar {\bf z}}) \rangle =} \\ && \int
\DD \tilde{\bf h} \, \exp \left\{ 
-\frac{1}{2} \int_{\bf x} \left[ \frac{1}{m_A^2}(\nabla \times \tilde{\bf h})^2 +
\tilde{\bf h}^2 \right] - i \int_{\bf x} \tilde{\bf h} \cdot {\bf B}^{\rm P}
\right\}. \nonumber 
\end{eqnarray} 
This shows that magnetic vortices described by a plastic field ${\bf B}^{\rm
P}$ couple to the fluctuating massive vector field $\tilde{\bf h}$, with a
coupling constant given by $g = \Phi_0 m_A = 2 \pi |\bar{\phi}|$ as in two
dimensions.  As the temperature approaches the critical temperature from
below, $\bar{\phi}$ tends to zero, so that the vortices decouple from
$\tilde{\bf h}$.  The finite penetration depth in the superconducting phase
is reflected by the mass term of the $\tilde{\bf h}$-field (\ref{Vdua}).

\begin{figure}
\begin{center}
\epsfxsize=4cm
\mbox{\epsfbox{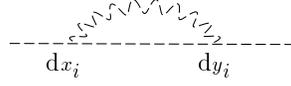}}
\end{center}
\vspace{-.5cm}
\caption{Biot-Savart interaction (wiggly line) between two line elements 
$\dd x_i$ and $\dd y_i$ of a magnetic vortex (straight line).
\label{fig:biotsavart} }
\end{figure}

After carrying out the integral over $\tilde{\bf h}$ in (\ref{Vdua}), we obtain
for the correlation function
\begin{eqnarray}  \label{vv*}
\langle V( L_{\bf z} ) V^*(L_{\bar {\bf z}}) \rangle = \exp \biggl\{ 
- \frac{1}{2} \int_{{\bf x}, {\bf y}} \bigl[ \!\!\!\!\!\!\!\!\!\!  &&
\rho_{\rm m} ({\bf x})\, G({\bf x} - {\bf y}) \, \rho_{\rm m} ({\bf y}) \\
\!\!\!\!\!\!\!\!\!\! && + m_A^2 B^{\rm P}_i ({\bf x}) \, G({\bf x} - {\bf
y}) \, B^{\rm P}_i ({\bf y}) \bigr] \biggr\}, \nonumber 
\end{eqnarray} 
where $\rho_{\rm m} ({\bf x}) = \Phi_0 [\delta ({\bf x} - {\bf z}) - \delta
({\bf x} - \bar{\bf z})]$ is the monopole density.  The first term in the
argument of the exponential function contains a diverging monopole
self-interaction for ${\bf x} = {\bf y}$.  This divergence is irrelevant and
can be eliminated by defining a renormalized operator
\begin{equation} 
\label{renopV}
V_{\rm r} (L_{\bf z}) = V(L_{\bf z}) \exp\left[ \tfrac{1}{2} \Phi_0^2 G(0)
\right].
\end{equation} 
The second term in the argument is the most important one for our purposes.
It represents a Biot-Savart interaction between two line elements $\dd x_i$
and $\dd y_i$ of the magnetic vortex (see Fig.~\ref{fig:biotsavart}).
For the renormalized operators we find
\begin{equation}  \label{correlation} 
\langle V_{\rm r}( L_{{\bf z}} ) V_{\rm r}^*( L_{\bar {\bf z}} )
\rangle = \exp(-M_V\left| L_{{{\bf z}}{\bar {\bf z}}} 
\right|) \exp\left(\frac{\Phi_0^2}{4 \pi} \frac{  
{\rm e}^{-m_A\left| L_{{{\bf z}}{\bar {\bf z}}} \right|} }{ \left|
L_{{{\bf z}}{\bar {\bf z}}} \right| } \right),
\end{equation} 
where $L_{{{\bf z}}{\bar {\bf z}}}$ is the flux tube connecting the
monopole at ${{\bf z}}$ with the antimonopole at ${\bar {\bf z}}$, and
$|L_{{{\bf z}}{\bar {\bf z}}}|$ is its length.  Initially, the two Dirac
strings may run to any point at infinity.  Due to the string tension,
however, they join on the shortest path $L_{{{\bf z}}{\bar {\bf z}}}$
between the monopoles.

The result (\ref{correlation}) is central to our line of arguments.  It
shows that the correlation function $\langle V_{\rm r}( L_{\bf z} )
V_{\rm r}^*( L_{\bar {\bf z}} )
\rangle$ behaves differently in the two phases \cite{KRE,Marino}.  In
the superconducting phase, where because of the Higgs mechanism $m_A \neq
0$, the first factor, which amounts to a confining linear potential between
the monopole and antimonopole, dominates.  As a result, the correlation
function decays exponentially for distances larger than $1/M_V$:
\begin{equation} 
\label{correlator}
\langle V_{\rm r}( L_{\bf z} ) V_{\rm r}^*(
L_{\bar {\bf z}}) \rangle \rightarrow 0.
\end{equation}
This behavior is typical for an operator in a phase without gapless
excitations.  On the other hand, in the high-temperature phase, where
$m_A=0$, the confinement factor in the correlation function
(\ref{correlation}) disappears, while the argument of the second exponential
turns into a pure Coulomb potential.  The correlation function remains,
consequently, finite for large distances:
\begin{equation}
\langle V_{\rm r}( L_{\bf z} ) V_{\rm r}^*(
L_{\bar {\bf z}} ) \rangle \rightarrow 1.
\end{equation} 
By the cluster property of correlation functions this implies that the
operator describing the finite vortex develops a vacuum expectation value.
This signals a proliferation of magnetic vortices.  Indeed, according to
(\ref{M}) the line tension $M_V$ of a vortex vanishes at the transition
point, where $\bar{\phi} \rightarrow 0$.  It should be noted that it is the
high-temperature phase and not the superconducting phase where $V_{\rm
r}(L_{\bf z})$ develops an expectation value.  Hence, it is indeed a
disorder parameter.

It is interesting to consider the limit $m_A \rightarrow 0$ in (\ref{Vdua}),
where the magnetic vortex decouples from the massive vector field.  This
limit yields the constraint $\nabla \times \tilde{\bf h} = 0$ which can be solved
by setting $\tilde{\bf h} = \nabla \gamma$.  The correlation function then takes
the simple form
\begin{equation} \label{dual:simple} 
\langle V_{\rm r}({\bf z}) V_{\rm r}^*(
{\bar {\bf z}}) \rangle = \int \DD \gamma \exp \left[ -\frac{1}{2} \int_{\bf
x} (\nabla \gamma)^2 + i \int_{\bf x} \gamma
\rho_{\rm m} \right].
\end{equation} 
In the absence of monopoles, the theory reduces to that of a free gapless
mode $\gamma$ that may be thought of as representing the magnetic scalar
potential.  This follows from combining the physical interpretation of the
vector field ${\bf h}$ (\ref{phys}) with the equation $\tilde{\bf h} = \nabla
\gamma$.  Specifically,
\begin{equation} 
\label{gammaid}
\nabla \gamma = i ( \nabla \times {\bf A} - {\bf B}^{\rm P} ).
\end{equation} 
Inserting the explicit form for the monopole density, which is given by
$\rho_{\rm m}({\bf x}) = \Phi_0 [\delta({\bf x}-{\bf z}) - \delta({\bf
x}-{\bar {\bf z}})]$, we see that in terms of the field $\gamma$, the
correlation function reads
\begin{equation}    \label{localrep}
\langle V_{\rm r}({\bf z} ) V_{\rm r}^*(
{\bar {\bf z}}) = \left\langle {\rm e}^{i \Phi_0 [\gamma ({\bf z})-
\gamma({\bar {\bf z}})] } \right\rangle.
\end{equation}  
This demonstrates that the operator $V_{\rm r}( L_{\bf z} )$ describing
the finite vortex, which was introduced in (\ref{dop}) via the singular
plastic field ${\bf B}^{\rm P}$ (\ref{subtr}), is now represented as an
ordinary field.  Since we are in the normal phase, where $V_{\rm r}$
develops a nonzero expectation value, the presence of the phase $\gamma$
indicates that this expectation value breaks a global U(1) symmetry,
with $\gamma$ the ensuing Goldstone field.  This point will be further
clarified below.

We note that Eq.\ (\ref{localrep}) reveals also that in the normal phase,
the Dirac string looses its physical relevance, the right-hand side
depending only on the end points ${\bf z}$ and ${\bar {\bf z}}$, not on the
path $L_{{\bf z}{\bar {\bf z}}}$ connecting these points.  The notion of a
magnetic vortex is of no relevance in this phase because the vortices
proliferate and carry no energy.  There is also no nontrivial topology to
assure their stability.  This is the reason for omitting any reference to
vortex lines in the argument of $V$ in Eqs.\ (\ref{dual:simple}) and
(\ref{localrep}).

We are now in a position to write down the dual theory of a
three-dimensional superconductor.  It features a grand-canonical ensemble of
fluctuating closed magnetic vortices of arbitrary shape and length having a
repulsive contact interaction.  The loop gas can be described by a complex
disorder $|\psi|^4$-theory.  In addition, as our study of a single external
vortex revealed, the magnetic vortices couple to the fluctuating vector
field $\tilde{\bf h}$ with a coupling constant $g$.  Hence, the dual theory is
given by
\cite{BS,Kawai,GFCM,KKR,MA}
\begin{equation}   \label{funcZ}
Z = \int \DD \tilde{\bf h} \DD \psi^{*} \DD \psi 
\,  \exp \left(- \int_{\bf x} {\cal H}_\psi\right)
\end{equation}  
with the Hamiltonian
\begin{equation}     \label{Hpsi}
{\cal H}_{\psi} =   \frac{1}{2 m_A^2} (\nabla \times 
\tilde{\bf h})^2 + \frac{1}{2} \tilde{\bf h}^2 + |(\nabla -i \Phi_0
\tilde{\bf h}) \psi|^2 + m_\psi^2 |\psi|^2 + u |\psi|^4 ,   
\end{equation} 
where the field $\psi$ is minimally coupled to the massive vector field
$\tilde{\bf h}$.  Equation (\ref{Hpsi}) is a description of the
superconducting state in terms of physical variables: the field ${\bf
n}$ describes the local magnetic induction, whereas $\psi$ accounts for
the loop gas of magnetic vortices.  There are no other physical entities
present in a superconductor.  The dual theory has no local gauge
symmetry because the vector field $\tilde{\bf h}$ is massive.

Although (\ref{funcZ}) was derived starting from the London limit, it is
also relevant near the phase transition.  The point is that integrating
out the size fluctuations of the scalar field $\phi$ would only generate
higher-order interaction terms.  But these modifications do not alter
the critical behavior of the theory.

The line tension $M_V$ (\ref{M}) appears in the dual theory as a one-loop
on-shell mass correction to the mass $m_\psi$ stemming from the graph
depicted in Fig.~\ref{fig:biotsavart}, which we now interpret as a Feynman
graph.  The straight and wiggly lines represent the $\psi$- and $\tilde{\bf
h}$-field correlation functions, respectively.  We have used dashed lines to
distinguish the Feynman graphs of the dual theory from those in the
Ginzburg-Landau model.

A measure for the interaction strength of a massive vector field in
three dimensions is given by the dimensionless parameter equal to the
square of the coupling constant multiplied by the range of the
interaction. For the dual theory this factor is $g^2/m_A \sim m_A/e^2$,
which is the inverse of the strength of the electromagnetic gauge field
${\bf A}$ in the superconducting phase.  This is a common feature of
theories which are dual to each other.

Another notable property of the dual theory is that in the limit $e
\rightarrow 0$ it changes into a local gauge theory \cite{GFCM},
\begin{equation}  
{\cal H}_{\psi} \rightarrow  \frac{1}{2} (\nabla \times
\tilde{\bf h})^2 + |(\nabla -i g \tilde{\bf h}) \psi|^2 + m_\psi^2
|\psi|^2 + u |\psi|^4 , 
\end{equation}  
as can be checked by rescaling the dual field $\tilde{\bf h}$ in the Hamiltonian
(\ref{Hpsi}).  By taking the limit $e \rightarrow 0$ in the Ginzburg-Landau
model, we obtain a $|\phi|^4$-theory, which describes a superfluid, with
a decoupled gauge field.

We next investigate what happens with the dual theory as we approach the
critical temperature.  Remember that $\bar{\phi}$ and therefore $m_A$ tends to
zero as $T$ approaches the critical temperature from below.  From the
first term in the Hamiltonian (\ref{Hpsi}) it again follows that $\nabla
\times \tilde{\bf h} \rightarrow 0$ in this limit, so that we can write once more
$\tilde{\bf h} = \nabla \gamma$, and (\ref{Hpsi}) becomes
\begin{equation} \label{Hpsi'} 
{\cal H}_{\psi} =  \tfrac{1}{2} (\nabla \gamma)^2 + |(\nabla -i
\Phi_0 \nabla \gamma) \psi|^2 + m_\psi^2 |\psi|^2 + u |\psi|^4 .
\end{equation} 
This equation shows that $\gamma$, representing the magnetic scalar
potential, cannot be distinguished from the phase of the disorder field.
Indeed, let $\Phi_0 \theta$ be this phase.  Then, the canonical
transformation $\theta \rightarrow \theta + \gamma$ absorbs the
scalar potential into the phase of $\psi$; the first term in
(\ref{Hpsi'}) decouples from the theory and yields a trivial
contribution to the partition function.  In this way, the dual theory
reduces to a pure $|\psi|^4$-theory
\begin{equation}  \label{Hpsi''}
{\cal H}_{\psi} =  |\nabla \psi|^2 + m_\psi^2 |\psi|^2
+ u |\psi|^4 .
\end{equation}   
At the transition temperature, the magnetic vortices proliferate and the
field $\psi$ develops an expectation value.  The transition is triggered
by a change in sign of $m_\psi^2$.  In the London limit, the Hamiltonian
(\ref{Hpsi''}) then takes the simple form
\begin{equation} \label{psisim}
{\cal H}_{\psi} = \tfrac{1}{2} |\bar{\psi}|^2 \Phi_0^2 (\nabla \gamma)^2 ,
\end{equation} 
with $|\bar{\psi}|$ the expectation value of the disorder field,
$|\bar{\psi}|/\sqrt{2} = \langle |\psi| \rangle$, and where we now represented
the phase of $\psi$ by $\Phi_0 \gamma$ to bring out the fact that $\gamma$
describes the magnetic scalar potential.  From a symmetry point of view
\cite{KRE}, $\gamma$ is the Goldstone mode of the spontaneously broken
global U(1) symmetry of the $|\psi|^4$-theory.  Whereas in the
Ginzburg-Landau formulation a magnetic vortex is described by a singular
plastic field ${\bf B}^{\rm P}$, in the dual formulation it is represented
by the Noether current associated with the global U(1) symmetry,
\begin{equation} 
{\bf j}_g = -i g\psi^* \stackrel{\leftrightarrow}{\nabla} \psi + 2 g^2
\tilde{\bf h} |\psi|^2.
\end{equation}   
This follows from comparing the terms coupling linearly to the fluctuating
$\tilde{\bf h}$ field.  In the normal phase, where $\psi$ develops a vacuum
expectation value, the Noether current becomes in the London limit ${\bf j}
= \nabla \gamma$.  This is the usual relation between the current of a
spontaneously broken symmetry and the ensuing Goldstone field. 

As we will demonstrate next, $\bar{\psi}$ has the value $|\bar{\psi}| =
1/\Phi_0$ \cite{KRE} of an inverse flux quantum, so that with our
normalization choice of the phase of the $\psi$-field, Eq.~(\ref{psisim})
takes the canonical form.  Let us introduce a closed vortex $L$ in the dual
theory (\ref{psisim}) by minimally coupling the magnetic scalar potential to
a vortex gauge field $\bbox{\varphi}^{\rm P}$ and consider the expectation
value of this configuration
\begin{equation} \label{Wilson} 
\langle W(L) \rangle = \int \DD \gamma \exp \left[ -\frac{1}{2} |\bar{\psi}|^2
\int_{\bf x} \left(\Phi_0 \nabla \gamma - \bbox{\varphi}^{\rm P} \right)^2
\right] 
\end{equation} 
We linearize the theory by introducing an auxiliary vector field ${\bf b}$
via
\begin{eqnarray}  
\lefteqn{\exp \left[ -\frac{1}{2} |\bar{\psi}|^2 \int_{\bf x}
\left(\Phi_0 \nabla \gamma - \bbox{\varphi}^{\rm P} \right)^2 \right] =}
\\ && \int  \DD {\bf b} \exp\left\{-\int_{\bf x}
\left[\frac{1}{2} {\bf b}^2 + i\, |\bar{\psi}| {\bf b} \cdot \left(\Phi_0\nabla
\gamma - \bbox{\varphi}^{\rm P}\right) \right]\right\}. \nonumber 
\end{eqnarray}  
This amounts to a duality transformation again.  The integral over $\gamma$
now yields the constraint $\nabla \cdot {\bf b} = 0$, demanding ${\bf b}$ to
be the rotation of a vector field, ${\bf b} =
\nabla \times {\bf A}$.  This gives
\begin{equation}   \label{sum}
\langle W(L) \rangle = \int \DD {\bf A} \, {\rm exp}\left\{-\int_{\bf x} \left[
\frac{1}{2} (\nabla \times {\bf A})^2 -  i  {\bf A} \cdot {\bf J}
\right] \right\},
\end{equation} 
where $J_i({\bf x}) := 2 \pi |\bar{\psi}| \oint_L \dd y_i \delta ({\bf x} - {\bf
y})$ describes the closed vortex.  It is natural to interpret the
fluctuating gapless gauge field ${\bf A}$ as the magnetic vector potential,
and the closed vortex as an electric current loop. This justifies the use of
the symbol $W$ in (\ref{Wilson}) which was first introduced in (\ref{scop}).
(The analogy between vortices and electric currents was first pointed out by
von Helmholtz.)  In this way, we get back the Ginzburg-Landau model with just
one external electric current loop.  The full model (\ref{gl:H}) is
recovered when we consider a loop gas of these defects and describe it by a
$|\phi|^4$-theory, provided we make the identification \cite{KRE}
\begin{equation} \label{rel}
2 \pi |\bar{\psi}| = 2 e,
\end{equation}
or $|\bar{\psi}| = 1/\Phi_0$.  It links the expectation value $\bar{\psi}$
of the disorder field $\psi$ describing the magnetic vortex loops to the
coupling constant $2e$ of the Ginzburg-Landau model.  It is the exact analog
of the relation between the expectation value $\bar{\phi}$ of the
$\phi$-field describing the electric current loops and the coupling constant
$g =
\Phi_0 m_A$ of the dual theory:
\begin{equation}  
2\pi |\bar{\phi}| = g.
\end{equation}  

Since $\bar{\psi}$ vanishes as $T$ approaches the critical temperature from
above, the coupling constant $2e$ goes to zero at the critical point and the
electric current loops decouple from ${\bf A}$.  Moreover, Eq.~(\ref{MW}),
giving the tension of electric current lines in the normal phase, shows that
indeed these lines carry no energy in the superconducting phase where they
proliferate.  This solves \label{pag:resol} the apparent contradiction we
mentioned on page \pageref{pag:contr}.  Precisely the same things happened
with the magnetic vortices: the coupling constant $g$ of the dual theory
vanishes as the critical temperature is approached from below, so that the
magnetic vortices decouple from the local magnetic induction and proliferate.
We thus have a complete duality here between magnetic vortices and electric
current loops in complementary phases.

At this stage we can also clarify the physical relevance
\label{pag:relevance} of the ultraviolet cutoff $|m_\psi|$ introduced in
the calculation of the tension (\ref{MW}) of a current loop.  This
calculation was performed in the London limit where current lines are
considered to be infinitely thin.  Outside the London limit, vortices of the
dual theory have a typical width of the order of the coherence length
$1/|m_\psi|$, which therefore provides a natural ultraviolet cutoff.

The line tension (\ref{MW}) we previously calculated in the framework of the
Ginzburg-Landau model can also be obtained in the dual theory directly from
(\ref{Wilson}).  The integration over $\gamma$ can be carried out by
substituting the field equation of the Goldstone field
\begin{equation} \label{fegm}
\gamma ({\bf x}) = - \frac{1}{2 \Phi_0} \int_{\bf y}  G_0({\bf x} - {\bf y})
\nabla \cdot  \bbox{\varphi}^{\rm P}_\mu ({\bf y}),
\end{equation} 
where $G_0$ is the Green function (\ref{Yuka}) with $m_A=0$.  This
yields an expression
\begin{equation}    \label{ww*}
\left\langle W( L ) \right\rangle = 
\exp \left\{ - \frac{1}{2 \Phi_0^2}  \int_{\bf x} \int_{\bf y} [\nabla 
\times \bbox{\varphi}^{\rm P}({\bf x})]_i  G_0
({\bf x} - {\bf y}) \, [\nabla \times \bbox{\varphi}^{\rm P}({\bf y})]_i
\right\},                                              
\end{equation}
very similar to the expression for the disorder parameter obtained in
Eq.\ (\ref{vv*}) with the monopole density set to zero since
\begin{equation} 
[\nabla \times \bbox{\varphi}^{\rm P}({\bf x})]_i = 2 \pi \int_L \dd y_i
\delta ({\bf x} - {\bf y}).
\end{equation} 
This immediately gives 
\begin{equation}     \label{Wil}
\langle W(L) \rangle = {\rm e}^{-M_W |L|},
\end{equation} 
with $|L|$ the vortex length and $M_W$ the line tension (\ref{MW}).

The field equation we derive from (\ref{sum}),
\begin{equation}
\nabla \times {\bf h} = i {\bf J}
\end{equation} 
is one of the two basic equations of magnetostatics.  It should be
noted, however, that an additional factor of $i$ shows up here.  As a
result, the Biot-Savart law yields opposite signs from what is usually
the case.  Two parallel currents repel instead of attract each other so
that a single current loop prefers to crumple.  It follows that a state
where these current loops proliferate has zero magnetization---as should
be the case in the superconducting phase of the Ginzburg-Landau model.

Because of the imaginary current, the local magnetic induction generated by
a current loop $L$ is also purely imaginary as follows from Ampere's law,
\begin{equation}  \label{Ampere}
{\bf h}({\bf x})= i \frac{1}{2 \Phi_0} \oint_L \dd {\bf y} \times \frac{{\bf
x}-{\bf y}}{ |{\bf x}-{\bf y}|^3} = i \frac{1}{2 \Phi_0} \nabla \Omega ({\bf
x}),
\end{equation}
where $\Omega({\bf x})$ is the solid angle that the loop subtends at ${\bf
x}$.  The same result can also be directly
derived from the dual theory.  Rewriting the field equation for $\gamma$
obtained from (\ref{Wilson}), we can relate the Goldstone field to the
solid angle in the following way \cite{GFCM}
\begin{equation}
\gamma({\bf x}) = \frac{1}{2\Phi_0} \int_S \dd^2 y_i \frac{({\bf x}-{\bf
y})_i}{|{\bf x}-{\bf y}|^3} = - \frac{1}{2\Phi_0}\Omega({\bf x}),
\end{equation}
where $\dd^2 y_i$ is an element of the surface $S$ spanned by the current
loop $L$.  Together with the observation that the local magnetic induction
can, apart from a factor $i$, be identified with the gradient of the phase
variable $\gamma$
\begin{equation} \label{hgamma}
{\bf h} = -i \nabla \gamma,
\end{equation}
[see (\ref{gammaid})], this yields the previous result (\ref{Ampere}).

As a side remark we note that the magnetic moment density, or magnetization,
is represented in the dual theory by $\bbox{\varphi}^{\rm P}$.  This follows
from (\ref{Wilson}) showing that an electric current loop couples to the
magnetic field $\nabla \gamma$ via $\bbox{\varphi}^{\rm P}$.

\begin{figure}
\begin{center}
\epsfxsize=6cm
\mbox{\epsfbox{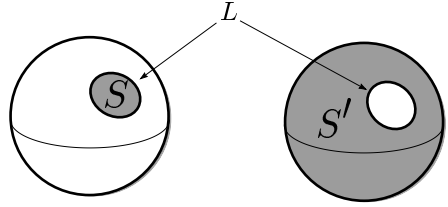}}
\end{center}
\caption{Two different surfaces $S$ and $S'$ spanning the loop $L$.  The flux
through the surfaces differ by a factor $4 \pi$.
\label{fig:monopolejail} } 
\end{figure}

With this physical identification of the Goldstone mode it follows that, in
the normal state, the disorder parameter $V_{\rm r}({\bf x})$ essentially
measures the solid angle $\Omega({\bf x})$
\begin{equation} 
\label{monometer}
V_{\rm r}({\bf x}) = {\rm e}^{i \Phi_0 \gamma ({\bf x})} = {\rm
e}^{i\Omega ({\bf x})/2}.
\end{equation} 
Since the operator $V({\bf x})$ was constructed by putting a magnetic
monopole at ${\bf x}$, we see that $\tfrac{1}{2} \Omega({\bf x})$ is the
magnetic flux emanated by the monopole through the electric current loop
$L$.  As a last remark we note that one can chose two topologically
different surfaces spanning the loop $L$ (see Fig. \ref{fig:monopolejail}).
Both lead, however, to the same phase factor because $\Omega$ differs only
by a factor of $4 \pi$.

Because the dual Ginzburg-Landau model involves only physical degrees of
freedom and a global U(1) symmetry, it is ideally suited as basis to study
the critical properties of the superconducting-to-normal phase transition
\cite{KKS}.  As has been argued above, at the mean-field level the massive
induction field decouples from the theory as the critical temperature is
approached from below.  The resulting theory is a $|\psi|^4$-theory which is
known to undergo a 2nd-order phase transition with $xy$-exponents.  This can
also be seen by integrating out the massive induction field in the
low-temperature phase.  This only leads to changes in the coefficients of
the $|\psi|^4$-theory.  Explicitly \cite{MA},
\begin{equation}  
{\cal H}_{\psi,{\rm eff}} =  |\nabla \psi|^2 +
\left(m_\psi^2 - g^2 \frac{m_A}{2 \pi} \right) |\psi|^2 + 
\left(u -  \frac{g^4}{4\pi m_A} \right) |\psi|^4.
\end{equation}  
In deriving this we used dimensional regularization and (irrelevant)
higher-order terms are omitted.  We note that all contributions
stemming from the vector field $\tilde{\bf h}$ vanish in the limit where $T$
approaches $T_{\rm c}$ from below, so that ${\cal H}_{\psi,{\rm eff}}$
reduces to a pure $|\psi|^4$-theory in this limit.

A one-loop renormalization-group analysis of the dual theory carried out in
Ref.\ \cite{KKS} also led to the conclusion that the disorder field behaves
as in a pure $|\psi|^4$-theory with $xy$-exponents.  In addition, the
exponents of the magnetic induction field where shown to retain their
Gaussian values because it decouples from the theory.  The $xy$-behavior of
the heat capacity was established experimentally in \cite{OHL}, while
experiments on YBaCuO \cite{Lin} seem to confirm the prediction of the
Gaussian value for the divergence of the magnetic penetration depth.  The
confirmation relies, however, on a delicate finite-size analysis, so that
there is still room for a reinterpretation of the data and the possibility
of suggesting a different critical behavior as has been done by Herbut
\cite{Herbut} who concluded that the penetration depth has an $xy$-exponent.
One recent lattice simulation directly of the Ginzburg-Landau model 
corroborated the Gaussian value for the divergence of the penetration depth
\cite{KKLP}, while an other found an $xy$-exponent \cite{OT}.

\chapter{Quantum Phase Transitions \label{chap:qpt}}
In this chapter, we discuss various continuous phase transitions at the
absolute zero of temperature---so-called quantum phase transitions.
Unlike in their classical counterparts taking place at finite
temperature and being an equilibrium phenomenon, time plays an important
role in quantum phase transitions.  Put differently, whereas the
critical behavior of classical 2nd-order phase transitions is governed
by thermal fluctuations, that of 2nd-order quantum transitions is
controlled by quantum fluctuations.  These transitions, which have
attracted much attention in recent years (for an introductory review,
see Ref.\
\cite{SGCS}), are triggered by varying not the temperature, but some
other parameter in the system, like the applied magnetic field or the
amount of disorder.  The natural language to describe these transitions
is quantum field theory.  In addition to a diverging correlation length
$\xi$, quantum phase transitions also have a diverging correlation time
$\xi_t$.  They indicate, respectively, the distance and time period over
which the order parameter characterizing the transition fluctuates
coherently.  The way the diverging correlation time relates to the
diverging correlation length, 
\begin{equation} \label{zcrit}
\xi_t \sim \xi^z, 
\end{equation} 
defines the so-called dynamic exponent $z$.  It is a measure for the
asymmetry between the time and space directions and  tells us how long
it takes for information to propagate across a distance $\xi$.  The
traditional scaling theory of classical 2nd-order phase transitions is
easily extended to include the time dimension \cite{Ma} because
relation (\ref{zcrit}) implies the presence of only one independent
diverging scale.  The critical behavior of a phase transition at finite
temperature is still controlled by the quantum critical point provided
$T < 1/\xi_t$.  This is what makes quantum phase transitions
experimentally accessible.

We start in the next section discussing the so-called
superfluid-to-Mott-insulating phase transition in the pure case, while in
Sec.\ \ref{sec:Dirt} we include (quenched) impurities.  In Sec.\
\ref{sec:CSGL} we discuss the effective theory describing the fractional
quantized Hall effect and argue that, in principle, it can be employed to
describe the quantum phase transitions in quantum Hall systems.  In Sec.\
\ref{sec:RG} we then apply renormalization-group analysis to this
theory.  In Sec.\ \ref{sec:scale} we discuss scaling and hyperscaling
theory applied to the systems under study and in Sec.\ \ref{sec:exp} we
discuss various experiments probing quantum phase transitions.
\section{Repulsively Interacting Bosons}
\label{sec:BT}
The first quantum phase transition we wish to investigate is the
superfluid-to-Mott-insulating transition of repulsively interacting bosons
in the absence of impurities \cite{FWGF}.  The transition is described by
the nonrelativistic $|\phi|^4$-theory (\ref{bec:Lagr}), which becomes
critical at the absolute zero of temperature at some (positive) value
$\mu_{\rm c}$ of the renormalized chemical potential.  The Mott insulating
phase is destroyed and makes place for the superfluid phase as $\mu$
increases.  Whereas in the superfluid phase the single-particle (Bogoliubov)
spectrum is gapless and the system compressible, the single-particle
spectrum of the insulating phase has an energy gap and the compressibility
$\kappa$ vanishes here.

\begin{figure}
\begin{center}
\epsfxsize=8.cm
\mbox{\epsfbox{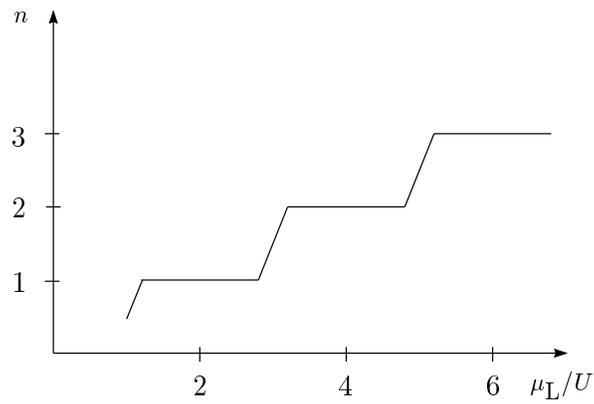}}
\end{center}
\caption{Schematic representation of the average number $n$ of particles per
site as function of the chemical potential $\mu_{\rm L}$ at some finite
value of the hopping parameter $t < t_{\rm c}$ \protect\cite{BSZ}.
\label{fig:occu}}
\end{figure}

The nature of the insulating phase can be best understood by putting the
theory on a lattice.  The lattice model is defined by the Hamiltonian
\begin{equation}    \label{BT:hu}
H_{\rm H} = - t  \sum_j (\hat{a}^{\dagger}_j \hat{a}_{j+1} + {\rm 
h.c.}) + \sum_j (- \mu_{\rm L} \hat{n}_j + U  \hat{n}_j^2),
\end{equation} 
where the sum $\sum_j$ is over all lattice sites.  The operator
$\hat{a}^{\dagger}_j$ creates a boson at site $j$ and $\hat{n}_j =
\hat{a}^{\dagger}_j \hat{a}_j$ is the particle number operator at that
site; $t$ is the hopping parameter, $U$ the interparticle repulsion, and
$\mu_{\rm L}$ is the chemical potential on the lattice.  The
zero-temperature phase diagram is as follows \cite{FWGF}.  In the limit
$t/U \rightarrow 0$, each site is occupied by an integer number $n$ of
bosons which minimizes the on-site energy (see Fig.\ \ref{fig:occu})
\begin{equation} 
\epsilon(n) = -\mu_{\rm L} n + U n^2.
\end{equation} 
It follows that within the interval $2n-1 < \mu_{\rm L}/U < 2n+1$, each
site is occupied by exactly $n$ bosons.  When the chemical potential is
negative, $n=0$.  The intervals become smaller when $t/U$ increases.
Within such an interval, where the particles are pinned to the lattice
sites, the single-particle spectrum has an energy gap, and the system is
in the insulating phase with zero compressibility, $\kappa =
n^{-2}\partial n/\partial \mu_{\rm L} =0$.  Outside these intervals, the
particles delocalize and can hop through the lattice.  Being at zero
temperature, the delocalized bosons condense in a superfluid state.  The
single-particle spectrum is gapless here and the system compressible
($\kappa \neq 0$).

\begin{figure}
\epsfxsize=10.cm
\mbox{\epsfbox{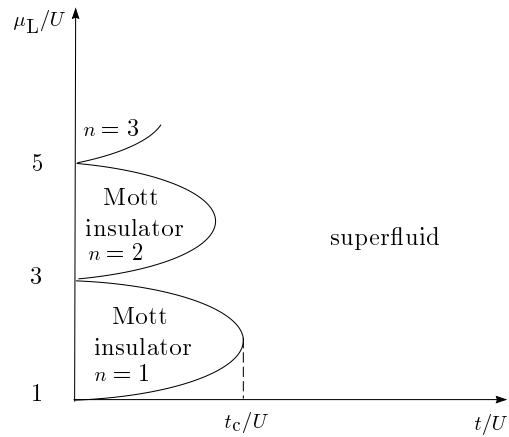}}
\caption{Schematic representation of the  phase diagram  of the lattice
model (\protect\ref{BT:hu}) at the absolute zero of temperature
\protect\cite{FWGF}.
\label{fig:qphase}}
\end{figure}

As $t/U$ increases, the gap in the single-particle spectrum as well as
the width of the intervals decrease and eventually vanish at some
critical value $t_{\rm c}$.  For values $t>t_{\rm c}$ of the hopping
parameter, the superfluid phase is the only phase present (see Fig.\
\ref{fig:qphase}).
The continuum model (\ref{bec:Lagr}), with renormalized chemical
potential $\mu > \mu_{\rm c}$ describes the condensed delocalized
lattice bosons which are present when the density deviates from integer
values (see Fig.\ \ref{fig:occu}).  In the limit $\mu
\rightarrow \mu_{\rm c}$ from above, the number of delocalized bosons
decreases and eventually becomes zero at the phase boundary $\mu=\mu_{\rm
c}$ between the superfluid and insulating phases.

Various quantum phase transitions belong to the universality class defined
by the zero-density transition of repulsively interacting bosons.  For
example, itinerant quantum antiferromagnets \cite{Hertz,Ian,KB}
as well as lower-dimensional (clean) superconductors belong to this
universality class.  As we have seen in Sec.\ \ref{sec:comp}, Cooper
pairs become tightly bound composite particles in the strong-coupling
limit, which are described by the nonrelativistic $|\phi|^4$-theory with
a weak repulsive interaction.  For $\mu > \mu_{\rm c}$, the field $\phi$
now describes the condensed delocalized Cooper pairs.  When the chemical
potential decreases, the condensate diminishes, and the system again
becomes insulating for $\mu < \mu_{\rm c}$ \cite{CFGWY}.  By continuity,
we expect also the superconductor-to-insulator transition of a (clean)
weakly interacting BCS superconductor to be in this universality class.
The restriction to lower dimensions is necessary for two different
reasons.  First, only for $d \leq 2$ the penetration depth is
sufficiently large [see, for example, below (\ref{2sc:mod})], so that it
is appropriate to work in the limit $\lambda_{\rm L} \rightarrow \infty$
with no fluctuating gauge field
\cite{FGG}.  Second, in lower dimensions, the energy gap which the fermionic
excitations face remains finite at the critical point, so that it is
appropriate to ignore these degrees of freedom.  Moreover, since also
the coherence length remains finite at the critical point, the Cooper
pairs look like point particles on the scale of the diverging
correlation length associated with the phase fluctuations,
even in the weak-coupling limit \cite{CFGWY}.

The nonrelativistic $|\phi|^4$-theory is also of importance for the
description of the fractional quantized Hall effect (FQHE) (see Sec.\
\ref{sec:CSGL}).  As function of the applied magnetic field, this
two-dimensional system undergoes a zero-temperature transition between a
so-called quantum Hall liquid, where the Hall conductance is quantized
in odd fractions of $e^2/2 \pi$, or, reinstalling Planck's constant,
$e^2/h$, and an insulating phase.  Here, the nonrelativistic
$|\phi|^4$-theory describes---after coupling to a Chern-Simons
term---the original electrons contained in the system bound to an odd
number of flux quanta.  The Hall liquid corresponds to the phase with
$\mu > \mu_{\rm c}$, while the other phase again describes the
insulating phase.  In this picture, the Hall liquid is characterized by
a condensate of composite particles.

It should be noted however that in most of these applications of the
nonrelativistic $|\phi|^4$-theory mentioned here, impurities play an
important role; this will be the subject of the succeeding section.

\begin{figure}
\begin{center}
\epsfxsize=2.cm
\mbox{\epsfbox{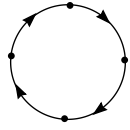}}
\end{center}
\caption{A closed oriented loop. \label{fig:oriented1}}
\end{figure}

\begin{figure}
\begin{center}
\epsfxsize=8.cm
\mbox{\epsfbox{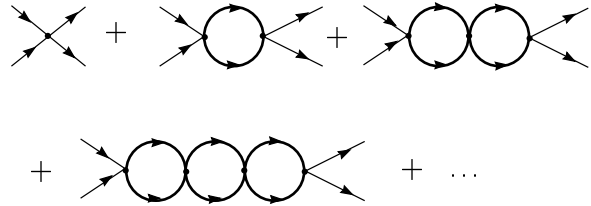}}
\end{center}
\caption{Ring diagrams renormalizing the vertex function of the neutral
$|\phi|^4$-theory. \label{fig:ring}}
\end{figure}

The critical properties of the zero-density transition of the
nonrelativistic $|\phi|^4$-theory were first studied by Uzunov
\cite{Uzunov}.  To facilitate the discussion let us make use of the fact
that in nonrelativistic theories the mass is---as far as critical phenomena
concerned---an irrelevant parameter which can be transformed away.  This
transformation changes, however, the scaling dimensions of the $\phi$-field
and the coupling constant which is of relevance to the renormalization-group
theory.  The engineering dimensions are
\begin{equation}  \label{BT:scale} 
[{\bf x}] = -1, \;\;\;\; [t] = -2, \;\;\;\; [\mu_0] = 2, \;\;\;\;
[\lambda_0] = 2-d,  \;\;\;\; [\phi] = \tfrac{1}{2}d,
\end{equation} 
with $d$ the number of space dimensions.  In two space dimensions the
coupling constant $\lambda_0$ is dimensionless, showing that the
$|\phi|^4$-term is a marginal operator, and $d_{\rm c}=2$ the upper critical
space dimension.  Uzunov showed that below the upper critical dimension
there appears a non-Gaussian IR fixed point.  He computed the corresponding
critical exponents to all orders in perturbation theory and showed them to
have Gaussian values, $\nu=\tfrac{1}{2}, \; z=2, \;
\eta=0$.  Here, $\nu$ characterizes the divergence of the
correlation length, $z$ is the dynamic exponent, and $\eta$ is the
correlation-function exponent which determines the anomalous dimension
of the field $\phi$.  The unexpected conclusion that a non-Gaussian
fixed point has nevertheless Gaussian exponents is rooted in the
analytic structure of the nonrelativistic propagator at zero (bare)
chemical potential ($\mu_0=0$):
\begin{equation} \label{BT:Green}
\raisebox{-0.3cm}{\epsfxsize=2.5cm
\epsfbox{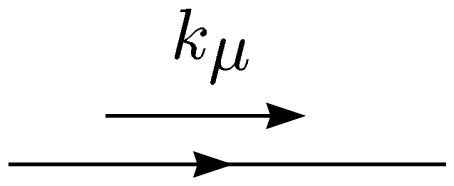}  }
= G(k) = \frac{i {\rm e}^{i k_0 \eta}}{k_0 - \tfrac{1}{2}{\bf k}^2 + i
\eta }, 
\end{equation}
where, as before, $\eta$ is a small positive constant that has to be taken
to zero after the loop integrations over the energies have been carried out.
By setting $\mu_0 = 0$, we are considering the system at criticality.  The
rule $k_0 \rightarrow k_0 + i \eta$ in (\ref{BT:Green}) expresses the fact
that in this nonrelativistic theory particles propagate only forward in
time.  In Feynman diagrams involving loops with more than one propagator,
the integrals over the loop energy are convergent and can be evaluated by
contour integration with the contour closed in either the upper or the lower
half plane.  If a diagram contains a loop which has all its poles in the
same half plane, it consequently vanishes.  Pictorially, such a loop has all
its arrows, representing the Green functions contained in the loop, oriented
in a clockwise or anticlockwise direction \cite{OB} (see Fig.\
\ref{fig:oriented1}).
We will refer to them as closed oriented loops.  Owing to this property most
diagrams are zero.  In particular, all self-energy diagrams vanish.  The
only surviving ones are the so-called ring diagrams which renormalize the
vertex (see Fig.\ \ref{fig:ring}).  
Because this class of diagrams constitute a geometric series, the one-loop
result is already exact.  The vertex renormalization leads to a non-Gaussian
fixed point in $d < 2$, while the vanishing of all the self-energy diagrams
asserts that the exponents characterizing the transition are not affected by
quantum fluctuations and retain their Gaussian values \cite{Uzunov}.  These
results have been confirmed by numerical simulations in $d=1$ \cite{BSZ} and
also by general scaling arguments \cite{FWGF}.  

We have seen that $d_{\rm c}=2$ is the upper critical dimension of the
nonrelativistic $|\phi|^4$-theory.  Dimensional analysis shows that for an
interaction term of the form
\begin{equation}
{\cal L}_{\rm int} = - g_0 |\phi|^{2k}
\end{equation}
the upper critical dimension is
\begin{equation}
\label{BT:dcnr}
d_{\rm c} = \frac{2}{k-1}.
\end{equation}
The two important physical cases are $d_{\rm c}=2$, $k=2$ and $d_{\rm c}=1$,
$k=3$, while $d_{\rm c} \rightarrow 0$ when $k \rightarrow \infty$.  For space
dimensions $d > 2$ only the quadratic term, $|\phi|^2$, is relevant so that
here the critical behavior is well described by a Gaussian theory. 

In the corresponding relativistic theory, the scaling dimensions of $t$ and
${\bf x}$ are, of course, equal $[t] = [{\bf x}] = -1$ and $[\phi] =
\tfrac{1}{2} (d-1)$.  This leads to different upper critical (space)
dimensions, viz., 
\begin{equation}
d_{\rm c} = \frac{k+1}{k-1} = \frac{2}{k-1} + 1,
\end{equation}
instead of (\ref{BT:dcnr}).  The two important physical cases are here $d_{\rm
c}=3$, $k=2$ and $d_{\rm c}=2$, $k=3$, while $d_{\rm c} \rightarrow 1$ when $k
\rightarrow \infty$.  On comparison with the nonrelativistic results, we see
that the nonrelativistic theory has an upper critical space dimension which
is one lower than that of the corresponding relativistic theory (see Table
\ref{table:1}).  Heuristically, this can be understood by noting that in a
nonrelativistic context the time dimension counts double in that it
has a scaling dimension twice that of a space dimension [see Eq.\
(\ref{BT:scale})], thereby increasing the {\it effective} spacetime
dimensionality by one.
\begin{table}
\caption{The upper critical space dimension $d_{\rm c}$ of a
nonrelativistic (NR) and a relativistic (R) quantum theory with a
$|\phi|^{2k}$ interaction term.}
\label{table:1}
\begin{center}
\vspace{.5cm}
\begin{tabular}{ccc|cccccccc} \hline \hline 
&  & & & &  & & &  & & \\[-.2cm] 
& $k$ & & & & $d_{\rm c}$(NR)&  & & $d_{\rm c}$(R) & & \\[.1cm]
\hline  
&  & & & &  & & &  & & \\[-.2cm] 
& 2 & & & & 2 & & & 3  & & \\ 
& 3 & & & & 1 & & & 2 & & \\ 
& $\infty$ & & & & 0 & & & 1 & &   \\[.1cm]
\hline \hline
\end{tabular}
\end{center}
\end{table}

From this analysis it follows that for a given number of space
dimensions the critical properties of a nonrelativistic theory are unrelated
to those of the corresponding relativistic extension.

In closing this section we recall that in a one-dimensional relativistic
theory---corresponding to the lowest upper critical dimension ($d_{\rm
c}=1)$---a continuous symmetry cannot be spontaneously broken.  However,
the theory can nevertheless have a phase transition of the so-called
Kosterlitz-Thouless type.  Given the connection between the relativistic
and nonrelativistic theories discussed above, it seems interesting to
study the nonrelativistic theory at zero space dimension ($d = 0$) to
see if a similar rich phenomenon as the Kosterlitz-Thouless transition
occurs in the quantum theory.  This may be of relevance to so-called
quantum dots.
\section{Including Quenched Impurities}
\label{sec:Dirt}
In the preceding section, we saw that in the absence of impurities
repulsively interacting bosons will undergo a 2nd-order quantum phase
transition.  As was pointed out there, this universality class is of
relevance to various condensed-matter systems.  However, in most of
these systems, as well in $^4$He in porous media, impurities play an
essential if not decisive role.  For example, the two-dimensional
superconductor-to-insulator transition investigated by Hebard and
Palaanen
\cite{HPsu1} is driven by impurities.  This means that, e.g., the correlation
length $\xi$ diverges as $|\hat{\Delta}^* - \hat{\Delta}|^{-\nu}$ when the
parameter $\hat{\Delta}$ characterizing the disorder approaches the critical
value $\hat{\Delta}^*$.  Hence, a realistic description of the critical
behavior of these systems should include impurities.

Some years ago, it has been argued that upon including quenched impurities,
the quantum critical behavior of the nonrelativistic $|\phi|^4$-theory
becomes unstable \cite{KU}.  Only after introducing an artificial
high-energy cutoff, an IR fixed point was found by Weichman and Kim
\cite{WK}.  However, as was pointed out by the authors, such a cutoff is
difficult to justify as it would imply that time is discrete.  So, it is
widely accepted that the random nonrelativistic $|\phi|^4$-theory has no
perturbatively accessible IR fixed point, if any at all \cite{FWGF}.  The
absence of an IR fixed point in the nonrelativistic $|\phi|^4$-theory theory
would imply that also the quantum critical behavior of the systems 
mentioned in the preceding section, is unstable with respect to impurity
influences.  Because of its implications for the description of the critical
behavior of these systems we have revisited the problem.  Below, it will be
shown that, contrary to general conviction, the random nonrelativistic
$|\phi|^4$-theory does have a new IR fixed point.  The calculations are
performed without introducing a (physically unnatural) high-energy cutoff.

To account for impurities, we add to the nonrelativistic $|\phi|^4$-theory
(\ref{bec:Lagr}) a term \cite{Ma}
\begin{equation} \label{Dirt:dis}
{\cal L}_{\Delta} = \psi({\bf x}) \, |\phi|^2,
\end{equation} 
with $\psi({\bf x})$ a random field.  As we have seen in the preceding
section, the theory becomes critical in the limit where the bare
chemical potential tends to zero, $\mu_0 \rightarrow 0$.  We shall study
the random theory in the symmetrical state where the chemical potential
is negative and the global U(1) symmetry unbroken.  We therefore set
$\mu_0 = - r_0$ again, with $r_0>0$.  We leave the number of space
dimensions $d$ unspecified for the moment.

The random field $\psi({\bf x})$ is assumed to be Gaussian distributed
\cite{Ma}:
\begin{equation} \label{random} 
P(\psi) = \exp \left[-\frac{1}{\Delta_0} \int_{\bf x} \, \psi^2({\bf x})
\right],
\end{equation}
characterized by the disorder strength $\Delta_0$.  The engineering
dimension of the random field is the same as that of the chemical potential
which is one, $[\psi]=1$, while that of the parameter $\Delta_0$ is
$[\Delta_0] = 2-d$ so that the exponent in (\ref{random}) is dimensionless.
The quantity
\begin{equation} \label{Dirt:Z}
Z[\psi] = \int \DD \phi^* \DD \phi \, \exp\left(i \int_x \, {\cal L}
\right),
\end{equation} 
where ${\cal L}$ now stands for the Lagrangian (\ref{bec:Lagr}) with the
term (\ref{Dirt:dis}) added, is the zero-temperature partition function
for a given impurity configuration $\psi$.  In the case of quenched
impurities, the average of an observable $O(\phi^*,\phi)$ is obtained as
follows
\begin{equation} 
\langle O(\phi^*,\phi) \rangle = \int \DD \psi P(\psi) \langle
O(\phi^*,\phi) \rangle_\psi,
\end{equation} 
where $\langle O(\phi^*,\phi) \rangle_\psi$ indicates the grand-canonical
average for a given impurity configuration, i.e., taken with respect to
(\ref{Dirt:Z}).  In other words, first the ensemble average is taken, and
only after that the averaging over the random field is carried out.

Since $\psi({\bf x})$ depends only on the $d$ spatial dimensions, the
impurities it describes should be considered as grains randomly distributed
in space.  When---as is required for the study of quantum critical
phenomena---time is included, the static grains trace out straight
worldlines.  That is to say, the impurities are line-like.  It has been shown
by Dorogovtsev \cite{Dorogovtsev} that the critical properties of systems
with extended defects must be studied in a double epsilon expansion,
otherwise no IR fixed point is found.  The method differs from the usual
epsilon expansion, in that it also includes an expansion in the defect
dimensionality $\epsilon_{\rm d}$.  To carry out this program in the present
context, where the defect dimensionality is determined by the dimensionality
of time, the theory has to be formulated in $\epsilon_{\rm d}$ time
dimensions.  The case of interest is $\epsilon_{\rm d}=1$, while in the
opposite limit, $\epsilon_{\rm d}\rightarrow 0$, the random nonrelativistic
$|\phi|^4$-theory reduces to the classical spin model with random
(point-like) impurities.  Hence, $\epsilon_{\rm d}$ is a parameter with which
quantum fluctuations can be suppressed.  An expansion in $\epsilon_{\rm d}$
is a way to perturbatively include the effect of quantum fluctuations on the
critical behavior.  Ultimately, we will be interested in the case
$\epsilon_{\rm d}=1$.

To calculate the quantum critical properties of the random theory, which
have been first studied in \cite{KU}, we will not employ the replica
method \cite{GrLu}, but instead follow Lubensky
\cite{Lubensky}.  In this approach, the averaging over impurities is carried
out for each Feynman diagram separately.  The upshot is that only those
diagrams are to be included which remain connected when $\Delta_0$, the
parameter characterizing the Gaussian distribution of the impurities, is set
to zero \cite{Hertzrev}.  To obtain the relevant Feynman rules of the random
theory we average the interaction term (\ref{Dirt:dis}) over the
distribution (\ref{random}):
\begin{eqnarray}  \label{Dirt:int}
\lefteqn{\int \DD \psi \, P(\psi)
\exp\left[i^{\epsilon_{\rm d}} \int
\dd^{\epsilon_{\rm d}} t \, \dd^d x \, \psi({\bf x}) \, |\phi(x)|^2  \right]
= } \nonumber \\ && \exp \left[\tfrac{1}{4} i^{2 \epsilon_{\rm d}} \Delta_0
\int \dd^{\epsilon_{\rm d}} t \, \dd^{\epsilon_{\rm d}} t' \, \dd^d x \,
|\phi(t,{\bf x})|^2 |\phi(t',{\bf x})|^2 \right].
\end{eqnarray} 
The randomness is seen to result in a quartic interaction term which is
nonlocal in time.  The factor $i^{\epsilon_{\rm d}}$ appearing in
(\ref{Dirt:int}) arises from the presence of $\epsilon_{\rm d}$ time
dimensions, each of which is accompanied by a factor of $i$.  The Feynman
rules of the random theory are now easily obtained
\begin{eqnarray} \label{Dirt:Feynart}
\raisebox{-0.3cm}{\epsfxsize=2.5cm
\epsfbox{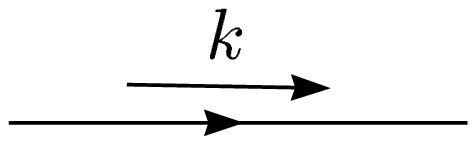}  }
 &=& \frac{-i^{- \epsilon_{\rm d}} {\rm e}^{i(\omega_1 + \omega_2 + \cdots +
\omega_{\epsilon_{\rm d}}) \eta}}{\omega_1 + \omega_2 + \cdots +
\omega_{\epsilon_{\rm d}} -{\bf k}^2 - r_0 + i \eta} \nonumber \\
\raisebox{-0.5cm}{\epsfxsize=2.5cm
\epsfbox{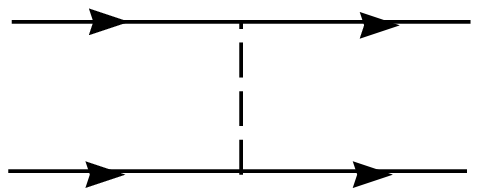}  }
&=& -4 i^{\epsilon_{\rm d}} \lambda_0  \\
\raisebox{-0.5cm}{\epsfxsize=2.5cm
\epsfbox{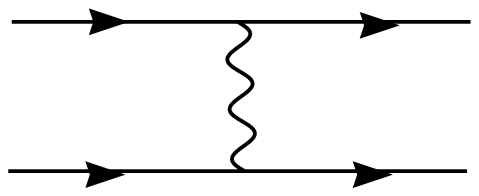}  } 
&=& i^{\epsilon_{\rm d}} (2 \pi)^{\epsilon_{\rm d}} \delta^{\epsilon_{\rm
d}}(\omega_1 + \omega_2 + \cdots + \omega_{\epsilon_{\rm d}})
\Delta_0, \nonumber 
\end{eqnarray} 
where we note that the Lagrangian in $\epsilon_{\rm d}$ time dimensions
involves instead of just one time derivative, a sum of $\epsilon_{\rm d}$
derivatives: $\partial_t \rightarrow \partial_{t_1} +
\partial_{t_2} + \cdots + \partial_{t_{\epsilon_{\rm d}}}$.  
The integral (\ref{Dirt:intb}) has an additional convergence
factor $\exp(i\omega \eta)$ for each of the $\epsilon_{\rm d}$ energy
integrals.  This factor, which is typical for nonrelativistic quantum
theories \cite{Mattuck}, is to be included in self-energy diagrams
containing only one $\phi$-propagator.

Following Weichman and Kim \cite{WK}, we evaluate the integrals over loop
energies assuming that all energies are either positive or negative.  This
allows us to employ Schwinger's propertime representation of propagators
\cite{proptime}, which is based on the integral representation of the gamma
function,
\begin{equation}  \label{Dirt:gamma}
\frac{1}{a^\alpha} = \frac{1}{\Gamma(\alpha)} \int_0^\infty \frac{\dd \tau}{\tau}
\tau^\alpha {\rm e}^{-\tau a}.
\end{equation}  
The energy integrals we encounter to the one-loop order can be carried out
with the help of the equations
\begin{eqnarray}  \label{Dirt:inta} 
\lefteqn{\int' \frac{\dd^{\epsilon_{\rm d}} \omega}{(2\pi)^{\epsilon_{\rm d}}}
\frac{1}{\omega_1 + \omega_2 + \cdots + \omega_{\epsilon_{\rm d}} -x
\pm i \eta} =} \nonumber \\ && -\frac{\Gamma(1-\epsilon_{\rm
d})}{(2\pi)^{\epsilon_{\rm d}}} {\rm sgn}(x) |x|^{\epsilon_{\rm d}-1}
\left({\rm e}^{\pm i \, {\rm sgn}(x) \pi
\epsilon_{\rm d}} + 1 \right),  \\ 
\lefteqn{\int' \frac{\dd^{\epsilon_{\rm d}} \omega}{(2\pi)^{\epsilon_{\rm d}}}
\frac{{\rm e}^{i(\omega_1 + \omega_2 + \cdots + \omega_{\epsilon_{\rm
d}})\eta}}{\omega_1 + \omega_2 + \cdots + \omega_{\epsilon_{\rm d}} -x + i x
\eta} =} \nonumber \\ &&  \frac{i \pi}{(2\pi)^{\epsilon_{\rm
d}}\Gamma(\epsilon_{\rm d})} (i|x|)^{\epsilon_{\rm d}-1} \left[
\sin(\tfrac{1}{2} \pi \epsilon_{\rm d}) - \frac{{\rm
sgn}(x)}{\sin(\tfrac{1}{2} \pi \epsilon_{\rm d} )} \right], 
\label{Dirt:intb} 
\end{eqnarray}  
where $\eta$ is again an infinitesimal positive constant which is to be
taken to zero after the energy integrals have been carried out.  The
prime on the integrals is to remind the reader that the energy integrals
are taken over only two domains with either all energies positive or
negative.  The energy integrals have been carried out by using again the
integral representation (\ref{Dirt:gamma}) of the gamma function.  In doing
so, the integrals are regularized and---as is always the case with analytic
regularizations---irrelevant divergences suppressed.

By differentiation with respect to $x$, Eq.\ (\ref{Dirt:inta}) can,
for example, be employed to calculate integrals involving integrands of the
form $1/(\omega_1 + \omega_2 + \cdots + \omega_{\epsilon_{\rm d}} -x + i
\eta)^2$.  Is is easily checked that in the limit $\epsilon_{\rm d}
\rightarrow 1$, where the energy integral can be performed with help of
contour integration, Eqs.\ (\ref{Dirt:inta}) and (\ref{Dirt:intb}) reproduce
the right results.  When considering the limit of zero time dimensions
($\epsilon_{\rm d} \rightarrow 0$), it should be remembered that the energy
integrals were taken over two separate domains with all energies either
positive or negative.  Each of these domains is contracted to a single point
in the limit $\epsilon_{\rm d} \rightarrow 0$, so that one obtains a result
which is twice that obtained by simply purging any reference to the time
dimensions.  

Before studying the random theory, let us briefly return to the
repulsively interacting bosons in the absence of impurities.  In this case,
there is no need for an $\epsilon_{\rm d}$-expansion and the formalism
outlined above should yield results for arbitrary time dimensions $0 \leq
\epsilon_{\rm d} \leq 1$, interpolating between the classical and quantum
limit.  After the energy integrals have been performed with the help of
Eqs.\ (\ref{Dirt:inta}) and (\ref{Dirt:intb}), the standard technique of
integrating out a momentum shell can be applied to obtain the
renormalization-group equations.  For the correlation-length exponent
$\nu$ we obtain in this way \cite{pla}
\begin{equation} \label{Dirt:nupure}
\nu = \frac{1}{2} \left[1 + \frac{\epsilon}{2} \frac{m+1}{(m+4) -
(m+3) \epsilon_{\rm d}} \cos^2( \tfrac{1}{2} \pi \epsilon_{\rm d}) \right].
\end{equation}   
Here, $\epsilon = 4-2\epsilon_{\rm d}-d$ is the deviation of the {\it
effective} spacetime dimensionality from 4, where it should be noted that
in (canonical) nonrelativistic theories, time dimensions have an engineering
dimension twice that of space dimensions.  (This property is brought out by
the Gaussian value $z=2$ for the dynamic exponent $z$.)  For comparison we
have extended the theory (\ref{bec:Lagr}) to include $m$ complex $\phi$
fields instead of just one field.  In the classical limit, Eq.\
(\ref{Dirt:nupure}) gives the well-known one-loop result for a classical spin
model with $2m$ real components \cite{Ma},
\begin{equation} 
\nu \rightarrow \frac{1}{2} \left(1 + \frac{\epsilon}{2}
\frac{m+1}{m+4} \right), 
\end{equation} 
while in the quantum limit it gives the result $\nu \rightarrow
\frac{1}{2}$, as required.  

The exponent (\ref{Dirt:nupure}), and also the location of the fixed point,
diverges when the number of time dimensions becomes $\epsilon_{\rm
d}\rightarrow (m+4)/(m+3)$.  Since this value is always larger than one, the
singularity is outside the physical domain $0 \leq \epsilon_{\rm d} \leq 1$.
This simple example illustrates the viability of the formalism developed
here to generate results interpolating between the classical and quantum
limit.

We continue with the random theory.  After the energy integrals
have been carried out, it is again straightforward to derive the
renormalization-group equations by integrating out a momentum shell
$\Lambda/b<k<\Lambda$, where $\Lambda$ is a high-momentum cutoff and
$b=\exp(l)$, with $l$ infinitesimal.  Defining the dimensionless variables
\begin{equation} 
\hat{\lambda} = \frac{K_d}{(2 \pi)^{\epsilon_{\rm d}}} \lambda 
\Lambda^{-\epsilon}; \;\;\;
\hat{\Delta} = K_d \Delta \Lambda^{d-4}; \;\;\;
\hat{r} = r \Lambda^{-2},
\end{equation} 
where $K_d$ given in (\ref{Kd}) is the area of a unit sphere in $d$
spatial dimensions divided by $(2\pi)^d$, we find \cite{pla}
\begin{eqnarray} \label{Dirt:reneq}
\frac{\dd \hat{\lambda}}{\dd l} &=&  \epsilon \hat{\lambda}  -8
\left[\Gamma(1-\epsilon_{\rm d}) + (m+3) \Gamma(2-\epsilon_{\rm d}) \right]
\cos(\tfrac{1}{2}\pi \epsilon_{\rm d}) 
\hat{\lambda}^2 + 6 \hat{\Delta} \hat{\lambda} \nonumber  \\
\frac{\dd \hat{\Delta}}{\dd l} &=&  (\epsilon + 2\epsilon_{\rm
d})\hat{\Delta }  + 4 \hat{\Delta}^2 - 16 (m+1)
\Gamma(2-\epsilon_{\rm d}) \cos(\tfrac{1}{2}\pi \epsilon_{\rm d} ) 
\hat{\lambda} \hat{\Delta} \nonumber  \\
\frac{\dd \hat{r}}{\dd l} &=& 2 \hat{r} + 4 \pi \frac{m+1}{
\Gamma(\epsilon_{\rm d})} \frac{\cos^2(\tfrac{1}{2} \pi \epsilon_{\rm d})}
{\sin(\tfrac{1}{2} \pi \epsilon_{\rm d} )}  \hat{\lambda} -
 \hat{\Delta}. 
\end{eqnarray} 
These results are to be trusted only for small values of $\epsilon_{\rm d}$.
For illustrative purposes we have, however, kept the full $\epsilon_{\rm d}$
dependence.  The set of equations yields the fixed point
\begin{eqnarray} \label{Dirt:fp}
\hat{\lambda}^* &=& \frac{1}{16 \cos(\tfrac{1}{2}\pi \epsilon_{\rm d} )
\Gamma(1-\epsilon_{\rm d})} \, \frac{\epsilon + 6 \epsilon_{\rm d}} 
{2m(1-\epsilon_{\rm d}) -1}  \\
\hat{\Delta}^* &=& \frac{1}{4} \frac{
m(1-\epsilon_{\rm d}) (2 \epsilon_{\rm d} -\epsilon) + 2 \epsilon_{\rm d}
(4-3\epsilon_{\rm d}) + \epsilon (2 -\epsilon_{\rm d})}{2m(1-\epsilon_{\rm d})
-1}, \nonumber 
\end{eqnarray}   
and the critical exponent 
\begin{equation} \label{Dirt:nufull} 
\nu = \frac{1}{2} + \frac{\epsilon +2 \epsilon_{\rm d}}{16} + 
\frac{m+1}{16} \frac{
(6\epsilon_{\rm d} + \epsilon )  [\epsilon_{\rm d}+\cos( \pi
\epsilon_{\rm d})]}{2m(1-\epsilon_{\rm d})-1}.
\end{equation} 
The dynamic exponent is given by $z = 2 + \hat{\Delta}^*$.
We see that both $\hat{\lambda}^*$ and $\hat{\Delta}^*$ diverge when
$\epsilon_{\rm d} \rightarrow 1-1/2m$.  At this point, the fixed point becomes
unphysical.  The singularity separates the quantum regime $\epsilon_{\rm
d}\lesssim 1$ from the classical regime $\epsilon_{\rm d} \gtrsim 0$ about
which perturbation theory is to be carried out.  When the equations are
expanded to first order in $\epsilon_{\rm d}$, we recover the IR fixed point
found by Weichman and Kim \cite{WK} using an high-energy cutoff:
\begin{equation} \label{dirt:WK}
\hat{\lambda}^*= \frac{1}{16} \frac{\epsilon + 6 \epsilon_{\rm d}}{2m-1};
\;\;\;  \hat{\Delta}^*= \frac{1}{4} \frac{(2-m)\epsilon + 2(m+4)
\epsilon_{\rm d}}{2m-1}, 
\end{equation}
with the critical exponent
\begin{equation} \label{Dirt:nuqm}
\nu = \frac{1}{2} \left[1 + \frac{1}{8} \frac{3m \epsilon + (5m +2)
2\epsilon_{\rm d}}{2m-1} \right].
\end{equation} 
We thereby provide support for the existence of this fixed point.  

The value of the critical exponent (\ref{Dirt:nuqm}) should be compared with
that of the classical spin model with $2m$ components in the presence of
random impurities of dimension $\epsilon_{\rm d}$ \cite{Dorogovtsev}:
\begin{equation} 
\nu = \frac{1}{2} \left[1 + \frac{1}{8} \frac{3m \epsilon + (5m +2)
\epsilon_{\rm d}}{2m-1} \right].
\end{equation}  
Taking into account that in a nonrelativistic quantum theory, time
dimensions count double as compared to space dimensions, we see that
both results are equivalent.  As to the dynamic exponent, we mention
that a remarkable simple scaling argument (see Sec.\ \ref{sec:scale})
using the quadratic terms in the effective theory (\ref{bec:Lef})
predicts the exact value $z=d$ for $\epsilon_{\rm d} =1$ \cite{FWGF}.
The perturbative result $z = 2 +
\hat{\Delta}^*$, with $\hat{\Delta}^*$ given by (\ref{dirt:WK}), is seen to
be far away from this.

The limit of interest to us, corresponding to $\epsilon_{\rm d} =1$, is
probably difficult to reach by low-order perturbation theory for the
quantum regime is separated by a singularity from the classical regime where
perturbation theory applies.  Although this might be an artifact of the
one-loop calculation, it is unlikely that by including a few more loops, the
quantum regime becomes accessible via the classical regime.  We note that
the singularity moves towards $\epsilon_{\rm d}=1$ when the number of field
components increases.

If the IR fixed point is to be of relevance to the systems mentioned in
Sec.\ \ref{sec:BT}, the impurities have to lead to localization in the
superfluid phase.  For the model at hand, albeit in $d=3$, this connection
has been established by Huang and Meng
\cite{HuMe}.  They showed that in this phase, the impurities give rise
to an additional depletion of the condensate as well as of the superfluid
mass density.  (The latter is defined, as we discussed in Sec.\
\ref{sec:ftbec}, by the response of the system to an
externally imposed velocity field as specified by the expression for the
momentum density, or mass current ${\bf g}$.)  They found that the
depletion of the superfluid mass density is larger than that of the
condensate, indicating that part of the zero-momentum states belongs to
the normal fluid rather than to the condensate.  They interpreted this
as implying that these states are trapped by impurities.

This situation should be contrasted to the one in the absence of
impurities, where the condensate is depleted only due to the
interparticle repulsion.  Despite the depletion, {\it all} the particles
were nevertheless seen in Sec.\ \ref{sec:ftbec} to participate in the
superflow motion at zero temperature.  In other words, the normal fluid
is dragged along by the condensate.  This conclusion was based on the
observation that the momentum density was given by ${\bf g} = m n {\bf
v}_{\rm s}$, where $n$ is the {\it total} particle number density and
${\bf v}_{\rm s}$ the velocity with which the condensate moves, i.e.,
the superfluid velocity.  (For clarity, we have reintroduced the mass
parameter $m$ here.)

In terms of the shifted field (\ref{bec:newfields}), the random term
(\ref{Dirt:dis}) reads
\begin{equation}  
{\cal L}_{\Delta} = \psi({\bf x}) (|\bar{\phi}|^2 + |\tilde{\phi}|^2 +
\bar{\phi} \tilde{\phi}^* + \bar{\phi}^* \tilde{\phi}  ).
\end{equation} 
The first two terms lead to an irrelevant change in the chemical
potential, so that we only have to consider the last two terms, which we
can cast in the form
\begin{equation}
{\cal L}_{\Delta} = \psi({\bf x}) \, \bar{\Phi}^\dagger \tilde{\Phi},
\;\;\;\;\;\;\;
\bar{\Phi} = \left(\begin{array}{l} \bar{\phi} \\ \bar{\phi}^*
\end{array} \right). 
\end{equation} 
The integral over $\tilde{\Phi}$ is Gaussian in the Bogoliubov
approximation and therefore easily performed to yield an additional term to the
effective action
\begin{equation} 
S_{\Delta} = -\frac{1}{2} \int_{x,y} \psi({\bf x}) \bar{\Phi}^\dagger \,
G_0(x-y) \bar{\Phi} \psi({\bf y}),
\end{equation} 
where the propagator $G_0$ is the inverse of the matrix $M_0$ introduced
in (\ref{bec:M}) with the field $U(x)$ set to zero.  Let us
first Fourier transform the fields,
\begin{eqnarray} 
G_0(x-y) &=& \int_k {\rm e}^{-i k \cdot (x-y)} \, G_0(k)    \\
\psi({\bf x}) &=& \int_{\bf k} {\rm e}^{i {\bf k} \cdot {\bf x}} \psi({\bf k}).
\end{eqnarray} 
The contribution to the effective action then appears in the form
\begin{equation} \label{S_d}
S_{\Delta} = -\frac{1}{2} \int_{\bf k} |\psi({\bf k})|^2
\bar{\Phi}^\dagger G(0,{\bf k}) \bar{\Phi}.
\end{equation} 
Since the random field is Gaussian distributed [see (\ref{random})], the
average over this field representing quenched impurities yields,
\begin{equation} 
\langle |\psi({\bf k})|^2 \rangle = \tfrac{1}{2} V \Delta_0.
\end{equation} 
The remaining integral over the loop momentum in (\ref{S_d}) is readily
carried out to yield in arbitrary space dimensions
\begin{equation} \label{L_D}
\langle {\cal L}_\Delta \rangle = \frac{1}{2} \Gamma(1-d/2)
\left(\frac{m}{2 \pi} \right)^{d/2} |\bar{\phi}|^2 (6 \lambda_0
|\bar{\phi}|^2 - \mu_0)^{d/2-1} \Delta_0. 
\end{equation} 
Because this is a one-loop result, we may to this order replace the bare
parameters with the (one-loop) renormalized ones.

In Sec.\ \ref{sec:bec} we saw that due to the interparticle repulsion, not
all the particles reside in the condensate.  We expect that the random field
causes an additional depletion of the condensate.  To obtain this, we
differentiate (\ref{L_D}) with respect to the chemical potential.  This
gives \cite{GPS,pla}
\begin{equation} 
\bar{n}_\Delta = \frac{\partial \langle {\cal L}_\Delta
\rangle}{\partial \mu} =
\frac{2^{d/2-5}\Gamma(2-d/2)}{\pi^{d/2}} m^{d/2} \lambda^{d/2-2}
\bar{n}_0^{d/2-1} \Delta_0,
\end{equation}   
where $\bar{n}_0$ denotes the number density of particles residing in the
condensate.  The divergence in the limit $\lambda \rightarrow 0$ for $d
<4$ signals the collapse of the system when the interparticle repulsion is
removed.

We next calculate the mass current ${\bf g}$ to determine superfluid mass
density, i.e., the mass density flowing with the superfluid velocity ${\bf
v}_{\rm s}$.  As we have seen in the preceding section, in the absence of
impurities and at zero temperature all the particles participate in the
superflow and move on the average with the velocity ${\bf v}_{\rm s}$.  We
expect this no longer to hold in the presence of impurities.  To determine
the change in the superfluid mass density due to impurities, we replace
$\mu_0$ with $\mu_{\rm eff}$ as defined in (\ref{bec:mureplacement}) and
$i\partial_0$ with $i\partial_0 - ({\bf u} - {\bf v}_{\rm s}) \cdot (-i
\nabla)$ in the contribution (\ref{S_d}) to the effective action, and
differentiate it with respect to the externally imposed velocity, $-{\bf
u}$.  We find to linear order in the difference ${\bf u}- {\bf v}_{\rm s}$:
\begin{equation} 
{\bf g} = \rho_{\rm s} {\bf v}_{\rm s} + \rho_{\rm n} {\bf u},
\end{equation} 
with the superfluid and normal mass density \cite{pla}
\begin{equation} 
\rho_{\rm s} = m\left(\bar{n} -  \frac{4}{d} \bar{n}_\Delta \right), \;\;\;\;
\rho_{\rm n} = \frac{4}{d} m \bar{n}_\Delta.
\end{equation}  
We see that the normal density is a factor $4/d$ larger than the mass
density $m\bar{n}_\Delta$ knocked out of the condensate by the impurities.
(For $d=3$ this gives the factor $\tfrac{4}{3}$ first found in Ref.\
\cite{HuMe}.)  Apparently, part of the zero-momentum states belongs for $d < 4$
not to the condensate, but to the normal fluid.  Being trapped by the
impurities, this fraction of the zero-momentum states are localized.  This
shows that the phenomenon of localization can be accounted for in the
Bogoliubov theory of superfluidity by including a random field.

An other realistic modification of the nonrelativistic $|\phi|^4$-theory
is to include a $1/r$ Coulomb repulsion.  Using very general scaling
arguments (see Sec.\ \ref{sec:scale}) in the context of the effective
theory, Fisher, Grinstein, and Girvin
\cite{FGG} predicted that in the presence of impurities the inclusion of
this interaction changes the value of the dynamic exponent to $z=1$, which
is exact again.  This prediction has been confirmed in experiments on the
superconductor-to-insulator transition in two-dimensional films
\cite{HPsu1}.  The same value $z=1$ has also been found in the quantum Hall
transitions, and---more recently---in metal-to-insulating transitions
observed in dilute two-dimensional electron systems in silicon MOSFETS
\cite{MIT,KSSMF,SKS,PFW}.  This suggests that the $1/r$ Coulomb
interaction plays an important role in these systems too.  However, let
us at this stage mention the renormalization-group results obtained by
Giamarchi and Schulz
\cite{GS} who studied fermions with {\it short-range} interactions in a
random one-dimensional system.  They found first of all that the
superconductor-to-insulator transition these fermions undergo is indeed in
the universality class of repulsively interacting bosons in the presence of
impurities.  Moreover, their result for the conductivity is consistent with
the value $z=1$ \cite{FGG}, implying that already a local repulsive
interaction as used in the nonrelativistic $|\phi|^4$-theory can lead to
this value for the dynamic exponent.
\section{CSGL Theory}
\label{sec:CSGL}
We now turn to the fractional quantized Hall effect (FQHE) which is the
hallmark of a new, intrinsically two-dimensional condensed-matter
state---the quantum Hall liquid.  Many aspects of this state are well
understood in the framework of the quantum-mechanical picture developed by
Laughlin \cite{Laughlin}.  Considerable effort has nevertheless been
invested in formulating an effective field theory which captures the
essential low-energy, small-momentum features of the liquid.  A similar
approach in the context of superconductors has proven most successful.
Initially, only the phenomenological model proposed by Ginzburg and Landau
\cite{GL} in 1950 was known here.  Most of the fundamental properties of the
superconducting state such as superconductivity---the property that gave
this condensed-matter state its name, Meissner effect, magnetic flux
quantization, Abrikosov flux lattice, and Josephson effect, can be explained
by the model.  The microscopic theory was given almost a decade later by
Bardeen, Cooper, and Schrieffer \cite{BCS}.  Shortly here after,
Gorkov \cite{Gorkov} made the connection between the two approaches by deriving
the Ginzburg-Landau model from the microscopic BCS theory, thus giving the
phenomenological model the status of an effective field theory.

A first step towards an effective field theory of the quantum Hall liquid
was taken by Girvin and MacDonald \cite{GMac} and has been developed further
by Zhang, Hansson and Kivelson \cite{ZHK}, who also gave an explicit
construction starting from a microscopic Hamiltonian.  Their formulation
incorporates time dependence which is important for the study of quantum
phase transitions.  This approach has proven very successful (for a review
see Ref.\ \cite{Zhang}).  In this and the following section, we shall argue
that the effective theory---the so-called Chern-Simons-Ginzburg-Landau
(CSGL) theory---can also be employed to describe the quantum phase
transition which a quantum Hall system undergoes as the applied magnetic
field changes.

We shall in this section first recall some basic properties of the CSGL
theory and show how it can be used to describe the field-induced
Hall-liquid-to-insulator transition of a Hall liquid.  In Sec.\ \ref{sec:RG} we
then apply renormalization group theory to the CSGL theory and study its
critical properties.

An important ingredient for obtaining an effective theory of the FQHE was
the identification by Girvin and MacDonald \cite{GMac} of a bosonic operator
$\phi$ which exhibits (algebraic) off-diagonal long-range order.  The
long-range order was found to be of a type known to exist in two-dimensional
bosonic superfluids.  They argued that this field should be viewed as an
order parameter in terms of which the effective field theory should be
formulated.  To account for the incompressibility of the quantum Hall liquid
they suggested to couple $\phi$ to a statistical gauge field $a_\mu$.  The
gapless spectrum of the neutral system then changes into one with an energy
gap \cite{ZHK}, thus rendering the charged system incompressible.

Girvin and MacDonald assumed that the statistical gauge field is governed
solely by a Chern-Simons term
\begin{equation}
\label{CSGL:CS}
{\cal L}_{\rm CS} = \frac{\theta}{2} \partial_0 {\bf a} \times {\bf a} -
\theta a_0 \nabla \times {\bf a}, 
\end{equation}
with $\nabla \times {\bf a}$ the statistical magnetic field and $\theta$ a
constant.  Because of the absence of a kinetic term (the usual Maxwell
term), the statistical gauge field does not represent a physical degree of
freedom.  In a relativistic setting, a Maxwell term is usually generated by
quantum corrections so that the statistical gauge field becomes dynamical at
the quantum level.  The quantum theory then differs qualitatively from the
classical theory.  On the other hand, as we shall see below, this need not
be the case in a nonrelativistic setting.  That is to say, the {\it Ansatz}
of the absence of a Maxwell term is here not necessarily obstructed by
quantum corrections.

The CSGL theory is described by the Lagrangian \cite{ZHK}
\begin{equation}
\label{CSGL:L}
{\cal L} = i \phi^* D_0 \phi -
\frac{1}{2m} |{\bf D} \phi|^2 + \mu_0 |\phi|^2 - \lambda_0 |\phi|^4 +
{\cal L}_{\rm CS}.
\end{equation}
The covariant derivatives $D_0 = \partial_0 + i e A_0 + i e a_0$ and
${\bf D} = \nabla - i e {\bf A} - i e {\bf a}$ give a minimal coupling
to the applied magnetic and electric field described by the gauge field
$A_\mu$ and also to the statistical gauge field.  For
definiteness we will assume that our two-dimensional sample is
perpendicular to the applied magnetic field, defining the $z$-direction,
and we choose the electric field to point in the $x$-direction.  The
charged field $\phi$ represents the Girvin-MacDonald order parameter
describing the original electrons bound to an odd number of flux
quanta.  To see that it indeed does, let us consider the field equation
for $a_0$:
\begin{equation} \label{CSGL:a0}
|\phi|^2 = - e \theta  \nabla \times {\bf a}.
\end{equation} 
The simplest solution of the CSGL Lagrangian is the uniform mean-field
solution
\begin{equation} 
|\phi|^2 = \bar{n}, \;\;\;\; {\bf a} = - {\bf A}, \;\;\;\; a_0 = - A_0 =
0,
\end{equation} 
where $\bar{n}$ indicates the constant fermion number density.  The
statistical gauge field is seen to precisely cancel the applied field.
The constraint equation (\ref{CSGL:a0}) then becomes
\begin{equation}  \label{CSGL:n}
\bar{n} = e \theta H,
\end{equation} 
with $H$ the applied magnetic field.  Now, if we choose $\theta^{-1} = 2
\pi (2l+1)$, it follows on integrating this equation that, as required,
with every electron there is associated $2l+1$ flux quanta:
\begin{equation} 
N = \frac{1}{2l+1} N_\otimes,
\end{equation} 
where $N_\otimes = \Phi/\Phi_0$, with $\Phi = \int_{\bf x} H$ the
magnetic flux, indicates the number of flux quanta.  Equation
(\ref{CSGL:n}) implies an odd-denominator filling factor $\nu_H$ which
is defined by
\begin{equation} 
\nu_H = \frac{\bar{n}}{H/\Phi_0}= \frac{1}{2l+1}.
\end{equation} 

The coupling constant $\lambda_0 \, (>0)$ in (\ref{CSGL:L}) is the
strength of the repulsive contact interaction between the composite
particles, and $\mu_0$ is a chemical potential introduced to account for
a finite number density of composite particles.  

It is well known from anyon physics that the inclusion of the
Chern-Simons term changes the statistics of the field $\phi$ to which
the statistical gauge field is coupled \cite{Wilczek}.  If one composite
particle circles another, it picks up an additional Aharonov-Bohm
factor, representing the change in statistics.  The binding of an odd
number of flux quanta changes the fermionic character of the electrons
into a bosonic one for the composite particles, allowing them to Bose
condense.  The algebraic off-diagonal long-range order of a quantum Hall
liquid can in this picture be understood as resulting from this
condensation.  Conversely, a flux quantum carries $1/(2l+1)$th of
an electron's charge \cite{Laughlin}, and also $1/(2l+1)$th of an
electron's statistics \cite{ASW}.

The defining phenomenological properties of a quantum Hall liquid are
easily shown to be described by the CSGL theory \cite{ZHK,Zhang}.  From
the lowest-order expression for the induced electromagnetic current one
finds
\begin{equation}
\label{CSGL:inducedji}
e j_i = \frac{\delta {\cal L}}{\delta A_i} = - \frac{\delta {\cal
L}_\phi}{\delta a_i} = \frac{\delta {\cal L}_{\rm CS}}{\delta a_i} = - e^2
\theta \epsilon_{ij} (\partial_0 a_j - \partial_j a_0) = e^2 \theta 
\epsilon_{ij} E_j,
\end{equation}
with ${\bf E}$ the applied electric field and where the Lagrangian
(\ref{CSGL:L}) is written as a sum ${\cal L} = {\cal L}_\phi + {\cal L}_{\rm
CS}$.  It follows that the Hall conductance $\sigma_{xy}$ is quantized in
odd fractions of $e^2/2 \pi$, or, reinstalling Planck's constant, $e^2/h$.
This result can also be understand in an intuitive way as follows.  Since
the composite particles carry a charge $e$, the applied electric field gives
rise to an electric current
\begin{equation} 
I = e \frac{\dd N}{\dd t}
\end{equation} 
in the direction of ${\bf E}$, i.e., the $x$-direction.  This is not the
end of the story because the composite objects carry in addition to
electric charge also $2l+1$ flux quanta.  When the Goldstone field
$\varphi$ encircles $2l+1$ flux quanta, it picks up a factor $2 \pi$ for
each of them
\begin{equation} 
\oint_\Gamma \nabla \cdot \varphi = 2 \pi (2l+1).
\end{equation} 
Now, consider two points across the sample from each other.  Let the
phase of these points initially be equal.  As a composite particle moves
downstream, and crosses the line connecting the two points, the relative
phase $\Delta \varphi$ between them changes by $2 \pi (2l+1)$.
This phase slippage \cite{PWA} leads to a voltage drop across the
sample given by 
\begin{equation} 
V_{\rm H} =  \frac{1}{e} \partial_0 \Delta \varphi = (2l+1) \Phi_0
\frac{\dd N}{\dd t}, 
\end{equation} 
where the first equation can be understood by recalling that due to minimal
coupling $\partial_0 \varphi \rightarrow \partial_0 \varphi + e A_0$.  For
the Hall resistance we thus obtain the expected value
\begin{equation} 
\rho_{xy} = \frac{V_{\rm H}}{I} = (2l+1) \frac{2 \pi}{e^2}.
\end{equation} 

If the CSGL theory is to describe an incompressible liquid, the spectrum
of the single-particle excitations must have a gap.  Without the
coupling to the statistical gauge field, the spectrum is given by the
gapless Bogoliubov spectrum (\ref{bec:bogo}).  To obtain the
single-particle spectrum of the coupled theory, we integrate out the
statistical gauge field.  The integration over $a_0$ was shown to yield
the constraint (\ref{CSGL:a0}) which in the Coulomb gauge $\nabla \cdot
{\bf a} = 0$ is solved by
\begin{equation}
\label{CSGL:solution}
a_i = \frac{1}{e \theta} \epsilon_{ij} \frac{\partial_j}{\nabla^2} |\phi|^2.
\end{equation}
The integration over the remaining components of the statistical gauge field
is now simply performed by substituting (\ref{CSGL:solution}) back into the
Lagrangian.  The only nonzero contribution arises from the term $- e^2
|\phi|^2 {\bf a}^2/2m$.  The spectrum of the charged system acquires as a
result an energy gap $\omega_{\rm c}$
\begin{equation}
E({\bf k}) = \sqrt{\omega_{\rm c}^2 + \epsilon^2({\bf k}) + 2 \mu_0
\epsilon( {\bf k}) },
\end{equation}
with $\omega_{\rm c} = \mu_0 /2\theta m\lambda_0$.  To lowest order, the gap
equals the cyclotron frequency of a free charge $e$ in a magnetic field $H$
\begin{equation}
\omega_{\rm c} = \frac{\bar{n}}{\theta m} = \frac{e H}{m}.
\end{equation}
The presence of this energy gap results in dissipationless flow with
$\sigma_{xx} =0$.

These facts show that the CSGL theory captures the essentials of a
quantum Hall liquid.  Given this success, it is tempting to investigate
if the theory can also be employed to describe the field-induced
Hall-liquid-to-insulator transitions.  It should however be borne in
mind that both the $1/|{\bf x}|$-Coulomb potential as well as impurities
should be incorporated into the theory in order to obtain a realistic
description of the FQHE.  The repulsive Coulomb potential is believed to
play a decisive role in the formation of the the composite particles,
while the impurities are responsible for the width of the Hall plateaus.
As the magnetic field moves away from the magic filling factor, magnetic
vortices will materialize in the system to make up the difference
between the applied field and the magic field value.  In the presence of
impurities, these defects get pinned and do not contribute to the
resistivities, so that both $\sigma_{xx}$ and $\sigma_{xy}$ are
unchanged.  Only if the difference becomes too large, the system reverts
to an other quantum Hall state with a different filling factor.

For positive bare chemical potential $\mu_0 >0$ the Girvin-MacDonald order
parameter $\phi$ has a nonvanishing expectation value given by $|\phi|^2 =
\mu_0/2\lambda_0$, implying that the composite particles are condensed.  When
$\mu_0 \rightarrow 0$, the condensate is drained of composite particles,
and at $\mu_0 = 0$, it vanishes altogether.  The system becomes critical
here and reverts to the insulating phase characterized by a negative
bare chemical potential.

In the spirit of Landau, we take a phenomenological approach towards the
field-induced phase transition of the CSGL theory.  And assume that when
the applied magnetic field $H$ is close to the upper critical field
$H^+_{\nu_H}$ at which the quantum Hall liquid with filling factor
$\nu_H$ is destroyed, the chemical potential of the composite particles
depends linearly on $H$, i.e., $\mu_0 \sim eH^+_{\nu_H}- eH$.  This
state can of course also be destroyed by lowering the applied field.  If
the system is near the lower critical field $H^-_{\nu_H}$, we assume
that the chemical potential is instead given by $\mu_0
\sim eH - eH^-_{\nu_H}$.  This is the basic postulate of our approach.

In the CSGL Lagrangian (\ref{CSGL:L}) we have again transformed the mass
$m$ of the nonrelativistic $|\phi|^4$-theory away.  In addition to the
engineering dimensions (\ref{BT:scale}), we have for the Chern-Simons
field
\begin{equation} \label{CSGL:dim}
[e a_i] = 1, \;\;\;\; [e a_0] = 2, \;\;\;\; [\theta] = 0.
\end{equation} 
In two space dimensions the coupling constant $\lambda_0$ was seen to be
dimensionless, implying that the $|\phi|^4$-term is a marginal operator.
We see from (\ref{CSGL:dim}) that also the Chern-Simons term is a
marginal operator.  Hence, the CSGL theory contains precisely those
terms relevant to the description of the Hall-liquid-to-insulator
transition in a quantized Hall system.
\section{Renormalization Group}
\label{sec:RG} 
We now turn our attention to the critical properties of the CSGL theory.  A
remaining open problem in this context is the nature of the
Hall-liquid-to-insulating transitions in the theory.  Taking place at the
absolute zero of temperature, these transitions are a pure quantum
phenomenon.  They are driven either by impurities, a periodic potential, or
by the applied magnetic field.  The impurity-driven phase transition has
been considered in Ref.\
\cite{KLZ_LWK}, while the transition driven by a periodic potential has
been studied in Refs.\ \cite{WW,PZ}.  The latter problem was mapped onto
a {\it relativistic} CSGL theory and it was concluded \cite{PZ} that the
transition induced by a periodic potential is 1st-order.  We in this
section shall discuss the transition induced by the magnetic field in
the absence of both impurities and a periodic potential.  We will be
working with the standard nonrelativistic CSGL theory.  As has been
eluded to above, there are essential differences between relativistic
and nonrelativistic Chern-Simons theories.  There are features special
to the nonrelativistic theory which in our view are important for
understanding the field-induced transition, and which cannot be
implemented in a relativistic formulation.

Experimentally, if the external field is changed so that the filling factor
$\nu_H$ moves away from an odd-denominator value, the system eventually
becomes critical and undergoes a transition to an insulating phase.  We will
argue that this feature is encoded in the CSGL theory.  We shall show that
the theory has an IR fixed point,  implying that the phase transition is
2nd-order, and determine the critical exponents.  They turn out to be
universal and independent of the filling fraction.

We will be working in the Coulomb gauge.  To this end we add a gauge-fixing
term $(\nabla \cdot {\bf a})^2/2 \alpha_{\rm gf}$ to the Lagrangian
(\ref{CSGL:L}), and take the limit $\alpha_{\rm gf} \rightarrow 0$.  In this
gauge, the propagator of the statistical gauge field is purely off-diagonal
\begin{equation}
\label{RG:statprop}
\raisebox{-0.4cm}{\epsfxsize=2.5cm
\epsfbox{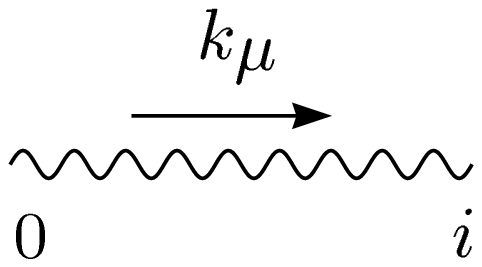} }
= G_{0 i}(k_) = - G_{i 0}(k) = i \alpha \epsilon_{ij}
\frac{k_j}{{\bf k}^2},
\end{equation}
with $\alpha = 1/\theta$.  The two propagators (\ref{BT:Green}) and
(\ref{RG:statprop}) of the CSGL theory are causing infrared divergences.  We
regularize these by introducing an infrared cutoff via the substitution
${\bf k}^2 \rightarrow {\bf k}^2 + 2r_0$ in both propagators.  In the scalar
propagator, this amounts to taking a negative chemical potential, $\mu_0 =
-r_0$, so that we are in the symmetrical state of the theory.  Because we
consider the theory in its upper critical dimension and at criticality, the
parameter $r_0$ is the only dimensional parameter present.  It is
necessary to introduce a dimensional parameter in a theory not having any,
not just to avoid infrared divergences, but also to be able to carry out the
renormalization-group program---this parameter playing the role of
renormalization-group scale parameter.  There are various ways to do so, our
proposal is one possibility.  Another would be to take the renormalization
point at nonzero external momenta, as has been done by, for example, Bergman
and Lozano \cite{LBL}.  The critical properties are, of course, independent
of the particular choice.

\begin{figure}
\begin{center}
\epsfxsize=8.cm
\mbox{\epsfbox{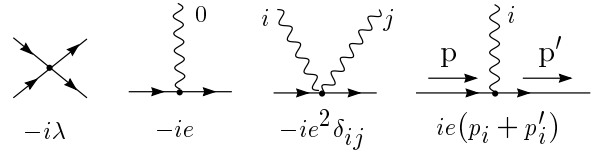}}
\end{center}
\caption{Feynman rules of the CSGL theory. \label{fig:feynman}}
\end{figure}

We evaluate Feynman diagrams according to the following computational scheme.
First, we carry out the discrete sums over repeated indices as well as the
integrations over loop energies.  Then, to handle ultraviolet
divergences, the resulting momentum integrals are analytically continued to
arbitrary space dimensions $d$.  It is to be noted that this step is taken
after the discrete sums have been carried out and the antisymmetric tensor
$\epsilon_{i j}$---which makes sense only in $d=2$---has disappeared from the
expressions.  Finally, the counter terms and the critical exponents are
determined by employing the minimal subtraction scheme in which only those
parts of the counter terms are considered which diverge in the limit $d
\rightarrow 2$.  The Feynman rules for the vertices are given in Fig.\ 
\ref{fig:feynman}.

\begin{figure}
\begin{center}
\epsfxsize=2.cm
\mbox{\epsfbox{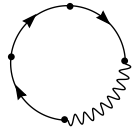}}
\end{center}
\caption{A closed oriented loop containing in addition a gauge
field propagator.   \label{fig:oriented2}}
\end{figure}

The reason we have chosen the Coulomb gauge is that in this gauge the
statistical gauge field propagator does not depend on the energy variable
$k_0$ [see Eq.\ (\ref{RG:statprop})].  This means that the analytic structure of
a given diagram is determined solely by its scalar propagators.  As a result,
the rule that closed oriented loops vanish is not corrupted if a loop contains
in addition gauge field propagators.  (See Fig.\ \ref{fig:oriented2} for an
example.)  Hence, also in the coupled theory most diagrams vanish, making the
quantum critical behavior of the CSGL theory tractable.

\begin{figure}
\begin{center}
\epsfxsize=8.cm
\mbox{\epsfbox{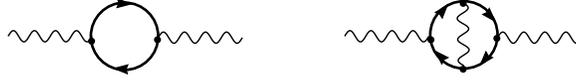}}
\end{center}
\caption{A one-loop and a two-loop example of vanishing diagrams which would
otherwise renormalize the charge $e$. \label{fig:noMaxwell}}
\end{figure}

In particular, all the diagrams which would renormalize the charge $e$ vanish.
(A one-loop and a two-loop example of this class of diagrams are depicted in
Fig.\ \ref{fig:noMaxwell}.)  Put differently, no Maxwell term for the statistical
gauge field is generated at the quantum level so that the statistical gauge
field propagator (\ref{RG:statprop}) remains purely off-diagonal.  This is an
important characteristic of the nonrelativistic CSGL theory not shared by the
corresponding relativistic extension where a Maxwell term is automatically
generated at the quantum level.  Since the Chern-Simons term is also not
renormalized, the statistical gauge field propagator (\ref{RG:statprop}) is not
effected at all by quantum corrections.

As in the neutral case, the boson self-energy $\Sigma$ also vanishes at
every loop order in the charged theory, and the only object that 
renormalizes is the self-coupling parameter $\lambda_0$.  Consequently, if an
IR fixed point is found in the CSGL theory, the 2nd-order
phase transition described by it has Gaussian exponents.

In order to examine the presence of an IR fixed point, we compute the beta
function using the scheme outlined in \cite{Ramond}.  To this end we evaluate
the diagrams:
\begin{eqnarray} \label{RG:1loopring}
\raisebox{-0.4cm}{\epsfxsize=2.cm 
\epsfbox{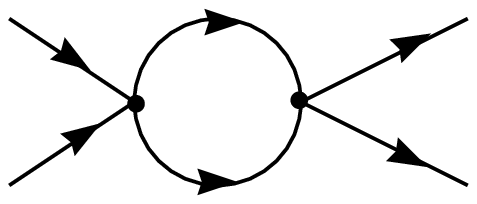}  }  \!\!\!\!
&=& \!\!\!\! 8 \lambda^2 \int_k \, \frac{1}{-k_0 - \tfrac{1}{2} {\bf k}^2 -
r + i \eta} \,\, \frac{1}{k_0 - \tfrac{1}{2} {\bf k}^2 -
r + i \eta} \nonumber \\  
\!\!\!\! &=& \!\!\!\!8 i\lambda^2 \int_{\bf k} \, \frac{1}{{\bf k}^2
+ 2 r}  = 8i \lambda^2 \, I_d,
\end{eqnarray} 
with $\lambda$ and $r$ denoting the renormalized parameters,
\begin{equation}
I_d = \frac{\Gamma(1-d/2)}{(4 \pi)^{d/2}} \frac{1}{(2r)^{1-d/2}} ,
\end{equation}
and 
\begin{eqnarray} \label{RG:1loop7UP}
\raisebox{-0.39cm}{\epsfxsize=2.7cm
\epsfbox{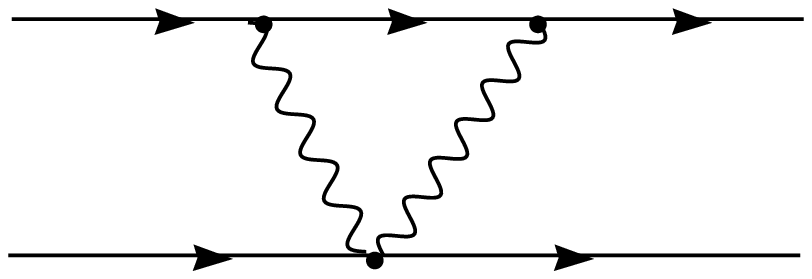}  } \!\!\!\!\!\!
&=& \!\!\!\! \alpha^2 \epsilon_{jl} \, \epsilon_{jm} \int_k \,
\frac{k_l}{{\bf k}^2 + 2r} \,\, \frac{1}{k_0 - \tfrac{1}{2} {\bf k}^2 -
r + i \eta} \,\, \frac{k_m}{{\bf k}^2 + 2r}
\nonumber \\ \!\!\!\!\!\!
&=& \!\!\!\! - \tfrac{1}{2} i \alpha^2 \int_{\bf k} \frac{{\bf k}^2}{({\bf
k}^2 + 2r)^2} \nonumber \\ \!\!\!\!\!\! &=& \!\!\!\! -
\tfrac{1}{2} i\alpha^2 \Omega_d \left[\Gamma\left(1-\frac{d}{2}\right) -
\Gamma\left(2-\frac{d}{2}\right) \right] ,
\end{eqnarray} 
where we introduced the abbreviation 
\begin{equation}
\Omega_d = \frac{1}{(4 \pi)^{d/2} } \frac{1}{(2r)^{1-d/2}}.
\end{equation}
There are three other diagrams of the form (\ref{RG:1loop7UP}), one having
the right external legs crossed and two having the triangle turned
upside-down.  To account for these, the result (\ref{RG:1loop7UP}) must be
included with an additional factor of 4.  The diagrams (\ref{RG:1loopring}) and
(\ref{RG:1loop7UP}) are found to diverge in the limit $d
\rightarrow 2$.  The divergences are cancelled by including the one-loop
counter term
\begin{equation}
\label{RG:1loopcounterterm}
\raisebox{-0.35cm}{\epsfxsize=1.8cm
\epsfbox{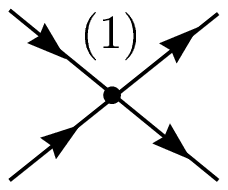} }
 = -i 4 \Omega_2 \, \frac{1}{\epsilon}
\left(4 \lambda^2 -  \alpha^2 \right),
\end{equation} 
with $\epsilon = 2-d$ the deviation from the upper critical dimension and
$\Omega_2 = 1/4 \pi$.  This leads in the usual way to the one-loop
beta function in the upper critical dimension ($d=2$ in our case)
\begin{equation}
\label{RG:beta1loop}
\beta(\lambda) = - \left( 1 - \lambda \frac{\partial}{\partial \lambda} -
\alpha \frac{\partial}{\partial \alpha} \right) a_1 = \frac{1}{\pi}
\left(4\lambda^2 - \alpha^2 \right), 
\end{equation}
where $a_1$ is ($i$ times) the residue of the simple pole in the counter term
(\ref{RG:1loopcounterterm}).  The result is in accordance with previous studies
by Lozano and Bergman \cite{LBL} who used a momentum cutoff to regularize the
ultraviolet divergences.

\begin{figure}
\begin{center}
\epsfxsize=8.cm
\mbox{\epsfbox{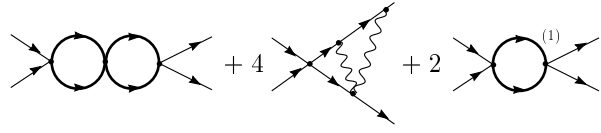}}
\end{center}
\caption{The relevant two-loop diagrams.  \label{fig:2loop}}
\end{figure}

We continue with the two-loop calculation.  The relevant diagrams are given in
Fig.\ \ref{fig:2loop}.  The factor 4 in the second diagram is to account
for the related diagram having the 4-vertex to the right rather than to the
left and for the diagrams having the triangles turned upside-down.  The factor
2 in the last diagram accounts for the fact that there are two vertices which
can be replaced with the counter term (\ref{RG:1loopcounterterm}).  
The first diagram is readily shown to yield
\begin{equation}
\label{RG:2loopring}
\raisebox{-0.3cm}{\epsfxsize=2.cm
\epsfbox{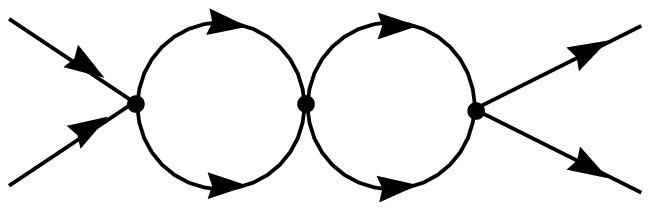} }
= - 16 i \lambda^3 I_d^2
\end{equation}
whereas the second gives
\begin{equation}
\label{RG:2loop7UP}
\raisebox{-.8cm}{\epsfxsize=1.8cm
\epsfbox{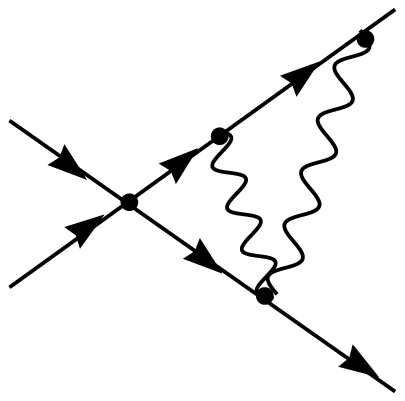} }
= - 2 i  \lambda \alpha^2 \int_{{\bf
k},{\bf l}} \, \frac{1}{{\bf k}^2 + 2r} \, \frac{1}{{\bf l}^2 +
2r} \,  \frac{({\bf k} - {\bf l}) \cdot {\bf l}}{({\bf k} - {\bf l})^2 +
2r}.
\end{equation}
To evaluate the remaining integrals, we employ the identity
\begin{equation}
\label{RG:id1}
\int_{\bf k} \, \frac{{\bf k}}{({\bf k}^2 + 2r) \, [({\bf k} - {\bf
l})^2 + 2r]} = \frac{1}{2}
\int_{\bf k} \, \frac{{\bf l}}{({\bf k}^2 + 2r)\,  [({\bf k} - {\bf
l})^2 + 2r]}
\end{equation}
which can be proven by introducing a new integration variable ${\bf k}' = -
{\bf k} +  {\bf l}$ in the left-hand side.  In this way, we obtain for the
second diagram
\begin{equation}
\label{RG:2loop7UPUP}
\raisebox{-.8cm}{\epsfxsize=1.8cm
\epsfbox{2loopb.eps} }
= i \lambda \alpha^2 \left( I_d^2 - J_d \right),
\end{equation}
where $J_d$ is the integral
\begin{equation}
\label{RG:jd}
J_d = 2r \int_{{\bf k},{\bf l}} \, \frac{1}{{\bf k}^2 + 2r} \,
\frac{1}{{\bf l}^2 + 2r} \,  \frac{1}{({\bf k} - {\bf l})^2 +
2r}
\end{equation}
which remains finite in the limit $d \rightarrow 2$.  Finally, the last
diagram in Fig.\ \ref{fig:2loop} is given by (\ref{RG:1loopring}) with $-i
\lambda$ replaced with the counter term (\ref{RG:1loopcounterterm}).

When the results are added, the sum of the three diagrams in Fig.\ 
\ref{fig:2loop} is shown to diverge again when the space dimension is given its
physical value $(d=2)$.  The theory can be rendered finite to this order by
introducing the two-loop counter term
\begin{equation}
\label{RG:2loopcounterterm}
\raisebox{-0.35cm}{\epsfxsize=1.8cm
\epsfbox{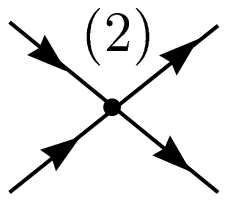} }
= - 16 i \, \Omega_2^2 \,
\frac{1}{\epsilon^2}  \lambda \left(4 \lambda^2 - \alpha^2 \right). 
\end{equation} 
Surprisingly, the counter term does not contain a simple pole in $\epsilon$.
Since the beta function is determined by the residue of the simple pole, it
follows that the one-loop result (\ref{RG:beta1loop}) is unchanged at the
two-loop level.  

The calculation of the beta function was extended to third order in the
loop expansion by Freedman, Lozano, and Rius \cite{FLR} using differential
regularization, and to fourth order by this author \cite{npfs} applying the
more conventional methods explained in this section.  We found that the
third and fourth order diagrams diverge like $1/\epsilon^3$ and
$1/\epsilon^4$, respectively.  In particular, the simple poles in $\epsilon$
dropped out so that the one-loop beta function is unaffected by three-
and four-loop corrections.  It is tempting to conjecture that this feature
persists to all orders in perturbation theory, implying that---just as in
the neutral system which corresponds to taking the limit $\theta\rightarrow
\infty$--- the one-loop beta function (\ref{RG:beta1loop}) is exact.

\begin{figure}
\vspace{-1.cm}
\begin{center}
\epsfxsize=5.cm
\mbox{\epsfbox{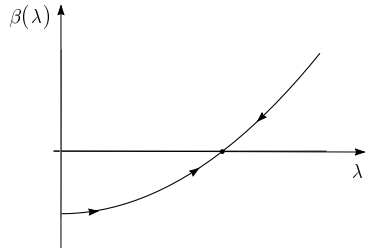}}
\end{center}
\vspace{-1.5cm}
\caption{Schematic representation of the $\beta$-function. \label{fig:beta}}
\end{figure}

The beta function is schematically represented in Fig.\
\ref{fig:beta}.  It yields an IR fixed point
$\lambda^{*^{\scriptstyle{2}}}= \tfrac{1}{4} \alpha^2$ determined by
$\theta = \alpha^{-1}$, or, equivalently, by the filling factor.  More
precisely, the strength of the repulsive coupling at the fixed point
$\lambda^{*} = \pi (2l +1)$ increases with the number $2l+1 \, (=1,3,5,
\ldots)$ of flux quanta bound to the electron.  The presence of the fixed
point shows that the CSGL theory undergoes a 2nd-order phase transition when
the chemical potential of the composite particles tends to zero.  Since the
self-energy $\Sigma$ is identically zero, the nontrivial fixed point has
nevertheless Gaussian exponents, $\nu =\tfrac{1}{2}, \; z=2, \; \eta=0$.
It should be noted that only the location of the fixed point depends on
$\theta$, the critical exponents---which in contrast to the strength of the
coupling at the fixed point are independent of the regularization and
renormalization scheme---are universal and independent of the filling factor.
This ``triviality'' is in accord with the experimentally observed
universality of the FQHE.  A dependence of the critical exponents on
$\theta$ could from the theoretical point of view
hardly be made compatible with the hierarchy construction \cite{HH}
which implies a cascade of phase transitions.  From this viewpoint
the present results are most satisfying: the CSGL theory is shown to
encode a new type of field-induced 2nd-order quantum phase
transition that is simple enough not to obscure the observed
universality of the FQHE.

Most of the properties found in the nonrelativistic theory are not shared by
the relativistic version of the CSGL theory.  On the contrary, by
calculating the effective potential, Pryadko and Zhang \cite{PZ} 
argued that the relativistic theory undergoes a fluctuation-induced
1st-order transition---similar to the one in massless scalar electrodynamics
in 3+1 spacetime dimensions first studied by Coleman and Weinberg
\cite{CW}.  From the standpoint of dimensional analysis the discrepancy in
critical behavior is not surprising.  Because the effective spacetime
dimensionality in the nonrelativistic theory is increased by one, the
scaling dimension of a given operator generally differs in both theories.

As a side remark we mention that the theory (\ref{CSGL:L}) with zero
chemical potential has also been successfully applied to the problem of
anyon scattering \cite{LBL}.  From it, the two-particle scattering
amplitude, which was known from the work of Aharonov and Bohm
\cite{AB,Wilczek} on the scattering of a charged particle from a magnetic
flux tube, could be calculated.  The result of Aharonov-Bohm could be
reproduced provided the coupling constant was given the specific value
$\lambda^* = \tfrac{1}{2} \alpha$ corresponding to the IR fixed point.  It
is sometimes presented as if taking this value is a mere choice.  However,
the scattering amplitudes considered here pertain to low-energy,
small-momentum phenomena, and these are governed by the IR fixed point.  In
other words, in describing these processes, there is no arbitrariness at
all; one must fix the value of the self-coupling $\lambda$ to correspond to
the IR fixed point.

Because of the necessity to introduce a dimensional parameter to avoid
infrared divergences and to be able to carry out the renormalization-group
program, the scale invariance of the classical theory (\ref{CSGL:L}) with
$\mu = 0$, first discussed in Ref.\ \cite{JP}, is broken at nonfixed points.
Only at the fixed point the theory is scale invariant.  From the discussion
above, it follows that the low-energy, small-momentum behavior of the system
is also scale invariant since it is governed by the fixed point.

The same remarks apply to the thermodynamic properties of an anyon gas which
can also be computed from the Lagrangian (\ref{CSGL:L}), as was done in Ref.\ 
\cite{EVB}.  These, too, are low-energy, small-momentum properties governed by
the IR fixed point so that one must set $\lambda^* = \tfrac{1}{2} \alpha$
when calculating them.

The results obtained so far are gratifying.  The CSGL theory, which captures
the essential properties of a quantum Hall liquid, was shown to be capable
of also describing the Hall-liquid-to-insulator transitions---at least in the
absence of impurities.  However, impurities plays an essential part in the FQHE,
so a realistic description of the critical behavior of a quantum Hall liquid
should account for impurities.  Including (quenched) impurities in the
superfluid state of repulsively interacting bosons was shown to lead
to the phenomenon of localization which is of paramount importance to the
FQHE.  Since upon invoking a statistical gauge field this superfluid state
represents a quantum Hall liquid, this is in principle a good starting point
to study the quantum critical behavior of the CSGL model in the presence of
impurities.  There is, however, a technical problem.  We have worked in a
double epsilon expansion thereby leaving physical spacetime.
Unfortunately, the Chern-Simons term governing the statistical gauge field
of the theory is defined only in $2+1$ dimensions.  This makes a study of
the critical behavior of the random CSGL theory in a double
epsilon expansion impossible, and a different method, possibly
nonperturbative in character, is required.  

Luckily, using general scaling arguments and information contained in
the effective theories considered in this report one is able to
acquire some additional understanding of the behavior of various quantum
phase transitions studied in experiment, although precise estimates for
critical exponents other than the dynamic one cannot be made on this basis.
\section{Scaling Theory}
\label{sec:scale}
The traditional scaling theory of classical 2nd-order phase transitions,
first put forward by Widom \cite{Widom}, is easily extended to include the
time dimension \cite{Ma} because relation (\ref{zcrit}) implies the presence
of only one independent diverging scale.  Let $\delta = K - K_{\rm c}$, with
$K$ the parameter that drives the phase transition, measure the distance
from the critical coupling $K_{\rm c}$.  A physical observable at the
absolute zero of temperature $O(k_0,|{\bf k}|,K)$ can in the critical region
close to the transition be written as
\begin{equation} \label{scaling0}
O(k_0,k,K) = \xi^{d_O} {\cal O}(k_0 \xi_t, |{\bf k}| \xi), \;\;\;\;\;\;\;\;
(T=0),
\end{equation} 
where $d_O$ is the dimension of the observable $O$.  The right-hand side
does not depend explicitly on $K$; only implicitly through $\xi$ and
$\xi_t$.  The closer one approaches the critical coupling $K_{\rm c}$, the
larger the correlation length and time become.

Since a physical system is always at some finite temperature, we have to
investigate how the scaling law (\ref{scaling0}) changes when the
temperature becomes nonzero.  The easiest way to include temperature in
a quantum field theory is to go over to imaginary time $\tau = it$, with
$\tau$ restricted to the interval $0 \leq \tau \leq \beta$.  The
temporal dimension becomes thus of finite extend.  The critical behavior
of a phase transition at finite temperature is, as we remarked before,
still controlled by the quantum critical point provided $\xi_t < \beta$.
If this condition is fulfilled, the system does not see the finite
extend of the time dimension.  Instead of the zero-temperature scaling
(\ref{scaling0}), we now have the finite-size scaling
\begin{equation} \label{scalingT}
O(k_0,|{\bf k}|,K,T) = \beta^{d_O/z} {\cal O}(k_0 \beta, |{\bf k}|
\beta^{1/z},\beta/\xi_t), \;\;\;\;\;\;\;\; (T \neq 0).
\end{equation} 
The distance to the quantum critical point is measured by the ratio
$\beta/\xi_t \sim |\delta|^{z\nu}/T$.

Let us continue to derive hyperscaling relations.  To this end, we consider
the two terms in the effective theory (\ref{bec:Lef}) quadratic in the
Goldstone field $\varphi$ with $m$ effectively set to 1 and write it in
the most general form \cite{FF}:
\begin{equation} \label{general} 
{\cal L}_{\rm eff}^{(2)} = - \tfrac{1}{2} \bar{\rho}_{\rm s} (\nabla
\varphi)^2 +  \tfrac{1}{2} \bar{n}^2 \kappa (\partial_0 \varphi)^2.
\end{equation}  
The coefficient $\rho_{\rm s}$ is the superfluid mass density which in
the presence of, for example, impurities does not equal $m
\bar{n}$---even at the absolute zero of temperature.  The other
coefficient,
\begin{equation} 
\bar{n}^2 \kappa = \frac{\partial \bar{n}}{\partial \mu} = \lim_{|{\bf k}|
\rightarrow 0} \Pi_{0 0} (0,{\bf k}) , 
\end{equation} 
with $\Pi_{0 0}$ the (0 0)-component of the full polarization tensor
(\ref{bec:Pi}), involves the full compressibility and particle number
density.  This is because the chemical potential is represented in the
effective theory by $\mu = -\partial_0 \varphi$ and
\begin{equation} 
\frac{\partial^2 {\cal L}_{\rm eff}}{\partial \mu^2} = \bar{n}^2 \kappa.
\end{equation} 
Equation (\ref{general}) leads to the general expression of the sound
velocity
\begin{equation} 
c^2 = \frac{\bar{\rho}_{\rm s}}{\bar{n}^2 \kappa}
\end{equation} 
at the absolute zero of temperature.

Let us next assume that the chemical potential is the control parameter, so
that $\delta \propto \mu - \mu_{\rm c}$ denotes the distance from the phase
transition, and $\xi \sim |\delta|^{-\nu}$.  Now, on the one hand, the
singular part of the free energy density $f_{\rm sing}$ arises from the
low-energy, long-wavelength fluctuations of the Goldstone field.  (Here, we
adopted the common practice of using the symbol $f$ for the density
$\Omega/V$ and of referring to it as the free energy density.)  The ensemble
averages give
\begin{equation} 
\langle (\nabla \varphi)^2 \rangle \sim \xi^{-2}, \;\;\;\;
\langle (\partial_0 \varphi)^2 \rangle \sim \xi_t^{-2} \sim \xi^{-2z} .
\end{equation} 
On the other hand, dimensional analysis shows that the singular part of
the free energy density scales near the transition as
\begin{equation} 
f_{\rm sing} \sim \xi^{-(d+z)}.
\end{equation} 
Combining these hyperscaling arguments, we arrive at the following
conclusions:
\begin{equation} \label{hyperrho} 
\rho_{\rm s} \sim \xi^{-(d+z-2)}, \;\;\;\; \bar{n}^2 \kappa \sim
\xi^{-(d-z)} \sim |\delta|^{(d-z)\nu}.
\end{equation} 
The first conclusion is consistent with the universal jump (\ref{jump})
predicted by Nelson and Kosterlitz \cite{NeKo} which corresponds to
taking $z=0$ and $d=2$.  Since $\xi \sim |\delta|^{-\nu}$, $f_{\rm
sing}$ can also be directly differentiated with respect to the chemical
potential to yield for the the singular part of the compressibility
\begin{equation} 
\bar{n}^2 \kappa_{\rm sing} \sim |\delta|^{(d+z)\nu -2}. 
\end{equation} 
Fisher and Fisher \cite{FF} continued to argue that there are two
alternatives.  Either $\kappa \sim \kappa_{\rm sing}$, implying $z \nu
=1$; or the full compressibility $\kappa$ is constant, implying $z=d$.
The former is consistent with the Gaussian values $\nu=\tfrac{1}{2}, \; z=2$ 
found by Uzunov \cite{Uzunov} for the pure case in $d < 2$.  The latter is
believed to apply to repulsively interacting bosons in a random media.  These
remarkable simple arguments thus predict the exact value $z=d$ for the
dynamic exponent in this case.

The above hyperscaling arguments have been extended by Fisher, Grinstein, and
Girvin \cite{FGG} to include the $1/|{\bf x}|$-Coulomb potential.  The
quadratic terms in the effective theory (\ref{effCoul}) may be cast in the
general form
\begin{equation} 
{\cal L}_{\rm eff}^{(2)} = \frac{1}{2} \left(\rho_{\rm s} {\bf k}^2 -
\frac{|{\bf k}|^{d-1}}{\hat{e}^2} k_0^2\right) |\varphi(k_0,{\bf k})|^2,
\end{equation}  
where $\hat{e}$ is the renormalized charge.  From (\ref{effCoul}) we find
that to lowest order:
\begin{equation} 
\hat{e}^2 = 2^{d-1} \pi^{(d-1)/2} \Gamma\left[\tfrac{1}{2}(d-1)\right] e_0^2.
\end{equation} 
The renormalized charge is connected to the (0 0)-component of the full
polarization tensor (\ref{bcs:cruc}) via
\begin{equation} 
\hat{e}^2 = \lim_{|{\bf k}| \rightarrow 0} \frac{|{\bf k}|^{d-1}}{\Pi_{0 0}
(0,{\bf k})} . 
\end{equation} 
A simple hyperscaling argument like the ones given in the preceding
paragraph shows that near the transition, the renormalized charge scales
as
\begin{equation} 
\hat{e}^2 \sim \xi^{1-z}.
\end{equation} 
They then argued that in the presence of random impurities this charge is
expected to be finite at the transition so that $z=1$.  This again is an
exact results which replaces the value $z=d$ of the neutral system.

The quantum phase transitions we are considering take place in charged
systems and are mainly probed by conductivity $\sigma$ or resistivity
$\rho$ measurements.  To see how the conductivity $\sigma$ relates to
the superfluid mass density $\rho_{\rm s}$, we minimally couple the
effective theory (\ref{general}) to an electromagnetic gauge field.  The
only relevant term for this purpose is the first one in (\ref{general})
with $\nabla \varphi$ replaced by $\nabla
\varphi - e {\bf A}$, where we allow the superfluid mass density to vary
in space and time.  The term in the action quadratic in ${\bf A}$ then
becomes in the Fourier representation
\begin{equation} 
S_\sigma = - \tfrac{1}{2} e^2 \int_{k_0,{\bf k}} {\bf A}(-k_0,-{\bf k})
\rho_{\rm s} (k_0,{\bf k}) {\bf A}(k_0,{\bf k}).
\end{equation} 
The electromagnetic current,
\begin{equation} 
{\bf j}(k_0,{\bf k}) = \frac{\delta S_\sigma}{\delta {\bf A}(-k_0,-{\bf k})}
\end{equation} 
obtained from this action can be written as
\begin{equation} 
{\bf j}(k_0,{\bf k}) = \sigma(k_0,{\bf k}) {\bf E}(k_0,{\bf k})
\end{equation} 
with the conductivity
\begin{equation} \label{conductivity}
\sigma(k) = -i e^2 \frac{\rho_{\rm s}(k)}{k_0}
\end{equation} 
essentially given by the superfluid mass density divided by $k_0$, where it
should be remembered that the mass $m$ is effectively set to 1 here.  By
virtue of the scaling relation (\ref{hyperrho}), it follows that $\sigma$
scales as \cite{AALR}
\begin{equation} 
\sigma \sim \xi^{-(d-2)}. 
\end{equation} 
In other words, the scaling dimension of the conductivity and therefore
that of the resistivity is zero in two space dimensions.  

Let us now consider a quantum phase transition triggered by changing
the applied magnetic field, i.e., $\delta \propto H - H_{\rm c}$.  
The critical field scales with $\xi$ as $H_{\rm c} \sim
\Phi_0/\xi^2$.  In fact, this expresses a more fundamental result, namely
that the scaling dimension $d_{\bf A}$ of ${\bf A}$ is one,
\begin{equation} 
d_{\bf A} = 1,
\end{equation} 
so that $|{\bf A}| \sim \xi^{-1}$.  From this it in turn follows that $E
\sim \xi_t^{-1} \xi^{-1} \sim \xi^{-(z+1)}$, and that the scaling dimension
$d_{A_0}$ of $A_0$ is $z$,
\begin{equation} 
d_{A_0} = z,
\end{equation} 
so that $A_0 \sim \xi_t^{-1} \sim \xi^{-z}$.  For the DC conductivities
in the presence of an external electric field we have on account of the
general finite-size scaling form (\ref{scalingT}) with $k_0=|{\bf k}|=0$:
\begin{equation}  \label{scalingE}
\sigma(H,T,E) = \varsigma(\delta^{\nu z}/T,\delta^{\nu (z+1)}/E).
\end{equation}
This equation shows that conductivity measurements close to the quantum
critical point collapse onto a single curve when plotted as function of
the dimensionless combinations $\delta^{\nu z}/T$ and $\delta^{\nu
(z+1)}/E$.  The best collapse of the data determines the values of $\nu
z$ and $\nu (z+1)$.  In other words, the temperature and electric-field
dependence determine the critical exponents $\nu$ and $z$ independently.
\section{Experiments}
\label{sec:exp}
\subsection{Superconductor-To-Insulator Transition}
The first experiments we wish to discuss are those performed by Hebard
and Paalanen on superconducting films in the presence of random
impurities \cite{HPsu1,HPsu2}.  It has been predicted by Fisher
\cite{MPAFisher} that with increasing applied magnetic field such
systems undergo a zero-temperature transition into an insulating state.
(For a critical review of the experimental data available in 1993, see
Ref.\ \cite{LG}.) 

Let us restrict ourselves for the moment to the $T\Delta$-plane of the
phase diagram by setting the applied magnetic field $H$ to zero.  For
given disorder strength $\Delta$, the system then undergoes a
Kosterlitz-Thouless transition induced by the unbinding of magnetic
vortex pairs at a temperature $T_{\rm KT}$ well below the bulk
transition temperature (see Sec.\ \ref{sec:2sc}).  The
Kosterlitz-Thouless temperature is gradually suppressed to zero when the
disorder strength approaches criticality $\hat{\Delta} \rightarrow
\hat{\Delta}_{\rm c}$.  The transition temperature scales with the
correlation length $\xi \sim |\hat{\Delta}_{\rm c} -
\hat{\Delta}|^{-\nu}$ as $T_{\rm KT} \sim \xi^{-z}$.

In the $H\Delta$-plane, i.e., at $T=0$, the situation is as follows.
For given disorder strength, there is now at some critical value $H_{\rm
c}$ of the applied magnetic field a phase transition from a
superconducting state of pinned vortices and condensed Cooper pairs to
an insulating state of pinned Cooper pairs and condensed vortices.  The
condensation of vortices disorder the ordered state as happens in
classical, finite temperature superfluid- and superconductor-to-normal
phase transitions \cite{GFCM}.  When the disorder strength approaches
criticality again, $H_{\rm c}$ is gradually suppressed to zero.  The
critical field scales with $\xi$ as $H_{\rm c} \sim \Phi_0/\xi^2$.
Together, the scaling results for $T_{\rm KT}$ and $H_{\rm c}$ imply
that \cite{MPAFisher}
\begin{equation}  \label{H-T}
H_{\rm c} \sim T_{\rm KT}^{2/z}.
\end{equation} 
This relation, linking the critical field to the Kosterlitz-Thouless
temperature, provides a direct way to measure the dynamic exponent $z$
at the $H=0$, $T=0$ transition.  This has been done first by Hebard and
Paalanen \cite{HPsu1,HPsu2}.  Their experimental determination of
$T_{\rm KT}$ and $H_{\rm c}$ for five different films with varying
amounts of impurities confirmed the relation (\ref{H-T}) with $2/z =
2.04 \pm 0.09$.  The zero-temperature critical fields were obtained by
plotting $\dd \rho_{xx}/\dd T|_H$ versus $H$ at the lowest accessible
temperature and interpolating to the the field where the slope is
zero. The resulting value $z= 0.98 \pm .04$ is in accordance with
Fisher's prediction \cite{MPAFisher}, $z=1$, for a random system with a
$1/|{\bf x}|$-Coulomb potential.

Hebard and Paalanen \cite{HPsu1} also investigated the field-induced
zero-temperature transition.  The control parameter is here $\delta
\propto H -H_{\rm c}$.  When plotted as function of $|H -H_{\rm
c}|/T^{1/\nu_H z_H}$ they saw their resistivity data collapsing onto two
branches; an upper branch tending to infinity for the insulating state,
and a lower branch bending down for the superconducting state.  The
unknown product $\nu_H z_H$ is experimentally determined by looking for
which value the best scaling behavior is obtained.  Further experiments
carried out by Yazdani and Kapitulnik \cite{YaKa} studying the
electric-field dependence of the resistivity also determined the product
$\nu_H (z_H+1)$.  The two independent measurements together fix the
critical exponents $\nu_H$ and $z_H$ separately.  From their best data,
Yazdani and Kapitulnik extracted the values \cite{YaKa}
\begin{equation} \label{zHnuH}
z_H = 1.0 \pm 0.1, \;\;\;\; \nu_H = 1.36 \pm 0.05.
\end{equation} 
\subsection{Quantum-Hall Systems}
We continue to discuss the field-induced quantum phase transitions in
quantum Hall systems.  Since an excellent discussion recently appeared in
the literature \cite{SGCS}, we shall be brief, referring the reader to
that review for a more thorough discussion and additional references.

One can image transitions from one Hall liquid to another Hall liquid
with a different (integer or fractional) filling factor, or to the
insulating state.  Experiments seem to suggest that all the quantum-Hall
transitions are in the same universality class.  The transitions are
probed by measuring the conductivities $\sigma_{xx}$ and $\sigma_{xy}$.
The scaling of the width of the transition regime with temperature as
predicted by Eq.\ (\ref{scalingE}) has been corroborated by DC
experiments on various transitions between integer quantum-Hall states
which were all found to yield the value $1/\nu z = 0.42 \pm 0.04$
\cite{WTPP}.  Also the scaling $\delta \sim E^{1/\nu (z+1)}$ has been
corroborated by experiment which yielded the value $\nu (z+1) \approx
4.6$ \cite{WET}.  Together with the previous result obtained from the
temperature scaling this gives
\begin{equation} 
z \approx 1, \;\;\;\; \nu \approx 2.3.
\end{equation} 
The value of the dynamic exponent strongly suggests that it is a result
of the presence of the $1/|{\bf x}|$-Coulomb potential.  The correlation
length exponent $\nu$ is seen to be large.
\subsection{$2d$ Electron Systems}
Recently, silicon MOSFET's at extremely low electron number densities
has been studied \cite{MIT,KSSMF,SKS,PFW}.  Earlier experiments at
higher densities seemed to confirm the general believe, based on the
work by Abrahams {\it et al.} \cite{AALR}, that such two-dimensional
electron systems do not undergo a quantum phase transition.  In that
influential paper, it was demonstrated that even weak disorder is
sufficient to localize the electrons at the absolute zero of temperature
thus excluding conducting behavior.  Electron-electron interactions were
however not included.  As we saw in Sec.\ \ref{sec:bec}, the $1/|{\bf
x}|$-Coulomb interaction becomes important at low densities and the
analysis of Abrahams {\it et al.} \cite{AALR} no longer applies.

The recent experiments have revealed a zero-temperature
conductor-to-insulator transition triggered by a change in the charge
carrier density $\bar{n}$.  That is, the distance to the critical point
is in these systems measured by $\delta \propto \bar{n} - \bar{n}_{\rm c}$.
Like in the quantum-Hall systems, these transitions are probed by
measuring the resistivity.  It scales with temperature near the
transition according to the scaling form (\ref{scalingE}) with $H$ set
to zero.  For $\bar{n} < \bar{n}_{\rm c}$, where the Coulomb interaction
is dominant and fluctuations in the charge carrier density are
suppressed, the electron system is insulating.  On increasing the
density, these fluctuations intensify and at the critical value
$\bar{n}_{\rm c}$, the system reverts to a conducting phase.  By plotting
their conductivity data as function of $T/\delta^{\nu z}$ with $\nu z =
1.6 \pm 0.1$, Popov\'{\i}c, Fowler, and Washburn \cite{PFW} saw it
collapse onto two branches; the upper branch for the conducting side of the
transition, and the lower one for the insulating side.  A similar
collapse with a slightly different value $1/\nu z = 0.83 \pm 0.08$ was found
in Ref.\ \cite{KSSMF}, where also the collapse of the data when plotted
as function of $\delta/E^{1/(z+1)\nu}$ was obtained.  The best collapse
resulted for $1/(z+1) \nu = 0.37 \pm 0.01$, leading to
\begin{equation} z = 0.8 \pm 0.1, \;\;\;\; \nu = 1.5 \pm 0.1.
\end{equation} 
The value for the dynamic exponent is close to the expected value $z=1$ for
a charged system with a $1/|{\bf x}|$-Coulomb interaction, while that of
$\nu$ is surprisingly close to the value (\ref{zHnuH}) found for the
superconductor-to-insulator transition.

A further experimental result for these two-dimensional electron systems
worth mentioning is the suppression of the conducting phase by an
applied magnetic field found by Simonian, Kravchenko, and Sarachik
\cite{SKS}.  They applied the field {\it parallel} to the plane of the
electrons instead of perpendicular as is done in quantum-Hall
measurements.  In this way, the field presumably couples only to the
spin of the electrons and the complications arising from orbital effects
do not arise.  At a fixed temperature, a rapid initial raise in the
resistivity was found with increasing field.  Above a value of about 20
kOe, the resistivity saturates.  It was pointed out that both the
behavior in a magnetic filed, as well as in zero field strongly
resembles that near the superconductor-to-insulator transition discussed
above, suggesting that the conducting phase might in fact be
superconducting.
\subsection{Conclusions}
We have seen that general scaling arguments combined with the effective
theories, can be employed to understand the scaling behavior observed in
various quantum phase transitions.  Most of the experiments seem to
confirm the expected value $z=1$ for a random system with a $1/|{\bf
x}|$-Coulomb interaction.  The number of different universality classes
present is yet not known.  Even if the conductor-to-insulator transition
observed in silicon MOSFET's at low electron number densities turns out
to be in the same universality class as the superconductor-to-insulator
transition, there are still the field-induced transitions in
quantum-Hall systems, which have a larger correlation-length exponent.

The paradigm provided by a repulsively interacting Bose gas, seems to be a
good starting point to describe the various systems.  However,
high-precision estimates calculated from this theory with impurities and
a $1/|{\bf x}|$-Coulomb interaction included are presently lacking.

\end{document}